\global\def\draftcontrol{0}

   \def\versionno{ ks s3 take 2}

\catcode`\@=11

\expandafter\ifx\csname draftcontrol\endcsname\relax\global\def\draftcontrol{0}
\fi

{\count255=\time\divide\count255 by 60
\xdef\hourmin{\number\count255}
\multiply\count255 by-60\advance\count255 by\time
\xdef\hourmin{\hourmin:\ifnum\count255<10 0\fi\the\count255}}
\def\draftdate{\number\month/\number\day/\number\year\ \ \ \hourmin }

\newcommand\makepapertitle{\par
  \begingroup
    \renewcommand\thefootnote{\@fnsymbol\c@footnote}%
    \def\@makefnmark{\rlap{\@textsuperscript{\normalfont\@thefnmark}}}%
    \long\def\@makefntext##1{\parindent 1em\noindent
            \hb@xt@1.8em{%
                \hss\@textsuperscript{\normalfont\@thefnmark}}##1}%
     \newpage
     \global\@topnum\z@   
     \@makepapertitle
     \thispagestyle{empty}\@thanks
  \endgroup
  \setcounter{footnote}{0}%
  \global\let\thanks\relax
  \global\let\makepapertitle\relax
  \global\let\@makepapertitle\relax
  \global\let\@thanks\@empty
  \global\let\@author\@empty
  \global\let\@date\@empty
  \global\let\@title\@empty
  \global\let\title\relax
  \global\let\author\relax
  \global\let\date\relax
  \global\let\and\relax
  \def\version{\let\version\@version\@gobble}
}
\def\@makepapertitle{%
  \newpage
   \ifnum\draftcontrol=1 {}
   \version\versionno
   \vskip 3em%
   \else
   \hfill\hbox to 3cm {\parbox{4cm}{\@pubnum}\hss}%
   \vskip 3em%
   \fi
   \begin{center}%
   \let \footnote \thanks
     {\LARGE {\@title}}%
     \vskip 1.5em%
     {\normalsize
       \lineskip .5em%
       \begin{tabular}[t]{c}%
         \@author
       \end{tabular}\par}%
     \vskip 1.5em%
     {\@bstract}%
     \end{center}%
     \vskip 1.5em
     \@date%
   \par
}

\gdef\@pubnum{}
\def\pubnum#1{%
  \gdef\@pubnum{#1}}

\gdef\@bstract{}
\def\Abstract#1{%
  \gdef\@bstract{%
   \parbox{\textwidth-0pc}{%
   \centerline{\bf Abstract}\penalty1000%
\kern.2cm%
\noindent
\renewcommand\baselinestretch{1.0}%
{#1}}}
}

\def\ps@paper{\let\@mkboth\@gobbletwo%
     \ifnum\draftcontrol=1
    \def\@oddfoot{\hbox to \textwidth{\tiny \versionno \hfil\tiny\draftdate}%
    \hskip -\textwidth \hbox to \textwidth{\hfil\rm\thepage\hfil}}%
     \else\def\@oddfoot{\hbox to \textwidth{\hfil\rm\thepage\hfil}}
     \fi
     \let\@evenfoot\@oddfoot
}

\def\body{\clearpage
          \pagestyle{paper}
    }

\def\@version#1{\ifnum\draftcontrol=1
\typeout{}\typeout{#1}\typeout{}
\vskip3mm\centerline{\hbox{\fbox{\normalsize{\tt DRAFT -- #1 -- }
                   {\draftdate}}}}\vskip3mm
\fi}
\let\version\@version
\long\def\eqlabel#1{\ifnum\draftcontrol=1
                    \tag@false  
                    \tag*{(\theequation) \hbox to -0.2cm{\hspace{0cm}\small{#1}\hss}}
                    \refstepcounter{equation}
                    \edef\@currentlabel{\theequation}
                    \ltx@label{#1}          
                    \else
                    \label{#1}
                    \fi
                    }
\let\st@bibitem\@bibitem
\let\st@lbibitem\@lbibitem
\ifnum\draftcontrol=1
  \def\@bibitem#1{%
    \st@bibitem{#1}\a@@label{#1}\ignorespaces}
  \def\@lbibitem[#1]#2{%
    \st@lbibitem[#1]{#2}\a@@label{#2}\ignorespaces}
  \def\a@@label#1{%
    \gdef\a@lab{\smash{\normalfont\small#1}}
    \ifvmode
      \if@inlabel
        \global\setbox\@labels\hbox{%
          \llap{\a@lab\let\a@lab\relax
                \kern\@totalleftmargin\kern\marginparsep}%
          \box\@labels}%
      \fi
    \fi}
\fi

\documentclass[12pt,letterpaper]{article}

\usepackage{amsmath,amssymb,array,calc,epsfig,rotating,bm,xcolor}
\usepackage[sort]{cite}
\usepackage{graphicx,esint,float}
\usepackage{psfrag,verbatim}
\usepackage[makeroom]{cancel}
\usepackage{xcolor}

\ifnum\draftcontrol=1
\fi

\renewcommand\baselinestretch{1.25}
\renewcommand\section{\@startsection {section}{1}{\z@}%
                                   {-3.5ex \@plus -1ex \@minus -.2ex}%
                                   {2.3ex \@plus.2ex}%
                                   {\normalfont\large\bfseries}}
\renewcommand\subsection{\@startsection{subsection}{2}{\z@}%
                                   {-3.25ex\@plus -1ex \@minus -.2ex}%
                                   {1.5ex \@plus .2ex}%
                                   {\normalfont\normalsize\bfseries}}
\renewcommand\subsubsection{\@startsection{subsubsection}{3}{\z@}%
                                   {-3.25ex\@plus -1ex \@minus -.2ex}%
                                   {1.5ex \@plus .2ex}%
                                   {\normalfont\normalsize\it}}
\renewcommand\paragraph{\@startsection{paragraph}{4}{\z@}%
                                   {-3.25ex\@plus -1ex \@minus -.2ex}%
                                   {1.5ex \@plus .2ex}%
                                   {\normalfont\normalsize\bf}}

\numberwithin{equation}{section}

\def\revise#1       {\raisebox{-0em}{\rule{3pt}{1em}}%
                     \marginpar{\raisebox{.5em}{\vrule width3pt\
                     \vrule width0pt height 0pt depth0.5em
                     \hbox to 0cm{\hspace{0cm}{%
                     \parbox[t]{4em}{\raggedright\footnotesize{#1}}}\hss}}}}

\newcommand\nxt[1]  {\\\fnxt#1}
\newcommand{\ie}{{\it i.e.,}\ }

\def\cala         {{\cal A}}

\def\calb         {{\cal B}}
\def\calc         {{\cal C}}

\def\cale         {{\cal E}}
\def\calf         {{\cal F}}

\def\calh         {{\cal H}}

\def\calk         {{\cal K}}
\def\call         {{\cal L}}
\def\calm         {{\cal M}}
\def\caln         {{\cal N}}
\def\calo         {{\cal O}}
\def\calp         {{\cal P}}
\def\calq         {{\cal Q}}
\def\calr         {{\cal R}}

\def\calt         {{\cal T}}

\def\calv         {{\cal V}}
\def\calw         {{\cal W}}

\def\reals        {{\mathbb R}}
\def\zet          {{\mathbb Z}}
\def\hw          {{\hat{\Omega}}}

\def\del          {\partial}

\def\tr           {\mathop{\rm Tr}}

\def\sqr#1#2{{\vcenter{\vbox{\hrule height.#2pt
 \hbox{\vrule width.#2pt height#1pt \kern#1pt
 \vrule width.#2pt}\hrule height.#2pt}}}}
\def\square{%
  \mathop{\mathchoice{\sqr{12}{15}}{\sqr{9}{12}}{\sqr{6.3}{9}}{\sqr{4.5}{9}}}}

\def\dd{\delta}

\def\e{\epsilon}

\def\ga{\gamma}

\def\rh{\hat{\rho}}

\def\hh{\hat{h}}

\def\hf{\hat{f}}
\def\hg{\hat{g}}
\def\hK{\hat{K}}
\def\aa1{\phi}
\def\cc1{\psi}
\def\hh{\hat{h}}

\def\Om{\Omega}
\def\om{\Omega}

\def\hr{\hat{\rho}}
\def\hf{\hat{f}}
\def\hK{\hat{K}}

\def\csb{{\chi\rm{SB}}}

\def\hmu{\hat{\mu}}

\def\va{\calv_A}
\def\vb{\calv_B}
\def\mucsb{\mu_{\chi{\rm SB}}}
\def\df{\delta f}
\def\dk{\delta k}

\catcode`\@=12

\begin{document}


\title{\bf A bestiary of black holes on the conifold with fluxes}

\date{March 27, 2021}

\author{
Alex Buchel\\[0.4cm]
\it $ $Department of Applied Mathematics\\
\it $ $Department of Physics and Astronomy\\ 
\it University of Western Ontario\\
\it London, Ontario N6A 5B7, Canada\\
\it $ $Perimeter Institute for Theoretical Physics\\
\it Waterloo, Ontario N2J 2W9, Canada
}

\Abstract{We present a comprehensive analysis of the black holes on warped
deformed conifold with fluxes in Type IIB supergravity.  These black
holes realize the holographic dual to thermal states of the ${\cal
N}=1$ supersymmetric $SU(N)\times SU(N+M)$ cascading gauge theory of
Klebanov et.al \cite{Klebanov:2000nc,Klebanov:2000hb} on round
$S^3$. There are three distinct mass scales in the theory: the strong
coupling scale $\Lambda$ of the cascading gauge theory, the
compactification scale $\mu=1/L_3$ (related to the $S^3$ radius $L_3$)
and the temperature $T$ of a thermal state. Depending on
$\Lambda\,,\, \mu$ and $T$, there is an intricate pattern of
confinement/deconfinement (Hawking-Page) and the chiral symmetry
breaking phase transitions. In the $S^3\to \reals^3$
decompactification limit, \ie $\mu\to 0$, we recover the
Klebanov-Tseytlin \cite{Aharony:2007vg} and the
Klebanov-Strassler \cite{Buchel:2018bzp} black branes.
}

\makepapertitle

\body

\version\versionno
\tableofcontents

\section{Introduction}\label{intro}

\begin{figure}[t]
\begin{center}
\psfrag{s}[cc][][1.0][0]{$S^3$}
\psfrag{t}[cc][][1.0][0]{$S^1$}
\psfrag{a}[cc][][1.0][0]{$S^2$}
\psfrag{q}[cc][][1.5][0]{$\calt_{con}$}
\psfrag{w}[cc][][1.5][0]{$\calt_{decon}$}
\includegraphics[width=6in]{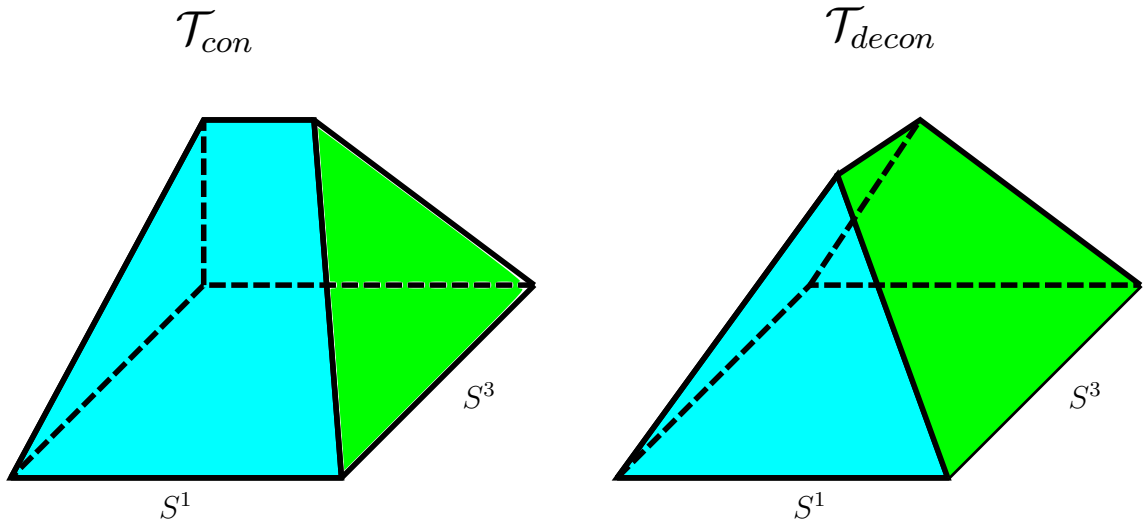}
\end{center}
  \caption{Geometrical phases in global $EAdS_5$: the cyan face represents Euclidean time direction $S^1$,
  compactified with the period $\frac 1T$, the green face represents the 3-sphere $S^3$. The bulk geometry
  with the $S^3$ smoothly shrinking to zero size represents the confined phase of thermal $\caln=4$ SYM
  (the left panel). The bulk geometry with the $S^1$ smoothly shrinking to zero size represents the
  deconfined phase of thermal $\caln=4$ SYM (the right panel). 
} \label{n4}
\end{figure}

An exciting aspect of the gauge theory/string theory correspondence \cite{Maldacena:1997re,Aharony:1999ti}
is that it geometrizes the confinement/deconfinement transitions of strongly
coupled gauge theories as topology changing transitions in dual supergravity backgrounds.
Following \cite{Witten:1998zw}, consider $\caln=4$ $SU(N)$ supersymmetric Yang-Mills (SYM) theory
on the 3-sphere $S^3$ of radius $L_3$. In the planar limit, $N\to \infty$ and $g_{YM}^2\to 0$ with $ g_{YM}^2 N$  fixed,
and for large 't Hooft coupling $N g_{YM}^2\gg 1$, this SYM  has a classical description as
Type IIB supergravity on $AdS_5\times S^5$. Thermal states of the theory have two scales: the temperature $T$
and the $S^3$ compactification scale $\mu=\frac {1}{L_3}$. There are two distinct phases of the theory:
the confined phase $\calt_{con}$ and the deconfined phase $\calt_{decon}$.
In the large-$N$ limit there is a sharp distinction
between the two phases: the confined phase has an entropy $s\propto O(N^0)$, while the deconfined phase
has an entropy $s\propto O(N^2)$. The former one is represented gravitationally
as $EAdS_5\times S^5$ with the Euclidean time direction $t_E$ compactified as
\begin{equation}
t_E\sim t_E+\frac 1T\,, 
\eqlabel{tecom}
\end{equation}
and the latter one as a black hole in global $AdS_5$. 
Which phase is the preferred one, depends on the ratio $T/\mu$: when $T\gg \mu$, the gravitational free energy
of the black hole solution is lower compare to $EAdS_5$, and the deconfined phase is the preferred one.
The confined phase is the preferred one at low temperatures. On the gravity side of the holographic duality,
the confinement/deconfinement
transition is the Hawking-Page thermal transition in anti-de Sitter space-time \cite{Hawking:1982dh}. 
Restricted to $EAdS_5$, the two thermal phases are shown in fig.~\ref{n4}. 
These are the only $SO(4)\times SO(6)$ symmetric thermal phases of the SYM.

\begin{figure}[t]
\begin{center}
\psfrag{s}[cc][][1.0][0]{$S^3$}
\psfrag{t}[cc][][1.0][0]{$\tilde{T}^{1,1}$}
\psfrag{a}[cc][][1.0][0]{$S^2$}
\psfrag{q}[cc][][1.5][0]{$\calv_A^s$}
\psfrag{w}[cc][][1.5][0]{$\calv_B$}
\psfrag{e}[cc][][1.5][0]{$\calv_A^b$}
\includegraphics[width=6in]{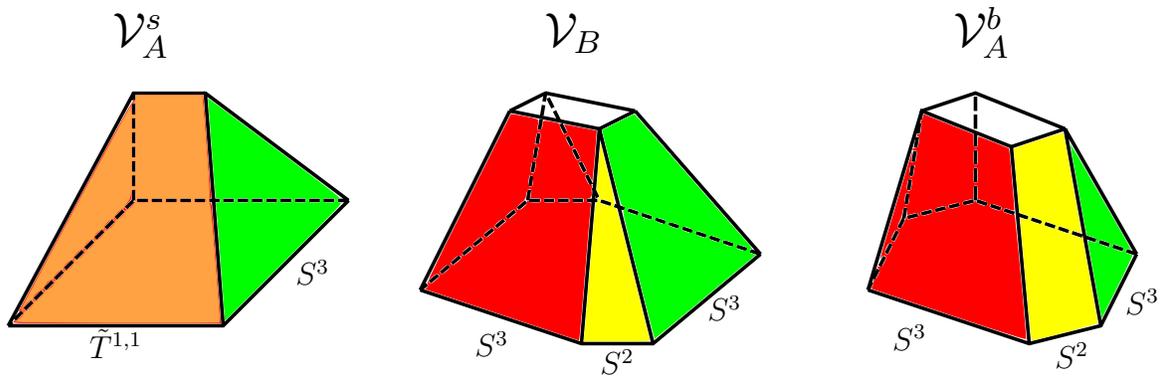}
\end{center}
  \caption{Vacua of the cascading gauge theory on $S^3$ (the green faces) are characterized by topologically distinct
  gravitational backgrounds: in the left panel the boundary $S^3$ shrinks to zero size,
  with the internal $\tilde{T}^{1,1}$ (orange face) of the warped conifold having unbroken $U(1)$ symmetry;
  in the middle panel the 2-cycle (the yellow faces) of the deformed conifold shrinks to zero size;
  in the right panel the boundary $S^3$ shrinks to zero size but the conifold is deformed ---
  the $U(1)$ symmetry of $\tilde{T}^{1,1}$ is spontaneously broken to $\zet_2$. 
} \label{vacuapic}
\end{figure}

In this paper we study vacua and black
holes\footnote{The black holes studied
are Schwarzschild-AdS-like, namely they carry no angular momentum or
electric charge.}  on the conifold \cite{Candelas:1989js}
with fluxes in Type IIB supergravity.
These gravitational backgrounds are holographic duals to the vacua and the thermal states of
$\caln=1$ $SU(N+M)\times SU(N)$ cascading gauge theory \cite{Klebanov:2000hb} on $S^3$.
Unlike $\caln=4$ SYM, the cascading gauge is not conformal, and has a strong coupling scale
$\Lambda$. Thus, already the vacua of the theory are characterized by two mass scales $\Lambda$ and $\mu$,
leading to a nontrivial phase structure. In fig.~\ref{vacuapic} we characterize distinct vacua of the
compactified cascading gauge theory according to the topology of the gravitational dual:
\begin{itemize}
\item $\calv_A^s$ --- vacua with the boundary $S^3$ (akin the 3-sphere in the $\caln=4$
discussion above) smoothly shrinking to zero size  (the left panel). In these vacua the chiral symmetry
of the gauge theory is unbroken.
\item $\calv_B$ --- vacua with the conifold 2-cycle $S^2$ smoothly shrinking to zero size (the middle panel).
In these vacua the chiral symmetry of the gauge theory is spontaneously broken.
\item $\calv_A^b$ --- vacua with the boundary $S^3$ (akin the 3-sphere in the $\caln=4$
discussion above) smoothly shrinking to zero size  (the right panel). In these vacua the chiral symmetry
of the gauge theory is spontaneously broken.
\end{itemize}
We discuss different vacua of the cascading gauge theory and the phase transitions in section \ref{vacua}.

\begin{figure}[t]
\begin{center}
\psfrag{s}[cc][][1.0][0]{$S^3$}
\psfrag{t}[cc][][1.0][0]{$\tilde{T}^{1,1}$}
\psfrag{a}[cc][][1.0][0]{$S^1$}
\psfrag{c}[cc][][1.0][0]{$S^2$}
\psfrag{q}[cc][][1.5][0]{$\calt_{con,A}^s$}
\psfrag{w}[cc][][1.5][0]{$\calt_{con,B}$}
\psfrag{e}[cc][][1.5][0]{$\calt_{con,A}^b$}
\psfrag{h}[cc][][1.5][0]{$\calt_{decon}^s$}
\psfrag{j}[cc][][1.5][0]{$\calt_{decon}^b$}
\includegraphics[width=6in]{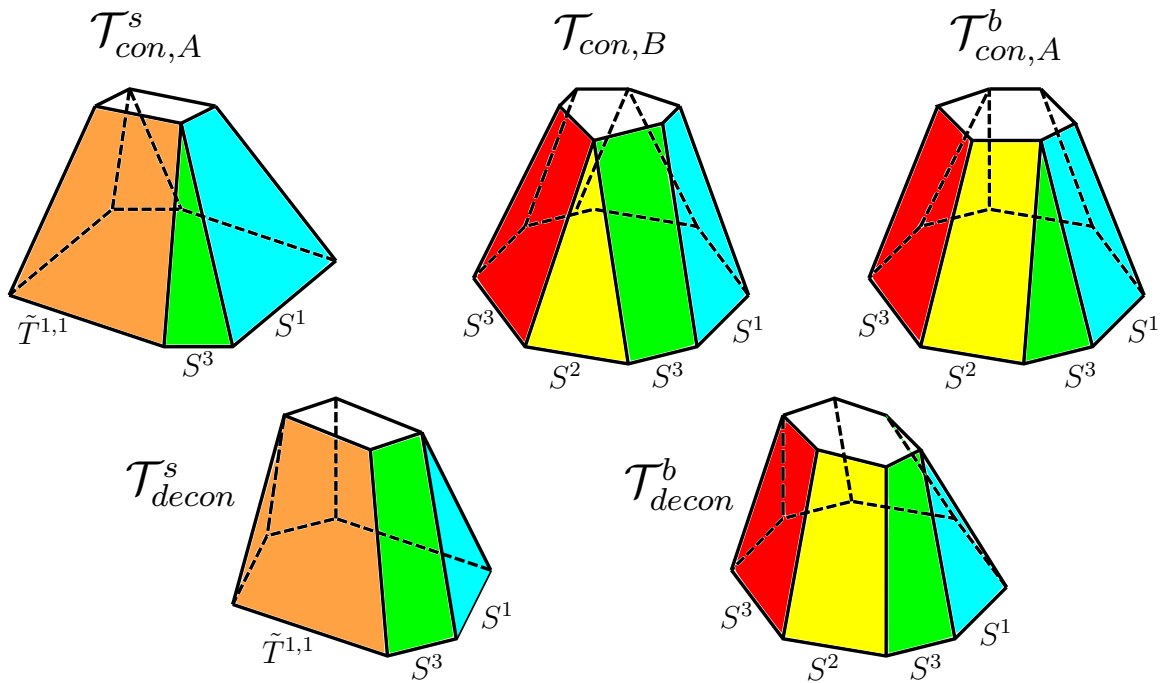}
\end{center}
  \caption{Top row: confined thermal states of the cascading gauge theory, obtained from the compactification of the
  Euclidean time direction as in \eqref{tecom} of the corresponding vacua, see fig.~\ref{vacuapic}. Bottom row,
  left panel: deconfined thermal states with the unbroken $U(1)$ symmetry --- the Klebanov-Tseytlin black holes.
  Bottom row,
  right panel: deconfined thermal states with the $U(1)$ symmetry spontaneously broken to $\zet_2$
  --- the Klebanov-Strassler black holes.
} \label{fig3}
\end{figure}

The set of possible thermal states of the cascading gauge theory is yet richer: in addition
to $\Lambda$ and $\mu$ we have the temperature $T$.  
In fig.~\ref{fig3} we characterize distinct vacua of the
compactified cascading gauge theory according to the topology of the gravitational dual:
\begin{itemize}
\item $\calt_{con,A}^s$ --- confined thermal states with the boundary $S^3$
smoothly shrinking to zero size  (the top row, left panel). In these states the chiral symmetry
of the gauge theory is unbroken. These thermal states are the Euclidean $\calv_{A}^s$ vacua with the compactified
time direction as in \eqref{tecom}. 
\item $\calt_{con,B}$ --- confined thermal states with the conifold 2-cycle $S^2$ smoothly shrinking to zero size.
In these vacua the chiral symmetry of the gauge theory is spontaneously broken
(the top row, middle panel). These thermal states are the Euclidean $\calv_{B}$ vacua with the compactified
time direction as in \eqref{tecom}. 
\item $\calt_{con,A}^b$ --- confined thermal states with the boundary $S^3$ 
smoothly shrinking to zero size  (the top row, right panel). In these states the chiral symmetry
of the gauge theory is spontaneously broken. These thermal states are the
Euclidean $\calv_{A}^b$ vacua with the compactified
time direction as in \eqref{tecom}.
\item $\calt_{decon}^s$ --- deconfined thermal states with the Euclidean time direction  $S^1$  (cyan faces) 
smoothly shrinking to zero size  (the bottom row, left panel). In these states the chiral symmetry
of the gauge theory is unbroken. We call these thermal states  the Klebanov-Tseytlin black holes. 
\item $\calt_{decon}^b$ ---  deconfined thermal states with the Euclidean time direction  $S^1$ 
smoothly shrinking to zero size  (the bottom row, right panel). In these states the chiral symmetry
of the gauge theory is spontaneously broken. We call these thermal states  the Klebanov-Strassler black holes. 
\end{itemize}
We discuss different thermal states of the cascading gauge theory and the phase transitions in section
\ref{bhsec}.

The black branes on the conifold were studied previously in \cite{Aharony:2007vg,Buchel:2018bzp} ---
these solutions can be obtained as the $\mu/\Lambda\to 0$, \ie the  $S^3$ decompactification, limit of the black holes
analyzed here. $\calv_A^s$ and $\calv_B$ vacua of the conifold with fluxes were analyzed earlier in
\cite{Buchel:2011cc}. $\calv_A^b$ vacua are constructed here for the first time.

The rest of the manuscript is organized as follows.
To have a self-contained presentation, in section \ref{ksreview} we review the cascading gauge theory
and its holographic dual \cite{Klebanov:2000hb}. We present an overview of how the cascading gauge theory
fits into the framework of the top-down holography, and what is precisely the gravitational dual
to the strong coupling scale $\Lambda$ of the theory. In section \ref{eareview} we review the consistent
truncation in $SU(2)\times SU(2)\times \zet_2$ invariant sector of Type IIB supergravity on
warped deformed conifold with fluxes to an arbitrary five dimensional
manifold $\calm_5$ \cite{Buchel:2010wp}. This effective five-dimensional gravitational action
is our holographic dual to cascading gauge theory on $\calm_4\equiv \del\calm_5$.  
In section \ref{kw} we use the framework of the effective action of section \ref{eareview}
to reproduce the known results about $\caln=1$ Klebanov-Witten superconformal gauge theory
holography \cite{Klebanov:1998hh}. The purpose is to gently familiarize the reader with the notations,
and the holographic renormalization of the conifold effective action. 
Sections \ref{vacua} and \ref{bhsec} collect the main results of this paper. For the most part, we tried
in these sections to make the presentation as free from the technicalities as practical. 
Interested reader can find technical guidance in section \ref{technical}, and more complete
assembly of relevant formulas in the appendices. We conclude, mainly listing the open problems,
in section \ref{conclude}. Extensive numerical tests conducted are covered in appendix \ref{numtests}.

Numerical work described in this paper has been done using Maplesoft$ ^{\rm \textregistered}$
(symbolic manipulations) and Wolfram {\it Mathematica}$ ^{\rm \textregistered}$ (numerics).
Unfortunately, it is not practical to publicly release the code and the data
produced, as it takes over 150MB.

\section{Review: KS cascading gauge theory and its holographic dual}\label{ksreview}

Following \cite{Herzog:2001xk,Bena:2019sxm,Buchel:2019pjb}, we present a brief review of the cascading gauge
theory and its holographic dual \cite{Klebanov:2000hb}.

The starting point is the original Maldacena duality \cite{Maldacena:1997re} between
$\caln=4$ $SU(N)$ supersymmetric Yang-Mills theory (SYM) and Type IIB supergravity on
$AdS_5\times S^5$:
\begin{equation}
ds_{10}^2=\frac{r^2}{L^2}\left(-dt^2 + dx_i^2\right)+\frac{L^2}{r^2}dr^2+L^2 \left(dS^5\right)^2\,; 
\eqlabel{ads5s5}
\end{equation}
there is also a constant dilaton $e^\Phi=g_s$, and a quantized self-dual R-R five form
$F_5=\calf_5+\star \calf_5$,
\begin{equation}
\calf_5=16\pi (\alpha')^2\ N\ {\rm vol}(S^5)\,,\qquad
\frac{1}{(4\pi^2\alpha')^2}\int_{S^5}\calf_5=N\in \zet\,.
\eqlabel{5formcft}
\end{equation}
The radii of $AdS_5$ and $S^5$ are
\begin{equation}
L^4=4\pi g_s N (\alpha')^2\,.
\eqlabel{ldef}
\end{equation}
The string coupling $g_s$ is related to the exactly marginal coupling $g_{YM}$ of $\caln=4$ SYM  as
\begin{equation}
g_{YM}^2=4\pi g_s\,.
\eqlabel{gymgs}
\end{equation}
This basic AdS/CFT correspondence, restricted as above to the supergravity approximation and to
leading order in $\alpha'$, implies the planar limit on the SYM size, namely $N\to \infty$
with $g_{YM}^2 N $ kept fixed, and for the large 't Hooft coupling constant, \ie   $g_{YM}^2 N\gg 1$.
In this paper we always work in the two-derivative Type IIB supergravity approximation. 

$\caln=4$ SYM is a conformal theory with central charges
$c=a$\footnote{In the supergravity approximation the two central charges are the same for any dual
pair in the gauge/gravity correspondence.}
\begin{equation}
c=a\bigg|_{\caln=4\ SYM}=\frac{N^2}{4}\,.\eqlabel{csym}
\end{equation}
The central charges characterize the quantum anomaly in the trace of the energy momentum tensor
of the theory on $\calm_4$,
\begin{equation}
\langle T_{\ \mu}^\mu\rangle= \frac{1}{(4\pi)^2} \biggl(c I_4-a E_4+ b\ \square R\biggr)\,,
\eqlabel{trace1}
\end{equation}
where the four-dimensional Euler density $E_4$ and the Weyl curvature $I_4$ are given by 
\begin{equation}
E_4=R_{\mu\nu\rho\lambda}R^{\mu\nu\rho\lambda}-4 R_{\mu\nu}R^{\mu\nu}+R^2\,,\qquad I_4=R_{\mu\nu\rho\lambda}R^{\mu\nu\rho\lambda}-2 R_{\mu\nu}R^{\mu\nu}+\frac 13 R^2\,,
\eqlabel{trace2}
\end{equation}
and $b$ is the renormalization scheme dependent constant. They 
 determine the Casimir energy $E_0$ of the vacuum state of the theory on $\calm_4$. 
The Casimir energy is renormalization scheme-dependent \cite{Assel:2015nca}
--- it depends on finite counterterms
in the effective action, which basis can be taken to be $E_4$, $I_4$, the Pontryagin
density $\tr(R\wedge R)$ and $R^2$. In this paper
we will be interested in gauge theories on $\calm_4=R\times S^3$, and it is only
the counterterm due to $R^2$ that is nonvanishing:
\begin{equation}
S_{ct}^{finite}=-\frac{b}{12(4\pi)^2}\int_{\calm_4}\ {\rm vol}_{\calm_4}\ R^2 \,,
\eqlabel{ct4}
\end{equation}
where the scheme-dependent coefficient $b$ correlates with the trace anomaly
ambiguity in \eqref{trace1}. The finite counterterm \eqref{ct4} leads to an ambiguity
$\delta E_0$ of the SYM on the $S^3$ of radius $\frac 1\mu$ as
\begin{equation}
\delta E_0= \frac{3b}{8}\ \mu\,.
\eqlabel{defde0}
\end{equation}
While we discussed here the Casimir energy ambiguity of the conformal theory,
it is clear that thermal states of the conformal theory on $S^3$ have ambiguous energy density
$\cale$ and the pressure $\calp$ as well. Specifically, corresponding to \eqref{ct4}, we have
renormalization scheme ambiguities leading to
\begin{equation}
\delta \cale=\frac{3b}{16\pi^2}\ \mu^4\,,\qquad \delta \calp= \frac13\ \delta \cale \,.
\eqlabel{dedp}
\end{equation}

The next step towards construction of the cascading gauge theory is to $\zet_2$ orbifold $\caln=4$
$SU(N)$ SYM. The resulting \cite{Douglas:1996sw}
$\caln=2$ $SU(N)\times SU(N)$ superconformal quiver gauge theory has hypermultiplets
transforming in $(N,\overline{N})\oplus(\overline{N},N)$ representation. From the $\caln=1$ point of view,
the hypermultiplets correspond to pairs of chiral multiplets $A_k$, $B_\ell$, $k,\ell=1,2$
in the $(N,\overline{N})$ and $(\overline{N},N)$ representations correspondingly. There are additional
$\caln=1$ chiral multiples $\Phi$ and $\tilde{\Phi}$ in the adjoint representations of the
two gauge group factors. The theory has a superpotential
\begin{equation}
g_{YM} \tr\ \Phi\left(A_1B_1+A_2B_2\right) +g_{YM} \tr \tilde{\Phi}\left(B_1A_1+B_2A_2\right)\,, 
\eqlabel{orbsuper}
\end{equation}
and $SU(2)\times SU(2)$ global flavor symmetry associated with rotations of $A_k$ and $B_\ell$. 
The holographic dual of the orbifold model was identified as $AdS_5\times S^5/\zet_2$ orbifold in
\cite{Kachru:1998ys}. Locally, the dual Type IIB background metric is as in \eqref{ads5s5};
because of the orbifolding the central charges are
\begin{equation}
c=a\bigg|_{S^5/\zet_2\ {\rm orbifold}}=\frac{N^2}{2}\,,
\eqlabel{n2ac}
\end{equation}
leading to a modified expression for the $AdS_5$ radius $L$:
\begin{equation}
L^4= 4\pi g_s N(\alpha')^2\ \frac{{\rm vol}(S^5)}{{\rm vol}(S^5/\zet_2)}=  8\pi g_s N(\alpha')^2\,.
\eqlabel{newl}
\end{equation}

In \cite{Klebanov:1998hh} (KW) the authors discussed an important holographic RG flow from the above
orbifold model in the UV. Specifically, they added to a superpotential \eqref{orbsuper} a relevant term
\begin{equation}
\frac{m}{2}\left(\tr \Phi^2-\tr \tilde{\Phi}^2\right)\,.
\eqlabel{mass}
\end{equation}
This mass term explicitly breaks the conformal invariance, and the supersymmetry to $\caln=1$. Below the
energy scale set by $m$, the adjoint chiral multiples can be integrated out leading to the superpotential
\begin{equation}
\frac{g_{YM}^2}{2m}\ \left[\tr\left(A_1B_1A_2B_2\right)-\tr\left(B_1A_1B_2A_2\right)\right]\,. 
\eqlabel{belowm}
\end{equation}
It was argued in  \cite{Klebanov:1998hh} that the mass-deformed orbifold model flows in the IR to $\caln=1$ superconformal
fixed point. Using the $U(1)_R$ symmetry of the superconformal fixed point, and the $SU(2)\times SU(2)$ global symmetry,
the anomalous dimensions of $A_k$ and $B_\ell$ chiral superfields can be computed as
\begin{equation}
\gamma_{A_k}=\gamma_{B_\ell}=-\frac14\,,
\eqlabel{andim}
\end{equation}
resulting in an exactly marginal superpotential in the IR:
\begin{equation}
\calw_{KW}=\frac{\lambda}{2}\ \epsilon^{ij}\epsilon^{k\ell} \ \tr A_iB_kA_jB_\ell\,.
\eqlabel{wir}
\end{equation}
The central charges $a|_{KW}$ and $c|_{KW}$ of the IR superconformal fixed point were determined
in \cite{Tachikawa:2009tt}; in the large-$N$ limit,
\begin{equation}
\frac{a|_{KW}}{a|_{S^5/\zet_2\ {\rm orbifold}}}=\frac{c|_{KW}}{c|_{S^5/\zet_2\ {\rm orbifold}}}=\frac{27}{32}\qquad \Longrightarrow\
\qquad c=a\bigg|_{KW}=\frac{27N^2}{64}\,.
\eqlabel{acflow}
\end{equation}
The holographic dual to the {\it full} gauge theory renormalization group flow described above is currently unknown.
The end point of the RG flow was argued to be holographically dual to a near-horizon geometry of D3 branes on a
singular conifold \cite{Candelas:1989js} --- this is commonly referred to as a Klebanov-Witten model \cite{Klebanov:1998hh}. 
Here, the background geometry is 
\begin{equation}
ds_{10}^2=\frac{r^2}{L^2}\left(-dt^2 + dx_i^2\right)+\frac{L^2}{r^2}dr^2+L^2 ds_{T^{1,1}}^2 \,,
\eqlabel{ads5t11}
\end{equation}
where 
\begin{equation}
ds_{T^{1,1}}^2=\frac{1}{9}\left(d\psi+\sum_{i=1}^2\cos\theta_i d\phi_i\right)^2+\frac 16\sum_{i=1}^2
\left(d\theta_i^2+\sin^2\theta_i d\phi_i^2\right)
\eqlabel{t11metric}
\end{equation}
is the metric on $T^{1,1}\equiv (SU(2)\times SU(2))/U(1)$ coset (the base of the conifold). The angular coordinates  range as  $0 \leq \psi \leq 4 \pi$, $0 \leq \theta_a \leq \pi$ and $0 \leq \phi_a \leq 2 \pi$ ($a=1,2$). Anticipating the generalization
to the deformed conifold  \cite{Candelas:1989js}, we introduce the following basis of one-forms on the compact
space \cite{Minasian:1999tt}:
\begin{equation}
\begin{split}
&g_1=\frac{\alpha^1-\alpha^3}{\sqrt 2}\,,\ \ \ g_2=\frac{\alpha^2-\alpha^4}{\sqrt 2}\,,\ \ \
g_3=\frac{\alpha^1+\alpha^3}{\sqrt 2}\,,\ \ \ g_4=\frac{\alpha^2+\alpha^4}{\sqrt 2}\,,\ \ \ g_5=\alpha^5\,,
\end{split}
\eqlabel{3form1}
\end{equation}
where 
\begin{equation}
\begin{split}
&\alpha^1=-\sin\theta_1\  d\phi_1\,,\qquad \alpha^2=d\theta_1\,,\qquad \alpha^3=\cos\psi\sin\theta_2\ d\phi_2-\sin\psi\ d\theta_2\,,\\
&\alpha^4=\sin\psi\sin\theta_2\ d\phi_2+\cos\psi\ d\theta_2\,,\qquad \alpha^5=d\psi+\cos\theta_1\ d\phi_1+\cos\theta_2\ d\phi_2\,,
\end{split}
\eqlabel{3form2}
\end{equation}
allowing to rewrite the metric \eqref{ads5t11} as
\begin{equation}
ds_{10}^2=\frac{r^2}{L^2}\left(-dt^2 + dx_i^2\right)+\frac{L^2}{r^2}dr^2+L^2\ \biggl(\frac 19 g_5^2
+\frac{1}{6}\left(g_3^2+g_4^2\right)+\frac{1}{6}\left(g_1^2+g_2^2\right)\biggr)\,.
\eqlabel{ads5t11b}
\end{equation}
Besides the metric, Type IIB supergravity background includes a constant string coupling $g_s$ and a self-dual R-R five form
$F_5=\calf_5+\star \calf_5$,
\begin{equation}
\calf_5=27\pi (\alpha')^2\ N\ {\rm vol}(T^{1,1})\,,\qquad \frac{1}{(4\pi^2\alpha')^2}\int_{T^{1,1}}\calf_5=N\in \zet\,.
\eqlabel{5formt11}
\end{equation}
The $AdS_5$ and $T^{1,1}$ radii $L$ are given by
\begin{equation}
L^4=4\pi g_s N(\alpha')^2\ \frac{{\rm vol}(S^5)}{{\rm vol}(T^{1,1})}=\frac{27}{4}\pi g_s N
(\alpha')^2\,.
\eqlabel{5formt112}
\end{equation}
The holographic computation of the boundary stress-energy trace anomaly reproduces the central charge \eqref{acflow}, see
\cite{Gubser:1998vd}.

The final step in constructing the cascading gauge theory is to {\it deform} the $\caln=1$
Klebanov-Witten superconformal gauge theory by shifting the rank of one of the gauge
groups \cite{Klebanov:2000hb}: $SU(N+M)\times SU(N)$, keeping pairs of chiral multiplets
$A_k$ and $B_\ell$, $k,\ell=1,2$,  in the bifundamental $(N+M,\overline{N})$ and
$(\overline{N+N},N)$ representations, and 
the superpotential as in \eqref{wir}. When $M\ne 0$, the theory is no longer conformal,
and the gauge couplings $g_1$ and $g_2$, of the gauge group factors $SU(N+M)$
and $SU(N)$ correspondingly, run with the renormalization group scale $\hat{\mu}$,
\begin{equation}
\begin{split}
&\frac{d}{d\ln(\hmu/\Lambda)}\ \frac{8\pi^2}{g_1^2}=3(N+M)-2 N(1-\gamma)\,,\\
&\frac{d}{d\ln(\hmu/\Lambda)}\ \frac{8\pi^2}{g_2^2}=3N-2 (N+M)(1-\gamma)\,,
\end{split}
\eqlabel{rgrunning}
\end{equation}
where $\gamma$ is the anomalous dimension of operators $\tr A_iB_j$ and $\Lambda$ is the strong coupling scale of
the cascading gauge theory\footnote{Below, we use the holographic dual to provide a precise
non-perturbative definition of $\Lambda$.}.
To leading order in $M/N$, $\gamma=-\frac12$ (see \eqref{andim}), so that
\begin{equation}
 \frac{8\pi^2}{g_1^2}- \frac{8\pi^2}{g_2^2}= 6 M \ln \frac{\hmu}{\Lambda}\ \times\ \biggl(1+\calo({M}/{N})\biggr)\,,
\eqlabel{rg1}
\end{equation}
while the sum of the gauge couplings is constant along the RG flow
\begin{equation}
 \frac{8\pi^2}{g_1^2}+ \frac{8\pi^2}{g_2^2}={\rm const}\equiv \frac{2\pi}{g_s}\,.
\eqlabel{rg2}
\end{equation}
An immediate consequence of \eqref{rg1} and \eqref{rg2} is that as one goes to the UV ($\hmu$ increases) $g_2$
diverges at some finite value of $\hmu=\hmu_{UV}\approx \Lambda e^{\frac{\pi}{3 M g_s}}$; while as one go to the IR ($\hmu$ decreases) $g_1$
diverges at some finite value of $\hmu=\hmu_{IR}\approx \Lambda e^{-\frac{\pi}{3 M g_s}}$. To continue the RG flow past $\hmu_{UV}$ or $\hmu_{IR}$, we must perform
$\caln=1$ gauge theory Seiberg duality \cite{Seiberg:1994pq}. For a cascading gauge theory discussed,
 Seiberg duality is a self-similar transformation of the theory leading to $N\to N+M$ for $\hmu> \hmu_{UV}$, and
$N\to N-M$ for $\hmu< \hmu_{IR}$ \cite{Klebanov:2000hb}. Effectively, the rank $N$ of the cascading gauge theory
runs along the RG flow as \cite{Buchel:2000ch}
\begin{equation}
N=N(\hmu)\ \propto\ +M^2\ \ln\frac{\hmu}{\Lambda}\,.
\eqlabel{rg3}
\end{equation}
Although the duality cascade extends indefinitely in the UV, it stops in the IR since the negative
values of $N(\hmu)$ are nonphysical. In general, the IR structure of the cascading gauge theory can
be rather involved \cite{Dymarsky:2005xt}; when $N$ is an integer multiple of $M$, the cascading gauge theory
ends up in the IR as the $\caln=1$ $SU(M)$ Yang-Mills theory. It confines with a spontaneous breaking of
the $U(1)_R$ chiral symmetry,
\begin{equation}
U(1)_R \to \zet_2\,.
\eqlabel{csb}
\end{equation}

The holographic dual of 'shifting the rank of one of the gauge groups' of the KW superconformal gauge theory is
to supplement the $N$ D3 branes at the tip of the conifold with $M$ D5 branes wrapping the vanishing
2-cycle of the conifold. The resulting Type IIB supergravity background takes form \cite{Klebanov:2000hb}
\begin{equation}
\begin{split}
&ds_{10}^2=H_{KS}^{-1/2}\left(-dt^2+dx_i^2\right)+\Omega_{1}^2 \left(d\tau^2+g_5^2\right)+
\Omega_{2}^2\left(g_3^2+g_4^2\right)+\Omega_{3}^2\left(g_1^2+g_2^2\right)\,,
\end{split}
\eqlabel{ks10}
\end{equation}
where
\begin{equation}
\begin{split}
&\Omega_i=\omega_{i,KS} H_{KS}^{1/4}\,,\qquad \omega_{1,KS}=\frac{\epsilon^{2/3}}{\sqrt{6}K}\,,\qquad
 \omega_{2,KS}=\frac{\epsilon^{2/3}K^{1/2}}{\sqrt{2}}\cosh\frac\tau2\,,\\
&\omega_{3,KS}=\frac{\epsilon^{2/3}K^{1/2}}{\sqrt{2}}\sinh\frac\tau2\,,\qquad K=\frac{(\sinh(2\tau)-2\tau)^{1/3}}
{2^{1/3}\sinh\tau}\,,\\
&\frac{d}{d\tau} H_{KS}=8 \frac{(2h_{2,KS}-M\alpha')h_{1,KS}-2 h_{2,KS}h_{3,KS}}{\epsilon^{8/3}K^2\sinh^2 \tau}\,,
\end{split}
\eqlabel{ks1}
\end{equation}
with
\begin{equation}
\begin{split}
&h_{1,KS}=\frac{M\alpha'g_s}{4}\frac{\cosh\tau-1}{\sinh\tau}\left(\frac{\tau\cosh\tau}{\sinh\tau}-1\right)\,,\qquad
h_{2,KS}=\frac{M\alpha'g_s}{4}\left(1-\frac{\tau}{\sinh\tau}\right)\,,
\\
&h_{3,KS}=\frac{M\alpha'g_s}{4}\frac{\cosh\tau+1}{\sinh\tau}\left(\frac{\tau\cosh\tau}{\sinh\tau}-1\right)\,,
\end{split}
\eqlabel{ks2}
\end{equation}
for the Einstein frame metric with a radial coordinate $\tau \in [0,\infty)$. The asymptotic boundary is
as $\tau\to +\infty$, while the deformed conifold 2-cycle collapsing as
\begin{equation}
\propto \epsilon^{4/3}\ \frac{\tau^2}{4}\ \left(g_1^2+g_2^2\right)\,,
\eqlabel{2cyclecollapse}
\end{equation}
as $\tau\to 0$, with the 3-cycle 
remaining finite
\begin{equation}
\propto \epsilon^{4/3}\ \left(\frac 12 g_5^2+g_3^2+g_4^2\right)\,.
\eqlabel{3cyclefinite}
\end{equation}
The dilaton is constant,
\begin{equation}
e^\Phi=g_s\,,
\eqlabel{fixeddil}
\end{equation}
R-R 3-form flux $F_3=F_3^{top}+dC_2$ and  NS-NS 3-form flux $H_3=dB_2$ are
\begin{equation}
\begin{split}
B_2  =&  h_{1,KS}\ g_1 \wedge g_2 + h_{3,KS}\ g_3 \wedge g_4,\qquad C_2=h_{2,KS}
\left(g_1\wedge g_3+g_2\wedge g_4\right)\,,\\
F_3^{top}=&\frac 12M\alpha'\ g_5\wedge g_3\wedge g_4\,,
\end{split}
\eqlabel{3fluxesks}
\end{equation}
and the self-dual R-R five-form $F_5=\calf_5+\star\calf_5$, 
\begin{equation}
g_s \calf_5=B_2\wedge F_3=\left(h_{2,KS} (h_{3,KS}-h_{1,KS}) +\frac 12 M\alpha' h_{1,KS}\right)\
g_1\wedge g_2\wedge g_3\wedge g_4\wedge g_5\,.
\eqlabel{f5ks}
\end{equation}
KS solution has 3 parameters:
\begin{itemize}
\item $g_s$ --- the string coupling, related to the RG invariant sum of the cascading gauge theory
coupling constants \eqref{rg2};
\item an integer  $M$ defining the topological part of the R-R 3-form flux  $F_3^{top}$ \eqref{3fluxesks}
(equivalently the number of D5 branes wrapping the conifold 2-cycle), related to the
difference of the ranks of the cascading gauge theory groups;
\item the conifold complex structure deformation parameter $\epsilon$, determining the strong coupling scale
$\Lambda$ of the cascading gauge theory and mass scale in the glueball spectrum
\begin{equation}
\Lambda=\frac{2^{5/12}e^{1/3}}{3^{3/2}}\ \frac{\epsilon^{2/3}}{Mg_s^{1/2}\alpha'}\,,\qquad m_{glueball}\equiv \frac{\epsilon^{2/3}}{Mg_s\alpha'} \,,
\eqlabel{lmg}
\end{equation}
where the specific numerical prefactor in the definition of the strong  coupling scale is chosen
to agree\footnote{In the previous studies of the cascading
gauge theory on $S^3$  \cite{Buchel:2011cc} or in $dS_4$ \cite{Buchel:2019pjb},
the strong coupling scale was defined
as $\Lambda_{S^3}=\Lambda_{dS_4}=2^{-1/4}\Lambda$, relative to $\Lambda$ in \eqref{lmg}.
} with the earlier computations of thermodynamics of the cascading gauge theory on $\reals^{3}$.  
\end{itemize}

To facilitate the  identification of the strong coupling scale in the numerical solutions representing
the cascading gauge theory on $S^3$ discussed in sections \ref{vacua}-\ref{bhsec}, we rewrite the
KS solution \eqref{ks1}-\eqref{f5ks} in a radial coordinate $\rho$ defined as in \eqref{5metrho}. 
Following \cite{Buchel:2019pjb} (see appendix B.3 there), from
\begin{equation}
\frac{(d\rho)^2}{\rho^4}=(\omega_{1,KS}({\tau}))^2 (d{\tau})^2\,,
\eqlabel{rhotau}
\end{equation}
and introducing
\begin{equation}
z\equiv e^{-{\tau}/3}\,,
\eqlabel{defz}
\end{equation}
we find 
\begin{equation}
\frac1\rho=\frac {\sqrt{6}\ (2\epsilon)^{2/3}}{4}\ \int_1^z\ du\  \frac{u^6-1}{u^2(1-u^{12}+12u^6 \ln u)^{1/3}}\,.
\eqlabel{solverho}
\end{equation}
In the UV, ${\tau}\to \infty$, $z\to 0$ and $\rho\to 0$ we have
\begin{equation}
\begin{split}
&e^{-{\tau}/3}\equiv z=
\frac{\sqrt{6}\ (2\epsilon)^{2/3}}{4} \rho \biggl(1+\calq \rho+\calq^2 \rho^2+\calq^3 \rho^3+\calq^4 \rho^4+\calq^5 \rho^5
+\biggl(\frac{27}{80} \epsilon^4 \ln 3+\calq^6\\
&+\frac{27}{800} \epsilon^4-\frac{9}{16} \epsilon^4 \ln 2+\frac{9}{20}
 \epsilon^4 \ln\epsilon+\frac{27}{40} \epsilon^4 \ln\rho\biggr) \rho^6+\biggl(
-\frac{63}{16} \epsilon^4 \calq \ln 2+\frac{189}{80} \epsilon^4 \calq \ln 3+\calq^7\\
&+\frac{729}{800} \calq \epsilon^4+\frac{63}{20} \epsilon^4 \calq \ln\epsilon
+\frac{189}{40} \calq \epsilon^4 \ln\rho\biggr) \rho^7+\biggl(\frac{2403}{400} \epsilon^4 \calq^2
-\frac{63}{4} \epsilon^4 \calq^2 \ln 2+\frac{189}{20} \epsilon^4 \calq^2 \ln 3\\
&+\frac{63}{5} \epsilon^4 \calq^2 \ln\epsilon+\calq^8
+\frac{189}{10} \epsilon^4 \calq^2 \ln\rho\biggr) \rho^8+\biggl(\frac{189}{5} \epsilon^4 \calq^3 \ln\epsilon+\frac{9729}{400}
 \epsilon^4 \calq^3-\frac{189}{4} \epsilon^4 \calq^3 \ln 2\\
&+\frac{567}{20} \epsilon^4 \calq^3 \ln 3+\calq^9+\frac{567}{10} \epsilon^4 \calq^3 
\ln\rho\biggr) 
\rho^9+\calo(\rho^{10}\ln\rho)\biggr)\,,
\end{split}
\eqlabel{rrhouv}
\end{equation} 
where 
\begin{equation}
\begin{split}
\calq=&\frac{\sqrt{6}\ (2\epsilon)^{2/3}}{4}\ \biggl\{
\int_0^1\ du\  \biggl(\frac{1-u^6}{u^2(1-u^{12}+12u^6 \ln u)^{1/3}}-\frac{1}{u^2}\biggr)-1\biggr\}\\
=&-\frac{\sqrt{6}\ (2\epsilon)^{2/3}}{4}\ \times\ 0.839917(9)\,.
\end{split}
\eqlabel{qdef}
\end{equation}
In the IR, ${\tau}\to 0$, $z\to 1_-$ and $y\equiv \frac1\rho\to 0$ we have
\begin{equation}
\begin{split}
\tau=\frac{\sqrt 6\ 2^{1/3}}{3^{1/3}\ \e^{2/3}}\ y\ \biggl(1-\frac{2^{2/3}\ 3^{1/3}}{15\ \e^{4/3}}\  y^2+
\frac{71\ 3^{2/3}\ 2^{1/3}}{2625\ \e^{8/3}}\  y^4
+\calo(y^6)\biggr)\,.
\end{split}
\eqlabel{rrhoir}
\end{equation}
Given an analytic Klebanov-Strassler solution \eqref{ks10}-\eqref{ks2}, \eqref{fixeddil}-\eqref{f5ks},
and the asymptotic correspondence between $\tau$ and $\rho$ (or $y$) coordinates \eqref{rrhouv} and
\eqref{rrhoir} it is straightforward to express all the KS background functions asymptotically
as $\tau\to \infty$ ($\rho\to 0$) and $\tau \to 0$ ($\rho\to\infty$). Of particular importance is
the UV asymptote of the NS-NS 3-form flux function $K_1$, to be defined below in \eqref{redef},
related to $h_{1,KS}$:
\begin{equation}
K_1\equiv 54 M\alpha'\ h_{1,KS}=\left(\frac 92 M\alpha'\right)^2 g_s\
\biggl\{\ln \biggl[\frac{2^{1/2}}{\Lambda^2} \left(\frac{2}{9M\alpha'}\right)^2 \frac{1}{g_s}\bigg]-2\ln \rho +\calo(\rho^3\ln \rho)\biggr\}\,,
\eqlabel{k1ass}
\end{equation}
where we used \eqref{rrhouv} and the definition of the strong coupling scale $\Lambda$ \eqref{lmg}.
The somewhat grotesque definition of $\Lambda$, leading to \eqref{k1ass} becomes natural
as we construct numerical solutions in sections \ref{vacua}-\ref{bhsec}, using  $P\propto M\alpha'$
as in \eqref{defpm} and the asymptotic parametrization of $K_1$ as in \eqref{as5}
\begin{equation}
K_1=K_0-2 P^2 g_s\ln \rho + \calo(\rho^3\ln\rho)\,.
\eqlabel{kointro}
\end{equation}
Indeed, we find in this case
\begin{equation}
\Lambda^2=\frac{\sqrt{2}}{P^2 g_s}\ e^{-\frac{K_0}{P^2g_s}}\,,
\eqlabel{defkolambda}
\end{equation}
where the factor $\sqrt{2}$ is introduced for historical reasons, to agree with the analysis
in \cite{Aharony:2007vg,Buchel:2010wp,Buchel:2018bzp}.

\section{Review: Type IIB supergravity on warped deformed conifold with fluxes}\label{eareview}

Consistent truncation in the $SU(2)\times SU(2)\times \zet_2$ invariant sector of Type IIB supergravity on
warped deformed conifold with fluxes  to a five dimensional manifold $\calm_5$  was derived
in \cite{Buchel:2010wp}:
\begin{equation}
\begin{split}
S_5\biggl[g_{\mu\nu},&\Omega_{i=1\cdots 3},\Phi,h_{i=1\cdots 3}\,,\,\{P,\Omega_0\}\biggr]= \frac{108}{16\pi G_5} 
\int_{\calm_5} {\rm vol}_{\calm_5}\ \Omega_1 \Omega_2^2\om_3^2\\
&\times \biggl\lbrace 
 R_{10}-\frac 12 \left(\nabla \Phi\right)^2
-\frac 12 e^{-\Phi}\left(\frac{(h_1-h_3)^2}{2\om_1^2\om_2^2\om_3^2}+\frac{1}{\om_3^4}\left(\nabla h_1\right)^2
+\frac{1}{\om_2^4}\left(\nabla h_3\right)^2\right)
\\
&-\frac 12 e^{\Phi}\left(\frac{2}{\om_2^2\om_3^2}\left(\nabla h_2\right)^2
+\frac{1}{\om_1^2\om_2^4}\left(h_2-\frac P9\right)^2
+\frac{1}{\om_1^2\om_3^4} h_2^2\right)
\\
&-\frac {1}{2\Omega_1^2\Omega_2^4\om_3^4}\left(4\Omega_0+ h_2\left(h_3-h_1\right)+\frac 19 P h_1\right)^2
\biggr\rbrace.
\end{split}
\eqlabel{5action}
\end{equation}
It is a functional of:
\begin{itemize}
\item a five-dimensional metric $g_{\mu\nu}$ on $\calm_5$,
\begin{equation}
ds_{5}^2 =g_{\mu\nu}(y) dy^{\mu}dy^{\nu}\,,
\eqlabel{5met}
\end{equation}
\item three scalars $\Omega_{i=1\cdots 3}(y)$ uplifting the metric on $\calm_5$ to a 10-dimensional metric of
Type IIB supergravity,
\begin{equation}
ds_{10}^2 = ds_{5}^2 + ds^2_{T^{1,1}}\,, \qquad ds^2_{T^{1,1}} = \Omega_1^2(y) g_5^2 + \Omega_2^2(y) (g_3^2 + g_4^2) + \Omega_3^2(y) (g_1^2 + g_2^2)\,,
\eqlabel{10dmetric}
\end{equation}
\item a dilaton $\Phi(y)$,
\item three scalars $h_{i=1\cdots 3}(y)$, uplifting to a R-R 3-form flux $F_3=F_3^{top}+dC_2$ and an NS-NS 3-form flux $H_3=dB_2$ as
\begin{equation}
\begin{split}
B_2  =&  h_1(y)\ g_1 \wedge g_2 + h_3(y)\ g_3 \wedge g_4,\qquad C_2=h_2(y) \left(g_1\wedge g_3+g_2\wedge g_4\right)\,,\\
F_3^{top}=&\frac 19 P\ g_5\wedge g_3\wedge g_4\,.
\end{split}
\eqlabel{3fluxes}
\end{equation}
\end{itemize}
The topological part of the R-R 3-from flux $F_3^{top}$ depends on 
a fixed parameter $P$, subject to quantization \cite{Aharony:2007vg,Herzog:2001xk}:
\begin{equation}
\frac{1}{4\pi^2\alpha'}\ \int_{\rm 3-cycle} F_3^{top}\qquad =\qquad \frac{2P}{9\alpha'}\ \in \zet\,,
\eqlabel{pquantization}
\end{equation}
leading to
\begin{equation}
P=\frac 92 M\alpha'\,,
\eqlabel{defpm}
\end{equation}
where $M$ is the number of fractional D3 branes (the number of D5 branes wrapping the 2-cycle of
the conifold base)\footnote{The $T^{1,1}$ 3-cycle
is parameterized by $\theta_2=\phi_2=0$, and the 2-cycle is parameterized by $\psi=0$, $\theta_1=-\theta_2$ and $\phi_1=-\phi_2$,
see \cite{Herzog:2001xk}.}.

The self-dual R-R five-form $F_5 = \calf_5+\star \calf_5$ Bianchi identity
\begin{equation}
d F_5=H_3\wedge F_3
\eqlabel{5formbianchi}
\end{equation}
can be integrated to obtain
\begin{equation}
g_s \calf_5 =\left(4\Omega_0+h_2(y) (h_3(y)-h_1(y)) +\frac P9 h_1(y)\right)\
g_1\wedge g_2\wedge g_3\wedge g_4\wedge g_5\,,
\eqlabel{5form}
\end{equation}
where $\Omega_0$ is an arbitrary constant. In the absence of the 3-form fluxes (the vanishing RHS of \eqref{5formbianchi}),
the 5-form is conserved and must be quantized \cite{Herzog:2001xk}
\begin{equation}
\frac{1}{(4\pi^2\alpha')^2}\ \ \int_{T^{1,1}} \calf_5\ =\ \frac{16\Omega_0}{\pi g_s(\alpha')^2}\ \in\ \zet\,,
\eqlabel{5formquantization}
\end{equation}
leading to
\begin{equation}
\Omega_0=\frac{\pi g_s(\alpha')^2 N}{16}\,,
\eqlabel{ncft}
\end{equation}
where $N$ is the number of D3 branes placed at the tip of the  singular conifold, and
$g_s$ is the asymptotic string coupling,
\begin{equation}
e^\Phi\ \longrightarrow\ g_s\,.
\eqlabel{defgs}
\end{equation}

Note that
 $R_{10}$ in \eqref{5action} is given by
\begin{equation}
\begin{split}
R_{10}=R_5&+\left(\frac{1}{2\om_1^2}+\frac{2}{\om_2^2}+\frac{2}{\om_3^2}-\frac{\om_2^2}{4\om_1^2\om_3^2}
-\frac{\om_3^2}{4\om_1^2\om_2^2}-\frac{\om_1^2}{\om_2^2\om_3^2}\right)-2\Box \ln\left(\om_1\om_2^2\om_3^2\right)\\
&-\biggl\{\left(\nabla\ln\om_1\right)^2+2\left(\nabla\ln\om_2\right)^2
+2\left(\nabla\ln\om_3\right)^2+\left(\nabla\ln\left(\om_1\om_2^2\om_3^2\right)\right)^2\biggr\}\,,
\end{split}
\eqlabel{ric5}
\end{equation}
and $R_5$ is the five-dimensional Ricci scalar of the metric \eqref{5met}.  
Furthermore, all the covariant derivatives $\nabla_\lambda$ in the effective action \eqref{5action} are
computed with respect to the metric \eqref{5met}. 
 Finally, $G_5$ is the five dimensional effective gravitational constant  
\begin{equation}
G_5\equiv \frac{G_{10}}{{\rm{vol}}_{T^{1,1}}}=\frac{27}{16\pi^3}\ G_{10}\,,
\eqlabel{g5deff}
\end{equation}
where $16 \pi G_{10}=(2\pi)^7g_s^2(\alpha')^4$ is  10-dimensional gravitational constant of 
Type IIB supergravity.

Our primary application of the effective action \eqref{5action} is in the context of black holes
on the warped deformed conifold, realizing the holographic dual to thermal states of the KS cascading gauge theory
on $S^3$.  For this goal, we find it convenient to introduce 
\begin{equation}
\begin{split}
h_1=&\frac 1P\left(\frac{K_1}{12}-36\Om_0\right)\,,\qquad h_2=\frac{P}{18}\ K_2\,,\qquad 
h_3=\frac 1P\left(\frac{K_3}{12}-36\Om_0\right)\,,\\
\Om_1=&\frac 13 f_c^{1/2} h^{1/4}\,,\qquad \Om_2=\frac {1}{\sqrt{6}} f_a^{1/2} h^{1/4}\,,\qquad 
\Om_3=\frac {1}{\sqrt{6}} f_b^{1/2} h^{1/4}\,,\qquad g=e^\Phi\,,
\end{split}
\eqlabel{redef}
\end{equation}
and following \cite{Aharony:2005zr} isolate the holographic coordinate $\rho$ in $\calm_5$ as:
\begin{equation}
ds_{5}^2 =g_{\mu\nu}(y) dy^{\mu}dy^{\nu}\equiv \frac{1}{h^{1/2}(x,\rho)\rho^{2}}\biggl(G_{ij}(x,\rho)dx^idx^j\biggr)+\frac{h^{1/2}(x,\rho)}{\rho^{2}}
g_{\rho\rho}(x,\rho)  (d\rho)^2\,.
\eqlabel{5metrho}
\end{equation}
The boundary of the space $\calm_5$, $\del\calm_5$, is taken at $\rho\to 0$, with   
\begin{equation}
\lim_{\rho\to 0}\ G_{ij}(x,\rho)dx^idx^j= -(dt)^2 + \frac{1}{\mu^2}\ (dS^3)^2\,,\qquad \lim_{\rho\to 0} g_{\rho\rho}(x,\rho)=1\,,
\eqlabel{delm5metric}
\end{equation}
where $(dS^3)^2$ is the unit size round metric on $S^3$, and $L_3=\frac 1\mu$ is its radius. Insisting that the $SO(4)$ symmetry of the boundary metric
\eqref{delm5metric} is the symmetry of the full $\calm_5$ metric \eqref{5metrho} implies that $G_{ij}$ and $g_{\rho\rho}$ are the functions
of the holographic radial coordinate $\rho$ only\footnote{Of course, the same implies for all
the scalars in the effective action \eqref{5action}.}. As emphasized in  \cite{Buchel:2011cc}, the parametrization \eqref{5metrho} is not
unique  --- the diffeomorphisms 
of the type 
\begin{equation}
\left( \begin{array}{c}
\rho  \\
h  \\
G_{ij}\\
g_{\rho\rho}\\
f_{a,b,c}\\
K_{1,2,3}\\
\Phi   \end{array} \right)\
\Longrightarrow \left( \begin{array}{c}
\hr  \\
\hh  \\
\hat{G}_{ij}\\
\hat{g}_{\rho\rho}\\
\hf_{a,b,c}\\
\hK_{1,2,3}\\
\hat{\Phi}   \end{array} \right)
=
\left( \begin{array}{c}
{\rho}/{(1+\alpha\ \rho)}  \\
(1+\alpha\ \rho)^4\ h \\
G_{ij}\\
g_{\rho\rho}\\
(1+\alpha\ \rho)^{-2}\ f_{a,b,c}\\
K_{1,2,3}\\
{\Phi}   \end{array} \right)\,,\qquad \alpha={\rm const}\,,
\eqlabel{leftover}
\end{equation}
preserve the general form of the metric. For\footnote{When $\mu\to 0$, \ie in the decompactification limit $S^3\to \reals^3$,
there is a residual rescaling of the radial coordinate $\rho\to \lambda \rho$, with the constant $\lambda$, where $\lambda$ is absorbed
into the rescaling of the time $t\to \lambda t$ and the $\reals^3$ spatial coordinates $x^i\to \lambda x^i$, with all the fields of the
effective action \eqref{5action} left invariant.} $\mu\ne 0$, we can completely fix \eqref{leftover}, \ie
the parameter $\alpha$ in \eqref{leftover},
requiring that for a geodesically complete $\calm_5$  the radial coordinate $\rho$
extends as 
\begin{equation}
\rho\in [0,+\infty)\,.
\eqlabel{extend}
\end{equation}

We choose different parameterizations of the metric \eqref{5metrho}
depending on whether or not it has a regular Schwarzschild horizon in the bulk:
\begin{equation}
\begin{split}
&{\rm (A)\ [no\ horizon]:}\qquad G_{ij} dx^idx^j\equiv -(dt)^2 +\frac{f_1(\rho)^2}{\mu^2}\ (dS^3)^2\,,\qquad g_{\rho\rho}=1\,;\\
&{\rm (B)\ [horizon]:}\qquad G_{ij} dx^idx^j\equiv -{f(\rho)}\ (dt)^2 + \frac{1}{\mu^2}\ (dS^3)^2\,,\qquad g_{\rho\rho}=  f^{-1}(\rho)\,.
\end{split}
\eqlabel{metpar}
\end{equation}
From \eqref{delm5metric}, at the $\calm_5$ boundary,
\begin{equation}
\lim_{\rho\to 0} f_1(\rho)=1\,,\qquad {\rm and}\qquad \lim_{\rho\to 0} f(\rho)=1\,,
\eqlabel{rho0}
\end{equation}
while the presence of the horizon as $\rho\to \infty$ in case (B) implies that 
\begin{equation}
\lim_{\rho\to \infty} f(\rho)=0\,.
\eqlabel{rhoi}
\end{equation}
Note that in case (A) --- no horizon in $\calm_5$ --- $f_1(\rho)\to 0$ as $\rho\to \infty$ is {\it one} of the possible
boundary condition dictated by the geodesic completeness of $\calm_5$; alternatively, we can require for a 2-cycle of the warped
deformed conifold to shrink to zero size, much like in the KS solution \cite{Klebanov:2000hb}, with $f_1$ nonvanishing in the
limit.

Given the radial coordinate as in \eqref{extend} and the metric ansatz \eqref{metpar},
there is an additional rescaling symmetry of the background geometry that effects the energy scale $\mu$:
\begin{equation}
\rho\to \frac{\rho}{\beta}\,,\qquad t\to \frac{t}{\beta} \,,\qquad \mu\to \beta {\mu}\,,\qquad
\beta={\rm const}\,,
\eqlabel{betatransform}
\end{equation}
with the rest of the background defining functions (\eqref{redef} and $h$, $f_1$ or $f$) unchanged.
Clearly, under the symmetry transformation \eqref{betatransform} any physical
observable $\calo$ of a mass-scaling dimension $\Delta$ would transform as
\begin{equation}
\calo\ \to\ \beta^\Delta \calo\,.
\eqlabel{brtadelta}
\end{equation}
In numerical analysis in sections \ref{vacua}-\ref{bhsec} we fix this symmetry either by fixing
$\mu$ or the rescaling of the radial coordinate --- the results of the computations will be presented
as dimensionless ratios, invariant under \eqref{betatransform}.

Finally, there are two rescaling symmetries acting on $P$ as $P\to \lambda P$:
\nxt with 
\begin{equation}
g\to \frac g\lambda \,,\ \{\rho,\mu,G_{ij},f_{a,b,c},h,K_{1,2,3}\}\to \{\rho,\mu,G_{ij},f_{a,b,c},h,K_{1,2,3}\}\,,
\eqlabel{kssym1}
\end{equation} 
\nxt or with
\begin{equation}
\rho\to \frac \rho\lambda\,,\ \{h,K_{1,3}\}\to \lambda^2\{h,K_{1,3}\}\,,\ 
\{\mu,G_{ij},f_{a,b,c},K_{2},g\}\to \{\mu,G_{ij},f_{a,b,c},K_{2},g\}\,.
\eqlabel{kssym2}
\end{equation} 
The former symmetry can be used to set the asymptotic string coupling as
\begin{equation}
\lim_{\rho\to 0} g=1\equiv g_s\,,
\end{equation}
while the latter can be used to either set $P=1$, or to set the asymptotic parameter $K_0$, see \eqref{kointro},
since under \eqref{kssym2} 
\begin{equation}
\frac{K_0}{P^2 g_s}\ \to\ {\frac{K_0}{P^2 g_s}} -2\ln\lambda\,.
\eqlabel{changek0}
\end{equation}

Generic solutions of the effective action \eqref{5action} have only $\zet_2$ $R$-symmetry; this chiral
symmetry is enhanced to $U(1)$ in solutions of the consistent truncation of the latter action
\cite{Aharony:2005zr}, constraining
\begin{equation}
K_1=K_3\,,\qquad f_a=f_b\,,\qquad  K_2\equiv 1\,.
\eqlabel{symsector}
\end{equation}

In appendix \ref{eomshol} we collect equations of motion derived from the effective action \eqref{5action}, produce the
corresponding asymptotic expansions, 
and review the holographic renormalization of the model. Additionally, in  appendix \ref{ksasvb},
we recast the supersymmetric solution of the Klebanov-Strassler \cite{Klebanov:2000hb}, reviewed in
section \ref{ksreview}, as
the spatial $S^3$-decompactification limit, \ie $\mu\to 0$, of the warped deformed conifold vacua
$\calv_B$. We collect explicit expressions for the boundary stress-energy tensor in vacua and
in thermal states of the model in appendix \ref{listtmunu}.

\section{Vacua and black holes in $AdS_5\times T^{1,1}$}\label{kw}

Prior to moving to a more technical example of Type IIB flux geometries on warped deformed conifold,
we consider the vacua and the black holes on a singular conifold, with the self-dual flux
$F_5$ only. For large values of the flux these solutions represent the vacua and the thermal states
of large-$N$ Klebanov-Witten theory \cite{Klebanov:1998hh} on $\calm_4=R\times S^3$.
There are not new results here: we just present this textbook example in the framework of
the effective action \ref{eareview}.

Note that the absence of the three form fluxes implies (see \eqref{redef})
\begin{equation}
P=0\,,\qquad K_1=K_3\equiv K_0=432 \Omega_0\,,\qquad  K_2\equiv 1\,,
\eqlabel{t11truncation}
\end{equation}
where $\Omega_0$ is quantized as in \eqref{ncft}.

\subsection{Conformal $\calv_A^s$ vacua}\label{conformalvas}

The equations of motion for horizonless bulk geometries of Type IIB supergravity \eqref{5action}, 
dual to Klebanov-Witten gauge theory vacua on $S^3$ are collected in appendix \ref{apa1}.
The limit $P\to 0$ of these equations is slightly subtle: first we need to set $K_i$ as in
\eqref{t11truncation}, followed by setting $P=0$. 
We find:
\begin{itemize}
\item from \eqref{a9}:
\begin{equation}
g(\rho)\equiv g_s\,;
\eqlabel{van1}
\end{equation}
\item from \eqref{a7}:
\begin{equation}
f_b(\rho)=f_a(\rho)\,.
\eqlabel{van2}
\end{equation}
\item Equations \eqref{a2} and \eqref{a3} become  identities once we set
\begin{equation}
f_c(\rho)=f_a(\rho)\,.
\eqlabel{van3}
\end{equation}
\item We expect that the warp factors $\Omega_i$ in \eqref{redef} are constant, thus we
set
\begin{equation}
f_a(\rho)=\frac{Q}{\sqrt{h(\rho)}}\,.
\eqlabel{fah}
\end{equation}
Consistency of \eqref{a2}, \eqref{a5} and \eqref{ac} implies that
\begin{equation}
Q=\frac 12 K_0^{1/2}\,.
\eqlabel{defqk0}
\end{equation}
Notice that the "size'' of the compact space $T^{1,1}$ is (see \eqref{t11truncation}
and  \eqref{ncft})
\begin{equation}
L^4\equiv Q^2=\frac{K_0}{4}=108 \Omega_0= \frac{27}{4} \pi g_sN (\alpha')^2 \,,
\eqlabel{van4}
\end{equation}
in agreement with \eqref{5formt112}. In what follows, we present the solution in
terms of $K_0$, rather than $L$ from \eqref{van4}.
\item We are left at this stage with the 3 equations, \ie \eqref{a1}, \eqref{a5} and
\eqref{ac}, for the remaining functions $f_1,h$ and the first-order constraint.
The most efficient way to proceed is to introduce
\begin{equation}
\tilde{h}\equiv \frac{h}{f_1^4}\,.
\eqlabel{van5}
\end{equation}
The new function must be defined on the interval \eqref{extend}, and from \eqref{as4}
and \eqref{as9}, and \eqref{ircaseaa}, must satisfy the boundary conditions:
\begin{equation}
\lim_{\rho\to 0} \tilde{h}=\frac{K_0}{4}\,,\qquad \lim_{\rho\to \infty}   \tilde{h}=\frac{1}{\mu^4 h_0^h}
={\rm finite}\,.
\eqlabel{van6}
\end{equation}
Eliminating $f_1'$ from \eqref{ac}, we find a decoupled equation for $\tilde{h}$:
\begin{equation}
0=\tilde{h}''-\frac{5(\tilde{h}')^2}{4\tilde{h}}\,.
\eqlabel{van7}
\end{equation}
The most general solution is
\begin{equation}
\tilde{h}=\frac{\calh_1}{(1+\calh_2\ \rho)^4}\,,
\eqlabel{van8}
\end{equation}
where $\calh_i$ are the integration constants.
Imposing the boundary conditions \eqref{van6} we determine
\begin{equation}
\tilde h\equiv \frac{K_0}{4}\,.
\eqlabel{van9}
\end{equation}
\item Given \eqref{van9}, we solve \eqref{a1}, subject to the boundary condition \eqref{rho0},
\begin{equation}
f_1= \frac{2}{(4+K_0\mu^2 \rho^2)^{1/2}}\,.
\eqlabel{van10}
\end{equation}
From \eqref{van5} we find
\begin{equation}
h=\frac{4K_0}{(4+K_0\mu^2 \rho^2)^{2}}\,.
\eqlabel{van11}
\end{equation}
\end{itemize}

Equipped with the analytic solution, we can extract the UV parameters \eqref{uvvevs}:
\begin{equation}
f_{a,1,0}=f_{a,3,0}=k_{2,3,0}=f_{a,4,0}=f_{c,4,0}=g_{4,0}=f_{a,6,0}=k_{2,7,0}=f_{a,8,0}=0\,,
\eqlabel{van12}
\end{equation}
and the IR parameters \eqref{irvevs}:
\begin{equation}
\begin{split}
&f_{a,0}^h=f_{b,0}^h=f_{c,0}^h=\frac{\mu^2 K_0}{4}\,,\qquad h_{0}^h=\frac{4}{\mu^4 K_0}\,,\\
&K_{1,0}^h=K_{3,0}^h=K_0\,,\qquad K_{2,0}^h=1\,,
\qquad g_{0}^h=g_s\,.
\end{split}
\eqlabel{van13}
\end{equation}

Using \eqref{epas} in the limit $P\to 0$ we find for the energy density $\cale_0$ and the
pressure $\calp$
\begin{equation}
\cale_0=3\calp=\frac{1}{8\pi G_5}\ \frac{K_0^2\mu^4}{32}=\frac{c}{2\pi^2}\ \mu^4\,,
\eqlabel{van14}
\end{equation}
where in the last equality we used the expression for $\Omega_0$ from \eqref{ncft}, $G_5$ from \eqref{g5deff},
and the expression
for the central charge $c$ of the large-$N$ Klebanov-Witten gauge theory \eqref{acflow},
resulting in the Casimir energy
\begin{equation}
E_0=\cale_0\ \times\ \frac{1}{\mu^3}\ {\rm vol}(S^3)= c\ \mu\,.
\eqlabel{van15}
\end{equation}
Note that \eqref{van15} obtained using the holographic renormalization
of \cite{Aharony:2005zr} is {\it different} from the standard result for the Casimir energy of
the holographic CFT \cite{Balasubramanian:1999re}:
\begin{equation}
E_{0}^{b=0}=\frac{3c}{4}\ \mu\,.
\eqlabel{casmin}
\end{equation}
We put the superscript $b=0$ in \eqref{casmin} to emphasize that this result was obtained in the
holographic renormalization scheme where the ambiguous coefficient $b$ multiplying $\square R$ of the conformal
theory boundary stress-energy tensor trace vanishes, see \eqref{trace1}. As detailed in \cite{Aharony:2005zr},
the minimal holographic renormalization of the flux geometries, reviewed in section \ref{apa3}, (dual to the
cascading gauge theory on generic $\calm_4$) {\it requires} the presence of $\calr_\gamma^2$ and
$\calr_{ab\ \gamma}\calr_\gamma^{ab}$ counterterms (see \eqref{lct}). These counterterms, in the conformal,
\ie $P\to 0$, limit produce the $\square R$ term in the boundary stress-energy tensor (see eq.(3.54) in
\cite{Aharony:2005zr}) as 
\begin{equation}
+\frac{1}{8\pi G_5}\ \square R\ \times \frac{1}{384}K_0^2\ =\ +\frac{c}{24\pi^2}\ \square R\ \equiv\
\frac{1}{(4\pi)^2}\ \times\ \underbrace{\frac{2c}{3}}_{\equiv b}\ \times\ \square R\,,
\eqlabel{boxrcft}
\end{equation}
which comparing with \eqref{trace1} implies that $b=\frac {2c}{3}$.   
Thus, following \eqref{defde0}, we expect that
\begin{equation}
E_0=E_0^{b=0}+\delta E_0^{b=2c/3}= \frac{3c}{4}\ \mu+\frac 38\times \frac{2c}{3}\ \mu=c\ \mu\,,
\eqlabel{e0inksframe}
\end{equation}
which is indeed the case.

\subsection{Conformal $\calt^s_{decon}$ thermal states}\label{conformalts}
The equations of motion for black hole geometries of Type IIB supergravity \eqref{5action},
dual to Klebanov-Witten gauge theory deconfined thermal states on $S^3$ are collected in appendix \ref{apa2}. 
To solve these equations we follow the same steps as in section \ref{conformalvas}. We find 
\begin{equation}
\begin{split}
&f=\frac{4 (f_{a,1,0} \rho+1) (K_0 \mu^2 \rho^2+2 f_{a,1,0}^2 \rho^2+4 f_{a,1,0} \rho+4)}
{(f_{a,1,0} \rho+2)^4}\,,\ f_a=f_b=f_c=\frac{(f_{a,1,0} \rho+2)^2}{4}\,,\\
&h=\frac{4K_0}{(f_{a,1,0}\rho+2)^4}\,,\qquad K_1=K_3\equiv K_0\,,\qquad K_2\equiv 1\,,\qquad g\equiv g_s\,,
\end{split}
\eqlabel{solvecftts}
\end{equation}
resulting in the UV parameters \eqref{uvvevsb}
\begin{equation}
\begin{split}
&f_{4,0}=\frac{1}{16}\ f_{a,1,0}^2 (2K_0\mu^2-f_{a,1,0}^2)\,,\\
&f_{a,3,0}=k_{2,3,0}=f_{c,4,0}=g_{4,0}=f_{a,6,0}=k_{2,7,0}=f_{c,8,0}=0\,,
\end{split}
\eqlabel{vant1}
\end{equation}
and the IR parameters  \eqref{irvevsbb}
\begin{equation}
\begin{split}
&f_{a,0}^h=f_{b,0}^h=f_{c,0}^h=\frac 14 f_{a,1,0}^2\,,\qquad h_{0}^h=\frac{4K_0}{f_{a,1,0}^4}\,,\qquad
K_{1,0}^h=K_{3,0}^h=K_0\,,\qquad K_{2,0}^h=1\,,\\
&g_{0}^h=g_s\,,\qquad f_1^h=\frac{4(K_0\mu^2+2f_{a,1,0}^2)}{f_{a,1,0}^3}\,.
\end{split}
\eqlabel{vant2}
\end{equation}
While $K_0$ is a fixed parameter, specifying the size of $T^{1,1}$,  see \eqref{van4},
the parameter $f_{a,1,0}$, which we express as
\begin{equation}
f_{a,1,0}\equiv K_0^{1/2}\ \kappa\,,
\eqlabel{redeffa10}
\end{equation}
is not fixed, and instead encodes the temperature of the state \eqref{thaw}:
\begin{equation}
T=\frac{\mu^2+2\kappa^2}{2\pi\kappa}\,.
\eqlabel{tcft}
\end{equation}
From section \ref{lvtb}, the remaining thermodynamic characteristics of the deconfined state are
\begin{equation}
\cale=3\calp=\frac{c}{2\pi^2}\ \biggl(\mu^4+3\kappa^2(\kappa^2+\mu^2)\biggr)\,,\qquad s=\frac{2c}{\pi}\ \kappa^3
\,,\qquad \calf=\frac{c}{2\pi^2}\biggl(\mu^4+\kappa^2(\mu^2-\kappa^2)\biggr)\,.
\eqlabel{therestcft}
\end{equation}

\begin{figure}[t]
\begin{center}
\psfrag{x}[cc][][1.5][0]{${\kappa}/{\mu}$}
\psfrag{e}[cc][][1.5][0]{$\cale/\cale_0-1$}
\includegraphics[width=6in]{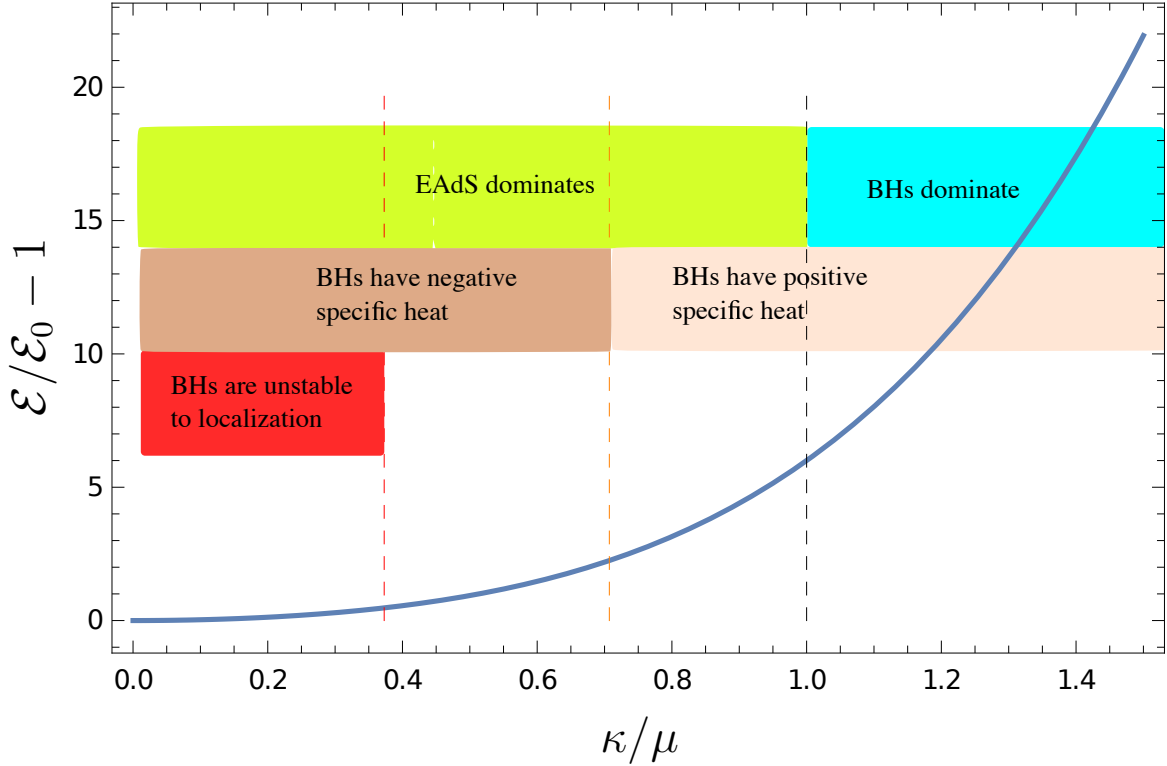}
\end{center}
  \caption{The phase diagram of black holes in $AdS_5\times T^{1,1}$. The energy density $\cale$ relative to the
  Casimir energy density $\cale_0$ is shown as a function of $\kappa/\mu$. $\kappa$ determines the Hawking
  temperature of the black hole, see \eqref{tcft}.
} \label{fig1}
\end{figure}

We end this section summarizing the well-known properties and the phase structure of the black holes in
$AdS_5\times T^{1,1}$: 
\nxt The first law of thermodynamics is satisfied:
\begin{equation}
d\cale=\frac{3c}{\pi^2}\ \kappa(2\kappa^2+\mu^2)\ \equiv\ T ds\,.
\eqlabel{detdscft}
\end{equation}
\nxt The specific heat density $c_V$ of small black holes, \ie the low entropy,
is negative:
\begin{equation}
c_V=-T\frac{\del^2 \calf}{\del T^2}=\frac{6c\kappa^3}{\pi}\ \frac{2\kappa^2+\mu^2}{2\kappa^2-\mu^2}<0 \qquad
\Longleftrightarrow\qquad \kappa<\frac{1}{\sqrt{2}}\mu\,.
\eqlabel{cvcft}
\end{equation}
\nxt The black holes discussed realize the deconfined thermal phase of the Klebanov-Witten gauge theory.
The confined thermal phase is represented by the $\calv_A^s$ vacua solutions of section \ref{conformalvas}
with the Euclidean time direction compactified with the appropriate inverse temperature period.  
The Hawking-Page \cite{Hawking:1982dh}
or the confinement phase transition \cite{Witten:1998zw} occurs when (compare \eqref{therestcft} and
\eqref{van14})
\begin{equation}
\calf\bigg|_{deconfined}\ge\  \calf=\cale\bigg|_{confined}\qquad \Longleftrightarrow\qquad  \kappa^2(\mu^2-\kappa^2)\ge 0 \,,
\eqlabel{condeconcft}
\end{equation}
leading to a critical temperature $T_{con/decon}$ of the confinement/deconfinement phase transition
\begin{equation}
T_{con/decon}=\frac{3\mu}{2\pi}\,.
\eqlabel{tcondecon}
\end{equation}
\nxt The black holes considered are 'smeared' over the compact space --- $T^{1,1}$ in this case. When the black
hole becomes sufficiently small, it becomes unstable with respect to localization on the compact
space \cite{Prestidge:1999uq,Hubeny:2002xn}.
From the dual boundary gauge theory perspective, the global $SU(2)\times SU(2)\times U(1)$ symmetry of the
Klebanov-Witten gauge theory is spontaneously broken. 
It was determined in \cite{Buchel:2015pla} that
the $AdS_5\times T^{1,1}$ black holes develop such an  instability when
\begin{equation}
\kappa < 0.37285(8)\ \mu\,.
\eqlabel{unstablecft}
\end{equation}

In fig.\ref{fig1} we collect the features of the black holes in $AdS_5\times S^5$. We plot the energy density
$\cale$ of the black holes relative to their Casimir energy density $\cale_0$  \eqref{van14} as a function of
$\kappa/\mu$. $\kappa$ is an auxiliary mass scale, which can be traded for the temperature $T$ following \eqref{tcft},
and $\mu$ is the boundary $S^3$ compactification scale. We choose $\kappa$ so that the Hawking-Page
(or confinement/deconfinement phase transition from the dual boundary perspective) is at
\begin{equation}
\frac{\kappa}{\mu}\bigg|_{\rm Hawking-Page}=1\,.
\eqlabel{hpkappa}
\end{equation}
Black holes dominate in the canonical ensemble for $\kappa>\mu$, and undergo the first-order phase transition
at $\kappa=\mu$. In the microcanonical ensemble, black holes remain thermodynamically stable for $\kappa > \mu/\sqrt{2}$,
and develop a perturbative instability as indicate in \eqref{unstablecft}.

\section{Round $S^3$ vacua of  the warped deformed conifold with fluxes}\label{vacua}

\begin{figure}[t]
\begin{center}
\psfrag{c}[cc][][1.0][0]{$\calv_A^s$}
\psfrag{a}[cc][][1.0][0]{$\calv_A^s$}
\psfrag{b}[cc][][1.0][0]{$\calv_B$}
\psfrag{x}[cc][][1.0][0]{$\mu/\Lambda$}
\psfrag{e}[cc][][1.0][0]{$\hat\cale$}
\psfrag{q}[cc][][1.0][0]{$\mu_u$}
\psfrag{p}[cc][][1.0][0]{$\mu_\csb$}
\includegraphics[width=6in]{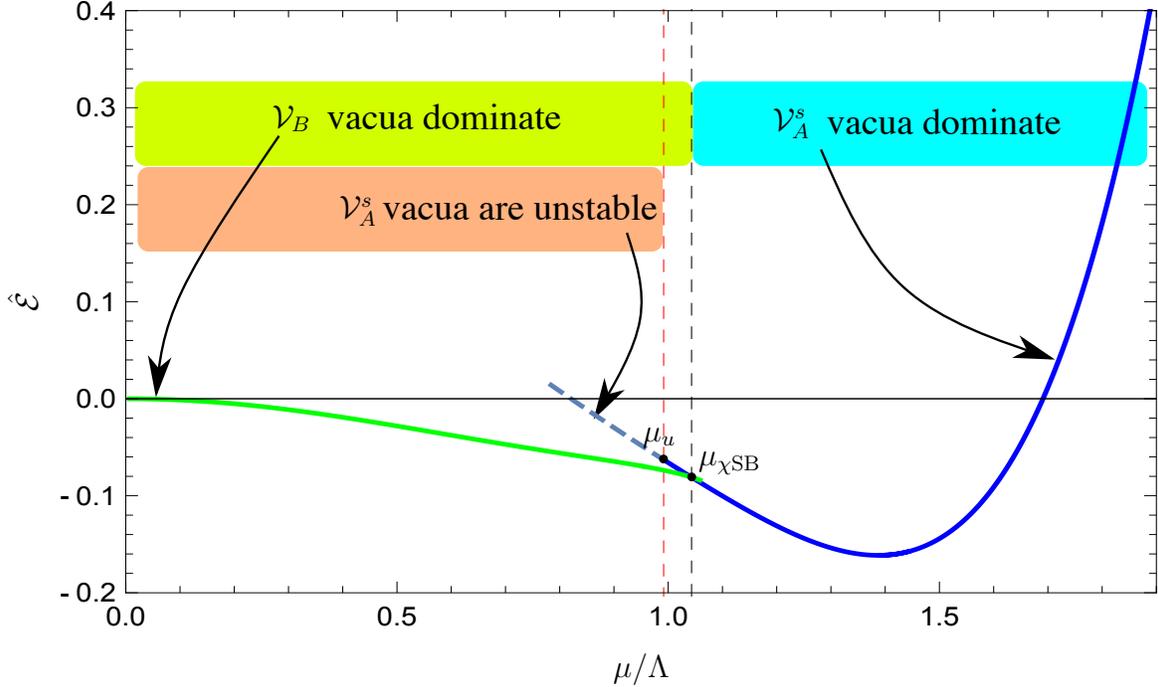}
\end{center}
  \caption{Round $S^3$ vacua of the warped deformed conifold with fluxes. The vertical
  dashed black line indicates the first-order transition between the chirally symmetric $\calv_A^s$
  (solid blue curve) and $\calv_{B}$
(solid green curve)  vacua with the spontaneously broken symmetry, see \eqref{mulambdablack}.
The vertical dashed red line indicates the onset of the perturbative chiral symmetry breaking
instability of the $\calv_A^s$ vacua (which are denoted with the grey dashed curve),
see \eqref{defmuunst}.
} \label{fluxvacuaI}
\end{figure}

\begin{figure}[t]
\begin{center}
\psfrag{c}[cc][][1.0][0]{$\calv_A^s$}
\psfrag{b}[cc][][1.0][0]{$\calv_B$}
\psfrag{a}[cc][][1.0][0]{$\calv_A^b$}
\psfrag{x}[cc][][1.0][0]{$\mu/\Lambda$}
\psfrag{e}[cc][][1.0][0]{$\hat\cale$}
\psfrag{p}[cc][][1.0][0]{$\mu_u$}
\includegraphics[width=6in]{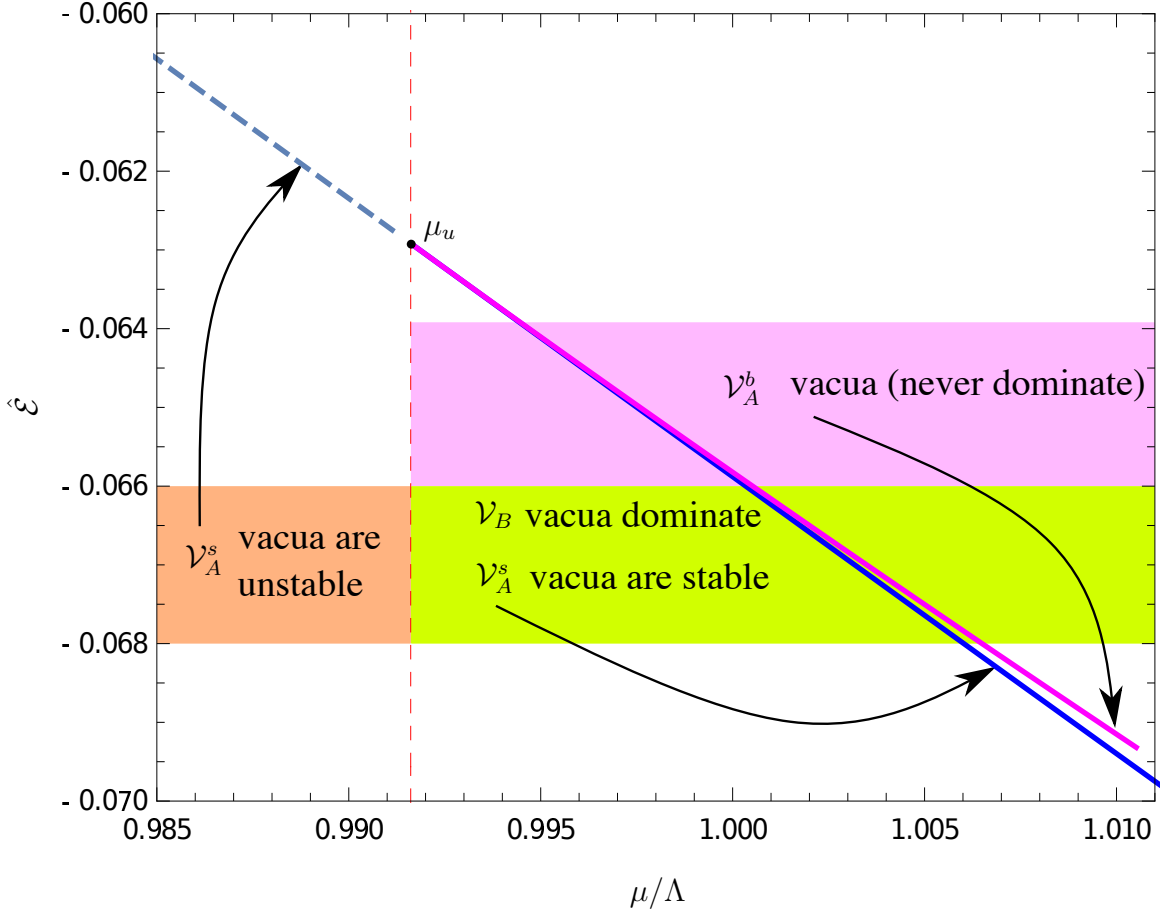}
\end{center}
  \caption{
The new branch of $\calv_A^b$ vacua (solid magenta curve) extends for $\mu> \mu_u$,
see \eqref{defmuunst}, represented by the vertical dashed red line. 
} \label{fluxvacuaII}
\end{figure}

We begin this section summarizing the results for the $S^3$ vacua structure of the warped deformed
conifold with fluxes. These vacua realize the holographic dual to the $SO(4)$ invariant
ground states of the cascading gauge theory \cite{Klebanov:2000hb} on $S^3$ of radius
$\frac 1\mu$. The theory has a strong coupling scale $\Lambda$ \eqref{lmg}.
In figs.~\ref{fluxvacuaI} and \ref{fluxvacuaII} we present the reduced Casimir energy density
$\hat\cale$, defined as
\begin{equation}
\hat\cale\equiv \frac{8\pi G_5}{P^4 g_s^2}\ \frac{\cale}{\Lambda^4}
=\frac{2^6\pi^4}{3^5 M^4}\ \frac{\cale}{\Lambda^4}\,,
\eqlabel{defhate}
\end{equation}
where we used \eqref{g5deff} and \eqref{defpm}, as a function of $\frac{\mu}{\Lambda}$.
Recall that there are 3 distinct types of vacua: $\calv_A^s$, $\calv_B$ and $\calv_A^b$,
see fig.~\ref{vacuapic}:
\nxt The $\calv_A^s$ vacua realize ground states of the cascading gauge theory
with unbroken chiral symmetry; in the bulk gravitational dual they are characterized topologically
by the presence of the trivial 3-cycle (the boundary $S^3$ can smoothly shrink to a zero size).
In the limit $\mu\gg \Lambda$ they resemble the ground states of the conformal
Klebanov-Witten gauge theory \cite{Klebanov:1998hh}, see section \ref{conformalvas}.
These vacua were originally constructed and analyzed in \cite{Buchel:2011cc}.
\nxt The $\calv_B$ vacua realize ground states of the cascading gauge theory
with the spontaneously broken  chiral symmetry;
in the bulk gravitational dual they are characterized topologically
by the presence of the trivial 2-cycle (the vanishing 2-cycle of the deformed conifold, see
\eqref{2cyclecollapse}). In the limit $\mu/\Lambda\to 0$ they represent the ground state of the
Klebanov-Strassler gauge theory on $\reals^3$. Note that, see \eqref{ksepk0},
\begin{equation}
\hat\cale\bigg|_{{\rm KS\ on}\ \reals^3}=0\,.
\eqlabel{hateks}
\end{equation}
These vacua were originally constructed and analyzed in \cite{Buchel:2011cc}.
\nxt  The $\calv_A^b$ vacua realize ground states of the cascading gauge theory
with the spontaneously broken chiral symmetry; in the bulk gravitational dual they are
characterized topologically
by the presence of the trivial 3-cycle (the boundary $S^3$ can smoothly shrink to a zero size).
These vacua are constructed and analyzed here for the first time. 

In fig.~\ref{fluxvacuaI} the dashed vertical black line denotes the first-order
(zero temperature) transition between $\calv_A^s$ (solid blue curve) and $\calv_B$
(solid green curve)
vacua:
\begin{equation}
\frac{\mucsb}{\Lambda}=1.043069(7)\,.
\eqlabel{mulambdablack}
\end{equation}
This is in perfect agreement with the corresponding transition point obtained in
\cite{Buchel:2011cc}\footnote{The critical value of  $\mu$ reported there, see
eq.~(5.76) in \cite{Buchel:2011cc}, is $2^{1/4}$ times bigger corresponding
to a $\sqrt{2}$ difference in the definition of the strong coupling scale $\Lambda^2$
\eqref{defkolambda} used in this work and in \cite{Buchel:2011cc}.}.
$\calv_A^s$ vacua dominate for  $\mu> \mucsb$,
implying that the ground state of the cascading gauge theory on sufficiently small $S^3$ has unbroken
chiral symmetry; while the $\calv_B$ vacua dominate for  $\mu<\mucsb$, implying
that the  ground state of the cascading gauge theory
on $S^3$ spontaneously breaks the chiral symmetry on  sufficiently large $S^3$.
Chirally symmetric $\calv_A^s$ states can be constructed for $\mu<\mucsb$, however,
they become perturbatively unstable for $\mu<\mu_u$,
\begin{equation}
\frac{\mu_u}{\Lambda}=0.991613(4)\,,
\eqlabel{defmuunst}
\end{equation}
represented by the dashed vertical red line in fig.~\ref{fluxvacuaI}. We represent $\calv_A^s$ vacua
for $\mu<\mu_u$ with a dashed grey curve. In \cite{Buchel:2011cc} the instability
\eqref{defmuunst} was identified\footnote{Again, up to the redefinition
of the strong coupling scale $\Lambda$ \eqref{defkolambda}, we find a perfect agreement
with the earlier computations.} analyzing the spectrum of the chiral symmetry breaking
fluctuations of the theory. Here we identify the instability studying the
linearized chiral symmetry breaking fluctuations about $\calv_A^s$ vacua
with with an explicit symmetry breaking source term \cite{Buchel:2019pjb}.
This latter approach would allow us to construct $\calv_A^b$ vacua, which escaped
the analysis in \cite{Buchel:2011cc}.

In fig.~\ref{fluxvacuaII} the solid magenta curve represent $\calv_A^b$ vacua --- much like $\calv_B$,
vacua, the dual states of the cascading gauge theory have $\zet_2$ chiral symmetry. Since their energy
density is {\it above} the energy density of $\calv_B$ states, they are not the true ground states. 
The scale of the figure does not allow to show this, but
(compare with fig.~\ref{fluxvacuaI})  for the range of $\mu/\Lambda$ reported,
\begin{equation}
\hat\cale\bigg|_{\calv_A^s}\ >\ \hat\cale\bigg|_{\calv_{B}}\,.
\end{equation}
What is exotic in the diagram~\ref{fluxvacuaII} is that
$\calv_A^b$ vacua exist only for $\mu>\mu_u$,
see \eqref{defmuunst}, represented by a vertical dashed red line, 
where $\calv_A^s$ vacua are perturbatively stable and
\begin{equation}
\hat\cale\bigg|_{\calv_A^b}\ >\ \hat\cale\bigg|_{\calv_A^s}\,,
\end{equation}
\ie the states with the spontaneously broken symmetry exist above (rather than below!)
the energy scale set by the transition point\footnote{This is not the first encounter of such a
phenomenon: see \cite{Buchel:2009ge} for the discovery and additional examples
\cite{Bosch:2017ccw,Buchel:2017map,Buchel:2018bzp,Buchel:2020thm,Buchel:2020jfs,Buchel:2020xdk}
in holography and  \cite{Chai:2020zgq} in QFTs.}. We report on the
perturbative stability of $\calv_A^b$ states elsewhere. We comment on the fate of the unstable
$\calv_A^s$ vacua in section \ref{bhsec}.

\begin{figure}[t]
\begin{center}
\psfrag{r}[bb][][1.0][0]{$\hat{\calk}/\hat{\calk}_{KS}$}
\psfrag{s}[tt][][1.0][0]{$\hat{R}^2_{S^3}/\hat{R}^2_{S^3,KS}$}
\psfrag{x}[cc][][1.0][0]{$\mu/\Lambda$}
\psfrag{e}[cc][][1.0][0]{$\hat\cale$}
\includegraphics[width=3in]{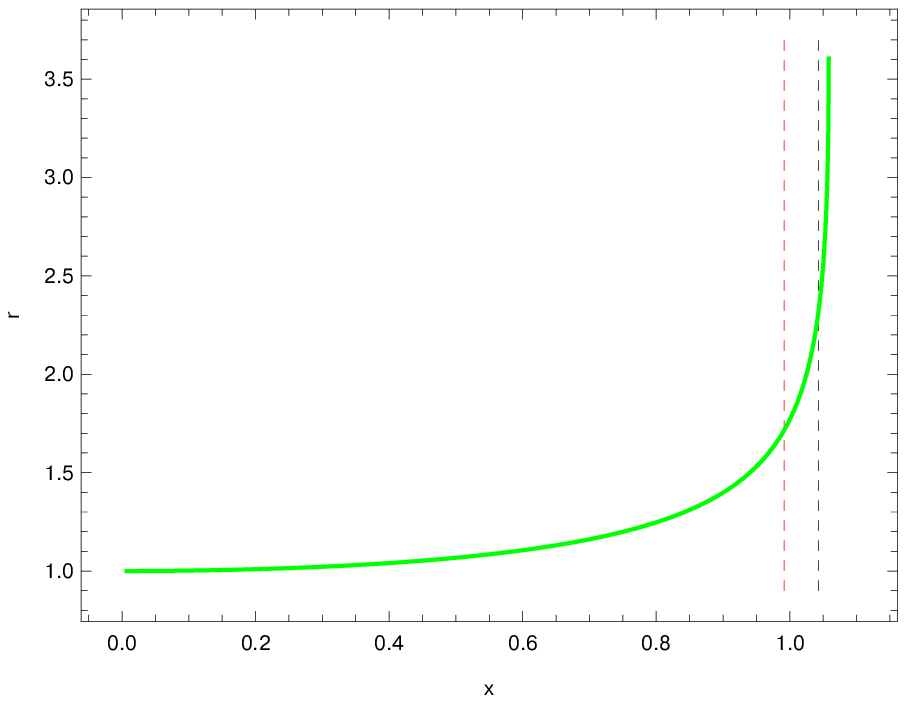}\ 
\includegraphics[width=3in]{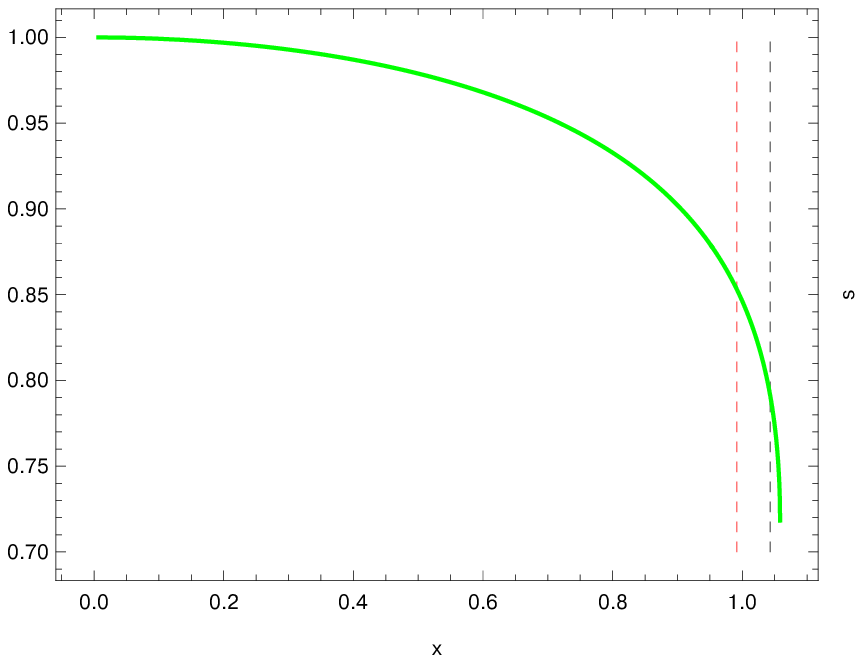}
\end{center}
  \caption{Left panel: the reduced Kretschmann scalar $\hat\calk$ of $\calv_B$ vacua in the deep
  IR of the bulk geometry, see \eqref{defhatk}. $\hat\calk_{KS}$ denotes the value of the
  scalar for the Klebanov-Strassler solution, \ie when $\mu/\Lambda=0$, see \eqref{defhatkks}.
  Right panel: the rapid growth of the scalar when $\mu\gtrsim \mucsb$
  (denoted by the vertical dashed black line) is caused by the collapsing 3-cycle of the conifold,
  see \eqref{r2s3van}. $\hat{R}^2_{S^3,KS}$ denotes the size of this 3-cycle for
  the Klebanov-Strassler solution, \ie when $\mu/\Lambda=0$, see \eqref{r2s3vanks}.
} \label{riem2b}
\end{figure}

\begin{figure}[t]
\begin{center}
\psfrag{r}[bb][][1.0][0]{$\hat{\calk}/\hat{\calk}_{KS}$}
\psfrag{t}[tt][][1.0][0]{$\hat{R}^2_{T^{1,1}}$}
\psfrag{x}[cc][][1.0][0]{$\mu/\Lambda$}
\psfrag{e}[cc][][1.0][0]{$\hat\cale$}
\includegraphics[width=3in]{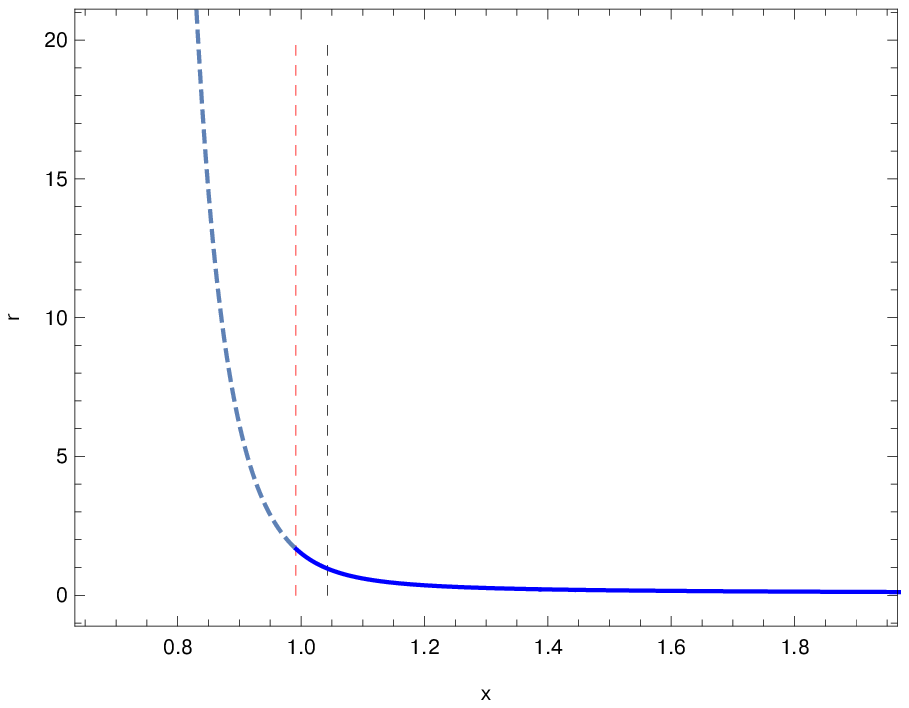}\ 
\includegraphics[width=3in]{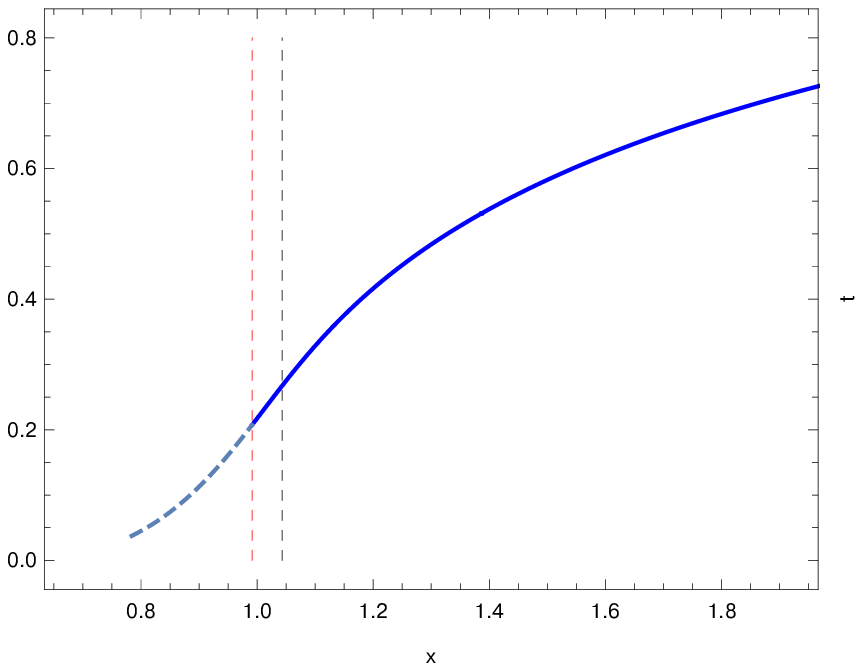}
\end{center}
  \caption{Left panel: the reduced Kretschmann scalar $\hat\calk$ of $\calv_A^s$ vacua in the deep
  IR of the bulk geometry, see \eqref{defhatk}. $\hat\calk_{KS}$ denotes the value of the
  scalar for the Klebanov-Strassler solution, \ie when $\mu/\Lambda=0$, see \eqref{defhatkks}.
  Right panel: the rapid growth of the scalar when $\mu\lesssim \mu_u$
  (denoted by the vertical dashed red line) is caused by the collapsing of the deformed
  $T^{1,1}$,
  see \eqref{defft11def} and \eqref{r2tt11}. 
} \label{riem2as}
\end{figure}

\begin{figure}[t]
\begin{center}
\psfrag{r}[bb][][1.0][0]{$\hat{\calk}/\hat{\calk}_{KS}$}
\psfrag{s}[tt][][1.0][0]{$\hat{R}^2_{S^2}$}
\psfrag{x}[cc][][1.0][0]{$\mu/\Lambda$}
\psfrag{e}[cc][][1.0][0]{$\hat\cale$}
\includegraphics[width=3in]{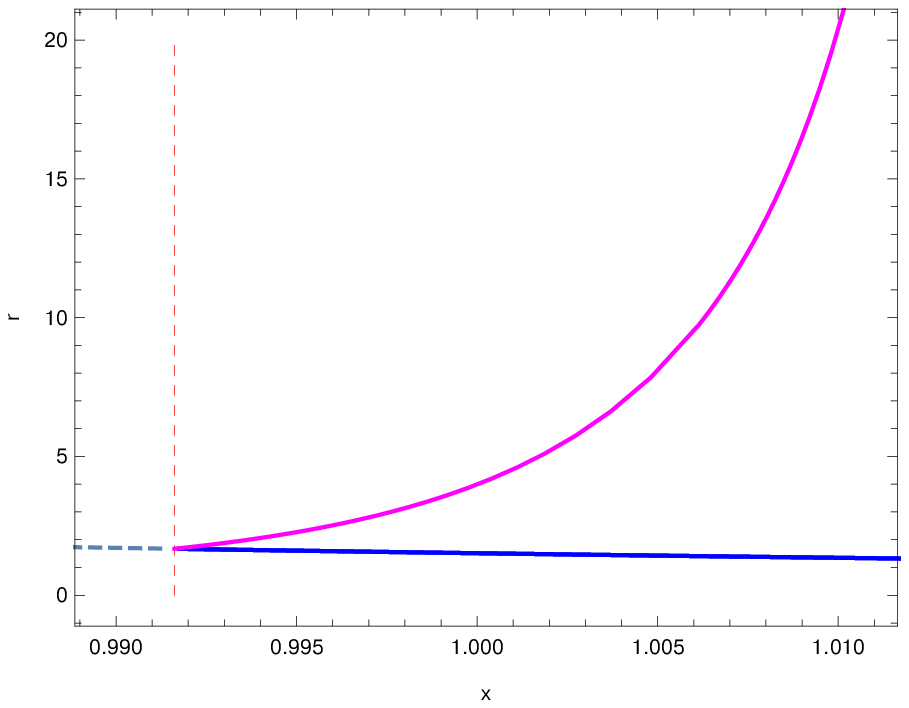}\ 
\includegraphics[width=3in]{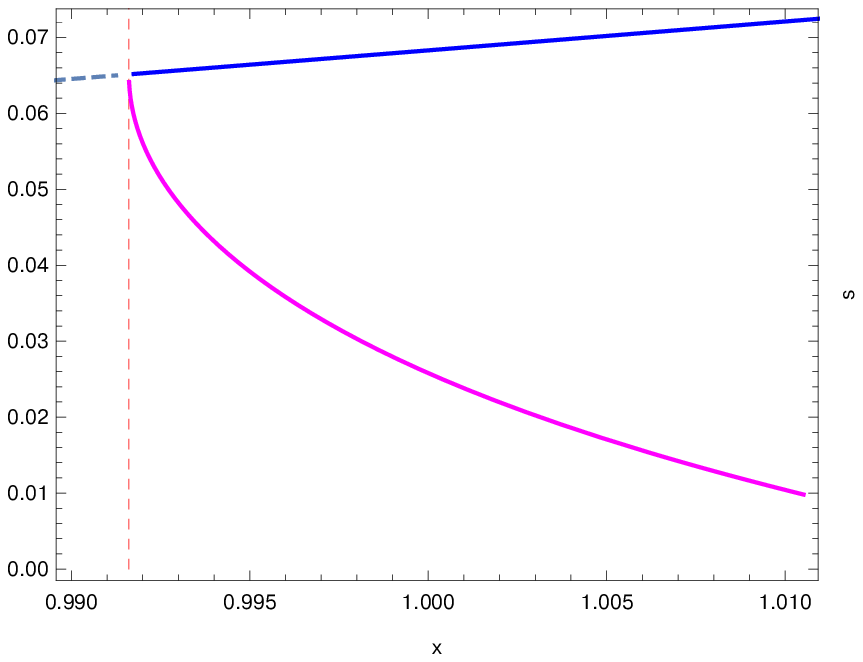}
\end{center}
  \caption{Left panel: the reduced Kretschmann scalar $\hat\calk$ of $\calv_A^b$ vacua (magenta curve)
  and $\calv_A^s$ vacua (blue/grey curves) in the deep
  IR of the bulk geometry, see \eqref{defhatk}. $\hat\calk_{KS}$ denotes the value of the
  scalar for the Klebanov-Strassler solution, \ie when $\mu/\Lambda=0$, see \eqref{defhatkks}.
  Right panel: the rapid growth of the scalar (magenta curve, $\calv_A^b$ vacua)
  when $\mu\gtrsim \mu_u$
  (denoted by the vertical dashed red line) is caused by the collapsing 2-cycle of the conifold,
  see \eqref{r2s2van}.
} \label{riem2ab}
\end{figure}

We encountered a (technical) obstruction of constructing $\calv_{B}$ and $\calv_A^b$ vacua
for  large values of $\mu$, and $\calv_A^s$ vacua for small values of $\mu$. The physical
reason for the obstruction is easy to understand focusing on the invariant
properties of the corresponding geometries. In the left panels of figs.~\ref{riem2b}-\ref{riem2ab}
we plot the reduced Kretschmann scalar $\hat\calk$, defined as
\begin{equation}
\hat\calk\equiv P^2 g_s\ \times\ \calr_{\mu\nu\lambda\rho}\calr^{\mu\nu\lambda\rho}\bigg|_{IR}\,,
\eqlabel{defhatk}
\end{equation}
where the ten-dimensional bulk Riemann tensor quadratic invariant is evaluated in the
deep IR of the geometry, \ie as $y\equiv \frac 1\rho\to 0$, as a function of $\mu/\Lambda$.
The quantity
\begin{equation}
\hat\calk_{KS}=\frac{64\times 3^{2/3} 2^{1/3}\times (110\times 3^{1/3}2^{2/3}+177147H_0^2)
}{295245H_0^3} 
\eqlabel{defhatkks}
\end{equation}
represents the reduced Kretschmann scalar of the Klebanov-Strassler geometry, \ie
when $\mu/\Lambda\to 0$.  $H_0\equiv 0.056288(0)$ is the numerical coefficient in the
IR asymptote of $h_0^h$, see \eqref{susyir}.
Interestingly, $\hat\calk_{KS}$, as defined, is independent of the strong coupling scale $\Lambda$
of the cascading gauge theory, equivalently the conifold deformation parameter $\epsilon$,
see \eqref{lmg}.
Clearly, the obstruction in extending the
construction of the flux vacua is associated with the rapid growth of the quadratic curvature
invariant in the deep IR of the geometry, leading to the breakdown of the supergravity
approximation. This is associated with an {\it additional} cycle\footnote{This is an additional 
cycle to the already collapsed one: the boundary $S^3$ for $\calv_{A}^s$ vacua,
the conifold 2-cycle for $\calv_{B}$ vacua, and the boundary $S^3$ for $\calv_{A}^b$ vacua ---
see fig.~\ref{vacuapic}.}
becoming small in the IR. As demonstrated in the right panels of figs.~\ref{riem2b}-\ref{riem2ab},
what is precisely this additional cycle, differs:
\nxt From the right panel in fig.~\ref{riem2b} for $\calv_B$ vacua, this is a conifold
3-cycle \eqref{3cycle}. We define the reduced radius square of the cycle (again evaluated
in the IR, $y\to 0$) as
\begin{equation}
\hat{R}^2_{S^3}\equiv \left(P^2 g_s\right)^{-1/2}\ \times\ \frac{f_{a,0}^h(h_0^h)^{1/2}}{3}\,.
\eqlabel{r2s3van}
\end{equation}
Note that the corresponding 3-cycle of the Klebanov-Strassler solution has a reduced size
(again independent of $\Lambda$ or $\epsilon$) 
\begin{equation}
\hat{R}^2_{S^3,KS}\equiv \left(P^2 g_s\right)^{-1/2}\ \times\ \left(\frac 23\right)^{1/3} H_0^{1/2}\,.
\eqlabel{r2s3vanks}
\end{equation}
\nxt From the right panel in fig.~\ref{riem2as} for $\calv_A^s$ vacua, this is (deformed)
$T^{1,1}$ (compare with \eqref{ads5t11b}):
\begin{equation}
(d{\tilde T}^{1,1})^2\bigg|_{IR}=f_{a,0}^h(h_0^h)^{1/2}\biggl(\frac{f_{c,0}^h}{f_{a,0}^h}\ \frac19\ g_5^2 +\frac 16(g_3^2+g_4^2)+\frac 16(g_1^2+g_2^2)\biggr)\,.
\eqlabel{defft11def}
\end{equation}
 We define the reduced radius square of the cycle (from \eqref{defft11def}) as
\begin{equation}
\hat{R}^2_{T^{1,1}}\equiv \left(P^2 g_s\right)^{-1/2}\ \times\ (f_{c,0}^h)^{1/5}(f_{a,0}^h)^{4/5}(h_0^h)^{1/2}\,.
\eqlabel{r2tt11}
\end{equation}
\nxt From the right panel in fig.~\ref{riem2ab} for $\calv_A^b$ vacua, this is a conifold
2-cycle \eqref{def2cycle}. We define the reduced radius square of the cycle (again evaluated
in the IR, $y\to 0$) as
\begin{equation}
\hat{R}^2_{S^2}\equiv \left(P^2 g_s\right)^{-1/2}\ \times\ \frac{f_{b,0}^h(h_0^h)^{1/2}}{3}\,.
\eqlabel{r2s2van}
\end{equation}

As before, the vertical dashed black line indicates \eqref{mulambdablack},
and the vertical dashed red line indicates \eqref{defmuunst}.

In the rest of this section we explain the technical details leading to the results reported above. 

\subsection{$\calv_B$ vacua}\label{calvbvac}

These vacua were originally constructed in \cite{Buchel:2011cc}.
Equations of motion describing $\calv_B$ vacua are collected in \eqref{a1}-\eqref{ac},
with the UV, \ie $\rho\to 0_+$,  asymptotics \eqref{as1}-\eqref{as6},  and the IR, \ie $y\equiv \frac 1\rho\to 0_+$,  asymptotics
\eqref{bredef}, \eqref{ircaseb}. There are 4 non-normalizable coefficients
\begin{equation}
K_0\,,\ \mu\,,\ P\,,\  g_s\,,
\eqlabel{nonvb}
\end{equation}
characterizing the cascading gauge theory \eqref{defuvcasea}, and 17 normalizable
coefficients \eqref{uvvevs} and \eqref{irvevsb} --- precisely as needed to solve
the system of 8 second-order equations  \eqref{a1}-\eqref{a3}, \eqref{a5}-\eqref{a9} and the first-order constraint
\eqref{ac}.

The numerical techniques for solving the required system of nonlinear ODEs were developed in
\cite{Aharony:2007vg}.  
The scaling symmetries \eqref{kssym1} and \eqref{kssym2} are used to set
\begin{equation}
g_s=1 \qquad {\rm and}\qquad P=1\,.
\eqlabel{setvb}
\end{equation}
Additionally, note that \eqref{betatransform} acts on $K_0$ as
\begin{equation}
\frac{K_0}{P^2 g_s}\ \to\ {\frac{K_0}{P^2 g_s}} -2\ln\beta\,,
\eqlabel{changgek0vb}
\end{equation}
which can be used to keep
\begin{equation}
\frac{K_0}{P^2 g_s}\ =\ {\rm constant}\ \equiv\ k_s\,.
\eqlabel{defks}
\end{equation}
It is important to keep in mind that the symmetry transformation \eqref{betatransform}
affects {\it all} the dimensionful observables, and so, once used, the numerical
results must be represented as dimensionless quantities.
At this stage the only variable non-normalizable coefficient is $\mu$, which is used to label
the sets of normalizable coefficients  \eqref{uvvevs} and \eqref{irvevsb}.
At $\mu=0$, $\calv_B$ vacuum is a Klebanov-Strassler solution, thus from \eqref{defk0ep},
\begin{equation}
\epsilon=2^{5/4} 3^{-3/4}  \ \exp\left(-\frac12-\frac34 k_s\right)\,,
\eqlabel{epks}
\end{equation}
leading to (see \eqref{susyuv})
\begin{equation}
\begin{split}
&f_{a,1,0}=2 \hat{\calq} \ \exp\left(-\frac13-\frac12 k_s\right)\,,\qquad
f_{a,3,0}=2 \ \exp\left(-1-\frac32 k_s\right)\,,\\
&k_{2,3,0}=-(2+3 k_s) \ \exp\left(-1-\frac32 k_s\right)\,,\qquad
f_{a,4,0}=-2 \hat{\calq} \ \exp\left(-\frac43-2 k_s\right)\,,\\
&f_{c,4,0}=0\,,\qquad g_{4,0}=0\,,\qquad
f_{a,6,0}=-\frac{2}{25} (25 \hat{\calq}^3+15 k_s-14) \ \exp\left(-2-3 k_s\right)\,,
\\ &k_{2,7,0}=\frac32 (37-30 k_s) \hat{\calq}^4 \ \exp\left(-\frac73-\frac72 k_s\right)\,,\\
&f_{a,8,0}=-2 \hat{\calq}^2 (\hat{\calq}^3+6 k_s-11) \ \exp\left(-\frac83-4 k_s\right)\,,
\end{split}
\eqlabel{susyuvks}
\end{equation} 
in the UV, with (see \eqref{qdef})
\begin{equation}
\hat{\calq}\equiv-\frac{4}{\sqrt{6}(2\epsilon)^{2/3}}\ \calq=0.839917(9) \,,
\end{equation}
and (see \eqref{susyir})
\begin{equation}
\begin{split}
&f_{a,0}^h=\frac43 3^{2/3} \ \exp\left(-\frac23-k_s\right)\,,\qquad
h_0^h=\frac{9}{16} 2^{2/3} \ \exp\left(\frac43+2 k_s\right)\ \times\ 0.056288(0)\,,\\
&g_0^h=1\,,\qquad
K_{1,3}^h=\ \exp\left(1+\frac32 k_s\right)\,,\qquad K_{2,2}^h=\frac12 3^{1/3} \
\exp\left(\frac23+k_s\right)\,,\\
&K_{2,4}^h=-\frac{11}{40} 3^{2/3} \ \exp\left(\frac43+2 k_s\right)\,,\qquad
K_{3,1}^h=\frac49 3^{2/3} \ \exp\left(\frac13+\frac12 k_s\right)\,,\qquad  f_{1,0}^h=1\,,
\end{split}
\eqlabel{susyirks}
\end{equation}
in the IR.

\begin{figure}[t]
\begin{center}
\psfrag{x}[cc][][1.0][0]{$\mu/\Lambda$}
\psfrag{e}[cc][][1.0][0]{$\hat\cale$}
\includegraphics[width=6in]{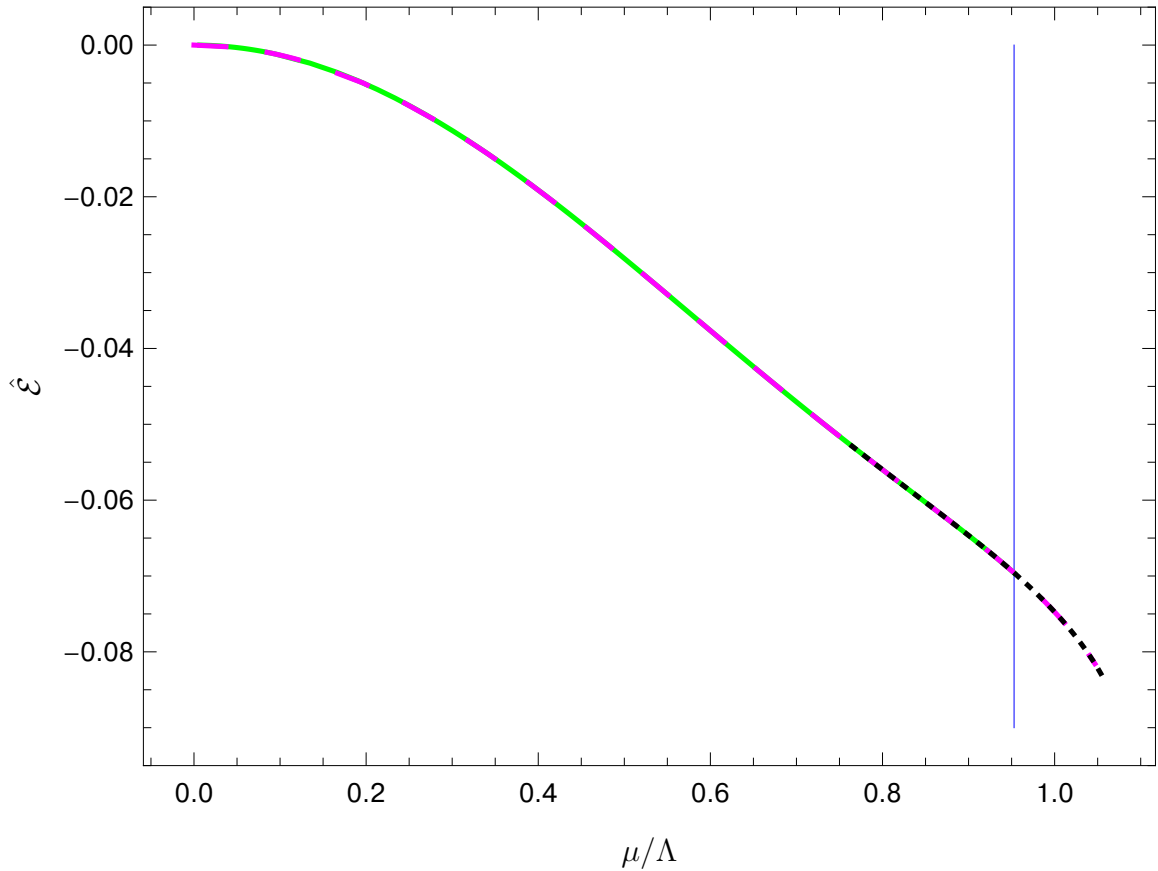}\ 
\end{center}
  \caption{Cross checks on numerics of $\calv_B$ vacua energy in the
  three different computational schemes: solid green curve, dashed magenta curve,
  and the dotted black curve.
  The vertical blue line denotes the common reference point in the green/black curve computational
  schemes, see
  \eqref{muk03}.
} \label{vbcompare}
\end{figure}

In practice, fixing $k_s$, we start with $\mu=0$ with \eqref{susyuvks} and
\eqref{susyirks},  and produce the data sets, incrementing $\mu$, for
\begin{equation}
\frac{\mu}{\Lambda}\equiv  2^{-1/4}\mu\ \exp\left(\frac 12 k_s\right)\,,
\eqlabel{defmul1}
\end{equation}
and $\hat{\cale}$ of \eqref{defhate}, where $\Lambda\equiv 2^{1/4} \exp\left(-\frac 12 k_s\right)$.
A powerful numerical check is that the  curves $\frac{\mu}{\Lambda}$---versus---$\hat\cale$ produced
for different values of $k_s$ must collapse in the overlapping range of $\mu/\Lambda$.
As fig.~\ref{vbcompare} demonstrates,
this is indeed what we find: 
the solid green curve is obtained with $k_s=\frac 14$ and the dashed magenta curve is obtained with
$k_s=1$ --- they differ at $\sim 10^{-5}$ fractional level or less. 
Yet a different computation scheme is to use the symmetry transformation \eqref{betatransform}
to keep $\mu=1$ fixed, and instead vary $K_0$, so that
\begin{equation}
\frac{\mu}{\Lambda}=2^{-1/4} \exp\left(\frac 12 K_0\right)\,.
\eqlabel{muk02}
\end{equation}
Of course, in this computational scheme we can not connect to the Klebanov-Strassler ($\mu=0$) solution,
but we can connect to, say, green curve at $K_0=\frac 14$ and $\mu=1$, \ie when 
\begin{equation}
\frac{\mu}{\Lambda}\bigg|_{\rm connection\ to\ green}=2^{-1/4} \exp\left(\frac 12\times \frac 14\right)= 0.952860(5)\,,
\eqlabel{muk03}
\end{equation}
which is represented by a vertical blue line. Such alternatively produced set of $\calv_B$ vacua
(dotted black curve in fig.~\ref{vbcompare}) must still agree with the green and magenta data sets.
Clearly, this is the case: we extend the data sets for the black curve for both $K_0< \frac 14$ and $K_0>\frac 14$ to show the overlaps with the green/magenta data sets.

\subsection{$\calv_A^s$ vacua}\label{calvas}

These vacua were originally constructed in \cite{Buchel:2011cc}.
Equations of motion describing $\calv_A^s$ vacua are collected in \eqref{a1}-\eqref{ac},
which  in the case of the unbroken chiral symmetry can be truncation
as
\begin{equation}
f_b(\rho)\equiv f_a(\rho)\equiv f_3(\rho)\,,\qquad K_3(\rho)=K_1(\rho)\equiv K(\rho)\,,\qquad
K_2(\rho)\equiv1 \,,
\eqlabel{chiraltrunc}
\end{equation}
leading  to nontrivial 5 second-order equations \eqref{a1}, \eqref{a2}, \eqref{a5}, \eqref{a6}, \eqref{a9} and a single
first order constraint \eqref{ac}. We have the UV, \ie $\rho\to 0_+$,  asymptotics \eqref{as1}-\eqref{as6},  and the IR, \ie $y\equiv \frac 1\rho\to 0_+$,  asymptotics \eqref{aredef},
\eqref{ircaseaa}.
There are 4 non-normalizable coefficients
\begin{equation}
K_0\,,\ \mu\,,\ P\,,\  g_s\,,
\eqlabel{nonvas}
\end{equation}
characterizing the cascading gauge theory \eqref{defuvcasea}, and $17-6=11$ normalizable
coefficients \eqref{uvvevs} and \eqref{irvevs} --- where the reduction in the number of the normalizable
coefficients reflects the constraints of the unbroken chiral symmetry,  see \eqref{uvirchiral}.

\begin{figure}[t]
\begin{center}
\psfrag{x}[cc][][1.0][0]{$\mu/\Lambda$}
\psfrag{e}[cc][][1.0][0]{$\hat\cale$}
\includegraphics[width=6in]{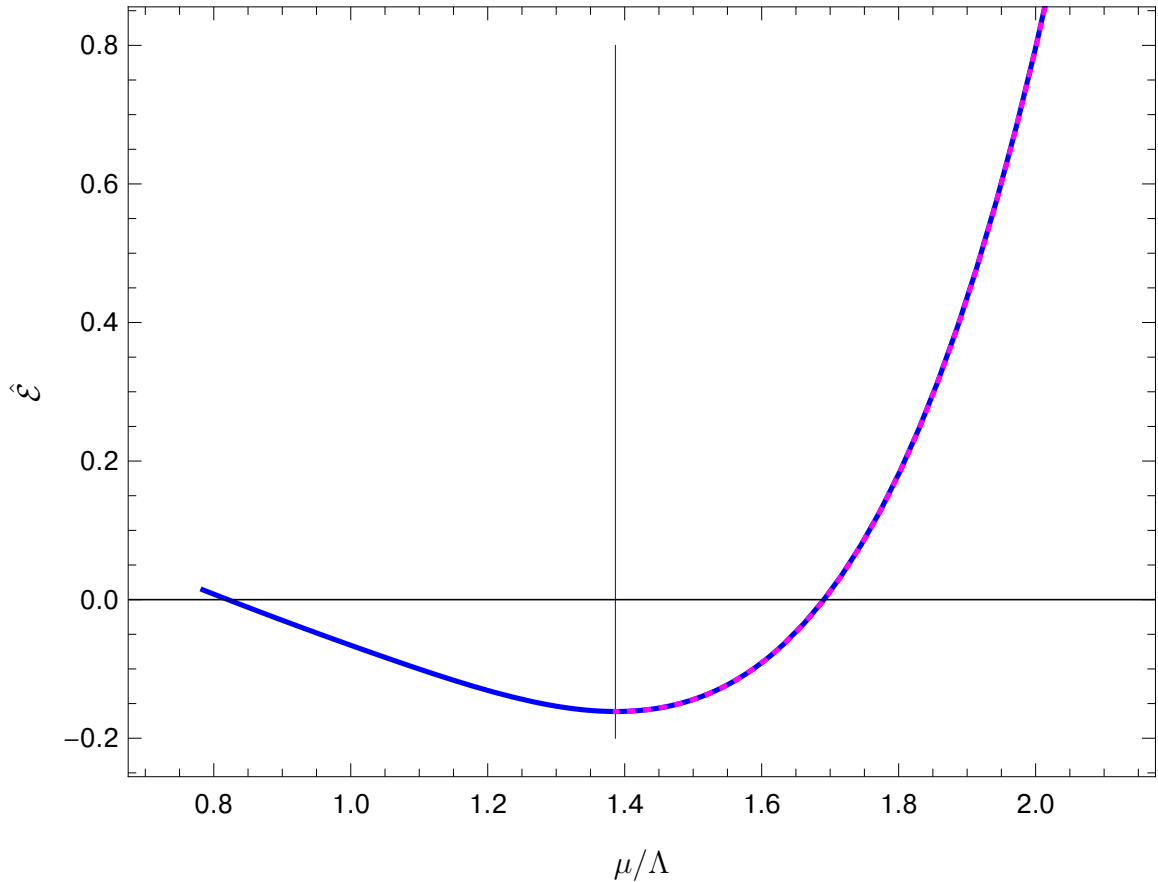}\ 
\end{center}
  \caption{Cross checks on numerics of $\calv_A^s$ vacua energy in the
  two different computational schemes: solid blue curve and the dotted magenta curve.
  The vertical blue line denotes the common reference point in the two computational
  schemes, see
  \eqref{connectas}.
} \label{vascompare}
\end{figure}

The numerical techniques for solving the required system of nonlinear ODEs were developed in
\cite{Aharony:2007vg}.  
We use the scaling symmetry \eqref{kssym1} to set
\begin{equation}
g_s=1 \,.
\eqlabel{setvas}
\end{equation}
We further use the symmetry transformation \eqref{betatransform}
to set\footnote{As before, this necessitates that  the numerical
results be represented as dimensionless quantities.}
\begin{equation}
\mu=1\,.
\eqlabel{muas}
\end{equation}
The final symmetry transformation \eqref{kssym2} is utilized to 
cross check the numerics:
\begin{itemize}
\item We can set $P=1$ and vary $K_0$, in which case
\begin{equation}
\frac{\mu}{\Lambda}\equiv 2^{-1/4} \exp\left(\frac 12 K_0\right)\qquad {\rm and}\qquad
 \Lambda=2^{1/4} \exp\left(-\frac 12 K_0\right)\,.
\eqlabel{mulambdavs}
\end{equation}
\item We can set $K_0=1$ and vary $P^2\equiv b$, in which case
\begin{equation}
\frac{\mu}{\Lambda}\equiv 2^{-1/4} b^{1/2}\exp\left(\frac {1}{2b}\right)\qquad {\rm and}\qquad
 \Lambda=2^{1/4} b^{-1/2}\exp\left(-\frac {1}{2b}\right)\,.
\eqlabel{mulambdavs2}
\end{equation}
\end{itemize}
Irrespectively which computational scheme is used, the produced curves
$\frac{\mu}{\Lambda}$---versus---$\hat\cale$ must collapse in the overlapping range of $\mu/\Lambda$.
As fig.~\ref{vascompare} demonstrates,
this is indeed what we find: 
the solid blue curve is obtained keeping fixed $P=1$,  and the dotted magenta curve is obtained
keeping fixed $K_0=1$. The two schemes connect at $K_0=1$ and $b=1$, \ie when
\begin{equation}
\frac{\mu}{\Lambda}\bigg|_{connection}=2^{-1/4}\exp\left(\frac 12\right)=1.38640(4) \,,
\eqlabel{connectas}
\end{equation}
which is represented by a vertical blue line\footnote{This is the smallest value of $\mu/\Lambda$
that can be reached in the computational scheme with $K_0=1$.}.

\subsection{$\calv_A^b$ vacua}\label{calvab}

Analyzing normal modes about $\va^s$ vacua it was established in \cite{Buchel:2011cc}
that they are perturbatively unstable
when $\mu< \mu_u$, see \eqref{defmuunst}. However, no new branch of vacua with spontaneously broken
symmetry, $\va^b$ in classification here, was identified. We use a different technique,
introduced in \cite{Buchel:2019pjb}, to independently reproduce the onset of the instability,
and construct $\va^b$ vacua.

\begin{figure}[t]
\begin{center}
\psfrag{x}[cc][][1.0][0]{$\mu/\Lambda$}
\psfrag{f}[bb][][1.0][0]{{${\color{blue} \frac{1}{\df_{3,0}}}\,,\, \frac{1}{\dk_{2,3,0}}\,,\,
{\color{orange} \frac{1}{\dk_{2,7,0}}}\qquad $}}
\psfrag{g}[tt][][1.0][0]{{${\color{blue} \frac{1}{\df_{0}^h}}\,,\, {\color{magenta}\frac{1}{\dk_{1,0}^h}}\,,\,
{ \frac{1}{\dk_{2,0}^h}}\qquad $}}
\includegraphics[width=3in]{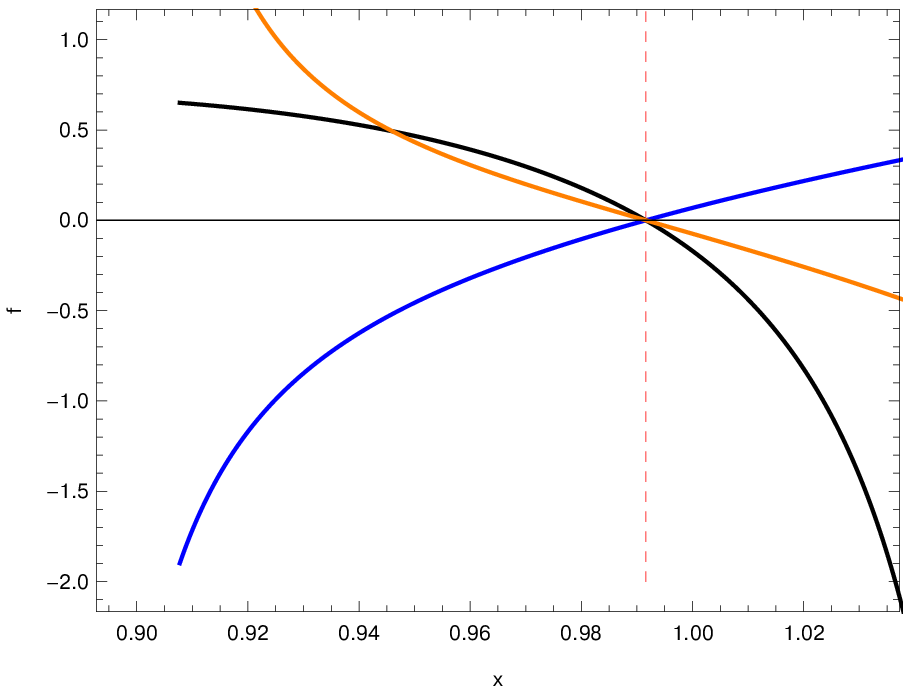}\
\includegraphics[width=3in]{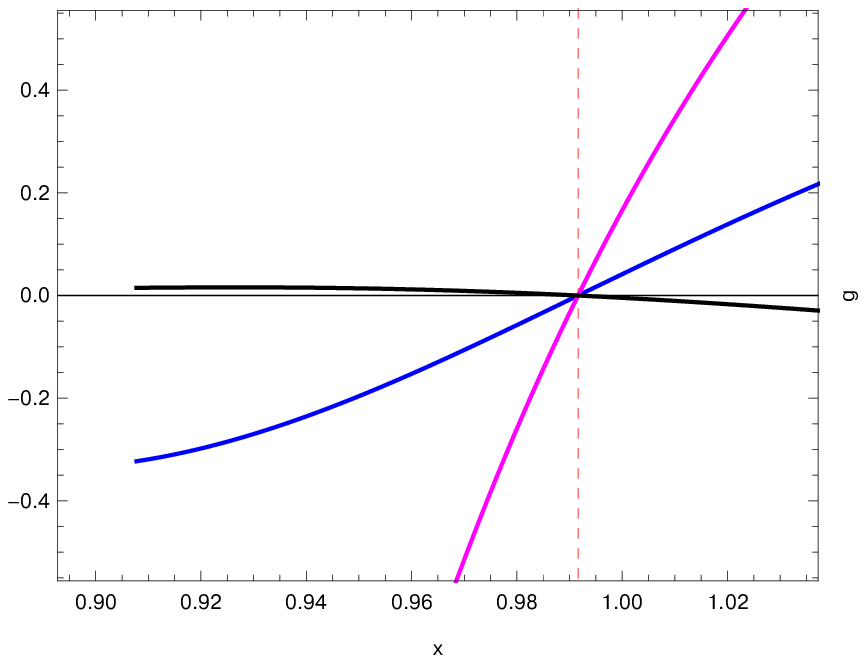}
\end{center}
  \caption{Parameters 
 $\{{\color{blue}\df_{3,0}}\,,\, \dk_{2,3,0}\,,\, {\color{orange}\dk_{2,7,0}}\,,\,
 {\color{blue}\df_{0}^h}\,,\, {\color{magenta} \dk_{1,0}^h}\,,\, {\dk_{2,0}^h}\}$ 
of the chiral symmetry breaking fluctuations over $\va^s$ vacua as a function of $\frac{\mu}{\Lambda}$,
evaluated at fixed explicit chiral symmetry breaking scale $m=1$, diverge
at $\mu=\mu_u$ \eqref{defmuunst}, indicated by a vertical red dashed line. $\mu_u$
identifies the bifurcation point of spontaneous symmetry broken $\va^b$ vacua off
chirally symmetric $\va^s$ vacua.
} \label{df30m1}
\end{figure}

Consider static,  chiral symmetry breaking fluctuations about $\va^s$ vacua:
\begin{equation}
\begin{split}
&f_a\equiv f_3+\df\,,\ f_b\equiv f_3-\df\,,\ K_1\equiv K+\dk_1\,,\ K_3\equiv K-\dk_1\,,\
K_2\equiv 1+\dk_2\,,
\end{split}
\eqlabel{deffluc1}
\end{equation}
with the remaining metric functions and the string coupling as in $\va^s$ vacua $\{f_1,f_c,h,g\}$.
It is straightforward to verify that truncation to $\{\df,\dk_{1}, \dk_2\}$
is consistent (at the linearized level).
Equations of motion for the fluctuations and their asymptotic expansions in the UV $(\rho\to 0)$ and the IR
($y=\frac 1\rho$) are collected in appendix \ref{flucas}. Once the non-normalizable coefficient 
(the explicit chiral symmetry breaking parameter, \ie the gaugino mass term) is fixed to 
$m=1$, the expansions are characterized by 6 UV/IR parameters 
\begin{equation}
\begin{split}
&{\rm UV:}\qquad \{\df_{3,0}\,,\, \dk_{2,3,0}\,,\, \dk_{2,7,0}\}\,;\\
&{\rm IR:}\qquad \{\df_{0}^h\,,\ \dk_{1,0}^h\,,\ \dk_{2,0}^h\}\,,
\end{split}
\eqlabel{uvirparslin}
\end{equation}
which is the correct number of parameters to find a unique solution of 3 second-order differential equations \eqref{vacasfl1}-\eqref{vacasfl3}
for $\{\df,\dk_{1},\dk_2\}$ on $\va^s$ background parameterized by $K_0$ (it is convenient to use
a computation scheme as in \eqref{mulambdavs}).

In fig.~\ref{df30m1} we assemble results for the fluctuation parameters
\eqref{uvirparslin} as $K_0$ label of $\va^s$ vacua is varied.
A signature of the spontaneous symmetry breaking is the divergence of all
the parameters, once the scale of the explicit chiral symmetry breaking,
the non-normalizable parameter $m$, is kept fixed. This occurs
at $\mu=\mu_u$ \eqref{defmuunst},
represented by vertical dashed red lines. The corresponding critical value of $K_0$ is
\begin{equation}
K_0^{critical}=0.329729(7)\,.
\eqlabel{k0crit}
\end{equation}

\begin{figure}[t]
\begin{center}
\psfrag{x}[cc][][1.0][0]{$\mu/\Lambda$}
\psfrag{a}[bb][][1.0][0]{${\dk_{2,3,0}/\df_{3,0}}$}
\psfrag{b}[tt][][1.0][0]{${\dk_{2,7,0}/\df_{3,0}}$}
\includegraphics[width=3in]{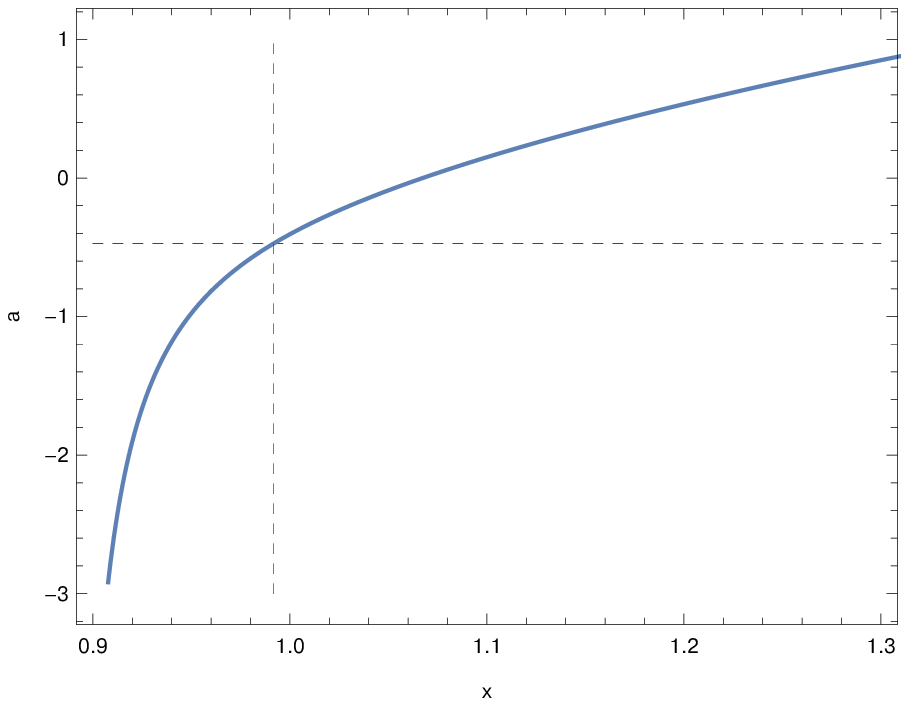}\ 
\includegraphics[width=3in]{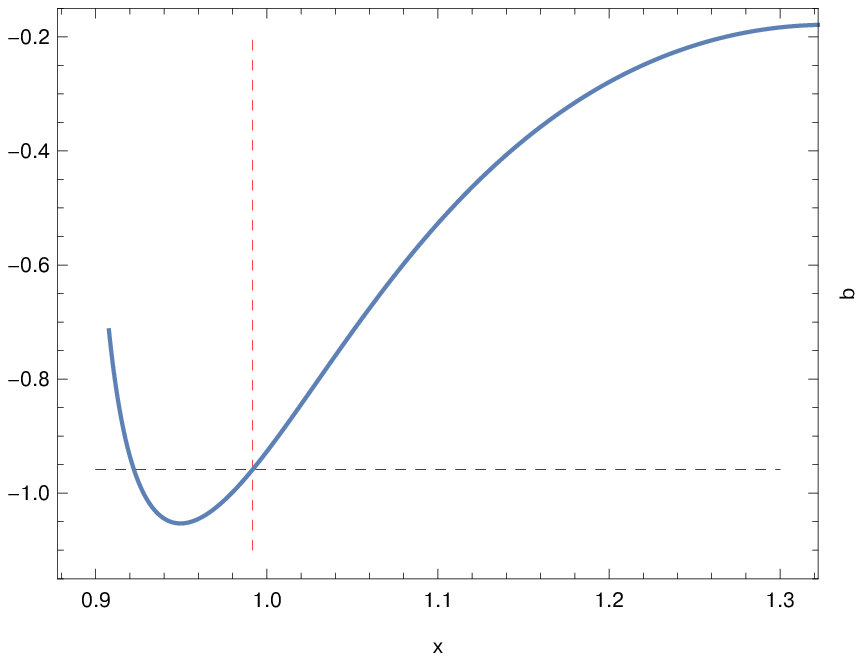}
\end{center}
  \caption{Susceptibilities of the UV parameters, see \eqref{defkappa}, of the linearized
  chiral symmetry breaking fluctuations. The red dashed vertical line
  denotes $\mu_u$, see \eqref{defmuunst}. The black horizontal dashed lines
  indicate the values of the susceptibilities at $\mu=\mu_{unstalble}$, see
  \eqref{numkappa}.
} \label{uvsuss}
\end{figure}

\begin{figure}[t]
\begin{center}
\psfrag{x}[tt][][1.0][0]{$\mu/\Lambda$}
\psfrag{c}[bb][][1.0][0]{${\df_0^h/\df_{3,0}}$}
\psfrag{d}[bb][][1.0][0]{${\dk_{1,0}^h/\df_{3,0}}$}
\psfrag{e}[tt][][1.0][0]{${\dk_{2,0}^h/\df_{3,0}}$}
\includegraphics[width=1.7in]{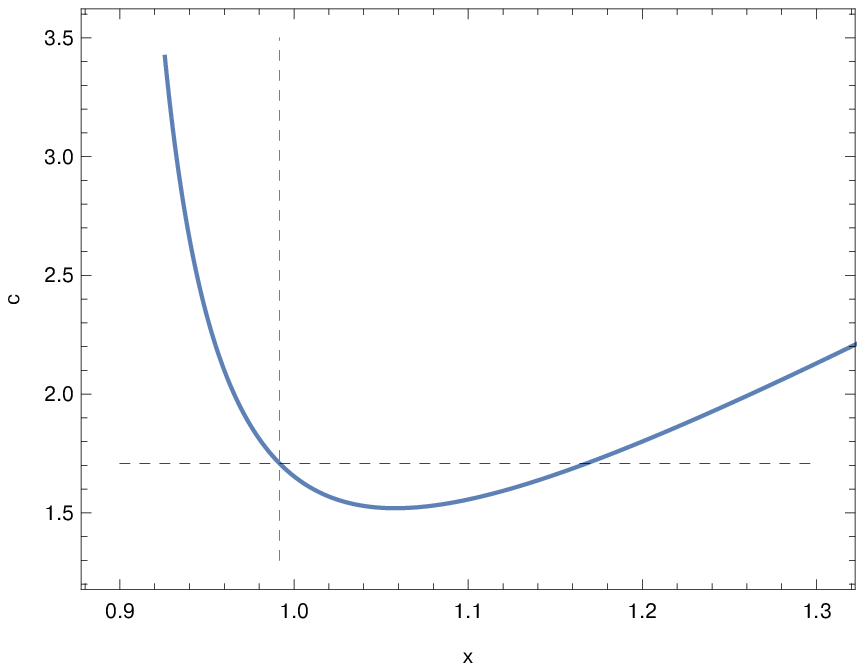}\qquad
\includegraphics[width=1.7in]{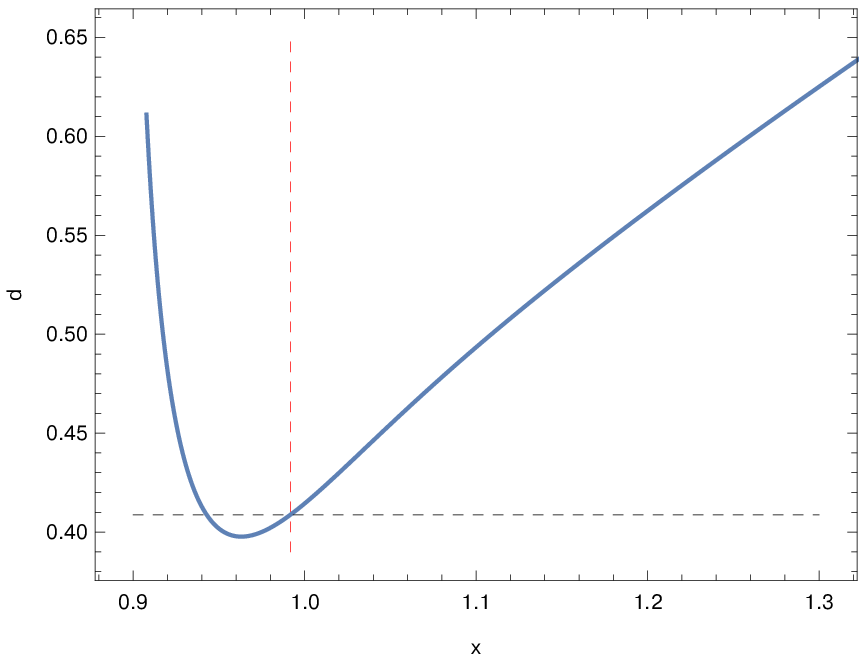}\qquad
\includegraphics[width=1.7in]{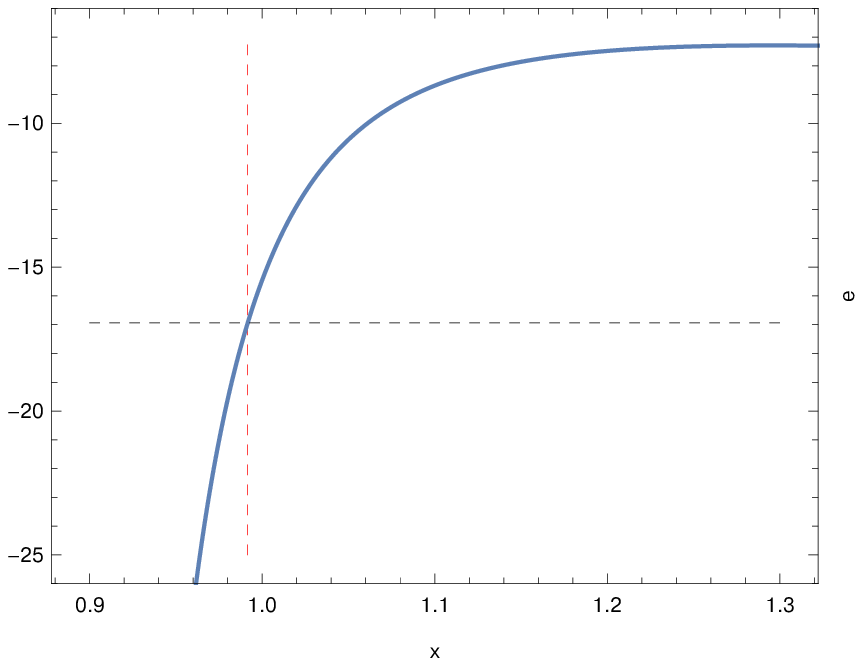}
\end{center}
  \caption{Susceptibilities of the IR parameters, see \eqref{defkappa}, of the linearized
  chiral symmetry breaking fluctuations. The red dashed vertical line
  denotes $\mu_u$, see \eqref{defmuunst}.  The black horizontal dashed lines
  indicate the values of the susceptibilities at $\mu=\mu_{unstalble}$, see
  \eqref{numkappa}.
} \label{irsuss}
\end{figure}

To use the critical fluctuations as a seed for $\va^b$ vacua, we need to know
the 'susceptibilities' 
\begin{equation}
\biggl\{\chi_{k_{2,3,0}}\,,\, \chi_{k_{2,7,0}}\,,\, \chi_{\hat{f}_0^h}\,,\, \chi_{k_{1,0}^h}\,,
\chi_{k_{2,0}^h} 
\biggr\} \equiv  \lim_{\mu\to \mu_u}\biggl\{
\frac{\dk_{2,3,0}}{\df_{3,0}}\,,\, \frac{\dk_{2,7,0}}{\df_{3,0}}\,,\,
\frac{\df_{0}^h}{\df_{3,0}}\,,\,
\frac{\dk_{1,0}^h}{\df_{3,0}}\,,\,\frac{\dk_{2,0}^h}{\df_{3,0}} 
\biggr\}\,.
\eqlabel{defkappa}
\end{equation}
In fig.~\ref{uvsuss} we present susceptibilities $\chi_{k_{2,3,0}}$ and  $\chi_{k_{2,7,0}}$ ---
notice that they are finite at $\mu_u$, represented by vertical dashed red lines.
The other susceptibilities (see fig.~\ref{irsuss}) are finite as well; we find:
\begin{equation}
\begin{split}
&\chi_{k_{2,3,0}}=-0.47398(9)\,,\qquad \chi_{k_{2,7,0}}=-0.95889(7)\,,\qquad \chi_{{f}_{0}^h}=1.7089(5)\,,\qquad\\
&\chi_{k_{1,0}^h}=0.40872(4)\,,\qquad \chi_{k_{2,0}^h}=-16.936(2)\,.
\end{split}
\eqlabel{numkappa}
\end{equation}

\begin{figure}[t]
\begin{center}
\psfrag{l}[cc][][1.0][0]{{$\lambda$}}
\psfrag{a}[cc][][1.0][0]{{$k_{2,3,0}$}}
\psfrag{b}[cc][][1.0][0]{{$k_{2,7,0}$}}
\includegraphics[width=2.5in]{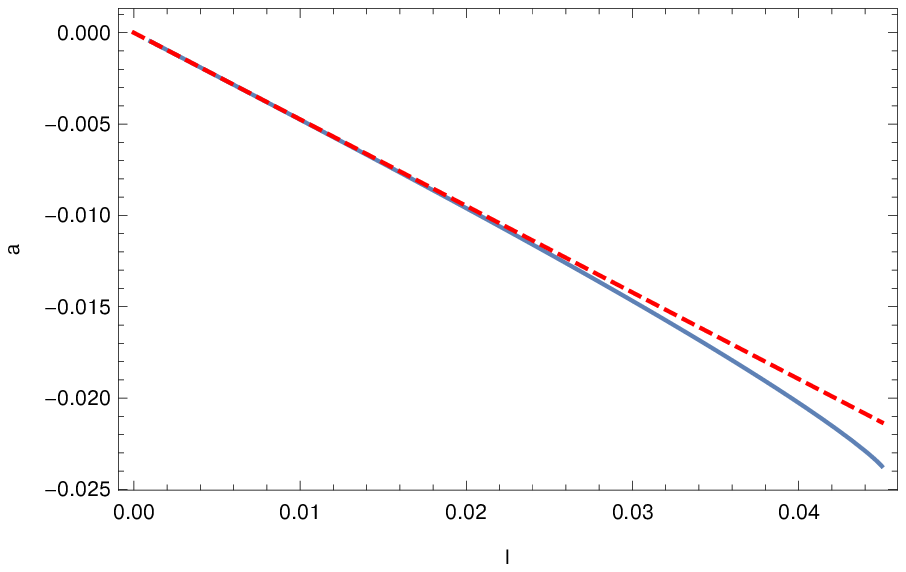}\qquad
\includegraphics[width=2.5in]{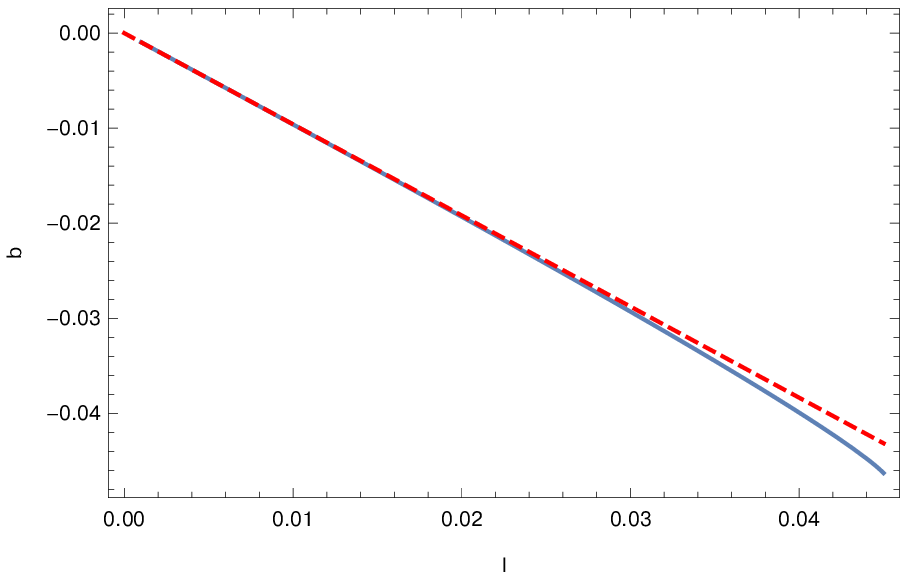}
\end{center}
  \caption{Sample of the UV parameters of $\va^b$ vacua constructed from
  the 'seed' \eqref{seedtypeab}. The linearized approximations in $\lambda$ are represented
  by dashed red lines. 
}\label{uvlambda}
\end{figure}

\begin{figure}[t]
\begin{center}
\psfrag{l}[cc][][0.7][0]{{$\lambda$}}
\psfrag{d}[bb][][0.7][0]{{$(K_{1,0}^h-K_{3,0}^h)/2$}}
\psfrag{c}[bb][][0.7][0]{{$(f_{a,0}^h-f_{b,0}^h)/2$}}
\psfrag{e}[bb][][0.7][0]{{$K_{2,0}^h-1$}}
\includegraphics[width=1.9in]{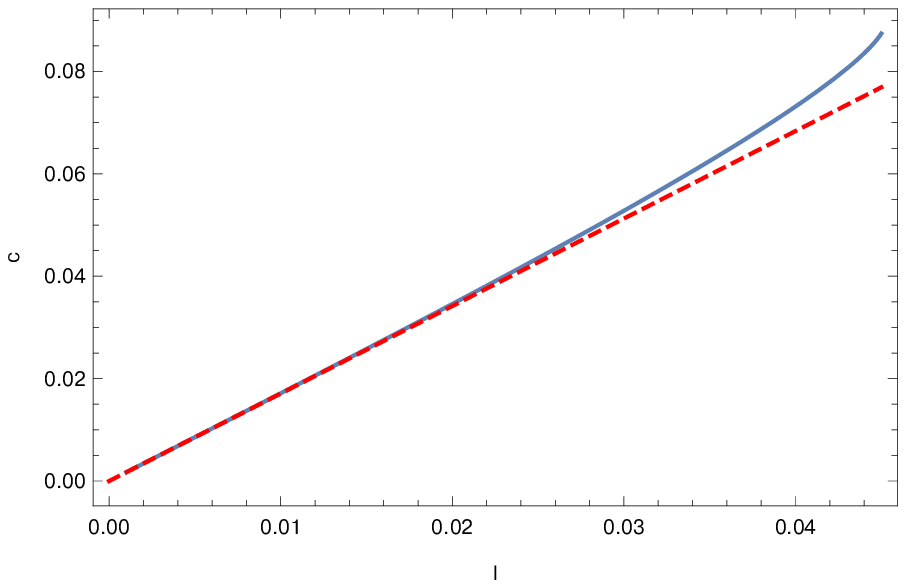}\ \ \ 
\includegraphics[width=1.9in]{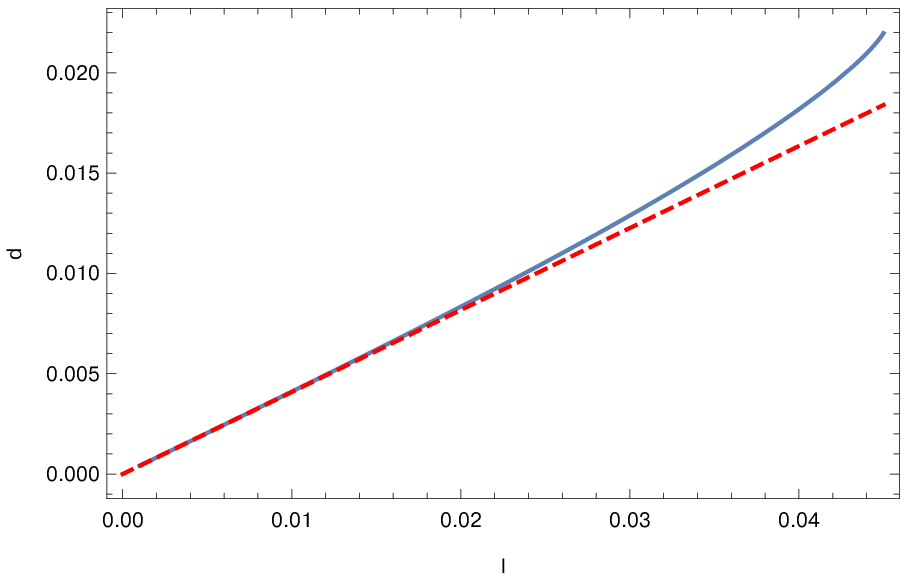}\ \ \
\includegraphics[width=1.9in]{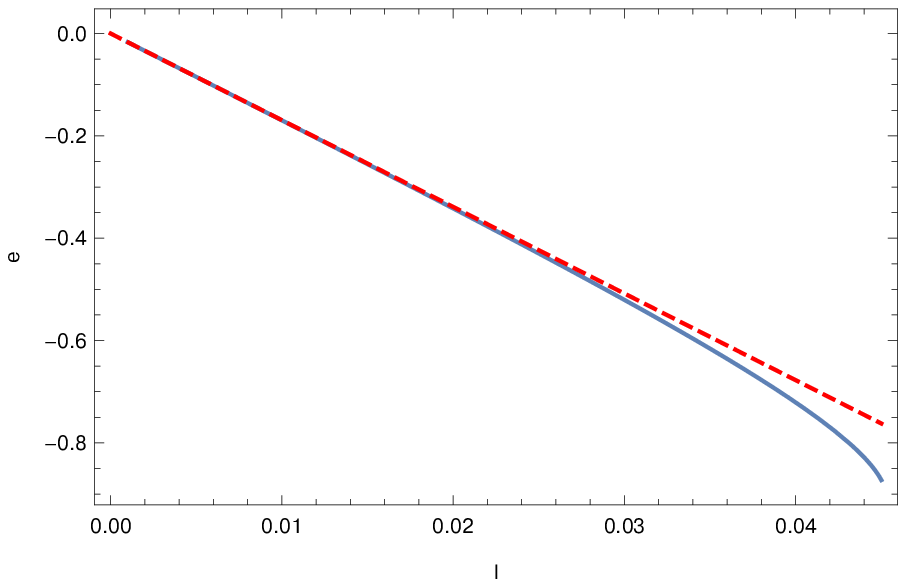}
\end{center}
  \caption{Sample of the IR parameters of $\va^b$ vacua constructed from
  the 'seed' \eqref{seedtypeab}. The linearized approximations in $\lambda$ are represented
  by dashed red lines. 
}\label{irlambda}
\end{figure}

Given \eqref{numkappa}, fully nonlinear $\va^b$ vacua, with $K_0$ close to $K_0^{crit}$,
can be constructed following the same procedure as the one employed in construction of
Klebanov-Strassler black brane in \cite{Buchel:2018bzp}. We highlight the main steps:
\begin{itemize}
\item We set $K_0=K_0^{crit}$ and compute the corresponding $\va^s$ vacuum.
This vacuum is characterized by (see \eqref{defuvcasea}, \eqref{uvvevs}, \eqref{irvevs}
and \eqref{uvirchiral})
\begin{equation}
\begin{split}
&{\rm UV}:\ \biggl\{K_0=K_0^{crit},\ \mu=1,\ P=1,\  g_s=1,\ f_{a,1,0}^{crit},\
f_{a,3,0}^{crit}=\frac 14 f_{a,1,0}^{crit},\ k_{2,3,0}^{crit}=0,\\
&\qquad\qquad f_{a,4,0}^{crit}\equiv f_{3,4,0}^{crit},\ f_{c,4,0}^{crit},\ g_{4,0}^{crit},\ f_{a,6,0}^{crit},\ k_{2,7,0}^{crit}=0,\ f_{a,8,0}^{crit}\biggr\}\,,
\\
&{\rm IR}:\ \biggr\{f_{a,0}^{h,crit}\equiv f_{3,0}^{h,crit},\ f_{b,0}^{h,crit}\equiv f_{3,0}^{h,crit},\
f_{c,0}^{h,crit},\ h_0^{h,crit},\ K_{1,0}^h\equiv K_{0}^{h,crit},\  K_{2,0}^h\equiv 1,\\
&\qquad\qquad K_{3,0}^h\equiv K_{0}^{h,crit},\
g_{0}^{h,crit}\biggl\}\,.
\end{split}
\eqlabel{crittypeas}
\end{equation}
\item Let's denote the amplitude of the symmetry breaking condensate (see \eqref{deffluc1})
\begin{equation}
\dd f_{3,0}\equiv \frac 12 \left(f_{a,3,0}-f_{b,3,0}\right)=\lambda\,.
\eqlabel{deflambdafluc}
\end{equation}
Then,
\begin{equation}
\left\{\dk_{2,3,0},\, \dk_{2,7,0},\,\df_{0}^h,\,\dd k_{1,0}^h,\,\dd k_{2,0}^h
\right\}=\lambda\ \{
\chi_{k_{2,3,0}},\, \chi_{k_{2,7,0}},\, \chi_{f_{0}^h},\, \chi_{k_{1,0}^h},\,
\chi_{k_{2,0}^h}\} +\calo(\lambda^2)\,.
\eqlabel{allotherveves}
\end{equation}
\item Using \eqref{deffluc1}, \eqref{crittypeas}-\eqref{allotherveves} we find to $\calo(\lambda^2)$:
\begin{equation}
\begin{split}
&K_0=K_0^{crit}+\calo(\lambda^2)\,,\qquad f_{a,1,0}=f_{a,1,0}^{crit}+\calo(\lambda^2)\,,\qquad f_{a,3,0}=f_{a,3,0}^{crit}
+\lambda+\calo(\lambda^2)\,,\\
&k_{2,3,0}=k_{2,3,0}^{crit}+\lambda\chi_{k_{2,3,0}}+\calo(\lambda^2)\,,\ \
f_{a,4,0}=f_{a,4,0}^{crit}+\calo(\lambda^2)\,,\ \ f_{c,4,0}=f_{c,4,0}^{crit}+\calo(\lambda^2)\,,\\
&g_{4,0}=g_{4,0}^{crit}+\calo(\lambda^2)\,,\ \ f_{a,6,0}=f_{a,6,0}^{crit}+\calo(\lambda^2)\,,\ \
k_{2,7,0}=k_{2,7,0}^{crit}+ \chi_{k_{2,7,0}}\lambda+\calo(\lambda^2)\,,\\
&f_{a,8,0}=f_{a,8,0}^{crit}+\calo(\lambda^2)\,,\ f_{a,0}^h=f_{a,0}^{h,crit}+\chi_{f_0^h}\ \lambda+\calo(\lambda^2)\,,\
\\&f_{b,0}^h=f_{b,0}^{h,crit}-\chi_{f_0^h} \lambda+\calo(\lambda^2)\,,\ \
f_{c,0}^h=f_{c,0}^{h,crit}+\calo(\lambda^2)\,,\ \ h_{0}^h=h_{0}^{h,crit}+\calo(\lambda^2)\,,\\
&K_{1,0}^h=K_{1,0}^{h,crit}+\chi_{k_{1,0}^h} \lambda+\calo(\lambda^2)\,,\qquad
K_{2,0}^h=K_{2,0}^{h,crit}+\chi_{k_{2,0}^h} \lambda+\calo(\lambda^2)\,,\\
&K_{3,0}^h=K_{3,0}^{h,crit}-\chi_{k_{1,0}^h} \lambda+\calo(\lambda^2)\,,\qquad
g_{0}^h=g_{0}^{h,crit}+\calo(\lambda^2)\,.
\end{split}
\eqlabel{seedtypeab}
\end{equation}
\item We construct fully nonlinear in $\lambda$ $\va^b$ vacua using
the linearized approximation \eqref{seedtypeab} as a seed. Select UV/IR parameters,
along with the corresponding linearized approximations (dashed red lines)
are shown in figs.~\ref{uvlambda}-\ref{irlambda}.
\end{itemize}

\section{Black holes on the warped deformed conifold with fluxes}\label{bhsec}

The physics of black holes on the warped deformed conifold with fluxes, correspondingly the dual deconfined
thermal states of the cascading gauge theory plasma on $S^3$ with a radius $\frac 1\mu$ is very rich.
There are interesting phase transitions
between them, as well as phase transitions towards confined thermal states. We set the stage by reviewing the
results in the decompactification limit $\frac{\mu}{\Lambda}\to 0$, followed by the general presentation
for $\frac{\mu}{\Lambda}\ne 0$. We highlight details in the following subsections, and delegate
the  numerical checks to the appendix \ref{numtests}. The $\mu=0$ results were obtained
over the years in \cite{Buchel:2000ch,Buchel:2001gw,Gubser:2001ri,Aharony:2005zr,Aharony:2007vg
,Buchel:2009bh,Buchel:2010wp,Buchel:2018bzp,Bena:2019sxm,Buchel:2020nqs}.
All the $\mu\ne 0$ results are new.  

\subsection{Conifold black branes and phase transitions at $\frac{\mu}{\Lambda}=0$}\label{mu0sec}

At $\mu=0$ we have (see fig.~\ref{fig3}):
\begin{itemize}
\item $\calt_{decon}^s:\qquad$ Klebanov-Tseytlin black branes, representing thermal deconfined states
of the cascading
gauge theory plasma with the unbroken chiral symmetry (blue / brown curves in the figures below);
\item $\calt_{decon}^b:\qquad$ Klebanov-Strassler black branes, representing thermal deconfined states of the cascading
gauge theory plasma with spontaneously broken  chiral symmetry (magenta curves in the figures below);
\item $\calt_{con,B}:\qquad$ confined states with spontaneously broken chiral symmetry (green curves
in the figures below).
\end{itemize}

Similar to \eqref{defhate}, it is convenient to introduce the reduced free energy density $\hat\calf$,
and the reduced entropy density $\hat s$:
\begin{equation}
\hat\calf\equiv \frac{8\pi G_5}{P^4 g_s^2} \frac{\cale}{\Lambda^4}
=\frac{2^6\pi^4}{3^5M^4} \frac{\cale}{\Lambda^4}\,,\qquad
\hat s\equiv \frac{8\pi G_5}{P^4 g_s^2} \frac{s}{\Lambda^3}
=\frac{2^6\pi^4}{3^5M^4} \frac{s}{\Lambda^3}\,.
\eqlabel{defhatf}
\end{equation}

\begin{figure}[ht]
\begin{center}
\psfrag{t}[cc][][1.5][0]{{$T/\Lambda$}}
\psfrag{f}[bb][][1.5][0]{{$\hat{\calf}$}}
\psfrag{a}[cc][][1.5][0]{{$\calt_{decon}^s$}}
\psfrag{b}[cc][][1.5][0]{{$\calt_{decon}^b$}}
\psfrag{g}[cc][][1.5][0]{{$\calt_{con,B}$}}
\psfrag{d}[cc][][1][0]{{$T_c$}}
\psfrag{c}[cc][][1][0]{{$T_\csb$}}
\psfrag{e}[cc][][1][0]{{$T_u$}}
\includegraphics[width=5.5in]{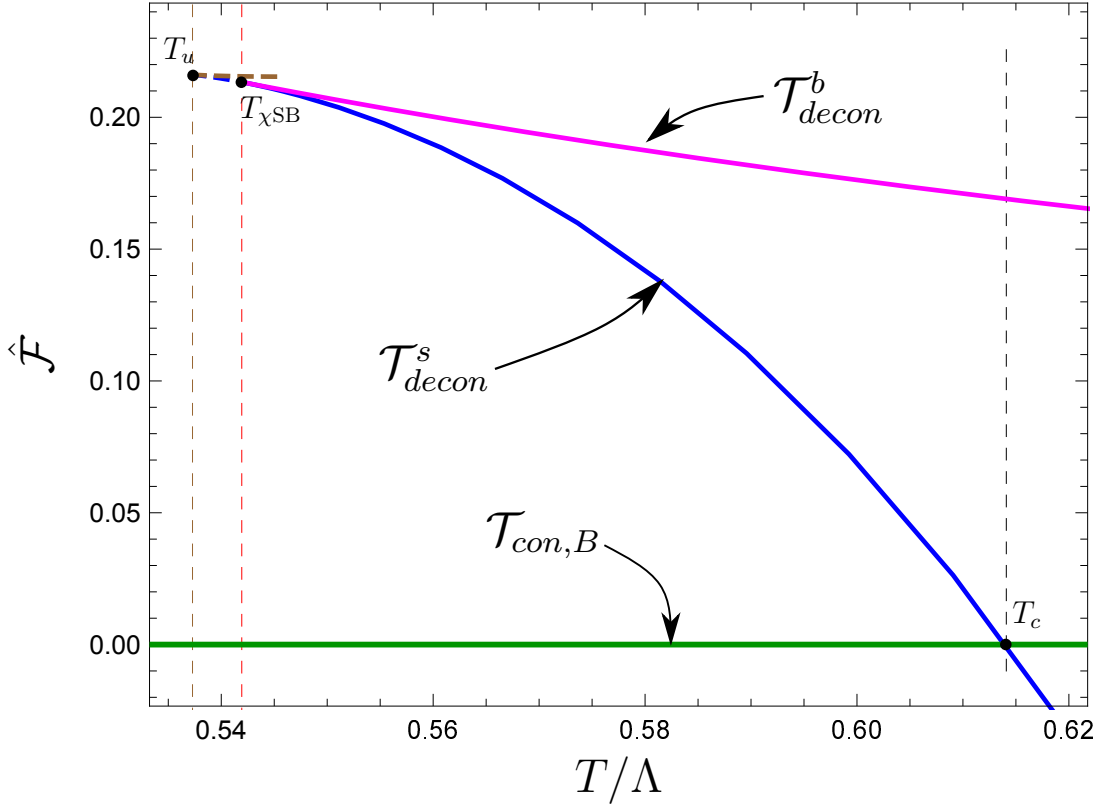}
\end{center}
  \caption{Phase diagram in the canonical ensemble at $\mu/\Lambda=0$: the reduced free energy density
$\hat\calf$, see \eqref{defhatf}, versus the reduced temperature $T/\Lambda$ for
different states in the theory.
Vertical dashed lines indicate critical temperatures $T_c$ (black) for the confinement-deconfinement phase transition,
$T_{\csb}$ (red) for the onset of the spontaneous chiral symmetry breaking, and $T_u$ (brown) for the
bifurcation point of the $\calt_{decon}^s$ states with positive/negative specific heat. 
}\label{canonical0}
\end{figure}

\begin{figure}[t]
\begin{center}
\psfrag{t}[cc][][0.7][0]{{$T/\Lambda$}}
\psfrag{f}[bb][][0.7][0]{{$\hat{\calf}$}}
\psfrag{a}[cc][][0.7][0]{{$\calt_{decon}^{s,+}$}}
\psfrag{e}[cc][][0.7][0]{{$\calt_{decon}^{s,-}$}}
\psfrag{b}[cc][][0.7][0]{{$\calt_{decon}^{b,-}$}}
\psfrag{d}[cc][][0.5][0]{{$T_\csb$}}
\psfrag{c}[cc][][0.7][0]{{$ _\csb$}}
\psfrag{g}[cc][][0.5][0]{{$T_u$}}
\psfrag{p}[cc][][0.7][0]{{$\calt_{con,B}$}}
\psfrag{v}[cc][][0.7][0]{{$\hat{c}_V$}}
\includegraphics[width=2.6in]{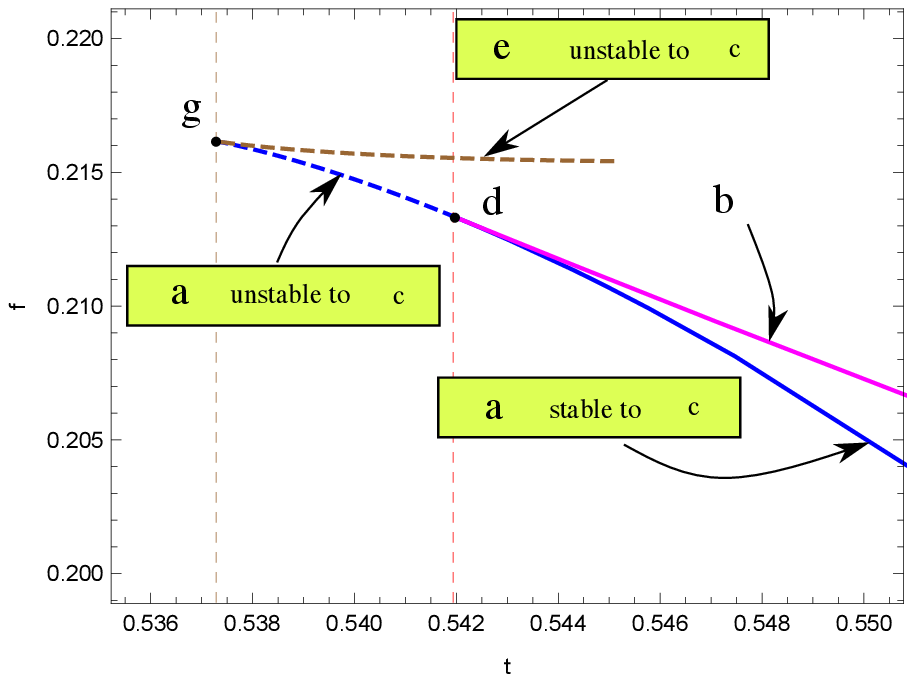}\ \ \
\includegraphics[width=2.6in]{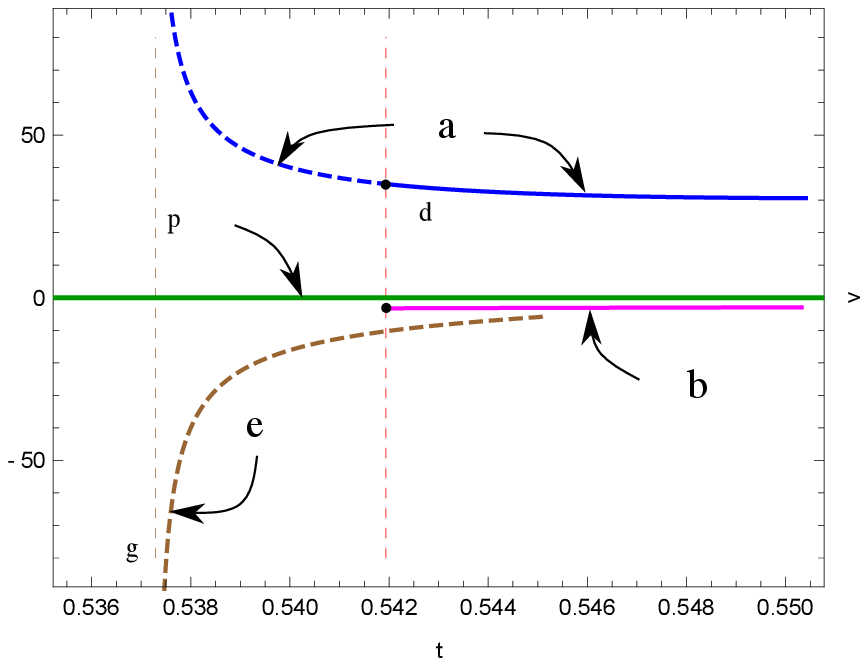}
\end{center}
  \caption{Left panel: detailed phase diagram in the canonical ensemble at $\mu/\Lambda=0$ for
  $T\gtrsim T_u$. $\calt_{decon}^s$ states unstable to spontaneous chiral symmetry breaking
  are denoted with dashed curves.
Right panel: specific heat $\hat{c}_V$ \eqref{defcvmu0} of different states.
 Brown curves denote $\calt_{decon}^s$ states
  with $\hat c_V<0$, and blue curves denote $\calt_{decon}^s$ states
  with $\hat c_V>0$. Note that the specific heat of $\calt_{decon}^b$ states (magenta curves),
  \ie the Klebanov-Strassler black branes, is negative. 
}\label{canonical0s}
\end{figure}

In fig.~\ref{canonical0} we recall the phase diagram in the canonical ensemble.
There are three critical temperatures
\begin{equation}
T_u\ < T_{\csb}\ < T_c\,.
\eqlabel{3t}
\end{equation}
\nxt $T_c$ is the temperature of the confinement-deconfinement phase transition
\cite{Aharony:2007vg},
\ie the transition $\calt_{decon}^s\leftrightarrow \calt_{con,B}$,
\begin{equation}
T_c=0.614(1)\Lambda\,.
\eqlabel{tcmu0}
\end{equation}
The confined phase with spontaneously broken chiral symmetry $\calt_{con,B}$ has a lower free
energy density for $T<T_c$, while the chirally symmetric deconfined phase $\calt_{decon}^s$ dominates when
$T>T_c$. Since the entropy density of $\calt_{decon}^s$ states  $\hat s\ne 0$,
while the reduced entropy density of $\calt_{con}^b$ states identically
vanishes\footnote{This is true in the supergravity, \ie the large-$M$ approximation.},
the phase transition at $T=T_c$ is of the first-order. 
\nxt $\calt_{decon}^s$ phase is stable with respect to linearized (perturbative)
chiral symmetry breaking fluctuations when $T>T_\csb$ \cite{Buchel:2010wp}
\begin{equation}
T_\csb=0.541(9)\Lambda\,,
\eqlabel{tcsbmu0}
\end{equation}
and is unstable (denoted by the dashed curves) when $T<T_\csb$.
$T=T_\csb$ is a bifurcation point of the phase diagram, where deconfined phase with spontaneously broken
chiral symmetry $\calt_{decon}^b$ joins the symmetric $\calt_{decon}^s$ phase \cite{Buchel:2018bzp}.
This phase, represented by the Klebanov-Strassler black branes,  exists for $T>T_{\csb}$, but
never dominates in the canonical ensemble as it has a higher free energy density than that of the chirally
symmetric phase $\calt_{decon}^s$ at the same temperature. 
\nxt $T_u$ \cite{Buchel:2009bh},
\begin{equation}
T_u=0.537(3) \Lambda\,,
\eqlabel{tumu0}
\end{equation} 
is the terminal temperature of the $\calt_{decon}^s$ states. It is the second bifurcation
point on the phase diagram, which separates deconfined chirally symmetric states
with positive specific heat $\hat c_V$,
\begin{equation}
\hat{c}_V\equiv \Lambda\ \frac{d\hat{\cale}}{d T}\,,
\eqlabel{defcvmu0}
\end{equation}
denoted as $\calt_{decon}^{s,+}$ (blue curves) in fig.~\ref{canonical0s}, from deconfined
chirally symmetric states with a negative specific heat $\calt_{decon}^{s,-}$
(brown curves). At a temperature where both $\calt_{decon}^{s,+}$ and  $\calt_{decon}^{s,-}$
states exist, the former ones always have a lower free energy density.
Note that the specific heat of the deconfined states with the spontaneously broken
chiral symmetry is also negative, thus these states are denoted as $\calt_{decon}^{b,-}$
(magenta curves) in fig.~\ref{canonical0s}.

Deconfined states of the cascading gauge theory plasma at $\mu/\Lambda=0$ are dual to  
black branes: the holographic geometries with a regular Schwarzschild horizon
with translational invariance. It was pointed out in \cite{Buchel:2005nt} that a
negative specific heat of the black branes immediately implies that the extended horizon is
perturbatively unstable to metric fluctuations, breaking the translational invariance.
The latter is a holographic dual to the fact that in an infinitely spatially extended media
with a negative specific heat the speed of the sound waves is purely imaginary:
\begin{equation}
c_s^2=\frac{s}{c_V}\ <0 \qquad {\rm if}\qquad c_V<0 \,.
\eqlabel{cs2}
\end{equation}
Thus, the states $\calt_{decon}^{s,-}$ and all the $\calt_{decon}^b$ states are perturbatively unstable
to inhomogeneous metric fluctuations.

\begin{figure}[t]
\begin{center}
\psfrag{a}[cc][][0.7][0]{{$(T-T_u)/\Lambda$}}
\psfrag{f}[tt][][0.7][0]{{$\sqrt{\Delta \hat\calf}$}}
\psfrag{k}[tt][][0.7][0]{{$\Delta \hat\calf\equiv\hat\calf[{\color{magenta}{\calt_{decon}^b}}]
-\hat\calf[{\color{blue}\calt_{decon}^s}]$}}
\psfrag{e}[bb][][0.7][0]{{$(\hat{\cale}-\hat\cale_u)^2$}}
\psfrag{c}[cc][][0.7][0]{{$\calt_{decon}^{s,+}$}}
\psfrag{d}[cc][][0.7][0]{{$\calt_{decon}^{s,-}$}}
\psfrag{b}[cc][][0.7][0]{{$(T-T_\csb)/\Lambda$}}
\includegraphics[width=2.6in]{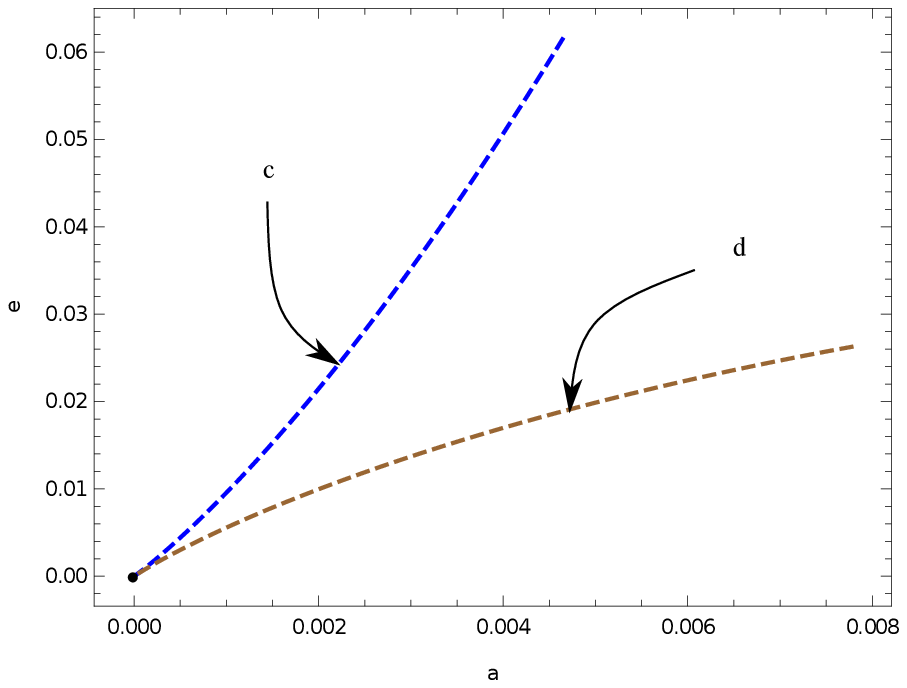}\ \ \
\includegraphics[width=2.6in]{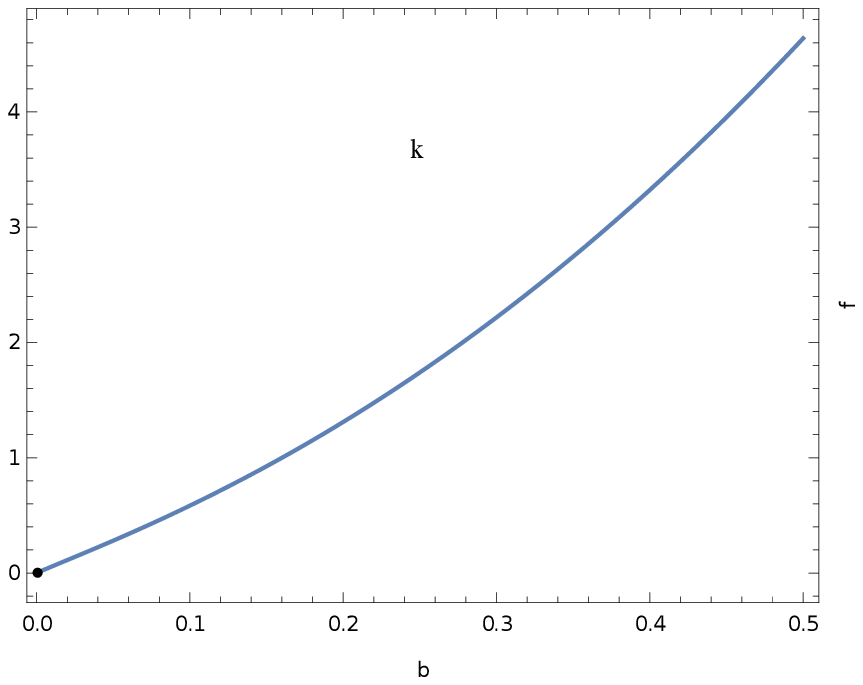}
\end{center}
  \caption{Critical behavior at $T=T_u$ (the left panel) and $T=T_\csb$ (the right panel).
   In the former case, the specific heat of $\calt_{decon}^{s,\pm}$ states diverge as $T\to T_{u+}$
  with the critical exponent $\alpha=\frac 12$. The scaling of the free energy density difference
  between the
  two phase as $T\to T_{\csb+}$ implies that the phase transition at $T=T_\csb$ is of the second order. 
}\label{criticalpic}
\end{figure}

In fig.~\ref{criticalpic} we discuss the criticality of $\mu/\Lambda=0$ states
at $T=T_u$ (the left panel) and $T=T_\csb$ (the right panel).
In the former case, both for  $\calt_{decon}^{s,-}$ and  $\calt_{decon}^{s,+}$ states
\begin{equation}
\hat\cale^{\pm}-\hat\cale_u\ \propto \sqrt{T-T_u}\qquad \Longrightarrow\qquad c_V^\pm\ \propto (T-T_u)^{-1/2} \,,
\eqlabel{cvdiv}
\end{equation}
implying that the specific heat diverges as $T\to T_u$ with the critical
exponent\footnote{In holographic models such critical behavior was identified first in
\cite{Buchel:2008uu}.}  
$\alpha=1/2$. Of course, this is consistent with the results
of the right panel in fig.~\ref{canonical0s}. In the vicinity of
$T=T_\csb$, the free energy density difference $\Delta\hat\calf$, between the
deconfined states with the spontaneously broken ($\calt_{decon}^b$) and the  unbroken
 ($\calt_{decon}^s$) chiral symmetry, scales as
 \begin{equation}
 \Delta\hat\calf\ \propto (T-T_\csb)^2\,,
 \eqlabel{delhatf}
 \end{equation}
as indicative of the second-order phase transition.

\begin{figure}[htb]
\begin{center}
\psfrag{e}[cc][][1.5][0]{{$\hat{\cale}$}}
\psfrag{s}[bb][][1.5][0]{{$\hat{s}$}}
\psfrag{a}[cc][][1.5][0]{{$\calt_{decon}^s$}}
\psfrag{b}[cc][][1.5][0]{{$\calt_{decon}^b$}}
\psfrag{c}[cc][][1][0]{{$c_s^2<0$}}
\psfrag{d}[cc][][1][0]{{$\ _\csb$}}
\psfrag{h}[cc][][1][0]{{$\hat{\cale}_\csb$}}
\psfrag{g}[cc][][1][0]{{$\hat{\cale}_u$}}
\includegraphics[width=5in]{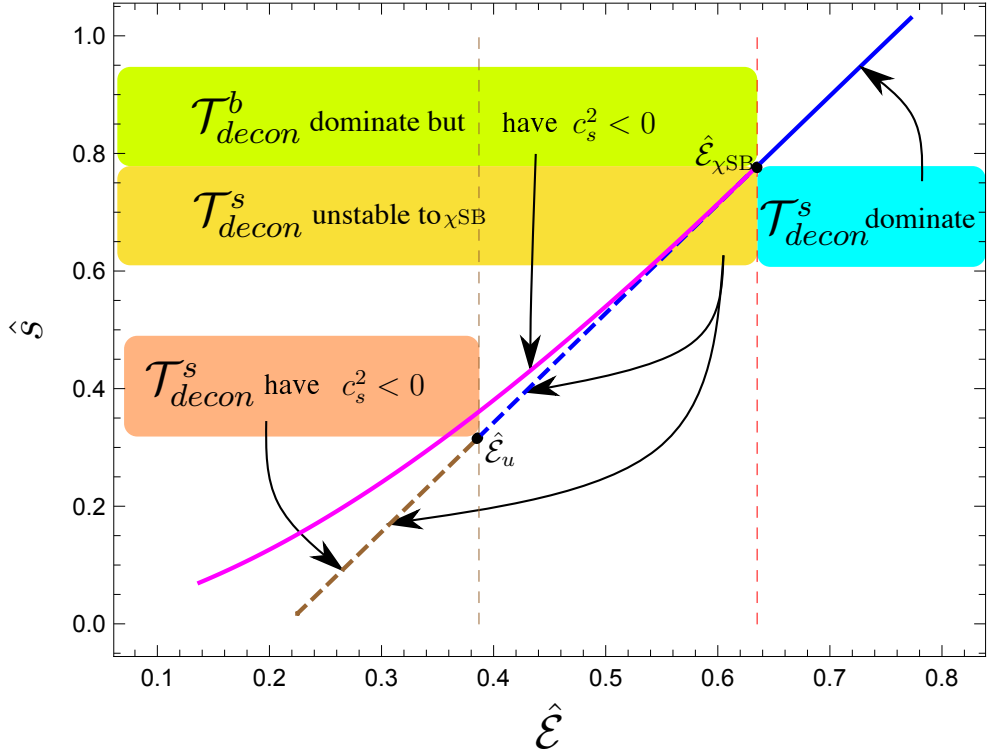}
\end{center}
  \caption{Phase diagram in the microcanonical ensemble at $\mu/\Lambda=0$: the reduced entropy
  density $\hat s$, see \eqref{defhatf}, versus the reduced energy density $\hat \cale$,
  see \eqref{defhate}, for different deconfined states in the theory. Vertical dashed
  lines indicate critical reduced energy densities $\hat\cale_{\csb}$ (red) for the
  of the spontaneous chiral symmetry breaking, and $\hat\cale_u$ (brown)
  for the onset of the spontaneous breaking of the spatial translational invariance in
  $\calt_{decon}^s$ states. 
}\label{microcanonical0}
\end{figure}

Fig.~\ref{microcanonical0} represents the phase diagram of the system at $\mu/\Lambda=0$
in the microcanonical ensemble.
Microcanonical ensemble is crucial in understanding the dynamics and the  equilibration of the
generic states in the theory. We will omit discussion of the confined states as the latter
have vanishing reduced entropy density $\hat s$ \eqref{defhatf} and thus never dominate
over the deconfined states in the microcanonical ensemble. There are two critical energy densities
(corresponding to critical temperatures $T_\csb$ and $T_u$ in the canonical ensemble)
\begin{equation}
\hat\cale_u\ <\ \hat\cale_\csb\,.
\eqlabel{calehier}
\end{equation}
\nxt $\hat\cale_\csb$, 
\begin{equation}
\hat\cale_\csb=0.635(1)\,,
\eqlabel{ecsbmu0}
\end{equation}
is the critical energy density below which $\calt_{decon}^s$  (represented by the dashed curves)
become unstable to perturbative chiral symmetry breaking instability. $\calt_{decon}^b$ states,
represented by the Klebanov-Strassler black branes \cite{Buchel:2018bzp} --- the magenta curve in
fig.~\ref{microcanonical0}, exist for $\hat\cale<\hat\cale_\csb$ and
are more entropic (and thus are the dominant ones) compare to  $\calt_{decon}^s$ states,
represented by the Klebanov-Tseytlin black branes
\cite{Buchel:2000ch,Buchel:2001gw,Gubser:2001ri,Aharony:2007vg}. While thermodynamic
instability, \ie the negative specific heat, in general does not affect the phase diagram in the
microcanonical ensemble, it does so when $\mu/\Lambda=0$ as in this case it is related to the
perturbative instability breaking the translational invariance of the thermal states ---
the purely imaginary speed of the sound waves in the cascading gauge theory interpretation.
Since $c_s^2<0$ for  $\calt_{decon}^b$ states (at all energy densities), they can not be the end point
of the evolution of the chiral symmetry breaking instability in  $\calt_{decon}^s$ states.
Neither can $\calt_{con,B}$ states be the end point of this instability, as these are the confined
states. Identification of the end point of the chiral symmetry breaking instability in
 $\calt_{decon}^s$ states remains an open problem. 
\nxt $\hat\cale_u$,
\begin{equation}
\hat\cale_u=0.386(9)\,,
\eqlabel{eumu0}
\end{equation}
is the critical energy density that separates  $\calt_{decon}^s$ states stable to spontaneous
breaking of the spatial translational invariance (blue curves with $\hat\cale>\hat\cale_u$),
from the states where it is
broken (brown curve).

\begin{figure}[t]
\begin{center}
\psfrag{k}[cc][][0.7][0]{{$\calk/\calk_\csb$}}
\psfrag{d}[cc][][0.7][0]{{$\calt_{decon}^b$}}
\psfrag{c}[cc][][0.7][0]{{$\calt_{decon}^s$}}
\psfrag{e}[tt][][0.7][0]{{$\hat\cale$}}
\psfrag{s}[bb][][0.7][0]{{$\hat{R}^2_{S^2}$}}
\psfrag{t}[tt][][0.7][0]{{$\hat{R}^2_{T^{1,1}}$}}
\psfrag{a}[cc][][0.5][0]{{$\hat\cale_\csb$}}
\psfrag{b}[cc][][0.5][0]{{$\hat\cale_u$}}
\includegraphics[width=1.9in]{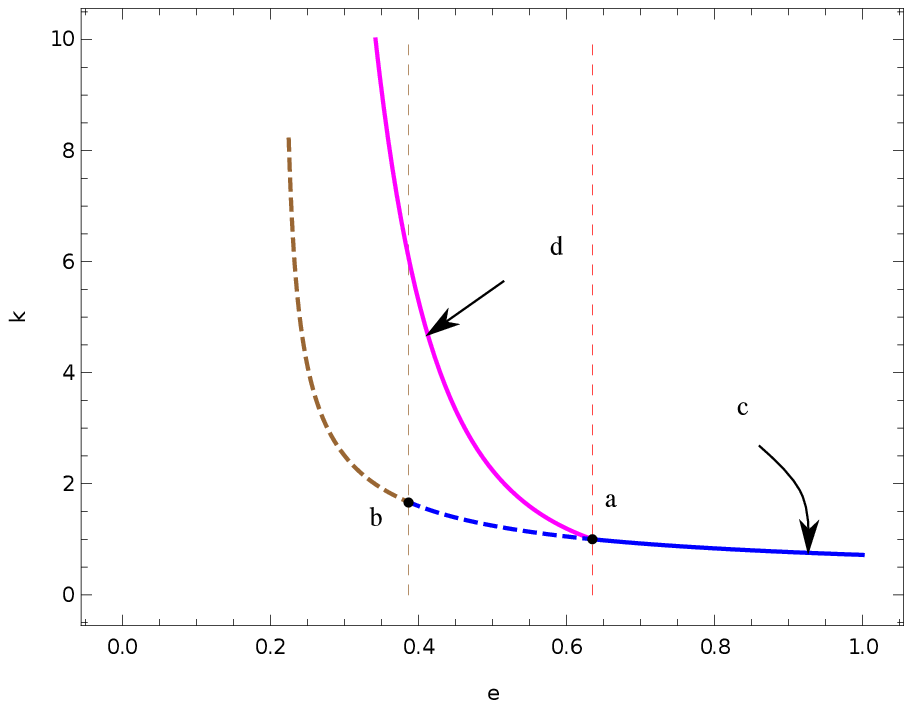}\ \ \ 
\includegraphics[width=1.9in]{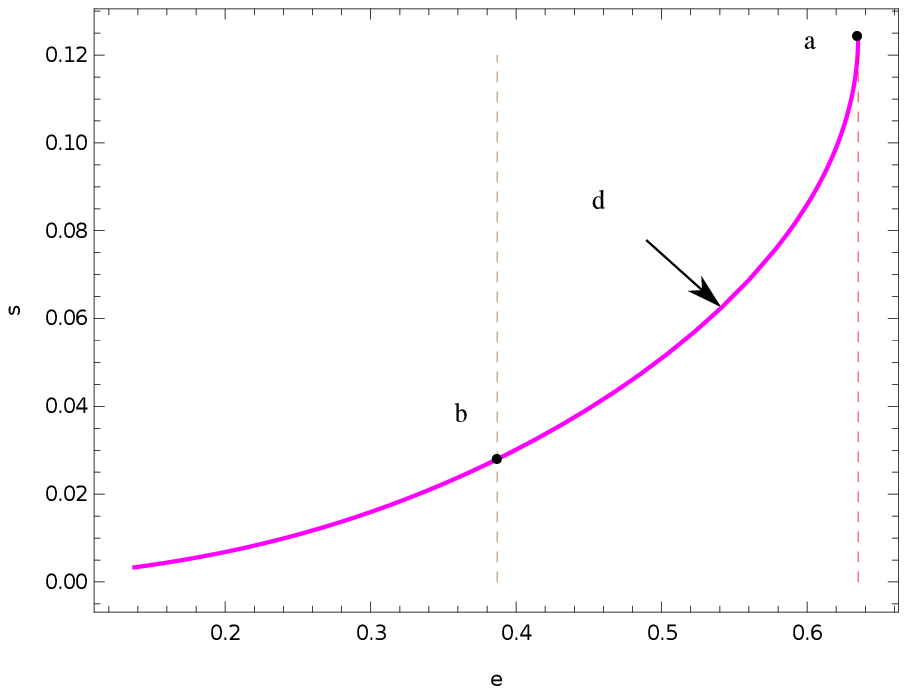}\ \ \
\includegraphics[width=1.9in]{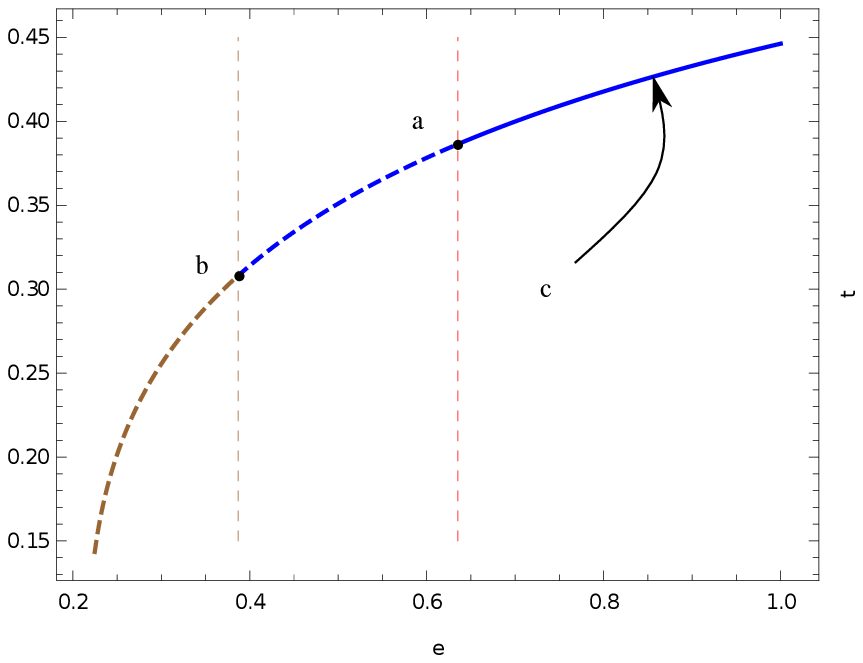}
\end{center}
  \caption{Left panel: the reduced Kretschmann scalar $\hat\calk$ \eqref{defhatk}
  of the holographic geometries
  dual to $\mu/\Lambda=0$ deconfined stated of the theory normalized to  $\hat\calk_\csb$.
  The small $\hat\cale$ growth of $\hat\calk$ of $\calt_{decon}^b$ states is associated
  with collapsing of the conifold 2-cycle (the central panel),
  and of $\calt_{decon}^s$ states is associated
  with collapsing of the deformed $T^{1,1}$ (the right panel).
}\label{mu0ks2t11}
\end{figure}

From fig.~\ref{microcanonical0}, $\calt_{decon}^b$ states exist only for $\hat\cale<\hat\cale_\csb$,
while  $\calt_{decon}^s$ extend to arbitrary high energy densities\footnote{We explicitly
construct these states for $\hat\cale\to \infty$
in section \ref{perkt}.}.  Neither $\calt_{decon}^s$ nor the $\calt_{decon}^b$ states
reach to the energy density of the true vacuum, see \eqref{ksepk0},
\begin{equation}
\hat\cale[\calv_{B}]\bigg|_{\mu/\Lambda=0}=0\,.
\eqlabel{vacbmu0}
\end{equation}
Fig.~\ref{mu0ks2t11} explains why that is the case.
In the left panel we present the Kretschmann scalar $\hat\calk$ \eqref{defhatk} for different
deconfined states in the theory normalized
to its value at $\hat\cale=\hat\cale_\csb$, \ie $\hat\calk_\csb$.
Note the sharp increase in $\hat\calk/\hat\calk_\csb$ as $\hat\cale <\hat\cale_\csb$.
For $\calt_{decon}^b$ states (magenta curve, the central panel) this growth is associated with the
collapse of the conifold 2-cycle, see \eqref{r2s2van}. 
For $\calt_{decon}^s$ states (blue/brown curves, the right panel) this growth is associated with the
collapse of the deformed $T^{1,1}$, see \eqref{defft11def} and \eqref{r2tt11}.

\subsection{Conifold black holes and phase transitions at $\frac{\mu}{\Lambda}\ne 0$}\label{meat}

In section \ref{cansec} we presents the results for the conifold black holes in the canonical
ensemble, followed by the microcanonical ensemble discussion in section \ref{microcansec}.
In section \ref{selsec} we dive into details of the black holes thermodynamics for select values
$\mu\ne 0$, highlighting similarities and differences with
the $\mu=0$ case covered in section \ref{mu0sec}.

The main conceptual difference with the black branes on the conifold,
reviewed in section \ref{mu0sec},  is that
the black hole horizon is compact; as the result, there is no simple
relation between the thermodynamic and the dynamic instabilities. While we do identify
branches of the black holes with the negative specific heat,
this does not imply that these black holes are unstable.
The stability analysis of the black holes with respect to perturbative metric fluctuations
breaking the spatial $SO(4)$ symmetry will be discussed elsewhere \cite{inprep1}.

\subsubsection{Canonical ensemble}\label{cansec}

The critical temperatures of the canonical ensemble phase diagram identified
in section \ref{mu0sec} become functions of $\mu/\Lambda$; the hierarchy of scales
\eqref{3t} is however preserved:
\begin{equation}
T_u(\mu)\ \le\ T_{\csb}(\mu)\ <\ T_c(\mu)\,.
\eqlabel{3t2}
\end{equation}
There is a special value $\mu^\star$, such that
\begin{equation}
T_u(\mu^\star)=T_\csb(\mu^\star)\qquad \Longrightarrow\qquad \mu^\star=0.848(0)\ \Lambda\,.
\eqlabel{defmustar}
\end{equation}
Similar to the black branes, at $\mu/\Lambda\ne 0$:
\begin{itemize}
\item The critical temperature
$T_u(\mu)$ identifies the terminal temperature of $\calt_{decon}^s$ states. This temperature
is a bifurcation point for the two branches of the Klebanov-Tseytlin black holes,
the $\calt_{decon}^s$ states, with the positive and the negative reduced specific heat
densities $\hat c_V$, see \eqref{defcvmu0}. We use blue curves to refer to   $\calt_{decon}^{s,+}$
states with  $\hat c_V>0$, and brown curves when the reduced specific heat density of the
$\calt_{decon}^{s,-}$ states is negative. As $T\to T_u(\mu)_+$, the reduced specific heat diverges
with the same critical exponent as for $\mu=0$, see \eqref{cvdiv}. We find that for all values of
$\mu/\Lambda$, at an appropriate\footnote{Both states $\calt_{decon}^{s,+}$ and $\calt_{decon}^{s,-}$
must exist.}  fixed temperature, $\calt_{decon}^{s,+}$ states have a smaller reduced
free energy densities $\hat\calf$ \eqref{defhatf} then that of the $\calt_{decon}^{s,-}$ states. 
\item The critical temperature $T_\csb(\mu)$ (red curve) denotes the onset of the perturbative
chiral symmetry breaking instability of the  $\calt_{decon}^{s}$ states.   $T_\csb(\mu)$
 is also the bifurcation point on the phase diagram, where $\calt_{decon}^s$ and $\calt_{decon}^b$
branches of states join with the second-order phase transition.
\item The reduced specific heat of the $\calt_{decon}^b$ states, or the Klebanov-Strassler black holes
(represented with magenta curves),
is always negative.
\end{itemize}
On the contrary to the black branes, at $\mu/\Lambda\ne 0$:
\begin{itemize}
\item $\calt_{decon}^{s,-}$ states, while having a negative specific heat, can be perturbatively
stable to the chiral symmetry breaking fluctuations, provided $\mu> \mu^\star$, see \eqref{defmustar}.
\item Related to above, while at $\mu=0$ the chiral symmetry is spontaneously broken
for $T< T_{\csb}$, for $\mu>\mu^\star$, the chiral symmetry breaking
occurs for $T>T_\csb(\mu)$ in $\calt_{decon}^{s,-}$ states.
In this regime, at a fixed temperature,
the reduced free energy density of the symmetry broken states $\calt_{decon}^b$ 
is lower than that of the  $\calt_{decon}^{s,-}$, however it is still larger than the  reduced free energy
density of the $\calt_{decon}^{s,+}$ states at the corresponding temperature, \ie
\begin{equation}
\mu> \mu^\star:\qquad \hat\calf[\calt_{decon}^{s,+}]\ <\
\hat\calf[\calt_{decon}^{b}]\ <\ \hat\calf[\calt_{decon}^{s,-}]\,.
\eqlabel{fhier}
\end{equation}
\item There is a conceptual difference associated with the critical temperature $T_c(\mu)$
for the confinement-deconfinement phase transitions,
albeit all these phase transitions are still of the first order.
At $\mu=0$ this transition is between
$\calt_{decon}^s$ and $\calt_{con,B}$ states. For $\mu/\Lambda\ne 0$, at a given value of
$\mu$, there can exist two or even three different confined states\footnote{
Figs.~\ref{fluxvacuaI} and \ref{fluxvacuaII} make this point obvious.}:
$\calt_{con,B}$ and $\calt_{con,A}^s$ when $\mu$ is large enough but smaller than
$\mu_\csb$, or  $\calt_{con,B}$, $\calt_{con,A}^s$ and  $\calt_{con,A}^b$ for $\mu\gtrsim \mu_{\csb}$.
As a result, we must augment the $T_c(\mu)$ label with the corresponding confined state
label. We use green curves to denote confinement-deconfinement transitions
$\calt_{decon}^s\leftrightarrow \calt_{con,B}$, blue curves to denote
transitions $\calt_{decon}^s\leftrightarrow \calt_{con,A}^s$, and
magenta curves to denote transitions $\calt_{decon}^s\leftrightarrow \calt_{con,A}^b$.
Much like $\calv_A^s$ vacua (see fig.~\ref{fluxvacuaI}),  $\calt_{con,A}^s$
states are perturbatively unstable to chiral symmetry breaking fluctuations for $\mu<\mu_u$.
We represent branches of $T_c(\mu)$ subject to such instabilities with dashed curves.
\end{itemize}

\begin{figure}[t]
\begin{center}
\psfrag{t}[cc][][1.5][0]{{$T_c/\Lambda$}}
\psfrag{m}[bb][][1.5][0]{{$\mu/\Lambda$}}
\psfrag{a}[cc][][1][0]{{$\mu_u$}}
\psfrag{g}[cc][][1][0]{{$\mu_\csb$}}
\psfrag{s}[cc][][1][0]{{$\calt_{decon}^s\leftrightarrow \calt_{con,A}^s$}}
\psfrag{h}[cc][][1][0]{{$\calt_{decon}^s\leftrightarrow \calt_{con,B}$}}
\includegraphics[width=5in]{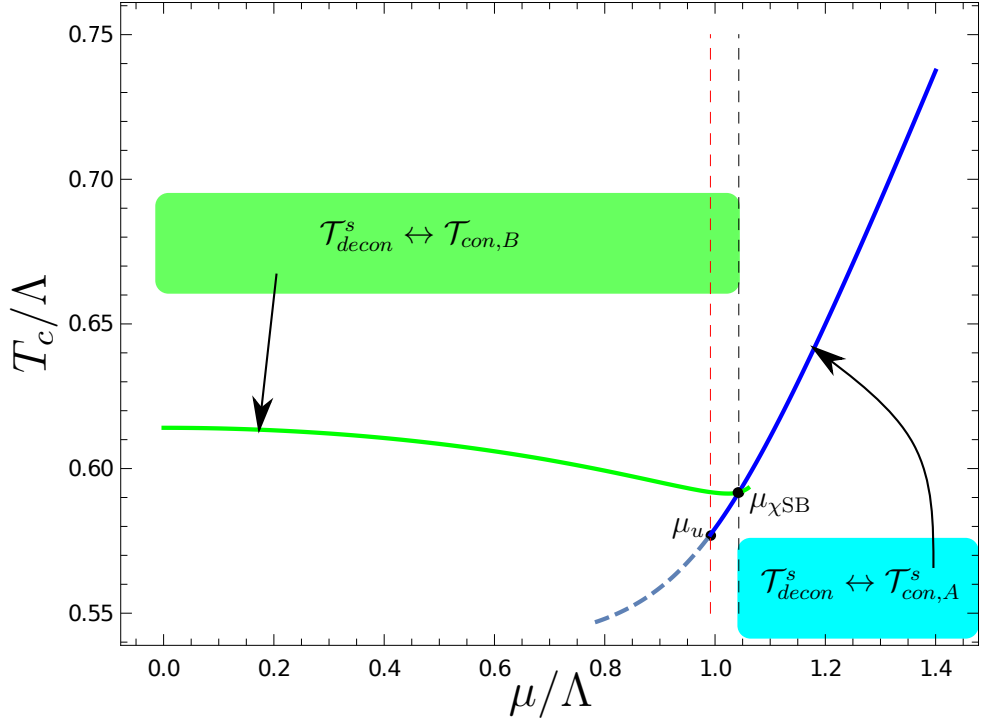}
\end{center}
  \caption{The temperature $T_c$ of the confinement/deconfinement phase transitions between the different states in the cascading gauge theory; equivalently, the analog of the Hawking-Page phase
  transition between the Klebanov-Tseytlin black holes and the different Euclidean vacua of the
  conifold flux geometries. The deconfined states are the preferred ones for $T> T_c$.
}\label{tclarge}
\end{figure}

In fig.~\ref{tclarge} we present the temperature of the first-order phase
confinement / deconfinement phase transition $T_c$ as a function of $\mu$ between $\calt_{decon}^s$
and $\calt_{con,B}$ states (the green curve) and between  $\calt_{decon}^s$
and $\calt_{con,A}^s$ states (the blue curves). The $\calt_{decon}^s\leftrightarrow \calt_{con,B}$
transition is the finite $\mu$ extension of the phase transition at $\mu=0$ identified
originally in \cite{Aharony:2007vg}. Note that for $\mu> \mu_\csb$ the confined states
$\calt_{con,A}^s$ are thermodynamically referred, compare to $\calt_{con,B}$
states\footnote{This reflects the $\csb$ phase transition between $\calv_{A}^s$ and $\calv_{B}$
vacua, see fig.~\ref{fluxvacuaI}.}. For $\mu<\mu_u$ the confined states $\calt_{con,A}^s$ are
perturbatively unstable with respect to chiral symmetry breaking fluctuations (represented
with the dashed curve). The blue curve extends to $\mu/\Lambda\to \infty$
reproducing the Hawking-Page transition of the $AdS_5\times T^{1,1}$ black holes,
see section \ref{conformalts}.

\begin{figure}[ht]
\begin{center}
\psfrag{t}[cc][][1.5][0]{{$T_c/\Lambda$}}
\psfrag{m}[bb][][1.5][0]{{$\mu/\Lambda$}}
\psfrag{a}[cc][][1][0]{{$\mu_u$}}
\psfrag{g}[cc][][1][0]{{$\mu_\csb$}}
\psfrag{s}[cc][][1][0]{{$\calt_{decon}^s\leftrightarrow \calt_{con,A}^s$}}
\psfrag{h}[cc][][1][0]{{$\calt_{decon}^s\leftrightarrow \calt_{con,A}^b$}}
\includegraphics[width=5in]{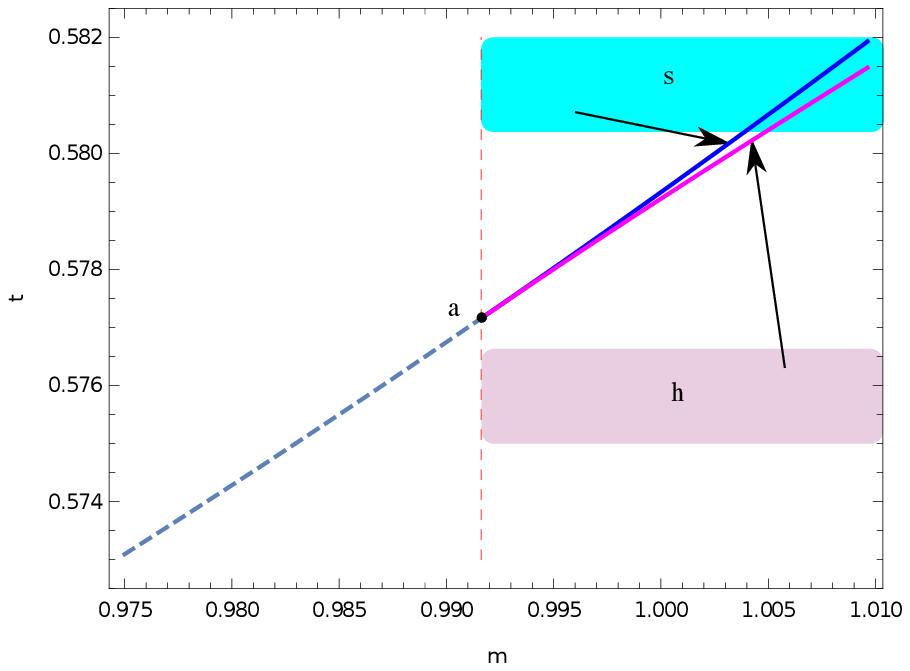}
\end{center}
  \caption{The temperature $T_c$ of the confinement/deconfinement phase transitions between the different states in the cascading gauge theory. Magenta curve represents the analog of
  the Hawking-Page phase
  transition between the Klebanov-Tseytlin black holes and the $\calv_A^b$ Euclidean vacua of the
  conifold flux geometries. The deconfined states are the preferred ones for $T> T_c$.
}\label{tcsmall}
\end{figure}

In fig.~\ref{tcsmall} we present the temperature of the first-order phase
confinement/deconfinement phase transition $T_c$ as a function of $\mu$ between $\calt_{decon}^s$
and $\calt_{con,A}^s$ states (the blue curve) and between  $\calt_{decon}^s$
and $\calt_{con,A}^b$ states (the magenta curve).
Note that these phase transitions are never realized in
practice since for the range of $\mu/\Lambda$ in question the preferred 
phase is always $\calt_{con,B}$, compare with fig.~\ref{tclarge}.

\begin{figure}[t]
\begin{center}
\psfrag{t}[cc][][1.5][0]{{$T_c/\mu$}}
\psfrag{m}[bb][][1.5][0]{{$\mu/\Lambda$}}
\psfrag{a}[cc][][1][0]{{$\mu_u$}}
\psfrag{g}[cc][][1][0]{{$\mu_\csb$}}
\psfrag{s}[cc][][1][0]{{$\calt_{decon}^s\leftrightarrow \calt_{con,A}^s$}}
\psfrag{h}[cc][][1][0]{{$T_c^{conformal}$}}
\includegraphics[width=6in]{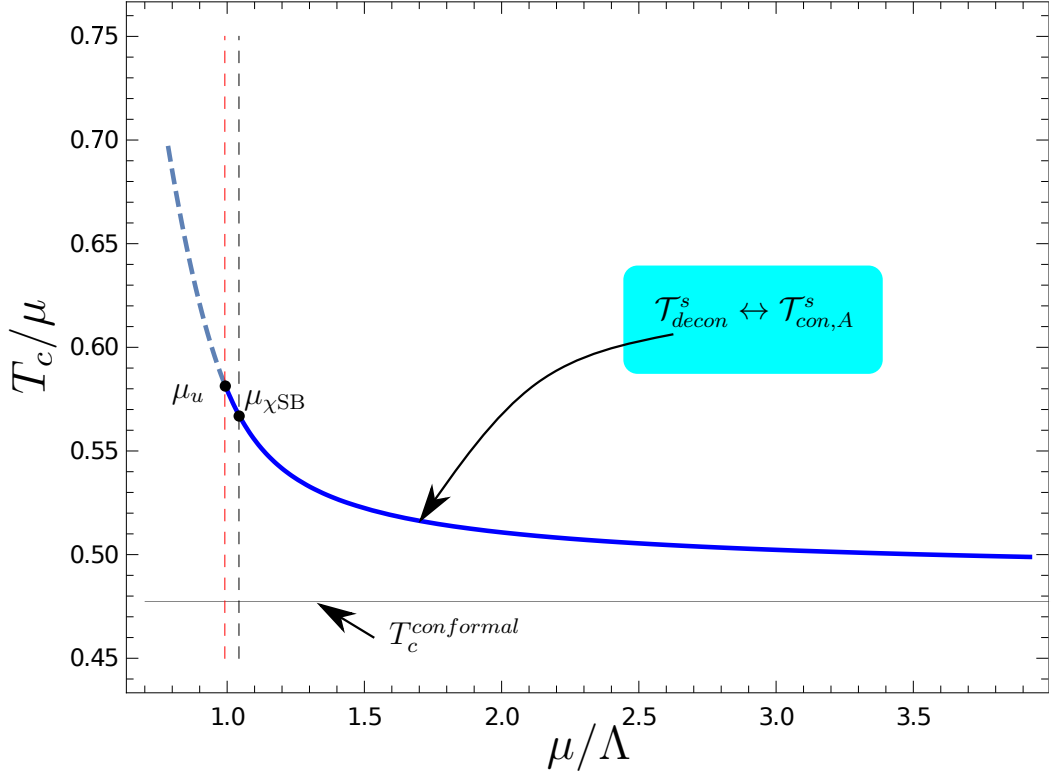}
\end{center}
  \caption{The temperature $T_c$ of the confinement/deconfinement phase transition 
  $\calt_{decon}^s\leftrightarrow \calt_{con,A}^s$ relative to $S^3$ compactification
  scale $\mu$. For large values of $\mu/\Lambda$ it approaches its conformal limit
   $T_c^{conformal}$, given by \eqref{tcondecon}. The deconfined states are the preferred
   ones for $T> T_c$.
}\label{tmularge}
\end{figure}

In fig.~\ref{tmularge} we present that temperature of the confinement/deconfinement transition
$\calt_{decon}^s\leftrightarrow \calt_{con,A}^s$ relative to the $S^3$ compactification scale $\mu$.
Here, we can take the limit $\mu/\Lambda\to \infty$, in which case the Klebanov-Tseytlin black holes
approach conformal black holes in $AdS_5\times T^{1,1}$, see section \ref{conformalts}. For the latter, the Hawking-Page transition
temperature $T_c^{conformal}$ is given by \eqref{tcondecon}, represented by the horizontal
black line.

\begin{figure}[t]
\begin{center}
\psfrag{e}[cc][][1.5][0]{{$\hat{\cale}$}}
\psfrag{m}[bb][][1.5][0]{{$\mu/\Lambda$}}
\psfrag{a}[cc][][1.5][0]{{$\calt_{con,B}$}}
\psfrag{b}[cc][][1.5][0]{{$\calt_{decon}^s$}}
\psfrag{n}[cc][][1.5][0]{{$\calt_{con,A}^s$}}
\psfrag{c}[cc][][1][0]{{$\csb$}}
\psfrag{f}[cc][][1][0]{{$\mu_\csb$}}
\psfrag{d}[cc][][1][0]{{$\mu_u$}}
\psfrag{p}[cc][][1][0]{{$\mu_{KS}$}}
\psfrag{z}[cc][][1][0]{{$\hat{\cale}_\csb(\mu)$}}
\psfrag{x}[cc][][0.9][0]{{$\hat{\cale}_{c}[\calt_{decon}^s
\leftrightarrow\calt_{con,B}]$}}
\psfrag{y}[cc][][0.9][0]{{$\hat{\cale}_{c}[\calt_{decon}^s
\leftrightarrow\calt_{con,A}^s]$}}
\includegraphics[width=5.5in]{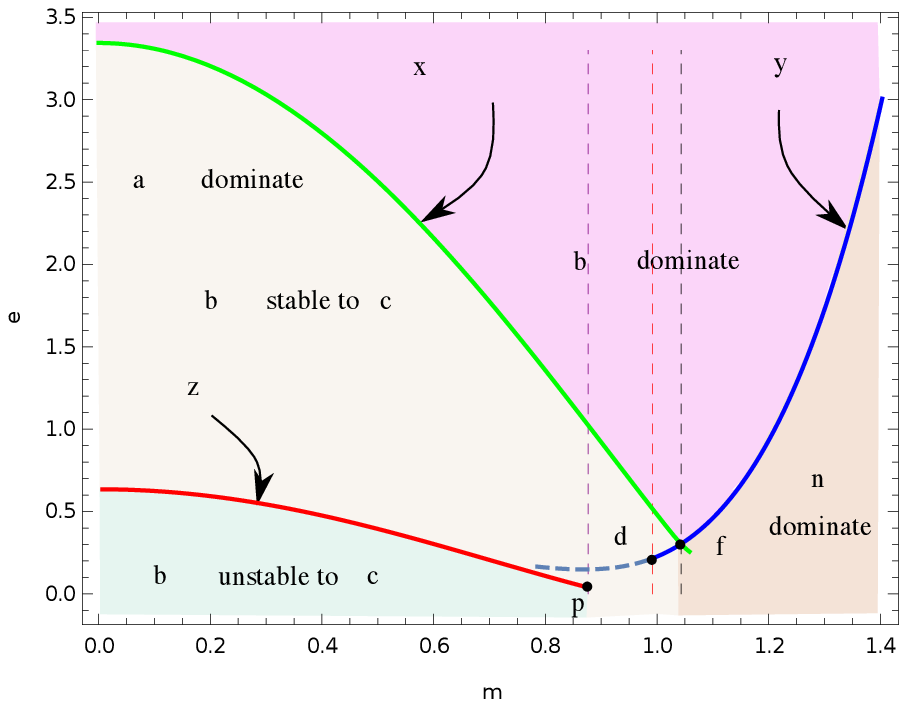}
\end{center}
  \caption{Critical energies $\hat\cale_c$ of $\calt_{decon}^s$ states (Klebanov-Tseytlin
  black holes) for different confinement/deconfinement phase
  transitions. The deconfined states 
  are preferred above green and blue curves. The Klebanov-Tseytlin black holes
  become unstable to $\csb$ fluctuations below the red curve $\hat{\cale}_\csb(\mu)$.
  The Klebanov-Strassler black holes exist only for $\hat\cale<\hat{\cale}_\csb$
  and only when $\mu<\mu_{KS}$ \eqref{defmuks}.
}\label{ecas}
\end{figure}

In fig.~\ref{ecas}, corresponding to the confinement/deconfinement temperatures $T_c$
we present the  reduced energy densities $\hat\cale_c$
(see \eqref{defhate}) of $\calt_{decon}^s$ states, \ie the Klebanov-Tseytlin black holes,
at the transition point. Once again, the green curve represents
$\calt_{decon}^s\leftrightarrow \calt_{con,B}$
transition, and the blue curves represents $\calt_{decon}^s\leftrightarrow \calt_{con,A}^s$ transition.
The deconfined states are the preferred ones above the corresponding $\hat\cale_c$.
The dashed curve indicates when the end-point of the phase transition, \ie the
$\calt_{decon}^s$ states, are perturbatively unstable to chiral symmetry breaking.
The red curve indicates the critical energy $\hat\cale_\csb$: the Klebanov-Tseytlin
black holes are unstable to chiral symmetry breaking for $\hat\cale<\hat\cale_\csb$.
The Klebanov-Strassler black holes exist only at energy densities $\hat\cale<\hat\cale_\csb$,
\ie below the red curve. Note that the red curve $\hat\cale_\csb$ terminates at $\mu=\mu_{KS}$,
\begin{equation}
\mu_{KS}=0.8766(6)\ \Lambda\,,
\eqlabel{defmuks}
\end{equation}
--- there are no Klebanov-Strassler black holes for $\mu>\mu_{KS}$.
We explain origin of $\mu_{KS}$ in section \ref{techtcsb}.

\begin{figure}[t]
\begin{center}
\psfrag{e}[cc][][1.5][0]{{$\hat{\cale}$}}
\psfrag{m}[tt][][1.5][0]{{$\mu/\Lambda$}}
\psfrag{a}[cc][][1.5][0]{{$\calt_{con,B}$}}
\psfrag{b}[cc][][1.5][0]{{$\calt_{decon}^s$}}
\psfrag{n}[cc][][1.5][0]{{$\calt_{con,A}^s$}}
\psfrag{f}[cc][][1.5][0]{{$\calt_{con,A}^b$}}
\psfrag{c}[cc][][1][0]{{$\csb$}}
\psfrag{d}[cc][][1][0]{{$\mu_u$}}
\psfrag{z}[cc][][1][0]{{$\hat{\cale}_\csb(\mu)$}}
\psfrag{l}[cc][][1][0]{{$\hat{\cale}_{c}[\calt_{decon}^s
\leftrightarrow\calt_{con,A}^b]$}}
\psfrag{y}[cc][][1][0]{{$\hat{\cale}_{c}[\calt_{decon}^s
\leftrightarrow\calt_{con,A}^s]$}}
\includegraphics[width=5.5in]{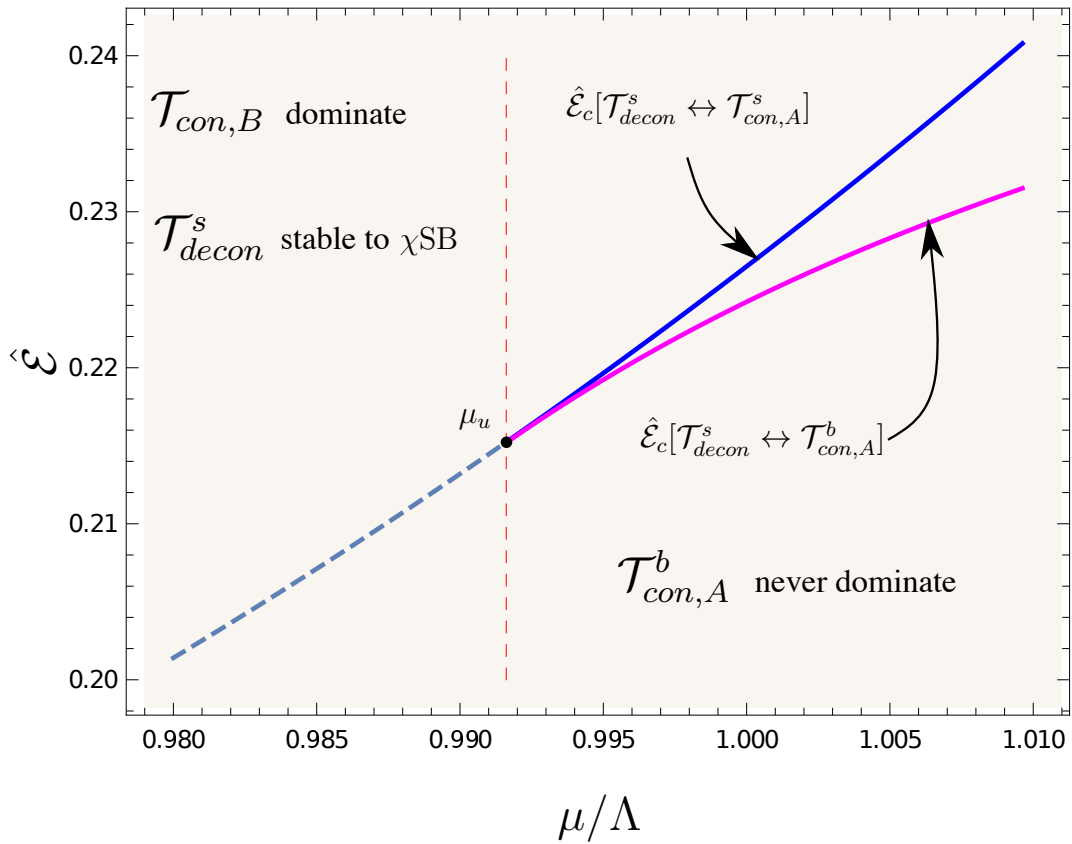}
\end{center}
  \caption{Critical energies $\hat\cale_c$ of $\calt_{decon}^s$ states (Klebanov-Tseytlin
  black holes) for different confinement/deconfinement phase
  transitions. These transitions do not happen in practice since the transition to a
  dominant confined phase $\calt_{con,B}$ occurs at  higher energy densities
  (compare with fig.~\ref{ecas}). 
}\label{ecab}
\end{figure}

In fig.~\ref{ecab} we present the reduced energy densities $\hat\cale_c$ of $T_{decon}^s$ states
at $\hat{\cale}_{c}[\calt_{decon}^s\leftrightarrow\calt_{con,A}^s$ confinement/deconfinement
transition (blue curve) and  $\hat{\cale}_{c}[\calt_{decon}^s\leftrightarrow\calt_{con,A}^b$
confinement/deconfinement transition (magenta curve). These transitions are never realized in
practice since the transition to a the preferred confined phase $\calt_{con,B}$
occurs at higher energy densities (compare with fig.~\ref{ecas}).

\begin{figure}[t]
\begin{center}
\psfrag{t}[bb][][1][0]{{$T_{\csb}/\Lambda$}}
\psfrag{m}[tt][][1][0]{{$\mu/\Lambda$}}
\psfrag{s}[cc][][0.7][0]{{$\mu^\star$}}
\psfrag{x}[cc][][0.7][0]{{$T_u(\mu^\star)=T_\csb(\mu^\star)$}}
\psfrag{c}[tt][][1][0]{{$1/\hat{c}_V(T_\csb)$}}
\includegraphics[width=2.6in]{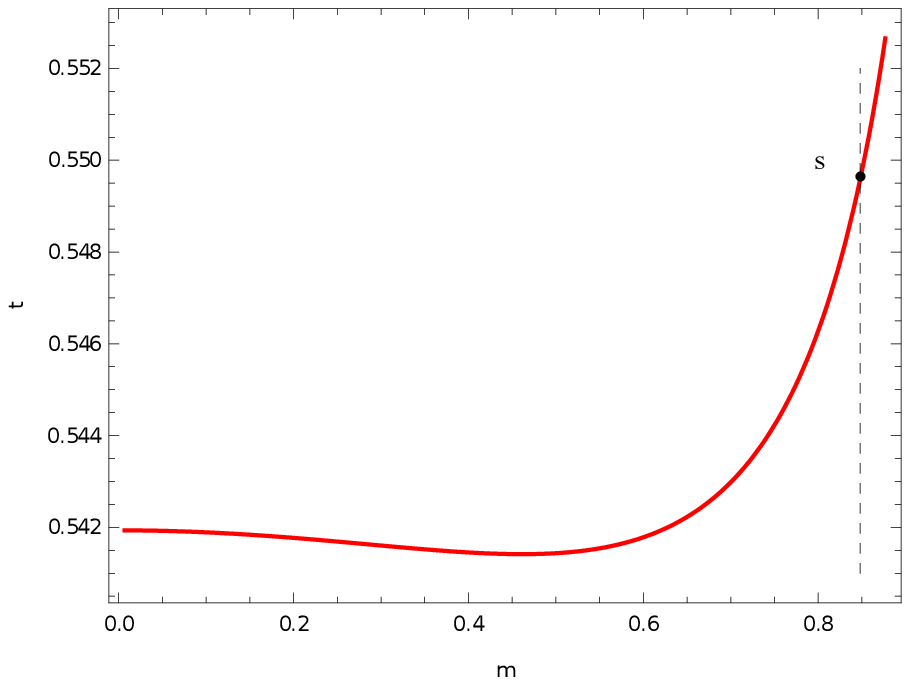}\ \
\includegraphics[width=2.6in]{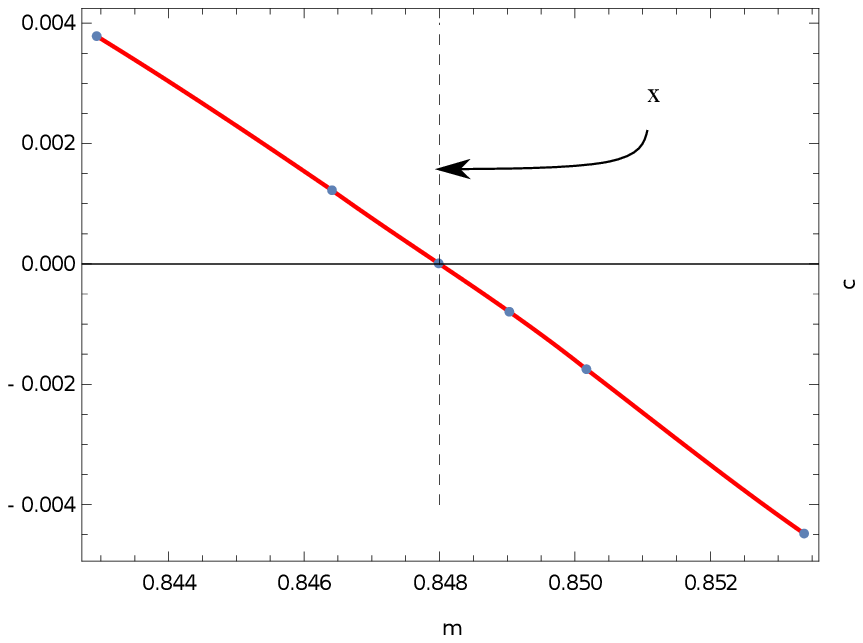}
\end{center}
  \caption{Left panel: the critical temperature $T_\csb$ of the spontaneous chiral symmetry breaking
  in $\calt_{decon}^s$ states, \ie the Klebanov-Tseytlin black holes, as a function of $\mu/\Lambda$.
  Right panel: $\mu^*$, see \eqref{defmustar}, identifies the $S^3$ compactification scale $\mu$
  with the divergent reduced specific heat $\hat{c}_V$ at $T_\csb(\mu^*)$. At $\mu=\mu^\star$
  the critical temperature $T_\csb$ coincides with the terminal temperature $T_u$ of $\calt_{decon}^s$ states. 
}\label{csbtcv}
\end{figure}

Recall that $T_u(\mu)$ is the terminal temperature for $\calt_{decon}^s$ states --- the Klebanov-Tseytlin black holes.
Thus $T_\csb(\mu)$, the critical temperature  at the onset of the chiral symmetry breaking,
can not exceed $T_u(\mu)$. In the left panel of fig.~\ref{csbtcv} we present results for $T_\csb$.
$\mu^\star$, see \eqref{defmustar}, identifies the $S^3$ compactification scale when both critical temperatures $T_u$
and $T_\csb$ coincide. In the right panel of fig.~\ref{csbtcv} we explain how to compute $\mu^\star$: we evaluate the
reduced specific heat density $\hat{c}_V$ \eqref{defcvmu0} along the $\hat\cale_\csb$ curve for $\calt_{decon}^s$ states
and identify the value of $\mu$ when it diverges.

Much like for $\mu=0$, for the  values of $\mu$ when both $\calt_{decon}^{s,\pm}$ and
$\calt_{decon}^b$ states exist, the free energy density of
 $\calt_{decon}^{s,+}$ states is always {\it lower} than that of  $\calt_{decon}^{b}$
 states\footnote{See section \ref{selsec} for some examples.},
 \ie whenever both Klebanov-Strassler and Klebanov-Tseytlin black holes exist,
 the latter ones (on the branch with the positive specific heat) always dominate
 in the canonical ensemble\footnote{As we will see in section \ref{microcansec},
 the story in the microcanonical ensemble is different: whenever both exist,
 Klebanov-Strassler black holes are always more entropic at a fixed energy density
 than the Klebanov-Tseytlin black holes. 
 }.
 This also implies that even if theoretically possible, the confinement/deconfinement
 phase transitions $\calt_{decon}^b\leftrightarrow \calt_{con,B}$, \
 \ie between the Klebanov-Strassler black holes and the $\calv_B$ Euclidean vacua would never
 occur in practice: these phase transitions would always be preceded by
 $\calt_{decon}^{s,+}\leftrightarrow \calt_{con,B}$ phase transitions. 

Some of the potential confinement/deconfinement phase transitions, are never realized
(even theoretically):
\begin{itemize}
\item $\calt_{decon}^b\leftrightarrow \calt_{con,A}^s$ --- the transition between
the Klebanov-Strassler black holes and the $\calv_A^s$ Euclidean vacua. These transitions
do not exist because there are no Klebanov-Strassler black holes for $\mu>\mu_u$, and there
are no stable  $\calv_A^s$ vacua (correspondingly $\calt_{con,A}^s$ confined thermal states) for
$\mu<\mu_u$, see  figs.~\ref{ecas} and \ref{fluxvacuaI}.
\item $\calt_{decon}^b\leftrightarrow \calt_{con,A}^b$ --- the transition between
the Klebanov-Strassler black holes and the $\calv_A^b$ Euclidean vacua. These transitions
do not exist because there are no Klebanov-Strassler black holes for $\mu>\mu_u$, and there
are $\calv_A^b$ vacua (correspondingly $\calt_{con,A}^b$ confined thermal states) for
$\mu<\mu_u$, see  figs.~\ref{ecas} and \ref{fluxvacuaII}.
\end{itemize}

\subsubsection{Microcanonical ensemble}\label{microcansec}

\begin{figure}[ht]
\begin{center}
\psfrag{e}[cc][][1.5][0]{{$\hat{\cale}$}}
\psfrag{m}[bb][][1.5][0]{{$\mu/\Lambda$}}
\psfrag{a}[cc][][1][0]{{$\mu=0$}}
\psfrag{b}[cc][][1][0]{{$\mu_1$}}
\psfrag{c}[cc][][1][0]{{$\mu_2$}}
\psfrag{f}[cc][][1][0]{{$\mu_\csb$}}
\psfrag{d}[cc][][1][0]{{$\mu_u$}}
\psfrag{p}[cc][][1][0]{{$\mu_{KS}$}}
\psfrag{o}[cc][][1][0]{{$\mu_1<\mu^\star<\mu_2$}}
\psfrag{z}[cc][][1.5][0]{{$\hat{\cale}_\csb(\mu)$}}
\psfrag{x}[cc][][1.5][0]{{$\hat{\cale}_{\calv_B}(\mu)$}}
\psfrag{n}[cc][][1.5][0]{{$\hat{\cale}_{\calv_A^s}(\mu)$}}
\psfrag{j}[cc][][1.5][0]{{$\color{orange}\calt_{decon}^b$}}
\psfrag{u}[cc][][1.5][0]{{$\color{blue}\calt_{decon}^s$}}
\includegraphics[width=6in]{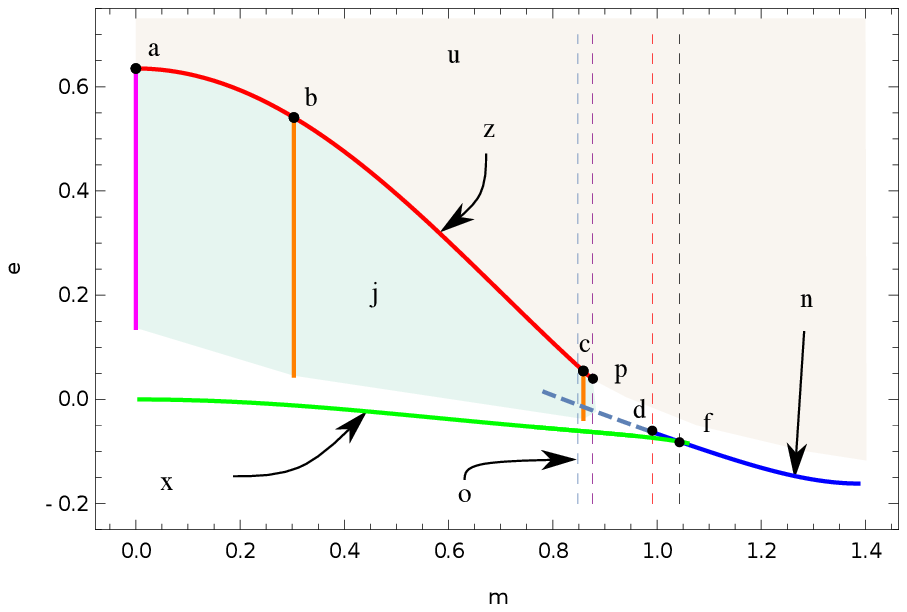}
\end{center}
  \caption{Landscape of the Klebanov-Tseytlin black holes, $\calt_{decon}^s$ states,  (the shaded region labeled by $\color{blue}\calt_{decon}^s$)
  and the Klebanov-Strassler black holes,  $\calt_{decon}^b$ states,  (the shaded region labeled by $\color{orange}\calt_{decon}^b$). Two
  $\calt_{decon}^b$ state spectra at $\mu=\mu_1$ and $\mu=\mu_2$ are represented by the vertical orange lines. The vertical magenta line represents
  the spectrum of the Klebanov-Strassler black branes, $\calt_{decon}^s$ states at $\mu=0$. 
}\label{ecsb}
\end{figure}

The phase diagram in the microcanonical ensemble for the conifold black branes ($\mu=0$)
morally repeats for black holes at $\mu/\Lambda>0$:
\begin{itemize}
\item Confined states have vanishing reduced entropy density $\hat s$ \eqref{defhatf} in the large
$M$ limit (the supergravity approximation of the dual geometry), and thus are never the preferred states
if any of the deconfined states $\calt_{decon}^s$, \ie the Klebanov-Tseytlin, or $\calt_{decon}^b$, \ie the Klebanov-Strassler
black holes, exist.
\item When both the $\calt_{decon}^s$ and $\calt_{decon}^b$ states exist, the latter ones are always more entropic,
and thus represent the preferred states.  
\end{itemize}

Since $\calt_{decon}^b$ states at $\mu=0$ have the
negative specific heat, the corresponding Klebanov-Strassler black branes are perturbatively unstable to
metric fluctuations breaking the translational invariance of the horizon \cite{Buchel:2005nt}.
This link is no longer true for the black holes, and the unstable hydrodynamic sound mode of the
black brane extended horizon could be lifted by the compactification \cite{inprep1}.

In fig.~\ref{ecsb} we discuss the landscape of $\calt_{decon}^s$ and $\calt_{decon}^b$ states. 
For reference, the green and the blue curves represent the vacua states $\calv_B$ and $\calv_A^s$
correspondingly. The red curves reproduces the critical energy density $\hat\cale_\csb$ for the range
$\mu\in[0,\mu_{KS}]$. $\mu_{KS}$ is the terminal point of the red curve. The Klebanov-Tseytlin black holes
are the preferred states in the microcanonical ensemble in the shaded region above the red the the blue curves,
labeled by $\color{blue}\calt_{decon}^s$. The Klebanov-Strassler black holes are the preferred states in the shaded region
labeled by $\color{orange}\calt_{decon}^b$. They exist below the red curve, and only for $\mu<\mu_{KS}$ \eqref{defmuks}.
The vertical magenta line, at $\mu=0$, represents the Klebanov-Strassler black brane constructed in \cite{Buchel:2018bzp}.
It extends along the $\hat\cale$ axis indicative to which reduced energy density, as the lower limit,
we have been able to construct this black brane numerically. Note that there
is likely a 
'gap' between the magenta curve and the green curve representing the vacuum $\calv_B$. 
Likewise, the sample of two vertical orange curves at $\mu=\mu_1$ and $\mu=\mu_2$
represents the energy spectra of the Klebanov-Strassler black holes, see section \ref{selsec} for additional details.
Once again, notice 
the potential gap in the spectrum of the Klebanov-Strassler black holes and the appropriate vacuum states $\calv_B$.
There is also a potential gap in the spectrum of Klebanov-Tseytlin black holes and the appropriate vacuum states $\calv_A^s$.
We emphasize this gap since it points out that the non-equilibrium energy states on $S^3$-compactified conifold geometries with fluxes, alternatively the low energy density states of the cascading gauge theory, might not thermalize
along $SO(4)$-invariant evolution trajectories\footnote{See further discussion of this in holography
in \cite{Buchel:2015lla,Buchel:2015sma}.}. Note that for our sample scales, $\mu_1<\mu^\star<\mu_2$ where 
$\mu_\star$ \eqref{defmustar} affects the sequence of the various phase transitions,
see \ref{selsec} for further details. Recall that $\calv_A^s$ vacua are perturbatively unstable to the spontaneous chiral symmetry breaking
when $\mu<\mu_u$, represented by the dashed curve. Interestingly, the Klebanov-Strassler black holes at $\mu=\mu_2$ reaches the energy densities of some of these unstable states ---
the corresponding vertical orange line intersects the dashed line of the unstable $\calv_A^s$ vacua. Thus, at least when $\mu<\mu_{KS}$, the unstable to $\csb$ fluctuations
 $\calv_A^s$ vacua are expected to thermalize as Klebanov-Strassler black holes. It is very interesting to understand the fate of the unstable $\calv_A^s$ vacua for
 $\mu\in(\mu_{KS},\mu_u)$ --- we do not have the answer to this question.

\subsubsection{Black holes thermodynamics for select values of $\mu/\Lambda\ne 0$}\label{selsec}

In this section we follow the discussion of section \ref{mu0sec} and presents results for the conifold black holes
thermodynamics at 
\begin{equation}
\mu_1=0.303(3)\Lambda \qquad {\rm and}\qquad  \mu_2= 0.858(5)\Lambda\,.
\eqlabel{mu1mu2}
\end{equation}
Note that $\mu_1<\mu^\star$ \eqref{defmustar} and $\mu_2>\mu^\star$ --- as we will see the relation of the compactification
scale $\mu$ relative to $\mu^\star$ affects the sequences of the phase transitions for the $\calt_{decon}^s$ states. 

At $\mu=\mu_1$ there is a single confinement-deconfinement phase transition: 
\begin{equation}
\calt_{decon}^s\leftrightarrow\calt_{con,B}:\qquad T_c(\mu_1)=0.612(1)\Lambda\,,\qquad \hat\cale_c(\mu_1)=3.025(2)\,.
\eqlabel{dettemu1}
\end{equation}
At $\mu=\mu_2$, theoretically there are two confinement-deconfinement phase transition: 
\begin{equation}
\begin{split}
&\calt_{decon}^s\leftrightarrow\calt_{con,B}:\qquad T_c(\mu_2)=0.596(9)\Lambda\,,\qquad \hat\cale_c(\mu_2)=1.103(5)\,,\\
&\calt_{decon}^s\leftrightarrow\calt_{con,A}^s:\qquad T_c(\mu_2)=0.553(4)\Lambda\,,\qquad \hat\cale_c(\mu_2)=0.148(8)\,.
\end{split}
\eqlabel{dettemu2}
\end{equation}
In practice only the transition at higher temperature/energy density occurs.
All these transitions are of the first order, with the deconfined phase being the preferred state at
$T>T_c$ or $\hat\cale>\hat\cale_c$.

\begin{figure}[t]
\begin{center}
\psfrag{t}[cc][][0.7][0]{{$T/\Lambda$}}
\psfrag{f}[cc][][0.7][0]{{$\hat{\calf}$}}
\psfrag{a}[cc][][0.7][0]{{$\calt_{decon}^s$}}
\psfrag{b}[cc][][0.7][0]{{$\calt_{decon}^b$}}
\psfrag{j}[cc][][1][0]{{$\mu=\mu_1$}}
\psfrag{k}[cc][][1][0]{{$\mu=\mu_2$}}
\psfrag{g}[cc][][0.5][0]{{$\csb$}}
\psfrag{p}[cc][][0.5][0]{{$\hat{c}_V>0$}}
\psfrag{n}[cc][][0.5][0]{{$\hat{c}_V<0$}}
\psfrag{c}[cc][][0.5][0]{{$T_\csb$}}
\psfrag{e}[cc][][0.5][0]{{$T_u$}}
\includegraphics[width=2.6in]{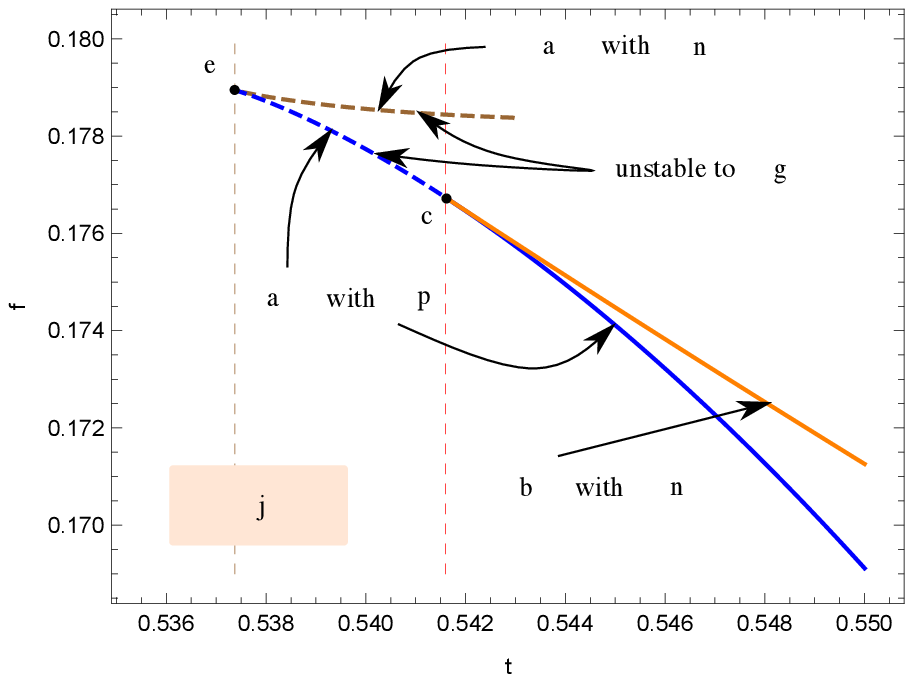}\ \
\includegraphics[width=2.6in]{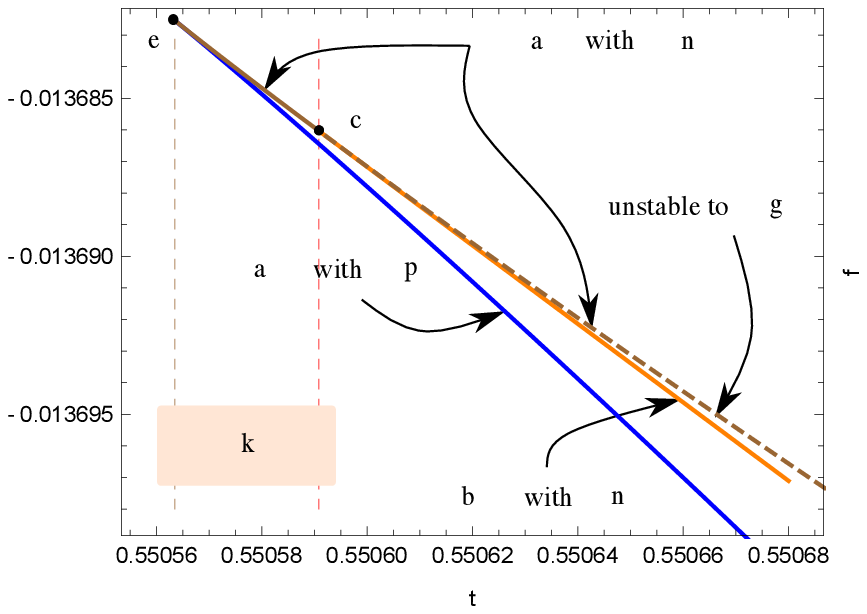}\ \
\end{center}
  \caption{Typical phase diagram in the canonical ensemble for deconfined states $\calt_{decon}^s$
  (the Klebanov-Tseytlin black holes) and  $\calt_{decon}^b$
  (the Klebanov-Strassler black holes) for $0<\mu=\mu_1<\mu^\star$ (see \eqref{defmustar})
  (the left panel), and  for $\mu=\mu_2>\mu^\star$ (the right panel).
}\label{can20500}
\end{figure}

In fig.~\ref{can20500} we highlight the similarities and the differences between the phase
diagrams at $\mu=\mu_1$ (the left panel)
and $\mu=\mu_2$ (the right panel) in the canonical ensemble. $T_u$ denotes the terminal point of $\calt_{decon}^s$
states, \ie the Klebanov-Tseytlin black holes; $T_\csb$ is the critical temperature for the onset of the perturbative
$\csb$ instability of $\calt_{decon}^s$ states.
\nxt $T_u$ is a bifurcation point of the two branches of $\calt_{decon}^s$ states: with the positive
specific heat $\calt_{decon}^{s,+}$ (blue), and the negative specific heat $\calt_{decon}^{s,-}$ (brown). The critical behavior as $T\to T_u$ is exactly the same as for $\mu=0$, see \eqref{cvdiv}.
\nxt While at $\mu=\mu_1$ $\calt_{decon}^{s,+}$ states are unstable to $\csb$ for $T<T_\csb$ (blue dashed curve),
at $\mu=\mu_2$ it is the $\calt_{decon}^{s,-}$ states that are unstable to $\csb$ for $T>T_\csb$ (brown dashed curve).
The critical behavior as $T\to T_\csb$ is exactly the same as for $\mu=0$, see \eqref{delhatf}.
\nxt $\calt_{decon}^b$ states (orange curves), \ie the Klebanov-Strassler black holes, exist for $T>T_{\csb}$.
These black holes always have a negative specific heat. Although they are thermodynamically unstable,
they are not necessarily dynamically unstable \cite{Buchel:2005nt,inprep1}.
\nxt When multiple phases present, $\calt_{decon}^{s,+}$ is always the preferred one:
\begin{equation}
\hat\calf[\calt_{decon}^{s,+}]\ <\
\hat\calf[\calt_{decon}^{b}]\ <\ \hat\calf[\calt_{decon}^{s,-}]\,.
\eqlabel{fhier2}
\end{equation}
\nxt The order of the transition from the $\calt_{decon}^b$ states to (dominant) $\calt_{decon}^{s,+}$ states differs:
it is of the second  order at $\mu=\mu_1$ and of the first-order at $\mu=\mu_2$.
\nxt The phase transition  from  the $\calt_{decon}^{s,-}$ states to (dominant) $\calt_{decon}^{s,+}$ states
is always of the first order.

\begin{figure}[t]
\begin{center}
\psfrag{e}[cc][][0.7][0]{{$\hat\cale$}}
\psfrag{s}[cc][][0.7][0]{{$\hat{s}$}}
\psfrag{a}[cc][][0.7][0]{{$\calt_{decon}^s$}}
\psfrag{b}[cc][][0.7][0]{{$\calt_{decon}^b$}}
\psfrag{j}[cc][][1][0]{{$\mu=\mu_1$}}
\psfrag{k}[cc][][1][0]{{$\mu=\mu_2$}}
\psfrag{g}[cc][][0.5][0]{{$\csb$}}
\psfrag{p}[cc][][0.5][0]{{$\hat{c}_V>0$}}
\psfrag{n}[cc][][0.5][0]{{$\hat{c}_V<0$}}
\psfrag{c}[cc][][0.5][0]{{$\hat{\cale}_\csb$}}
\psfrag{w}[cc][][0.5][0]{{$\hat{\cale}_u$}}
\includegraphics[width=2.6in]{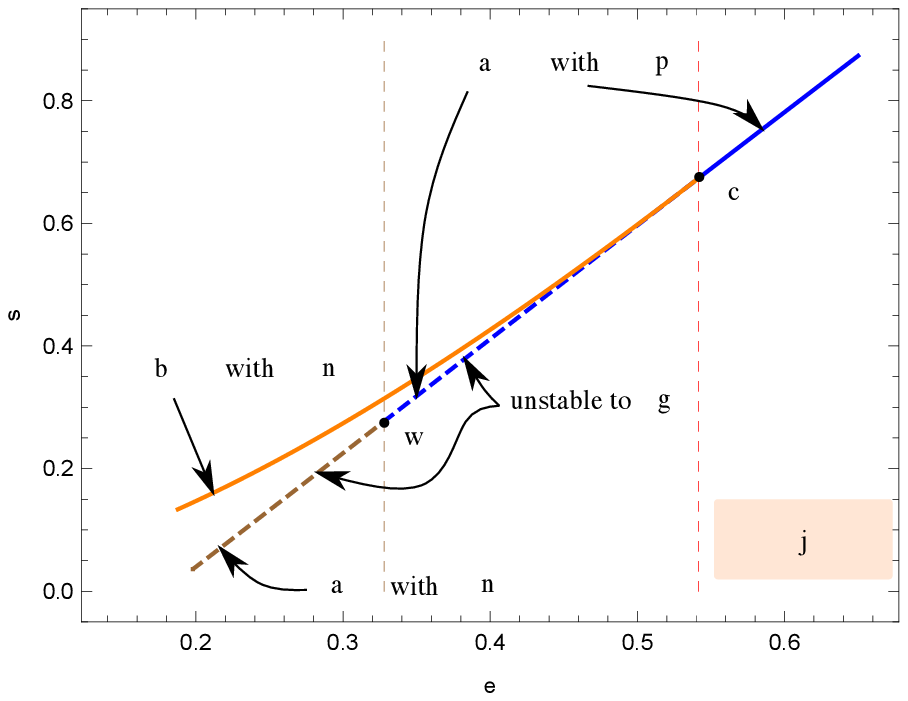}\ \
\includegraphics[width=2.6in]{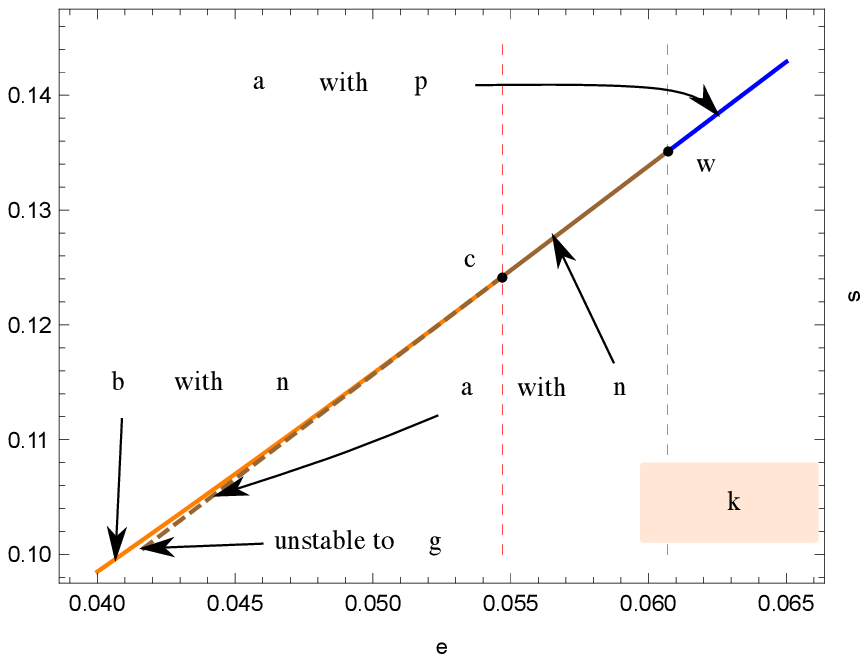}\ \
\end{center}
  \caption{Typical phase diagram in the microcanonical ensemble for deconfined states $\calt_{decon}^s$
  (the Klebanov-Tseytlin black holes) and  $\calt_{decon}^b$
  (the Klebanov-Strassler black holes) for $0<\mu=\mu_1<\mu^\star$ (see \eqref{defmustar})
  (the left panel), and  for $\mu=\mu_2>\mu^\star$ (the right panel).
}\label{micro20500}
\end{figure}

In fig.~\ref{micro20500} we present the typical phase diagram in the microcanonical ensemble for the
black holes on the conifold.
\nxt $\calt_{decon}^b$ (the orange curves; the Klebanov-Strassler black holes) are always the preferred states,
when they exist along with   $\calt_{decon}^s$ (blue/brown curves, the Klebanov-Tseytlin black holes) states. 
\nxt  $\calt_{decon}^b$ states exist for $\hat\cale<\hat\cale_\csb$, in which regime the Klebanov-Tseytlin
black holes are perturbatively unstable to $\csb$ fluctuations (dashed curves).
\nxt Note that $\hat\cale_\csb> \hat\cale_u$ at $0<\mu=\mu_1<\mu^\star$, but  $\hat\cale_\csb< \hat\cale_u$
at $\mu=\mu_2>\mu^\star$.

\begin{figure}[t]
\begin{center}
\psfrag{e}[cc][][1.5][0]{{$\hat\cale$}}
\psfrag{k}[bb][][1.5][0]{{$\hat\calk_\csb/\hat\calk$}}
\psfrag{a}[cc][][1][0]{{$\calt_{decon}^s$}}
\psfrag{b}[cc][][1][0]{{$\calt_{decon}^b$}}
\psfrag{j}[cc][][1][0]{{$\mu=\mu_1$}}
\psfrag{c}[cc][][1][0]{{$\hat{\cale}_\csb$}}
\psfrag{w}[cc][][1][0]{{$\hat{\cale}_u$}}
\psfrag{d}[cc][][1][0]{{$\hat{\cale}_{\calv_B}$}}
\includegraphics[width=5in]{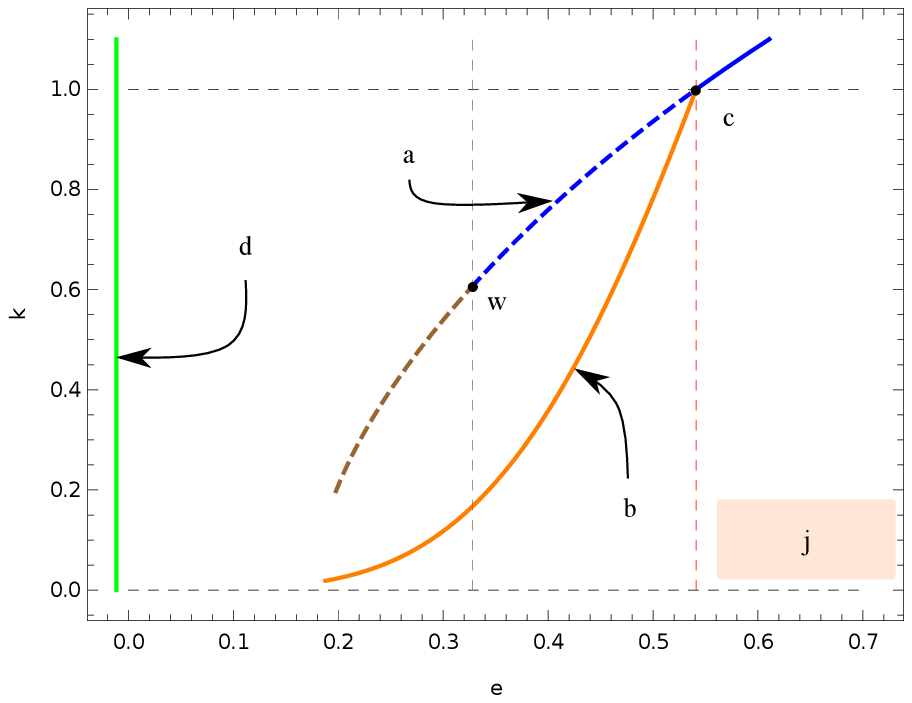}
\end{center}
  \caption{Inverse Kretschmann scalar $\hat\calk$ normalized to its value at
  $\hat\cale=\hat\cale_\csb$ for $\calt_{decon}^s$ states
(blue/drown curves) and
 $\calt_{decon}^b$ states
(orange curve) at $\mu=\mu_1$.  Vertical green line denotes the energy density of the
vacuum $\hat\calv_B$ at $\mu=\mu_1$.
}\label{rim2ist20}
\end{figure}

In fig.~\ref{rim2ist20} we present the inverse of the reduced Kretschmann scalar
$\hat\calk$ \eqref{defhatk}, normalized to its value at the
$\hat\cale=\hat\cale_\csb$ for $\calt_{decon}^s$ states
(blue/drown curves, the Klebanov-Tseytlin black holes) and
 $\calt_{decon}^b$ states
(orange curve, the Klebanov-Strassler black holes) at $\mu=\mu_1$.
The vertical green line indicates the energy density of the
$\hat\calv_B$ vacuum state at $\mu=\mu_1$. Our numerics
suggests\footnote{A definite conclusion requires more precise
analysis with improved numerical codes.}
that $\hat\calk$ likely diverges in deconfined states prior to reaching the
vacuum energy --- there is potentially a gap in the spectrum of
black holes on the
conifold.

\subsection{Technical details on constructing the black holes,
computing  $T_\csb(\mu)$ and $T_c(\mu)$ }\label{technical}

In constructing black holes on the conifold with fluxes it is convenient
to start at $\mu=0$ and slowly increase the $S^3$ curvature, compare to the strong coupling
scale $\Lambda$ of the theory. At $\mu=0$ we have black branes, thus we would like to
recycle  the numerical results obtained in 
\cite{Aharony:2007vg,Buchel:2010wp,Buchel:2018bzp}. In the latter references
the "universal'' radial coordinate $x$,
\begin{equation}
1-x\equiv -\frac{G_{tt}}{G_{x^ix^i}}\,,\qquad x\in [0,1]\,,
\eqlabel{defxold}
\end{equation}
was used, with $x\to 0_+$ being the asymptotic boundary, and $x\to 1_-$ being the
regular Schwarzschild horizon. It would appear to  be more
natural to use the same radial coordinate
$x$, instead of $\rho$ as defined in \eqref{5metrho} and \eqref{metpar}.
The problem is that the radial coordinate $x$ is simply not defined when $\mu\ne 0$.
Indeed, the definition \eqref{defxold} is sensible only that the ratio of the
warp factors $G_{tt}$ and $G_{x^ix^i}$ is a monotonic function from the
boundary to the horizon. When $\mu\ne 0$, this function is {\it not} monotonic.
Indeed, even in the conformal limit, see \eqref{solvecftts},
\begin{equation}
-\frac{G_{tt}}{G_{x^ix^i}}=\frac{4 (f_{a,1,0} \rho+1) (K_0 \mu^2 \rho^2+2 f_{a,1,0}^2 \rho^2+4 f_{a,1,0} \rho+4)}
{(f_{a,1,0} \rho+2)^4}\,,
\eqlabel{fconf}
\end{equation}
has a local maximum at
\begin{equation}
\rho=\rho_{max}\equiv \frac{2 \mu ((2\kappa^2+2\mu^2)^{1/2}+\mu)}{(2 \kappa^2+\mu^2) K_0^{1/2} \kappa}\,,
\eqlabel{xmax}
\end{equation}
with $\kappa$ defined in as in \eqref{redeffa10}, and
\begin{equation}
-\frac{G_{tt}}{G_{x^ix^i}}\bigg|_{\rho=0}=1\,,\qquad -\frac{G_{tt}}{G_{x^ix^i}}\bigg|_{\rho=\rho_{max}}>1\,,
\qquad -\frac{G_{tt}}{G_{x^ix^i}}\bigg|_{\rho\to +\infty}=0\,.
\eqlabel{maxratio}
\end{equation}
Of course, at $\mu=0$, we can fully map the results for the black branes on the conifold in $x$ radial coordinate
to those in $\rho$: from eq.~(A.11) and (A.17) of \cite{Buchel:2018bzp}
\begin{equation}
\begin{split}
&K_1^{[x]}=P^2 g_0 k_s-\frac 12 P^2 g_0 \ln x+\calo\left(x^{3/4}\ln x\right)\,,\qquad g^{[x]}=g_0 \biggl[1+\calo(x\ln x)\biggr]\,,
\\
&h^{[x]}=\frac{P^2 g_0}{a_0^2} \left(\frac 18+\frac{k_s}{4}\right)-\frac{P^2g_0}{8a_0^2}\ln x +\calo(x^{1/2}\ln x)\,,
\end{split}
\eqlabel{khold}
\end{equation}
and using
\begin{equation}
-\frac{G_{tt}}{G_{x^ix^i}}\equiv (1-x)^2\equiv f(\rho)\qquad \Longrightarrow\qquad  x=1-\sqrt{f}\,,
\eqlabel{rhox}
\end{equation}
with the asymptotic expansions \eqref{bs5}, \eqref{bs8} and \eqref{bs9} we find
\begin{equation}
g_s=g_0\,,\qquad K_0=P^2 g_0\left(k_s+\frac12\ln 2-\frac12\ln(-f_{4,0})\right)\,.
\eqlabel{match1}
\end{equation}
Additionally, matching the $g_{tt}$ components of the metric on $\calm_5$ \eqref{5met} in different radial coordinates,
\begin{equation}
g_{tt}\equiv \frac{(1-x)^2}{(h^{[x]})^{1/2}(2x-x^2)^{1/2}}\equiv \frac{f}{h^{1/2}\rho^2}\qquad \Longrightarrow\qquad f_{4,0}=-a_0^2\,,
\eqlabel{match2}
\end{equation}
where $a_0^2\propto s T$ of the thermal states of the black branes on the conifold.
The scaling symmetries of the equations of motion in \cite{Aharony:2007vg,Buchel:2010wp,Buchel:2018bzp} were always used to
set $a_0=1$, with varying $k_s$. This means, effectively,  keeping the temperature scale 'fixed', while varying the
strong coupling scale $\Lambda$ of the theory. The results were presented in terms of a single dimensionless ratio
$\frac{T}{\Lambda}$. Setting $a_0=1$ implies that, see \eqref{match2},
\begin{equation}
f_{4,0}=-1\,.
\eqlabel{match3}
\end{equation}
While \eqref{match3} was established for $\mu=0$, we keep it even for  $\mu\ne 0$. This means that out of
three scales $\mu,\Lambda,T$ relevant to the black holes on the conifold, we fixed one of them --- the temperature. As in
\cite{Aharony:2007vg,Buchel:2010wp,Buchel:2018bzp}, this is consistent  as long as we use dimensionless observables.  

We now outline the steps  used to produced the data reported in section \ref{meat}.

\subsubsection{Klebanov-Tseytlin black holes at $\mu\ne 0$}\label{techkt}

In addition to \eqref{match3}, we use the scaling symmetries \eqref{kssym1} and \eqref{kssym2} to fix
\begin{equation}
P=1\,,\qquad g_s=1\,.
\eqlabel{fix1}
\end{equation}
For the data files of \cite{Buchel:2018bzp} we identify the black brane solution corresponding to
(see \eqref{defkolambda})
\begin{equation}
K_0=4\qquad \Longleftrightarrow\qquad \Lambda=2^{1/4} e^{-2}\,,
\eqlabel{fix2}
\end{equation}
\ie when $k_s=4-\ln2/2$. We use the coordinate transformation $x\leftrightarrow \rho$ of \eqref{rhox}
to map the corresponding black brane data to \eqref{uvvevsb}, \eqref{irvevsbb}, remembering the
constraints of the unbroken chiral symmetry \eqref{uvirchiralt}. We thus obtained our first entry
line for the Klebanov-Tseytlin black hole data  file at $\mu=0$. We populate this  {\it KT-master} file, with each
line labeled by $\mu$, keeping fixed $f_{4,0},P,g_s,K_0$ as in \eqref{match3}-\eqref{fix2}.
Each entry line in the file represents a black hole at a single value of $T/\Lambda$, labeled by
\begin{equation}
\frac{\mu}{\Lambda}=2^{-1/4} e^{2}\ \mu\equiv \hat\mu\,.
\eqlabel{mulambdakt}
\end{equation}

Given this KT-master file, we can produce Klebanov-Tseytlin black holes at fixed $\mu/\Lambda$, for different
values of $T/\Lambda$ as follows\footnote{We use a separate file for distinct values of $\mu/\Lambda$.}:
\nxt Pick an entry from the KT-master file at required value of $\hat\mu_r$. This is a Klebanov-Tseytlin
black holes at {\it some} temperature $T$, see \eqref{thaw}.
\nxt We continue to keep \eqref{match3} and \eqref{fix1}, but we now vary both $K_0$ and $\mu$ along the line
\begin{equation}
\left\{K_0,\mu\right\}\ \equiv\ \left\{K_0, \hat\mu_r 2^{1/4} e^{-K_0/2}\right\} \,.
\eqlabel{k0muline}
\end{equation}
Since we vary $K_0$, we vary the strong coupling scale of the theory
\begin{equation}
\Lambda=2^{1/4} e^{-K_0/2}\,,
\eqlabel{varyk0}
\end{equation}
and thus, the ratio $T/\Lambda$. The corresponding change in $\mu$ as per \eqref{k0muline}
guarantees that during this $T/\Lambda$ variation
\begin{equation}
\frac{\mu}{\Lambda}=\hat\mu_r 2^{1/4} e^{-K_0/2}\ \times\ 2^{-1/4} e^{K_0/2}\ =\ \hat{\mu}_r
={\rm constant}\,.
\eqlabel{mulvar}
\end{equation}

Each Klebanov-Tseytlin black hole data file (at a given value of $\hat\mu_r$) is verified
with respect to the first law of thermodynamics  \eqref{firstlaw}. For example, the left panel of 
fig.~\ref{fl20} verifies this for the Klebanov-Tseytlin black hole at $\mu=\mu_1$ \eqref{mu1mu2} ---
the accuracy is $\sim 10^{-7}$ and better. This is a typical accuracy achieved for the Klebanov-Tseytlin
black holes at $\mu/\Lambda \ne 0$.

\subsubsection{Computation of $T_\csb(\mu)$}\label{techtcsb}

$T_{\csb}$ is the temperature of the onset of the perturbative chiral symmetry breaking
instability of $\calt_{decon}^s$ states, \ie the Klebanov-Tseytlin black holes. To identify it we study
$\csb$ fluctuations about $\calt_{decon}^s$ states and search for the normalizable mode.

As in \eqref{deffluc1}, we set
\begin{equation}
f_a\equiv f_3+\df\,,\ f_b\equiv f_3-\df\,,\ K_1\equiv K+\dk_1\,,\ K_3\equiv K-\dk_1\,,\
K_2\equiv 1+\dk_2\,,
\eqlabel{defflt}
\end{equation}
with the remaining metric functions and the string coupling as in $\calt_{decon}^s$ thermal states $\{f,f_c,h,g\}$.
It is straightforward to verify that truncation to $\{\df,\dk_{1}, \dk_2\}$
is consistent (at the linearized level).
Equations of motion for the fluctuations and their asymptotic expansions in the UV $(\rho\to 0)$ and the IR
($y=\frac 1\rho$) are collected in appendix \ref{flucts}. The expansions are characterized by 6 UV/IR parameters 
\begin{equation}
\begin{split}
&{\rm UV:}\qquad \{\df_{3,0}\,,\, \dk_{2,3,0}\,,\, \dk_{2,7,0}\}\,;\\
&{\rm IR:}\qquad \{\df_{0}^h\,,\ \dk_{1,0}^h\,,\ \dk_{2,0}^h\}\,,
\end{split}
\eqlabel{uvirparslin2}
\end{equation}
Without the loss of generality, we fix the overall normalization of the linearized $\csb$ fluctuations  setting
\begin{equation}
\df_{0}^h\equiv 1\,.
\eqlabel{fh0fixed}
\end{equation}
Note that we have only 5 parameters in \eqref{uvirparslin2}, specifying a solution to
3 second-order differential equations \eqref{tsfl1}-\eqref{tsfl3}
for $\{\df,\dk_{1},\dk_2\}$ on $\calt_{decon}^s$ background parameterized by $K_0$, with given $\mu/\Lambda$.
For a generic value of $K_0$ (equivalently $T/\Lambda$) there is no solution: given $\mu/\Lambda$,
$K_0$ must be tuned to identify the $\csb$ normalizable mode. This tuned value of $K_0$, for the
corresponding background Klebanov-Tseytlin black hole, would determine the temperature $T$ \eqref{thaw}, which is
precisely $T_{\csb}$ at a given value $\mu/\Lambda$:
\begin{equation}
\frac{T_\csb}{\Lambda}=\frac{f_1^h}{4\pi \sqrt{h_0^h}}\ \times\ 2^{-1/4} e^{K_0/2}\,.
\eqlabel{tcsbl}
\end{equation}
The access to the numerical data for the chiral symmetry breaking for the black branes on the conifold
\cite{Buchel:2010wp} greatly simplifies our task:
\nxt Using the map \eqref{rhox}, with \eqref{match3} and \eqref{fix1},
we identify from the data in \cite{Buchel:2010wp} the onset of the chiral symmetry breaking
at $\mu=0$ to occur at
\begin{equation}
K_0=-0.43170(7)\,,\qquad f_1^h=3.8074(5)\,,\qquad h_0^h=0.14353(3)\,,
\eqlabel{k0mu0}
\end{equation}
leading to (from \eqref{tcsbl})
\begin{equation}
\frac{T_\csb}{\Lambda}\bigg|_{\mu=0}=0.54193(5)\,.
\eqlabel{tcsbl1}
\end{equation}
\nxt We continue to keep \eqref{match3} and \eqref{fix1}, and be vary $\mu$, 
simultaneously solving for the Klebanov-Tseytlin black hole $\{f,f_c,f_3,h,K,g\}$
and the $\csb$ fluctuations $\{\df,\dk_{1},\dk_2\}$. Note that in total we have
8 second-order ODEs\footnote{Recall that for the Klebanov-Tseytlin  black hole
$f_a=f_b=f_3$, $K_1=K_2=K$ and $K_2=1$.} \eqref{b1}, \eqref{b2}, \eqref{b5}, \eqref{b6},
\eqref{b9},  \eqref{tsfl1}-\eqref{tsfl3}, and a single  first order equation \eqref{bc}.
To specify the solution we need to adjust $8\times 2 +1=17$ parameters.
These are (see \eqref{uvvevsb},  \eqref{irvevsbb}  and \eqref{uvirflucts}):
\begin{equation}
\begin{split}
&{\rm UV}:\qquad \{K_0,f_{a,1,0},f_{c,4,0},g_{4,0},f_{a,6,0},f_{c,8,0}\}\qquad {\rm and}\qquad
\{\df_{3,0},\dk_{2,3,0},\dk_{2,7,0}\}\,;\\
&{\rm IR}:\qquad \{f_{a,0}^h,f_{c,0}^h,h_0^h,K_{1,0}^h,g_0^h,f_1^h\}\qquad {\rm and}\qquad \{\dk_{1,0}^h,\dk_{2,0}^h\}\,.
\end{split}
\eqlabel{parcsb1}
\end{equation}
Given \eqref{parcsb1} we compute from \eqref{tcsbl} $T_\csb(\mu)/\Lambda$ as a function of
$\mu/\Lambda$.

\begin{figure}[t]
\begin{center}
\psfrag{m}[cc][][1][0]{{$\mu_{num}^2$}}
\psfrag{k}[bb][][1][0]{{$1/\dk_{2,3,0}$}}
\psfrag{l}[tt][][1][0]{{$\mu/\Lambda$}}
\includegraphics[width=2.6in]{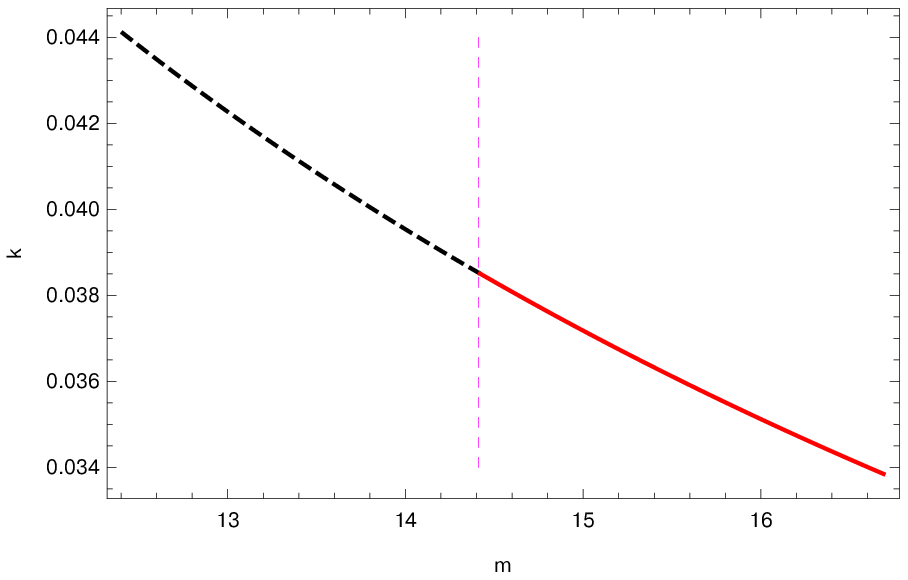}\
\includegraphics[width=2.6in]{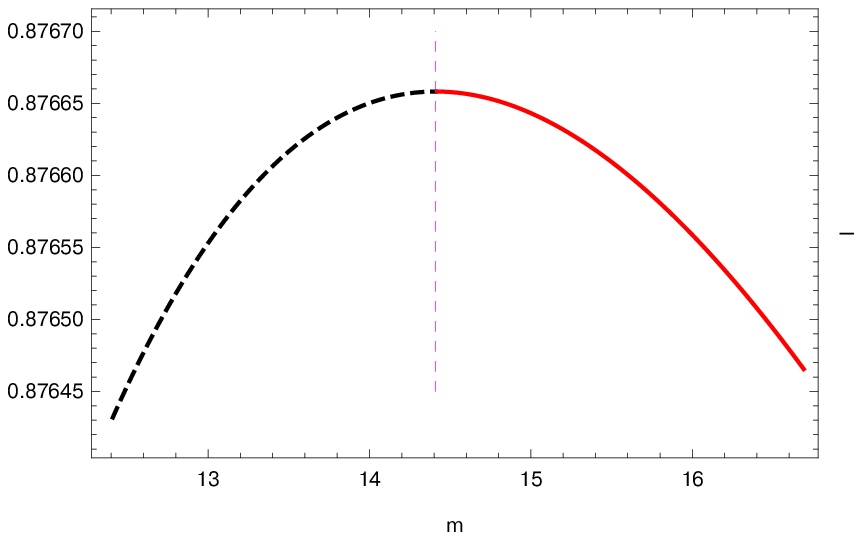}
\end{center}
  \caption{Variation of parameters \eqref{parcsb1} of the critical Klebanov-Tseytlin black hole
  with $\mu_{num}$: while $\dk_{2,3,0}$ monotonically increases with $\mu_{num}$ (left panel),
  $\mu/\Lambda$
  (right panel) reaches a maximum at $\mu_{num}=\mu_{num,max}$ \eqref{mumax},
  represented by the vertical dashed
  magenta line. 
}\label{nonmon}
\end{figure}

\begin{figure}[t]
\begin{center}
\psfrag{t}[bb][][1][0]{{$T_\csb/\Lambda$}}
\psfrag{e}[tt][][1][0]{{$\hat\cale_\csb$}}
\psfrag{m}[cc][][1][0]{{$\mu/\Lambda$}}
\includegraphics[width=2.6in]{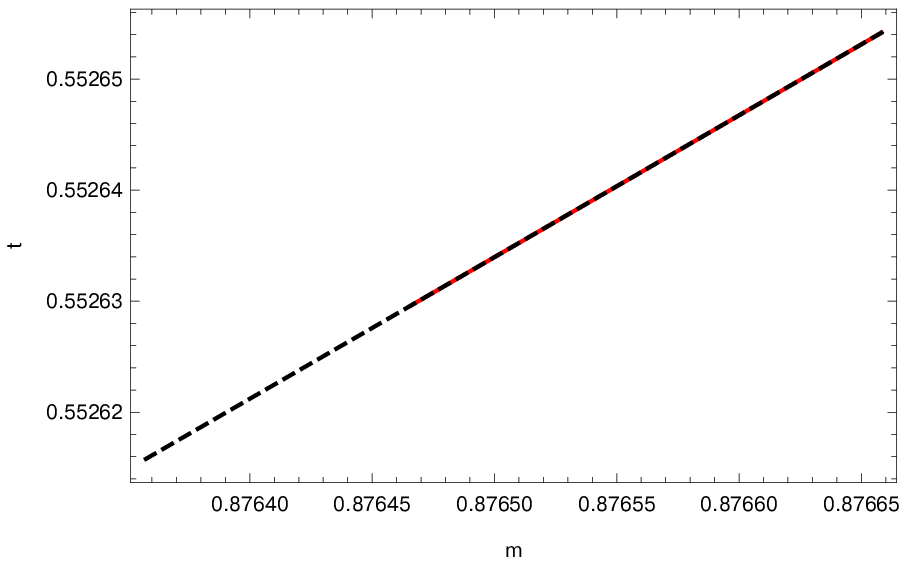}\
\includegraphics[width=2.6in]{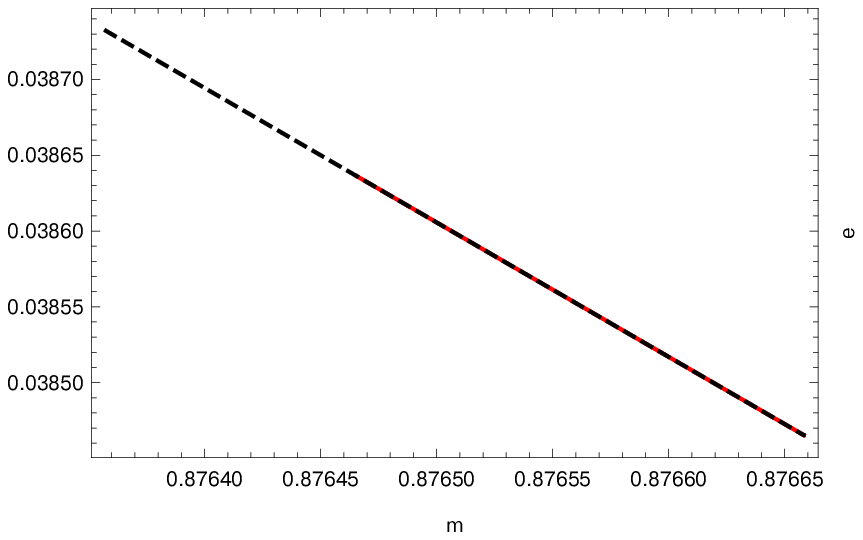}
\end{center}
  \caption{The critical Klebanov-Tseytlin black holes for $\mu_{num}<\mu_{num,max}$
  (dashed black curves in fig.~\ref{nonmon}) and $\mu_{num}>\mu_{num,max}$
  (red curves in fig.~\ref{nonmon}) have identical values of $T_\csb/\Lambda$ (left panel)
  and $\hat\cale_\csb$ (right panel) as a function of $\mu/\Lambda$.
}\label{nonmon2}
\end{figure}

We now explain the origin of $\mu_{KS}$  \eqref{defmuks} --- the terminal point of $T_\csb(\mu)$
or $\hat\cale_\csb(\mu)$. Parameters \eqref{parcsb1}, characterizing the critical
Klebanov-Tseytlin black holes along with $\csb$ fluctuations, are functions of $\mu_{num}$ ---
we use the subscript $\ _{num}$ to highlight that these are "numerical'' values of $\mu$ in the
computation scheme with \eqref{match3} and \eqref{fix1}.  
However, these parameters are
{\it not all}  monotonic functions of $\mu_{num}$.  In  fig.~\ref{nonmon} we present the
variation of some parameters in \eqref{parcsb1} of the critical Klebanov-Tseytlin black hole
  with $\mu_{num}$: while $\dk_{2,3,0}$ monotonically increases with $\mu_{num}$, $\mu/\Lambda$ reaches a
  maximum at
  \begin{equation}
  \mu_{num}^2=\mu_{num,max}^2\equiv 14.41(0)\,,
\eqlabel{mumax}
\end{equation}
represented by the vertical dashed magenta line. This maximal value of
  $\mu/\Lambda$ is precisely  $\mu_{KS}/\Lambda$ --- there is no instability to $\csb$ fluctuations
  of Klebanov-Tseytlin black holes for $\mu/\Lambda> \mu_{KS}/\Lambda$. A remarkable test on our numerics, the
  critical Klebanov-Tseytlin black holes for $\mu_{num}<\mu_{num,max}$
  (black dashed curves in fig.~\ref{nonmon})
  and for $\mu_{num}>\mu_{num,max}$ (red curves in fig.~\ref{nonmon}) are physically identical:
  in fig.~\ref{nonmon2} we plot $T_{\csb}/\Lambda$ (left panel) and $\hat\cale_{\csb}$ (right panel)
  of these two branches as a function of $\mu/\Lambda$ --- there is a perfect overlap.

It is possible to set up numerics in such a way that we vary directly $\hat\mu\equiv \mu/\Lambda$ and thus set
\begin{equation}
\mu\equiv 2^{1/4} e^{-K_0/2}\ \hat\mu \,.
  \eqlabel{fixmu}
  \end{equation}
In this numerical scheme we have been unable to extend the critical line of the Klebanov-Tseytlin black holes
past $\hat\mu={\mu_{KS}}/{\Lambda}$
as well.

\subsubsection{Klebanov-Strassler black holes at $\mu\ne 0$}\label{techks}

In section \ref{techtcsb} we constructed the set of critical Klebanov-Tseytlin black holes, labeled
by $\mu/\Lambda \in [0,\mu_{KS}/\Lambda]$, unstable to spontaneous chiral symmetry breaking.
Each point of $T_\csb(\mu)$ (or $\hat\cale_\csb(\mu)$ ) of the phase diagrams discussed in
sections \ref{mu0sec} and \ref{meat} is a bifurcation point for the branch of Klebanov-Strassler
black holes, equivalently $\calt_{decon}^b$ --- the thermal deconfined states of the cascading gauge theory
with the spontaneously broken chiral symmetry.

Given a bifurcation point on the critical Klebanov-Tseytlin black hole branch, the construction of the Klebanov-Strassler black hole branch is straightforward.
In fact the procedure is identical to the one employed for the black branes in \cite{Buchel:2018bzp}, or
to the construction of $\calv_A^b$ vacua in section \ref{calvab}. An important subtlety is that we need
to keep $\mu/\Lambda$ fixed, allowing for the variation of the temperature/energy density
of the Klebanov-Strassler black hole. This is done as in section \ref{techkt}: we vary both $K_0$
and $\mu$ along the line \eqref{k0muline}, where the KS black hole branch label $\hat\mu$ 
is picked from the range $\hat\mu \in [0,\mu_{KS}/\Lambda]$. Specific procedural steps are as follows.
\nxt From \eqref{parcsb1} we identify the susceptibilities (compare with \eqref{defkappa}) as 
\begin{equation}
\biggl\{\chi_{f_{3,0}}\,,\, \chi_{k_{2,3,0}}\,,\, \chi_{k_{2,7,0}}\,,\,  \chi_{k_{1,0}^h}\,,
\chi_{k_{2,0}^h} 
\biggr\} \equiv  \biggl\{\frac{\df_{3,0}}{\df_0^h}\,,\,
\frac{\dk_{2,3,0}}{\df_0^h}\,,\, \frac{\dk_{2,7,0}}{\df_0^h}\,,\,
\frac{\dk_{1,0}^h}{\df_0^h}\,,\,\frac{\dk_{2,0}^h}{\df_0^h} 
\biggr\}\,,
\eqlabel{defkappabh}
\end{equation}
remembering \eqref{fh0fixed}.
\nxt Similar to \eqref{deflambdafluc}, we denote the amplitude of the chiral symmetry breaking
condensate  (see \eqref{defflt})
\begin{equation}
\df_{0}^h\equiv \frac 12 \left(f_{a,0}^h-f_{b,0}^h\right)=\lambda\,.
\eqlabel{deflambdaflucbh}
\end{equation}
Then,
\begin{equation}
\left\{\df_{3,0}\,\, \dk_{2,3,0},\, \dk_{2,7,0},\, \dd k_{1,0}^h,\,\dd k_{2,0}^h
\right\}=\lambda\ \{\chi_{f_{3,0}},\,
\chi_{k_{2,3,0}},\, \chi_{k_{2,7,0}},\,  \chi_{k_{1,0}^h},\,
\chi_{k_{2,0}^h}\} +\calo(\lambda^2)\,.
\eqlabel{allothervevesbh}
\end{equation}
\nxt Using \eqref{defflt}, keeping \eqref{match3} fixed, and correlating $\mu$ and $K_0$ variation
as in \eqref{k0muline}, we find to $\calo(\lambda^2)$:
\begin{equation}
\begin{split}
&K_0=K_0^{crit}+\calo(\lambda^2)\,,\qquad f_{a,1,0}=f_{a,1,0}^{crit}+\calo(\lambda^2)\,,\qquad f_{a,3,0}=
\chi_{f_{3,0}} \lambda+\calo(\lambda^2)\,,\\
&k_{2,3,0}=\lambda\chi_{k_{2,3,0}}+\calo(\lambda^2)\,,\ \
f_{c,4,0}=f_{c,4,0}^{crit}+\calo(\lambda^2)\,,\\
&g_{4,0}=g_{4,0}^{crit}+\calo(\lambda^2)\,,\ \ f_{a,6,0}=f_{a,6,0}^{crit}+\calo(\lambda^2)\,,\ \
k_{2,7,0}= \chi_{k_{2,7,0}} \lambda+\calo(\lambda^2)\,,\\
&f_{c,8,0}=f_{c,8,0}^{crit}+\calo(\lambda^2)\,,\ f_{a,0}^h=f_{a,0}^{h,crit}+\lambda+\calo(\lambda^2)\,,\
\\&f_{b,0}^h=f_{a,0}^{h,crit}-\lambda+\calo(\lambda^2)\,,\ \ 
f_{c,0}^h=f_{c,0}^{h,crit}+\calo(\lambda^2)\,,\ \  h_{0}^h=h_{0}^{h,crit}+\calo(\lambda^2)\,,\\
&K_{1,0}^h=K_{1,0}^{h,crit}+\chi_{k_{1,0}^h} \lambda+\calo(\lambda^2)\,,\qquad
K_{2,0}^h=1+\chi_{k_{2,0}^h} \lambda+\calo(\lambda^2)\,,\\
&K_{3,0}^h=K_{1,0}^{h,crit}-\chi_{k_{1,0}^h} \lambda+\calo(\lambda^2)\,,\ \ 
g_{0}^h=g_{0}^{h,crit}+\calo(\lambda^2)\,,\ \ f_{1}^h=f_{1}^{h,crit}+\calo(\lambda^2)\,,
\end{split}
\eqlabel{seedtypeabbh}
\end{equation}
where the superscript $\ ^{crit}$ stands for the values  of the corresponding parameters
in \eqref{parcsb1} along the critical line of the Klebanov-Tseytlin black holes. 
\nxt We construct fully nonlinear in $\lambda$ $\calt_{decon}^b$ states, \ie the Klebanov-Strassler black holes,
using the linearized approximation \eqref{seedtypeabbh} as a seed.

An important check on our numerics is the validity of the first law of thermodynamics 
\eqref{firstlaw}.  For example, the right panel of 
fig.~\ref{fl20} verifies this for the Klebanov-Strassler black hole at $\mu=\mu_1$ \eqref{mu1mu2} ---
the accuracy is $\sim 10^{-5}$ and better. This is a typical accuracy achieved for the Klebanov-Strassler
black holes at $\mu/\Lambda \ne 0$.

\subsubsection{Computation of $T_c(\mu)$}

$T_c(\mu)$ is the temperature of the confinement deconfinement phase transition.
There are distinct confined states: $\calt_{con,B}$ and $\calt_{con,A}^s$ (see fig.~\ref{tclarge})
and   $\calt_{con,A}^b$ (see fig.~\ref{tcsmall}). Correspondingly, there are distinct curves
$T_{c}(\mu)$ for the transition between the deconfined chirally symmetric states $\calt_{decon}^s$,
\ie the Klebanov-Tseytlin black holes, and these confined states. Computation of the deconfined
temperature is a numerically consuming procedure,
albeit very straightforward:
\nxt We follow section \ref{techkt} to generate spectra of Klebanov-Tseytlin black holes
labeled by $\mu/\Lambda$.
\nxt At fixed $\hat\mu\equiv \mu/\Lambda$, we compute the reduced
free energy density $\hat\calf_{\calt_{decon}^s}(T;\hat\mu)$ (see \eqref{defhatf}) of $\calt_{decon}^s$ states,
using the expressions in appendices \ref{lvtb} and  \ref{lvts}. 
\nxt The reduced free energy density of the confined states is temperature independent, and
is simply the reduced vacuum energy density of $\calv_{B}$, $\calv_A^s$ or $\calv_{A}^b$ \eqref{fise}
at the corresponding value of $\hat\mu$.
\nxt For example, for $\calt_{decon}^s\leftrightarrow \calt_{con,B}$ transition reported in  fig.~\ref{tclarge},
$T_c(\mu)$ is determined from
\begin{equation}
\hat\calf_{\calt_{decon}^s}(T;\hat\mu)\bigg|_{T=T_c(\mu)}\ =\ \calv_{B}(\hat\mu)\,.
\eqlabel{tcconfb}
\end{equation}
\nxt We replicated \eqref{tcconfb} for the transitions $\calt_{decon}^s\leftrightarrow \calt_{con,A}^s$ and
$\calt_{decon}^s\leftrightarrow \calt_{con,A}^b$, and extract the corresponding confinement/deconfinement
temperatures $T_c$.

\subsection{Conifold black holes at $\{T,\mu\}\gg \Lambda$}\label{perkt}

In this section we extend the work on black branes on the conifold with fluxes in the limit
$T\gg \Lambda$ of \cite{Aharony:2005zr,Gubser:2001ri,Buchel:2009bh} to Klebanov-Tseytlin
black holes, \ie $\calt_{decon}^s$ states. We work to the leading order in
the limit $\{T,\mu\}\gg \Lambda$, without
any particular hierarchy between $T$ and $\mu$. The main motivation for this analysis is to have yet another
check on the numerics and the holographic renormalization of the model.  

Chiral symmetry is unbroken in the limit $T\gg \Lambda$ \cite{Buchel:2000ch}, so our starting point
are the conformal $\calt_{decon}^s$ thermal states of section \ref{conformalts}.
As in \cite{Buchel:2009bh}, we use a perturbative in $P^2 g_s/\hK_0$ 
ansatz to solve  \eqref{b1}-\eqref{bc}:
\begin{equation}
\begin{split}
&f=\frac{4 (\hf_{a,1,0} \rho+1) (\hK_0 \mu^2 \rho^2+2 \hf_{a,1,0}^2 \rho^2+4 \hf_{a,1,0} \rho+4)}
{(\hf_{a,1,0} \rho+2)^4}\ \times\  
\biggl[1+\sum_{n=1}^\infty \left(\frac{P^2 g_s}{\hK_0}\right)^2\ f_{n}(\rho)\biggr]\,,\\
&f_a=f_b=\frac{(\hf_{a,1,0}\rho+2)^2}{4}\ \times\ \biggl[1+\sum_{n=1}^\infty \left(\frac{P^2 g_s}{\hK_0}\right)^2\ f_{3,n}(\rho)\biggr]\,,\\
&f_c=\frac{(\hf_{a,1,0}\rho+2)^2}{4}\ \times\ \biggl[
1+\sum_{n=1}^\infty \left(\frac{P^2 g_s}{\hK_0}\right)^2\ f_{c,n}(\rho)\biggr]\,,\\
&h=\frac{4\hK_0}{(\hf_{a,1,0}\rho+2)^4}\ \times\ \biggl[1+\sum_{n=1}^\infty \left(\frac{P^2 g_s}{\hK_0}\right)^2\ h_{n}(\rho)\biggr]\,,\\
&K_1=K_3= \hK_0\ \times\ \biggl[1+\sum_{n=1}^\infty \left(\frac{P^2 g_s}{\hK_0}\right)^2\ k_{1,n}(\rho)\biggr]\,,\qquad
K_2\equiv 1\,,\\
&g=g_s\ \times\  \biggl[1+\sum_{n=1}^\infty \left(\frac{P^2 g_s}{\hK_0}\right)^2\ g_{n}(\rho)\biggr]\,,
\end{split}
\eqlabel{persolbh}
\end{equation}
were we used the conformal solution \eqref{solvecftts}  in the limit $\Lambda/T\to 0$.
It is convenient to introduce 
\begin{equation}
x\equiv \frac{\hf_{a,1,0}}{2}\rho\,,\qquad q\equiv \frac{\mu^2 \hK_0}{\hf_{a,1,0}^2}\,,\qquad z\equiv
\frac{P^2g_s}{\hK_0}\,.
\eqlabel{defqx}
\end{equation}
All the functions $f_n$, $f_{3,n}$, $f_{c,n}$, $h_n$, $k_{1,n}$, $g_n$ become
the functions of the new radial coordinate $x$, with the parametric dependence on $q$.
We present the explicit equations and the asymptotic expansions for these functions for $n=1$
in appendix \ref{perbh}.

Using \eqref{defqx}, the asymptotic expansions \eqref{uvper1}-\eqref{irper}, we
identify\footnote{Note that the relation for $K_0$ is exact to all orders in $z$, provided
the additive integration constants are set to zero for all $k_{1,n}$ functions, see
\eqref{defconst} for $k_{1,1}$. }
(compare \eqref{uvvevsb} and \eqref{irvevsbb})
\begin{equation}
\begin{split}
&K_0=\hK_0 \biggl(1-2\ln \frac{\hf_{a,1,0}}{2}\ z\biggr)\,,\qquad
f_{a,1,0}=\hf_{a,1,0}\left(1+\frac12\hf_{c,1.0}\ z+\calo(z^2)\right)\,,\\
\end{split}
\eqlabel{somepert1}
\end{equation}
\begin{equation}
\begin{split}
&f_{4,0}=\frac{1}{16} \hf_{a,1,0}^4 (2 q-1)+\frac{1}{96} \hf_{a,1,0}^4
\biggl(q (8 q\ln2 +36 \ln2+3 q)-4 q (2 q+9) \ln\hf_{a,1,0} \\&+6 \hf_{1,4,0}\biggr)\ z+\calo(z^2)
\,,\qquad f_{c,4,0}=-\frac{1}{144}\hf_{a,1,0}^4 \left(5q^2+3q-3\right)\ z+\calo(z^2)\,,\\
\end{split}
\eqlabel{somepert2}
\end{equation}
\begin{equation}
\begin{split}
&f_{a,0}^h=\frac 14\hf_{a,1,0}^2
\left(1+f_{3,1,0}^h\ z+\calo(z)^2\right)\,,\qquad f_{c,0}^h=\frac 14\hf_{a,1,0}^2
\left(1+f_{c,1,0}^h\ z+\calo(z)^2\right)\,,
\end{split}
\eqlabel{somepert3}
\end{equation}
\begin{equation}
\begin{split}
&h_{0}^h=\frac{4\hK_0}{\hf_{a,1,0}^4}\left(1+\hh_{1,0}^h\ z
+\calo(z^2)\right)\,,\qquad f_{1}^h=\frac{4(q+2)}{\hf_{a,1,0}} \left(1+\hf_{1,0}^h\ z+\calo(z^2)\right)\,,
\end{split}
\eqlabel{somepert4}
\end{equation}
where we identified only the parameters necessary to compute the energy
density $\cale$ \eqref{eptb},
the entropy density $s$ \eqref{enttb} and the temperature $T$ \eqref{thaw}.

Using \eqref{defkolambda}, \eqref{somepert1} and \eqref{defqx}
we identify
\begin{equation}
\frac{\mu^2}{\Lambda^2}=2^{3/2} P^2 g_s\ \frac{q}{\hK_0}\ e^{\frac{\hK_0}{P^2g_s}}\,.
\eqlabel{mulambpert}
\end{equation}
We produce the numerical data for Klebanov-Tseytlin black holes using $q$ as a label
(see appendix \ref{perbh}) ---
to discuss thermodynamics of a  {\it fixed} theory we must keep $\{P,g_s,\mu,\Lambda\}$
constant as the temperature $T$ varies\footnote{Here, the temperature variation is induced
by variation of $q$.}.  This obviously necessitates that
\begin{equation}
\hK_0=\hK_0(q)\qquad \Longrightarrow\qquad \frac{d\hK_0}{dq}=-\frac{P^2g_s}{q}\ \times\ \frac{\hK_0 }{\hK_0-P^2g_s}\,.
\eqlabel{k0q}
\end{equation}
The fact that $\hK_0$ {\it must} depend on temperature, compactification scale is not new
and was first observed in \cite{Buchel:2000ch}.

Notice that at $q=0$, \ie the limit of the Klebanov-Tseytlin
black brane,
\begin{equation}
f_{4,0}\bigg|_{q=0}=-\frac{1}{16}\hf_{a,1,0}^4\biggl(1-\hf_{1,4,0}\ z+\calo(z^2)\biggr)\,.
\eqlabel{f40q0}
\end{equation}
As discussed earlier, see \eqref{match3}, $f_{4,0}$ sets the overall mass scale, and
can be adjusted at will, without affecting physics, provided we express observables
as dimensionless quantities. In \eqref{match3} $f_{4,0}$ was fixed matching to
(fully nonlinear) numerics used for black branes, \ie at $q=0$ and finite $T/\Lambda$.
Here, we do the same, matching $f_{4,0}$ to the {\it perturbative} black brane numerics of  
\cite{Aharony:2007vg}. We omit the technical details and simply present the answer:
\begin{equation}
\hf_{a,1,0}=2\,,\qquad \hf_{1,4,0}=\frac 13\,.
\eqlabel{matchpert}
\end{equation}

Using \eqref{defqx}, \eqref{matchpert}, \eqref{somepert1}-\eqref{somepert4},
and the thermodynamic expressions of appendices \ref{lvtb} and \ref{lvts} we find
\begin{equation}
\frac{8\pi G_5\ \cale}{\mu^4}=\frac{\hK_0^2}{32} \biggl(
\frac{q^2+3 q+3}{q^2}+\frac{54 \hf_{c,1,0}-2 q-69}{6q}\ \frac{P^2g_s}{\hK_0}+\calo(z^2)
\biggr)\,,
\eqlabel{epert}
\end{equation}
\begin{equation}
\frac{8\pi G_5\ s}{\mu^3}=\frac{\pi \hK_0^2}{8q^{3/2}} \biggl(
1+\frac{\hh_{1,0}^h+4 \hf_{3,1,0}^h+\hf_{c,1,0}^h}{2}\ \frac{P^2g_s}{\hK_0}+\calo(z^2)
\biggr)\,,
\eqlabel{spert}
\end{equation}
\begin{equation}
\frac{T}{\mu}= \frac{q+2}{2\pi q^{1/2}} \biggl(
1+\frac{2 \hf_{1,0}^h-\hh_{1,0}^h}{2}\ \frac{P^2g_s}{\hK_0}+\calo(z^2)
\biggr)\,.
\eqlabel{tpert}
\end{equation}
The first law of thermodynamics in the form $d\cale = T ds$, keeping
$\{P,g_s,\mu,\Lambda\}$ constant, and using \eqref{k0q}, is automatically
satisfied at $\calo(z^0)$, while at $\calo(z)$ it leads to the constraint:
\begin{equation}
\begin{split}
&0\ =\ \calf\call_{[1],2}(q)\ q^2+\calf\call_{[1],1}(q)\ q+\calf\call_{[1],0}(q)\ \equiv\
\calf\call_{[1]}(q)\,,\\
\end{split}
\eqlabel{flpert1}
\end{equation}
with
\begin{equation}
\begin{split}
&\calf\call_{[1],2}=-\frac43 (\hf_{3,1,0}^h)'-\frac13 (\hf_{c,1,0}^h)'+3 (\hf_{c,1,0})'
-\frac13 (\hh_{1,0}^h)'-\frac23\,,\\
&\calf\call_{[1],1}=2 \hf_{3,1,0}^h+\frac12 \hf_{c,1,0}^h-\frac83 (\hf_{3,1,0}^h)'
-\frac23 (\hf_{c,1,0}^h)'-\frac23 (\hh_{1,0}^h)'+\hf_{1,0}^h-3 \hf_{c,1,0}+\frac{19}{6}\,,\\
&\calf\call_{[1],0}=4 \hf_{3,1,0}^h+\hf_{c,1,0}^h+2 \hf_{1,0}^h+\frac23\,,
\end{split}
\eqlabel{flpert2}
\end{equation}
where $ '\equiv d/dq$.
In fig.~\ref{flpert} we check the first law of thermodynamics \eqref{flpert1} presenting
\begin{equation}
\calc_{[1]}\ \equiv\ \frac{\calf\call_{[1]}(q)}{{\rm max} [ \calf\call_{[1],2}(q)\,,\,\calf\call_{[1],1}(q)
\,,\,\calf\call_{[1],0}(q)]}\,, 
\eqlabel{cdeef}
\end{equation}
as a function of $q$. We find that $\calc_{[1]}\sim 10^{-8}$ for the range $q\in[0,0.8]\bigcup[1.2,2]$.
We deliberately excluded the region around $q=1$ for an important reason: it turns out that almost
all parameters in \eqref{pertpar} diverge as $q\to 1$ (both from above and below). 
As we demonstrate now, this fact has interesting implications for the leading corrections
to the confinement/deconfinement
temperature $T_c$ for the transition $\calt_{decon}^s\leftrightarrow\calt_{con}^s$ in the limit
$\{T,\mu\}\gg \Lambda$.

\begin{figure}[ht]
\begin{center}
\psfrag{w}[bb][][1][0]{{$1/\calc_1$}}
\psfrag{r}[tt][][1][0]{{$1/\calc_2$}}
\psfrag{q}[tt][][1][0]{{$q$}}
\includegraphics[width=2.6in]{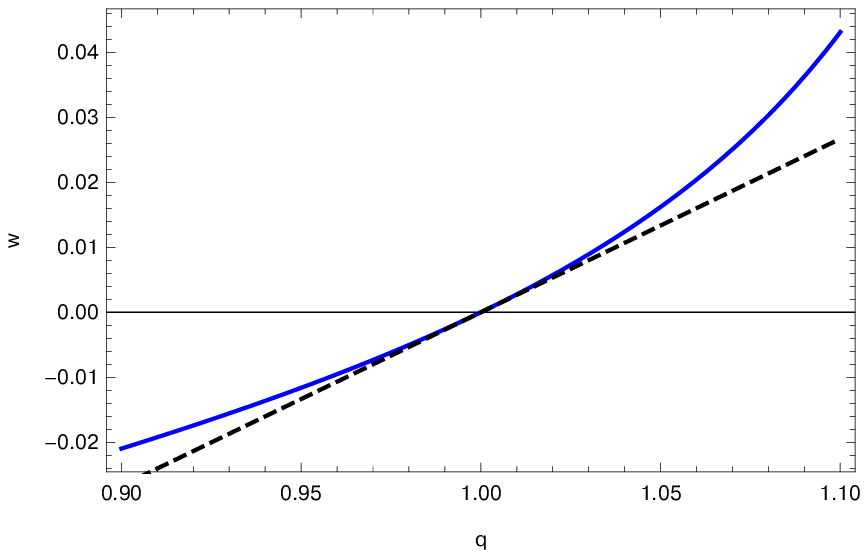}\
\includegraphics[width=2.6in]{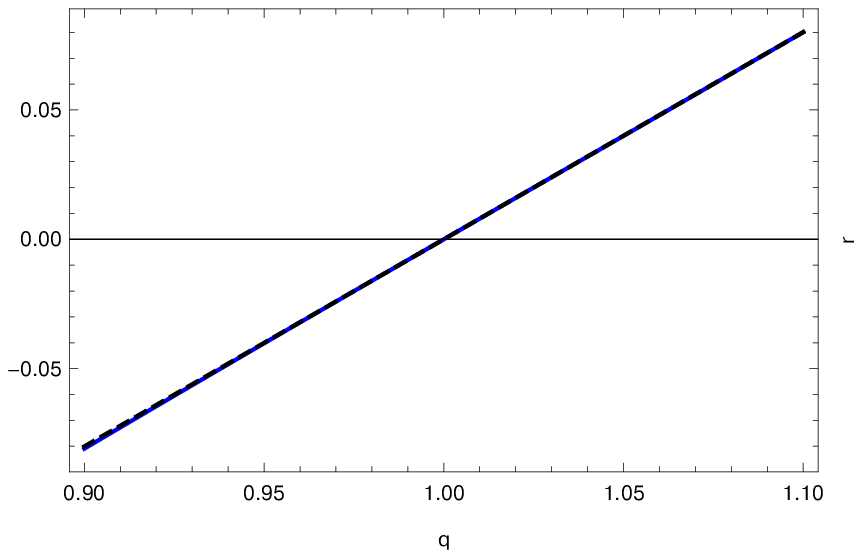}
\end{center}
  \caption{Certain linear combinations of the parameters  in \eqref{pertpar},
  $\calc_1$ (the left panel, \eqref{defcc1}) and $\calc_2$ (the right panel, \eqref{defcc2}),
  diverge as $q\to 1$. The dashed black lines are the tangents to the curves $1/\calc_1$
  and  $1/\calc_2$ at $q=1$.
}\label{c1c2}
\end{figure}

We will not presents results for the divergence of all the parameters  in \eqref{pertpar},
and focus instead on two combinations $\calc_1$ and $\calc_2$, which will be relevant for
the computation of the leading in $P^2g_s/\hK_0$ corrections to the Hawking-Page temperature
\eqref{tcondecon}:
\begin{equation}
\calc_1(q) \equiv\ \left(2 \hf_{1,0}^h
-9 \hf_{c,1,0}+\hf_{c,1,,0}^h+4 \hf_{3,1,0}^h\right)\ q+4 \hf_{1,0}^h+2 \hf_{c,1,0}^h+8 \hf_{3,1,0}^h \,,
\eqlabel{defcc1}
\end{equation}
\begin{equation}
\calc_2(q) \equiv 2 \hf_{1,0}^h-\hh_{1,0}^h\,.
\eqlabel{defcc2}
\end{equation}
In fig.~\ref{c1c2} the blue curves represent  $\calc_1^{-1}$ (the left panel) and $\calc_2^{-1}$
(the right panel) as a function of $q$. The dashed black lines represent the slopes to the
curves at $q=1$:
\begin{equation}
\begin{split}
&\calc_1^{-1}=\delta_1\ (q-1)+\calo((q-1)^2)\,,\qquad \delta_1=0.26695(6)\,,\\
&\calc_1^{-2}=\delta_2\ (q-1)+\calo((q-1)^2)\,,\qquad \delta_2=0.80086(7)\,.
\end{split}
\eqlabel{alphaspert}
\end{equation}
Note a serendipitous numerical fact:
\begin{equation}
3 \delta_1-\delta_2=-1.0(5)\times 10^{-9}\,,
\eqlabel{ddm}
\end{equation}
\ie with very good numerical accuracy  $3\delta_1=\delta_2$. 
We are now ready to compute the correction to $T_c$ as $\mu/\Lambda\to \infty$ for Klebanov-Tseytlin
black holes. From \eqref{epert}-\eqref{tpert} we have
\begin{equation}
\begin{split}
\frac{8\pi G_5\ \calf\left[\calt_{decon}^s\right]}{\mu^4}=&
\frac{8\pi G_5\ \cale}{\mu^4}-\frac{8\pi G_5\ s}{\mu^3}\times
\frac{T}{\mu}\\
=&\frac{\hK_0^2}{32}\biggl(\frac{q^2+q-1}{q^2}-\frac{(6\calc_1+2q^2+69q)}{6q^2}\
\frac{P^2 g_s}{\hK_0}
+\calo(z^2)
\biggr)\,,
\end{split}
\eqlabel{fpertq}
\end{equation}
where $\calc_1$ is given \eqref{defcc1}. For the confined state,
\begin{equation}
\frac{8\pi G_5\ \calf\left[\calt_{con}^s\right]}{\mu^4}=\frac{8\pi G_5\
\cale\left[\calv_{A}^s\right]}{\mu^4}=\frac{\hK_0^2}{32}\biggl(1+\gamma
\frac{P^2g_s}{\hK_0}+\calo(z^2)\biggr)\,,
\eqlabel{fpertq1}
\end{equation}
where the coefficient $\gamma$ was computed in\footnote{We independently verified
that result here.} \cite{Buchel:2011cc}
\begin{equation}
\gamma=-2.2725(9)\,.
\eqlabel{valgamma}
\end{equation}
Confinement/deconfinement transition occurs when
\begin{equation}
\calf\left[\calt_{con}^s\right]=\calf\left[\calt_{decon}^s\right]\,.
\eqlabel{perttrans}
\end{equation}
To leading order
in $P^2g_s/\hK_0$ it happens at
\begin{equation}
q\bigg|_{\rm leading\ order}=1\,.
\eqlabel{qlead}
\end{equation}
Precisely because $\calc_1$ diverges as $\propto {1}/{(q-1)}$ as $q\to 1$, the leading
correction to \eqref{qlead} in the limit $\mu/\Lambda\to \infty$ are nonanalytic in $P^2g_s/\hK_0$.
Using \eqref{alphaspert}, we find from \eqref{perttrans}
\begin{equation}
q\bigg|_{con/decon}=1\pm \delta_1^{-1/2}\ \left(\frac{P^2g_s}{\hK_0}\right)^{1/2}+\calo\left(\frac{P^2g_s}{\hK_0}\right)\,.
\eqlabel{qsublead}
\end{equation}
To compute the correction to the transition temperature \eqref{tpert} due to \eqref{qsublead}
we need to remember that $\calc_2(q)$ \eqref{defcc2} diverges as $q\to 1$ as well ---
see \eqref{alphaspert}. We find
\begin{equation}
\begin{split}
\frac{T_c}{\mu}=&\frac{3}{2\pi}
\pm \frac{3\delta_1-\delta_2}{4\delta_1^{1/2}\delta_2\pi}\ \left(\frac{P^2g_s}{\hK_0}\right)^{1/2}+\calo\left(\frac{P^2g_s}{\hK_0}\right)\\
=&\frac{3}{2\pi} + 0\times \left(\frac{1}{{2\ln\frac{\mu}{\Lambda}}}\right)^{1/2}
+\calo\left(\frac{1}{2\ln\frac{\mu}{\Lambda}}\right)\,,
\end{split}
\eqlabel{tcondeonpert}
\end{equation}
where in the second equality we used $3\delta_1-\delta_2=0$ (motivated by \eqref{ddm})
and \eqref{mulambpert} at $q=1$ and to leading order in $\mu/\Lambda\to\infty $.
Thus, to compute the leading in the limit $\mu/\Lambda\to \infty$ nonvanishing correction to
the confinement/deconfinement 
transition temperature  $T_c^{conformal}$, and compare with
results reported in fig.~\ref{tmularge} we need to develop perturbative
solution to the Klebanov-Tseytlin black hole \eqref{persolbh} to order $n=2$.
We will not pursue this computation here.

\section{Conclusion and open questions}\label{conclude}

In this paper we studied the vacua and the black holes on the conifold of Type IIB supergravity
with fluxes. These background geometries realize the holographic dual to vacua and
thermal states of the cascading gauge theory \cite{Klebanov:2000hb} on a 3-sphere. We uncovered 
rich phase diagrams in the canonical and microcanonical ensembles, identified distinct
confinement/deconfinement phase transitions, spontaneous breaking of the chiral symmetry
in vacua and in thermal states. Further details are provided in sections \ref{vacua}
and \ref{bhsec}.

We list now open question for the future work.
\begin{itemize}
\item In a review section \ref{conformalvas}, see fig.~\ref{fig1} in particular,
we recalled that $AdS_5\times T^{1,1}$
black holes we unstable to localization on $T^{1,1}$ \cite{Prestidge:1999uq,Hubeny:2002xn,Buchel:2015pla}.
There are no studies in the literature
on the localization instabilities in the presence of internal (3-form) fluxes on $T^{1,1}$.
\item We still do not know what is the end point of the chiral symmetry breaking
instability of Klebanov-Tseytlin black branes, \ie when $\mu=0$. Although Klebanov-Strassler black
branes are the preferred states in the microcanonical ensemble \cite{Buchel:2018bzp},
their translational invariant
horizon is unstable to gravitational perturbations, the sound modes in the dual cascading
gauge theory plasma. 
\item In holography, there is a simple relation between the thermodynamic and the dynamical
instabilities of the black branes \cite{Buchel:2005nt}. In this work we established that
Klebanov-Strassler black holes, as well as some branches of the Klebanov-Tseytlin
black holes, have a negative specific heat, \ie are thermodynamically unstable.
Whether or not they are dynamically unstable (when $\mu\ne 0$) is an open question.
\item In this paper we used holography to study the vacua and the thermal states of the interesting strongly coupled
quantum field theory theory which has a confinement and a spontaneous breaking of a symmetry.
It would be very interesting to understand how to enlarge the holographic duality used here,
and include the background magnetic field.\footnote{See
\cite{Buchel:2020vkv} for a recent related discussion in  $\caln=2^*$ model.}.
This would open
the possibility to explore confinement/deconfinement and chiral symmetry breaking
in the presence of the magnetic field in a top-down holographic model.
A challenge is to identify a simple enough consistent truncation including a
bulk gauge field for the class of models in \cite{Cassani:2010na,Buchel:2014hja}.  
\item In this paper we identified certain conifold vacua  which are unstable to the spontaneous
breaking of the chiral symmetry, and could 'thermalize' into Klebanov-Strassler black holes.
It would be interesting to simulate this dynamics.
\item There is an interesting interval in the $S^3$ compactification scale
of the cascading gauge theory, namely $\mu\in [\mu_{KS},\mu_u]$. In this range, chirally symmetric vacua
of the conifold with fluxes are unstable to symmetry breaking fluctuations, yet, there are no
Klebanov-Strassler black holes, see fig.~\ref{ecsb}. What is the fate of these vacua instabilities? 
\end{itemize}

\section*{Acknowledgments}
This research is supported in part by Perimeter Institute for Theoretical Physics.
Research at Perimeter Institute is supported in part by the Government
of Canada through the Department of Innovation, Science and Economic Development
Canada and by the Province of Ontario through the Ministry of Colleges and
Universities. This work was further supported by
NSERC through the Discovery Grants program.

\appendix
\section{Equations of motion, asymptotic expansions, and the holographic renormalization}\label{eomshol}

\subsection{Case (A): horizonless warped deformed conifold with fluxes}\label{apa1}

Using the metric ansatz \eqref{metpar} we derive from \eqref{5action} the following equations of motion: 
\begin{equation}
\begin{split}
&0=f_1''+\frac{2 (f_1')^2}{f_1}-\frac{f_1'}{2} \biggl(
\frac6\rho-\frac{2 f_b'}{f_b}-\frac{2 f_a'}{f_a}-\frac{f_c'}{f_c}
\biggr)-\frac{2 h \mu^2}{f_1}\,,
\end{split}
\eqlabel{a1}
\end{equation}
\begin{equation}
\begin{split}
&0=f_a''-\frac{3f_a}{2f_1^2} (f_1')^2-\frac{1}{8 f_a} (f_a')^2
-\frac{f_a}{8f_b^2} (f_b')^2+\frac{f_a}{8h^2} (h')^2+\frac{f_a}{8g^2}(g')^2
+\frac{g P^2}{36h f_b} (K_2')^2\\
&+\frac{5}{32g h f_a P^2} (K_3')^2-\frac{3f_a}{32g h f_b^2 P^2} (K_1')^2
+\biggl(
\frac{3f_a'}{2f_1}-\frac{3h' f_a}{4h f_1}-\frac{3f_c' f_a}{4f_1 f_c}
-\frac{3f_b' f_a}{2f_b f_1}-\frac{3f_a}{2r f_1}
\biggr) f_1'\\
&+\biggl(
\frac{f_a'}{4f_c}-\frac{f_b' f_a}{4f_b f_c}
\biggr) f_c'
+\frac{f_a' f_b'}{2f_b}+\frac{h'f_a}{h \rho}-\frac{3 f_a'}{\rho}
-\frac{9(K_1-K_3)^2}{64f_b h g f_c \rho^2 P^2}
-\frac{K_1^2}{8f_b^2 h^2 f_a f_c \rho^2}+\frac{5 f_a}{\rho^2}
\\&
-\frac{g (K_2^2 (5 f_a^2-3 f_b^2)+12 f_b^2 (K_2-1)) P^2}{8h f_b^2 f_a f_c \rho^2}
-\frac{K_2 (K_1-K_3) (K_2 (K_1-K_3)-4 K_1)}{32f_b^2 h^2 f_a f_c \rho^2}
\\&-\frac{45f_a^2}{16f_b f_c \rho^2}
+\frac{9f_a}{8f_c \rho^2}+\frac{3 f_a}{f_b \rho^2}+\frac{27f_b}{16f_c \rho^2}
-\frac{9}{\rho^2}+\frac{3 f_c}{f_b \rho^2}+\frac{3h f_a\mu^2}{2f_1^2} \,,
\end{split}
\eqlabel{a2}
\end{equation}
\begin{equation}
\begin{split}
&0=f_b''-\frac{3f_b}{2f_1^2} (f_1')^2-\frac{f_b}{8f_a^2} (f_a')^2
-\frac{1}{8 f_b} (f_b')^2+\frac{f_b}{8h^2} (h')^2+\frac{f_b}{8g^2} (g')^2
+\frac{g P^2 (K_2')^2}{36h f_a}\\&-\frac{3f_b (K_3')^2}{32g h f_a^2 P^2}
+ \frac{5(K_1')^2}{32g h f_b P^2}
+\bigg(
 \frac{3f_b'}{2f_1}- \frac{3h' f_b}{4h f_1}-\frac{3f_b f_c'}{4f_1 f_c}
 -\frac{3f_b f_a'}{2f_a f_1}-\frac{3f_b}{2\rho f_1}\biggr) f_1'
+\biggl(
\frac{f_b'}{4f_c}\\
&- \frac{f_b f_a'}{4f_a f_c}
\biggr) f_c'+\frac{f_b' f_a'}{2f_a}-\frac{3 f_b'}{\rho}
+\frac{f_bh'}{h \rho} -\frac{9(K_1-K_3)^2}{64h f_a g f_c \rho^2 P^2}
-\frac{K_1^2}{8h^2 f_a^2 f_b f_c \rho^2}+\frac{5 f_b}{\rho^2}
+ \frac{27f_a}{16f_c \rho^2}\\
&+\frac{g (K_2^2 (3 f_a^2-5 f_b^2)+20 f_b^2 (K_2-1)) P^2}{8h f_c f_b f_a^2 \rho^2}
-\frac{K_2 (K_1-K_3) (K_2 (K_1-K_3)-4 K_1)}{32f_c f_b h^2 f_a^2 \rho^2}
\\&+\frac{9f_b}{8f_c \rho^2}-\frac{9}{\rho^2}-\frac{45f_b^2}{16f_a f_c \rho^2}
+\frac{3 f_b}{f_a \rho^2}+\frac{3 f_c}{f_a \rho^2}+\frac{3h f_b\mu^2}{2f_1^2} \,,
\end{split}
\eqlabel{a3}
\end{equation}
\begin{equation}
\begin{split}
&0=f_c''-\frac{f_c}{8f_a^2} (f_a')^2-\frac{f_c}{8f_b^2} (f_b')^2
-\frac{1}{2 f_c} (f_c')^2-\frac{g f_c P^2 (K_2')^2}{12h f_b f_a}
-\frac{3f_c (K_3')^2}{32g h f_a^2 P^2}-\frac{3 (K_1')^2 f_c}{32h g f_b^2 P^2}
\\&+\frac{f_c}{8g^2} (g')^2+\frac{f_c}{8h^2} (h')^2
-\frac{3f_c}{2f_1^2} (f_1')^2+\biggl(
\frac{9f_c'}{4f_1}-\frac{3h' f_c}{4h f_1}-\frac{3f_c f_b'}{2f_b f_1}
-\frac{3f_c f_a'}{2f_a f_1}-\frac{3f_c}{2\rho f_1}
\biggr) f_1'
\\&+\biggl(\frac{3f_b'}{4f_b}+ \frac{3f_a'}{4f_a}-\frac3\rho
\biggr) f_c'
-\frac{f_c f_b' f_a'}{2f_b f_a}+\frac{f_ch'}{h \rho}+\frac{27(K_1-K_3)^2}{64h f_b f_a g \rho^2 P^2}
-\frac{K_1^2}{8f_b^2 h^2 f_a^2 \rho^2}+\frac{5 f_c}{\rho^2}+\frac{63f_a}{16f_b \rho^2}
\\&+\frac{3g (K_2^2 (f_a^2+f_b^2)-4 f_b^2 (K_2-1)) P^2}{8h f_b^2 f_a^2 \rho^2}
-\frac{K_2 (K_1-K_3) (K_2 (K_1-K_3)-4 K_1)}{32f_b^2 h^2 f_a^2 \rho^2}
-\frac{63}{8 \rho^2}\\
&+\frac{3 f_c}{f_b \rho^2}+\frac{63f_b}{16f_a \rho^2}
+\frac{3 f_c}{f_a \rho^2}-\frac{9 f_c^2}{f_a f_b \rho^2}+\frac{3h f_c\mu^2}{2f_1^2} \,,
\end{split}
\eqlabel{a4}
\end{equation}
\begin{equation}
\begin{split}
&0=h''-\frac{9}{8 h} (h')^2-\frac{h}{8g^2} (g')^2+\frac{3h}{2f_1^2} (f_1')^2+\frac{h}{8f_a^2} (f_a')^2
+\frac{h}{8f_b^2} (f_b')^2+\frac{g P^2 (K_2')^2}{12f_b f_a}
\\&+\frac{3(K_3')^2}{32g f_a^2 P^2}+\frac{3(K_1')^2}{32g f_b^2 P^2}
+\biggl(
\frac{3f_b' h}{2f_b f_1}+\frac{15h'}{4f_1}+\frac{3f_a' h}{2f_a f_1}+\frac{3f_c' h}{4f_1 f_c}
+\frac{15h}{2\rho f_1}
\biggr) f_1'
+\biggl(
\frac{h'}{2f_c}+\frac{h}{\rho f_c}
\\&+\frac{f_b' h}{4f_b f_c}
+\frac{f_a' h}{4f_a f_c}
\biggr) f_c'
+\bigg(
\frac{f_b'}{f_b}+\frac{f_a'}{f_a}
-\frac{4}{\rho}\biggr) h'
+\frac{h f_b'f_a'}{2f_b f_a}+\frac{2 h f_b'}{f_b \rho}+\frac{2 h f_a'}{f_a \rho}
+\frac{45(K_1-K_3)^2}{64f_b f_a g f_c \rho^2 P^2}
\\&+ \frac{5g (K_2^2 (f_a^2+f_b^2)-4 f_b^2 (K_2-1)) P^2}{8f_b^2 f_a^2 f_c \rho^2}
+\frac{9K_2 (K_1-K_3) (K_2 (K_1-K_3)-4 K_1)}{32f_b^2 h f_a^2 f_c \rho^2}-\frac{13 h}{\rho^2}
\\&+\frac{9K_1^2}{8f_b^2 h f_a^2 f_c \rho^2}
+\frac{h f_c}{f_a f_b \rho^2}-\frac{3 h}{f_b \rho^2}-\frac{3 h}{f_a \rho^2}
+\frac{9h f_a}{16f_b f_c \rho^2}-\frac{9h}{8f_c \rho^2}+\frac{9h f_b}{16f_a f_c \rho^2}
-\frac{3h^2\mu^2}{2f_1^2} \,,
\end{split}
\eqlabel{a5}
\end{equation}
\begin{equation}
\begin{split}
&0=K_1''+\biggl(
\frac{f_c'}{2f_c}-\frac{h'}{h}-\frac{f_b'}{f_b}+\frac{f_a'}{f_a}+\frac{3 f_1'}{f_1}
-\frac{g'}{g}-\frac3\rho
\biggr) K_1'
-\frac{9f_b (K_1-K_3)}{2f_a f_c \rho^2}\\&-\frac{g (K_2 (K_2 (K_1-K_3)-4 K_1+2 K_3)+4 K_1) P^2}{h f_a^2 f_c \rho^2}\,,
\end{split}
\eqlabel{a6}
\end{equation}
\begin{equation}
\begin{split}
&0=K_2''+\biggl(
\frac{3 f_1'}{f_1}-\frac{h'}{h}+\frac{f_c'}{2f_c}+\frac{g'}{g}-\frac3\rho
\biggr) K_2'
- \frac{9(K_1-K_3) (K_2 (K_1-K_3)-2 K_1)}{8g h f_a f_b f_c P^2 \rho^2}
\\&-\frac{9(K_2 (f_a^2+f_b^2)-2 f_b^2)}{2f_a f_b f_c \rho^2}\,,
\end{split}
\eqlabel{a7}
\end{equation}
\begin{equation}
\begin{split}
&0=K_3''+\left(
\frac{f_c'}{2f_c}-\frac{h'}{h}+\frac{f_b'}{f_b}-\frac{f_a'}{f_a}
+\frac{3 f_1'}{f_1}-\frac{g'}{g}-\frac3\rho
\right) K_3'
+\frac{9f_a (K_1-K_3)}{2f_b f_c \rho^2}\\
&+\frac{K_2 g (K_2 (K_1-K_3)-2 K_1) P^2}{h f_b^2 f_c \rho^2}\,,
\end{split}
\eqlabel{a8}
\end{equation}
\begin{equation}
\begin{split}
&0=g''-\frac1g (g')^2+
\biggl(
\frac{f_c'}{2f_c}+\frac{f_b'}{f_b}+\frac{f_a'}{f_a}+\frac{3 f_1'}{f_1}-\frac3\rho\biggr) g'
-\frac{P^2 g^2 (K_2')^2}{9h f_a f_b}+\frac{(K_3')^2}{8h f_a^2 P^2}
+\frac{(K_1')^2}{8h f_b^2 P^2}\\&+\frac{9(K_1-K_3)^2}{16f_a f_b h f_c \rho^2 P^2}
-\frac{g^2 (K_2 (K_2 (f_a^2+f_b^2)-4 f_b^2)+4 f_b^2) P^2}{2h f_a^2 f_b^2 f_c \rho^2}\,.
\end{split}
\eqlabel{a9}
\end{equation}
Additionally we have the first order constraint
\begin{equation}
\begin{split}
&0=\frac{(K_1')^2}{P^2}+\frac{8g^2 P^2 (K_2')^2 f_b}{9f_a}
+\frac{f_b^2 (K_3')^2}{f_a^2 P^2}+\frac{4 f_b^2 h}{g} (g')^2
-\frac{48 h f_b^2 g}{f_1^2} (f_1')^2-8 g f_b h\biggl(
\frac{f_b'}{f_c}+\frac{f_b f_a'}{f_a f_c}\\
&+\frac{3 f_1' f_b}{f_1 f_c}-\frac{4 f_b}{\rho f_c}
\biggl) f_c'
-24 g f_b \biggl(
\frac{2 f_b' h}{f_1}+\frac{h' f_b}{f_1}+\frac{2 f_a' f_b h}{f_1 f_a}-\frac{6 f_b h}{f_1 \rho}
\biggr) f_1'
-4 g h (f_b')^2\\
&-\frac{16 g h f_b' f_a' f_b}{f_a}+\frac{4 g (h')^2 f_b^2}{h}-\frac{4 g h (f_a')^2 f_b^2}{f_a^2}+\frac{64 g h f_b' f_b}{\rho}
+\frac{32 g h' f_b^2}{\rho}+\frac{64 g h f_a' f_b^2}{\rho f_a}\\
&-\frac{9f_b (K_1-K_3)^2}{2f_a \rho^2 f_c P^2}-\frac{4 g^2 (K_2^2 (f_a^2+f_b^2)-4 f_b^2 (K_2-1)) P^2}{\rho^2 f_c f_a^2}
-\frac{4 g K_1^2}{h \rho^2 f_c f_a^2}
\\&-\frac{g K_2 (K_1-K_3) (K_2 (K_1-K_3)-4 K_1)}{h \rho^2 f_c f_a^2}
-2 g f_b h \biggl(
\frac{48 f_b}{\rho^2}+\frac{16 f_c}{f_a \rho^2}-\frac{48}{\rho^2}-\frac{48 f_b}{f_a \rho^2}
+\frac{9 f_a}{f_c \rho^2}\\&-\frac{18 f_b}{f_c \rho^2}+\frac{9 f_b^2}{f_a f_c \rho^2}
\biggr)+\frac{48 h^2 f_b^2 g\mu^2}{f_1^2}\,.
\end{split}
\eqlabel{ac}
\end{equation}
We explicitly verified that the constraint \eqref{ac} is consistent with \eqref{a1}-\eqref{a9}. Thus the
second-order equation for $f_c$ \eqref{a4} can be eliminated in favor of the constraint equation
\eqref{ac} where $f_c'$ enters linearly. In total, we expect that a solution is
specified by $8\times 2+ 1\times 1=17$ parameters (8 second-order equations for
$f_{a,b}, K_{1,2,3}, h,g,f_1$ and a single first-order equation for $f_c$). 
 
The general UV (as $\rho\to 0$ ) asymptotic solution of \eqref{a1}-\eqref{ac} describing the vacua
of the warped deformed conifold with fluxes and $S^3$ spatial boundary, \ie the vacua of the cascading
gauge theory on $S^3$, takes the form
\begin{equation}
\begin{split}
&f_a=1+f_{a,1,0} \rho+\rho^2 \biggl(
\frac{5}{16} \mu^2 P^2 g_s+\frac14 \mu^2 K_0+\frac14 f_{a,1,0}^2-\frac12 \mu^2 P^2 g_s \ln\rho\biggr)+\rho^3 f_{a,3,0}
\\&+\calo(\rho^4\ln^2 \rho)\,,
\end{split}
\eqlabel{as1}
\end{equation}
\begin{equation}
\begin{split}
&f_b=1+f_{a,1,0} \rho+\rho^2 \biggl(
\frac{5}{16} \mu^2 P^2 g_s+\frac14 \mu^2 K_0+\frac14 f_{a,1,0}^2-\frac12 \mu^2 P^2 g_s
\ln\rho\biggr)\\&+\rho^3 \biggl(
\frac12 \mu^2 f_{a,1,0} P^2 g_s-f_{a,3,0}\biggr)
+\calo(\rho^4\ln^2 \rho)\,,
\end{split}
\eqlabel{as2}
\end{equation}
\begin{equation}
\begin{split}
&f_c=1+f_{a,1,0} \rho+\rho^2 \biggl(
\frac 38 \mu^2 P^2 g_s+\frac14 \mu^2 K_0+\frac14 f_{a,1,0}^2-\frac12 \mu^2 P^2 g_s \ln\rho\biggr)
+\frac14 \mu^2 f_{a,1,0} P^2 g_s \rho^3\\&+\calo(\rho^4\ln^2 \rho)\,,
\end{split}
\eqlabel{as3}
\end{equation}
\begin{equation}
\begin{split}
&h= \frac18 P^2 g_s+\frac14 K_0-\frac12 P^2 g_s \ln\rho+\rho \biggl(
-\frac12 f_{a,1,0} K_0+f_{a,1,0} P^2 g_s \ln\rho\biggr)
+\rho^2 \biggl(
\frac{23}{288} \mu^2 P^4 g_s^2\\
&-\frac16 P^2 g_s \mu^2 K_0-\frac14 P^2 g_s f_{a,1,0}^2
-\frac18 \mu^2 K_0^2+\frac58 f_{a,1,0}^2 K_0+\frac{1}{96} P^2 g_s (32 P^2 g_s \mu^2
+48 K_0 \mu^2\\&-120 f_{a,1,0}^2) \ln\rho
-\frac12 \mu^2 P^4 g_s^2 \ln^2\rho\biggr)
+\rho^3 \biggl(
\frac14 P^2 g_s \mu^2 f_{a,1,0} K_0-\frac{13}{32} P^4 g_s^2 f_{a,1,0} \mu^2
\\&+\frac{11}{24} P^2 g_s f_{a,1,0}^3
+\frac38 \mu^2 f_{a,1,0} K_0^2-\frac58 K_0 f_{a,1,0}^3+\biggl(
\frac54 P^2 g_s f_{a,1,0}^3-\frac12 P^4 g_s^2 f_{a,1,0} \mu^2
\\&-\frac32 P^2 g_s \mu^2 f_{a,1,0} K_0\biggr)
\ln\rho+\frac32 P^4 g_s^2 f_{a,1,0} \mu^2 \ln^2\rho\biggr)+\calo(\rho^4\ln^3 \rho)\,,
\end{split}
\eqlabel{as4}
\end{equation}
\begin{equation}
\begin{split}
&K_1=K_0-2 P^2 g_s \ln\rho+P^2 g_s f_{a,1,0} \rho+\rho^2 \biggl(
\frac{1}{4} P^2 g_s (3 P^2 g_s \mu^2+K_0 \mu^2-f_{a,1,0}^2)
\\&-\frac12 \mu^2 P^4 g_s^2 \ln\rho\biggr)
+\rho^3 \biggl(
\frac{1}{12} P^2 g_s (8 k_{2,3,0}-8 P^2 f_{a,1,0} g_s \mu^2-3 K_0 f_{a,1,0} \mu^2+f_{a,1,0}^3+8 f_{a,3,0})
\\&+2 P^2 g_s f_{a,3,0} \ln\rho\biggr)+\calo(\rho^4\ln^2 \rho)\,,
\end{split}
\eqlabel{as5}
\end{equation}
\begin{equation}
\begin{split}
&K_2=1+\rho^3 \biggl(k_{2,3,0}+\left(3 f_{a,3,0}-\frac34 \mu^2 f_{a,1,0} P^2 g_s\right)
\ln\rho\biggr)+\calo(\rho^4\ln \rho)\,,
\end{split}
\eqlabel{as6}
\end{equation}
\begin{equation}
\begin{split}
&K_3=K_0-2 P^2 g_s \ln\rho+P^2 g_s f_{a,1,0} \rho+\rho^2 \biggl(
\frac14P^2g_s(3P^2g_s\mu^2+K_0\mu^2-f_{a,1,0}^2)
\\&-\frac12 \mu^2 P^4 g_s^2 \ln\rho\biggr)
+\rho^3 \biggl(\frac{1}{12}P^2g_s(f_{a,1,0}^3-4P^2g_sf_{a,1,0}\mu^2-3K_0f_{a,1,0}\mu^2-8f_{a,3,0}-8k_{2,3,0})
\\&+P^2 g_s (P^2 g_s f_{a,1,0}  \mu^2-2 f_{a,3,0}) \ln\rho\biggl)+\calo(\rho^4\ln^2 \rho)\,,
\end{split}
\eqlabel{as7}
\end{equation}
\begin{equation}
\begin{split}
&g=g_s \biggl(1-\frac14 \mu^2 P^2 g_s \rho^2+\frac14 \mu^2 f_{a,1,0} P^2 g_s \rho^3+\calo(\rho^4\ln \rho)\biggr)\,,
\end{split}
\eqlabel{as8}
\end{equation}
\begin{equation}
\begin{split}
&f_1=1+\rho^2 \biggl(-\frac{1}{16} \mu^2 P^2 g_s-\frac18 \mu^2 K_0+\frac14 \mu^2 P^2 g_s \ln\rho\biggr)
+\rho^3 \biggl(-\frac{1}{16} \mu^2 f_{a,1,0} P^2 g_s\\&+\frac18 \mu^2 f_{a,1,0} K_0
-\frac14 \mu^2 f_{a,1,0} P^2 g_s \ln\rho\biggr)+\calo(\rho^4\ln^2 \rho)\,.
\end{split}
\eqlabel{as9}
\end{equation}
It is characterized by the cascading gauge theory defining parameters
\begin{equation}
K_0\,,\ \mu\,, \ P\,,\ g_s \,,
\eqlabel{defuvcasea}
\end{equation}
correspondingly related to the strong coupling scale $\Lambda$ \eqref{defkolambda}, the $S^3$ compactification scale
$\mu$ \eqref{delm5metric}, the rank difference of the gauge group factors $M$ \eqref{defpm}, and the
renormalization group flow invariant sum of the gauge couplings \eqref{rg2}.  Additionally, there are
9 normalizable coefficients related to the diffeomorphism parameter $\alpha$ \eqref{leftover} and the expectation values of the various operators
in the boundary theory
\begin{equation}
\{f_{a,1,0}\,,\ f_{a,3,0}\,,\ k_{2,3,0}\,,\ f_{a,4,0}\,,\ f_{c,4,0}\,,\ g_{4,0}\,,\ f_{a,6,0}\,,\ k_{2,7,0}\,,\ f_{a,8,0}
\}\,.
\eqlabel{uvvevs}
\end{equation}

The IR (as $y\equiv \frac 1\rho\to 0$) asymptotics of \eqref{a1}-\eqref{ac} differ depending on the
topology of the background geometry. We call vacua of the cascading gauge theory theory with the
boundary spatial $S^3$ smoothly
shrinking to zero size $\va$, and vacua with the 2-cycle of the warped deformed conifold smoothly shrinking to zero
size $\vb$. Note that the 3-cycle of the warped deformed conifold can not vanish without producing a naked singularity
since it  supports nonzero (when $P\ne 0$) RR 3-form flux \eqref{pquantization}.

\begin{itemize}
\item $\calv_A$ vacua of the cascading gauge theory. To identify smooth geometries with vanishing $S^3$ as $y\to 0$
we introduce
\begin{equation}
f_1^h\equiv y^{-1}\ f_1\,,\qquad f_{a,b,c}^h\equiv y^2\ f_{a,b,c}\,,\qquad h^h\equiv y^{-4}\ h\,.
\eqlabel{aredef}
\end{equation}
The IR asymptotic expansion
\begin{equation}
\begin{split}
&f_{a,b,c}^h=\sum_{n=0} f_{a,b,c,n}^h\ y^{2n}\,,\qquad h^h=h^h_0+\sum_{n=1} h^h_n\ y^{2n}\,,\ \ K_{1,2,3}=\sum_{n=0} K_{1,2,3,n}^h\ y^{2n}\,,
\\
&g=\sum_{n=0} g_{n}^h\ y^{2n}\,,\qquad f_1^h=\mu  \sqrt{h^{h}_0}\ \biggl(
1+\sum_{n=1} f_{1,n}^h\ y^{2n}\biggr)\,,
\end{split}
\eqlabel{ircaseaa}
\end{equation}
is characterized by 8 parameters:
\begin{equation}
\{f_{a,0}^h\,,\, f_{b,0}^h\,,\,f_{c,0}^h\,,\,h_{0}^h\,,\,K_{1,0}^h\,,\,K_{2,0}^h\,,\,K_{3,0}^h\,,\,g_{0}^h\}\,.
\eqlabel{irvevs}
\end{equation}
Note that given given \eqref{ircaseaa},
\begin{equation}
\begin{split}
&\frac{1}{h^{1/2}\rho^2}\biggl(-dt^2+\frac{f_1^2}{\mu^2}\ \left(dS^3\right)^2\biggr)+\frac{h^{1/2}}{\rho^2}\ (d\rho)^2=
\frac{1}{\sqrt{h^h}}\biggl(-dt^2+\frac{(f_1^h)^2}{\mu^2}\ y^2\left(dS^3\right)^2 \biggr)\\&+
\sqrt{h^h} (dy)^2\qquad \underbrace{\longrightarrow}_{y\to 0}\qquad -\frac{1}{\sqrt{h_0^h}}\ dt^2+
\sqrt{h_0^h}\ \biggl(y^2\ \left(dS^3\right)^2+(dy)^2\biggr)\,,
\end{split}
\eqlabel{irlimitaa}
\end{equation}
\ie $S^3$ indeed smoothly shrinks to zero size as $y\to 0$. It is important to emphasize that
$\calv_A$ vacua defined by \eqref{ircaseaa} have either $U(1)$ or $\zet_2$ chiral symmetry ---
the chiral symmetry is unbroken in the former ($\calv_A^s$ vacua), and is
spontaneously broken in the
latter ($\calv_A^b$ vacua). Specifically, unbroken chiral symmetry dictates \eqref{symsector}, leading to 
\begin{equation}
\begin{split}
&{\rm UV:}\qquad f_{a,3,0}=\frac14\mu^2f_{a,1,0} P^2g_s\,,\qquad k_{2,3,0}=0\,,\qquad k_{2,7,0}=0\,;
\\
&{\rm IR:}\qquad f_{b,0}^h=f_{a,0}^h\,,\qquad K_{3,0}^h=K_{1,0}^h\,,\qquad K_{2,0}^h=1\,.
\end{split}
\eqlabel{uvirchiral}
\end{equation}
\item $\calv_B$ vacua of the cascading gauge theory. To identify smooth geometries with vanishing $S^2$ as $y\to 0$
we introduce
\begin{equation}
f_{a,b,c}^h\equiv y^2\ f_{a,b,c}\,,\qquad h^h\equiv y^{-4}\ h\,.
\eqlabel{bredef}
\end{equation}
The IR asymptotic expansion
\begin{equation}
\begin{split}
&f_{a}^h=f_{a,0}^h+\sum_{n=1} f_{a,n}\ y^{2n}\,,\ \ f_{b}^h=3y^2+\sum_{n=2} f_{b,n}\ y^{2n}\,,\ \
f_{c}^h=\frac 34f_{a,0}^h+\sum_{n=1} f_{c,n}\ y^{2n}\,,\\
&K_1=K_{1,3}^h y^3+\sum_{n=2}K_{1,n}y^{2n+1}\,,\qquad K_2=K_{2,2}^h y^2+\sum_{n=2}K_{2,n}y^{2n}\,,\\
&K_3=K_{3,1}^h y+\sum_{n=1}K_{3,n}y^{2n+1}\,,\qquad h^h=h^h_0+\sum_{n=1} h^h_n\ y^{2n}\,,\\
&g=\sum_{n=0} g_{n}^h\ y^{2n}\,,\qquad
f_1^h= \sum_{n=0} f_{1,n}^h\ y^{2n}\,,
\end{split}
\eqlabel{ircaseb}
\end{equation}
is characterized by 8 parameters:
\begin{equation}
\{f_{a,0}^h\,,\, h_{0}^h\,,\,K_{1,3}^h\,,\,K_{2,2}^h\,,\,K_{2,4}^h\,,\,K_{3,1}^h\,,\,g_{0}^h\,,\,f_{1,0}^h\}\,.
\eqlabel{irvevsb}
\end{equation}
Note that given given \eqref{ircaseb},
\begin{equation}
\begin{split}
&\frac{h^{1/2}}{\rho^2}\ (d\rho)^2+\frac{f_b h^{1/2}}{6}(g_1^2+g_2^2)=
\sqrt{h^h} (dy)^2+\frac{f_b^h(h^h)^{1/2}}{3}\ \frac12(g_1^2+g_2^2)\bigg|_{\rm 2-cycle}
\\
&\underbrace{\longrightarrow}_{y\to 0}\qquad 
\sqrt{h_0^h}\ \biggl(y^2\ \left(dS^2\right)^2+(dy)^2\biggr)\,,
\end{split}
\eqlabel{irlimitb}
\end{equation}
where $\bigg|_{\rm 2-cycle}$ means restriction to the $T^{1,1}$ 2-cycle. Following \cite{Herzog:2001xk},
this means setting $\psi=0$, $\phi_2=-\phi_1$, $\theta_2=-\theta_1$ in one-forms $\{g_i\}$ (see \eqref{3form1}) on $T^{1,1}$:
\begin{equation}
(g_1^2+g_2^2)\bigg|_{\rm 2-cycle}=2 \biggl((d\theta_1)^2+\sin^2\theta_1\ (d\phi_1)^2 \biggr)=2
\left(dS^2\right)^2\,.
\eqlabel{def2cycle}
\end{equation}
On the other hand, the 3-cycle supporting RR flux remains finite, provided $f_{a,0}^h h_0^h\ne 0$:
\begin{equation}
\begin{split}
&\frac{f_ch^{1/2}}{9}g_5^2+\frac{f_ah^{1/2}}{6}(g_3^2+g_4^2)
=\frac{f_c^h(h^h)^{1/2}}{9}g_5^2+\frac{f_a^h(h^h)^{1/2}}{6}(g_3^2+g_4^2)\\
&\underbrace{\longrightarrow}_{y\to 0}\qquad \frac{f_{a,0}^h(h_0^h)^{1/2}}{6}\
\left(\frac 12 g_5^2+g_3^2+g_4^2\right)\biggl|_{{\rm3-cycle}:\ \theta_2=\phi_2=0,\theta_1=2\eta, \psi=\xi_1+\xi_2,\phi_1=\xi_1-\xi_2}\\
&=\frac{f_{a,0}^h(h_0^h)^{1/2}}{6}\ 2\left((d\eta)^2+\cos^2\eta (d\xi_1)^2+\sin^2\eta (d\xi_2)^2\right)
=\frac{f_{a,0}^h(h_0^h)^{1/2}}{3}\ \left(dS^3\right)^2\,.
\end{split}
\eqlabel{3cycle}
\end{equation}
From \eqref{irlimitb}, $S^2$ indeed smoothly shrinks to zero size as $y\to 0$. Because
$f_a\ne f_b$ as $y\to 0$,  
$\calv_B$ vacua defined by \eqref{ircaseb}  have only $\zet_2$ chiral symmetry --- the chiral
symmetry is spontaneously broken. 
\end{itemize}

\subsubsection{Klebanov-Strassler solution \cite{Klebanov:2000hb} as $\mu\to 0$
limit of $\calv_B$ conifold vacua}\label{ksasvb}

The supersymmetric Klebanov-Strassler solution \cite{Klebanov:2000hb} is a decompactification
limit $\mu\to 0$ of $\calv_B$ conifold vacua discussed above (see section \ref{ksreview} for a review).
Indeed,
\begin{itemize}
\item in the UV, using \eqref{rrhouv} we identify (see \eqref{uvvevs}):
\begin{equation}
\begin{split}
&f_{a,1,0}=-2 \calq\,,\qquad f_{a,3,0}=\frac{3\sqrt{6}}{4}\ \epsilon^2\,,\qquad k_{2,3,0}=\frac{3 \sqrt{6}}{8} \epsilon^2 (3 \ln3-5 \ln2+4 \ln\epsilon)\,,\\
&f_{a,4,0}=\frac{3\sqrt{6}}{4}\ \calq\epsilon^2\,,\qquad f_{c,4,0}=0\,,\qquad g_{4,0}=0\,,\\
&f_{a,6,0}=\biggl(-\frac{27}{16} \ln2
+\frac{81}{50}+\frac{81}{80} \ln3+\frac{27}{20} \ln\epsilon\biggr) \epsilon^4+\frac{3\sqrt{6}}{4} \calq^3  \epsilon^2\,,
\\ &k_{2,7,0}=\frac{45\sqrt{6}}{8} \biggl(
\frac{57}{10}-5\ln2+3\ln3+4\ln\epsilon\biggr)\ \calq^4\epsilon^2\,,\\
&f_{a,8,0}=\left(\frac{27}{2}\ln\epsilon-\frac{135}{8}\ln2+\frac{81}{8}\ln 3+\frac{405}{16}\right) \calq^2\epsilon^4
+\frac{3\sqrt{6}}{4} \calq^5  \epsilon^2\,,
\end{split}
\eqlabel{susyuv}
\end{equation} 
\item in the IR, using \eqref{rrhoir} we identify (see \eqref{irvevsb}):
\begin{equation}
\begin{split}
&f_{a,0}^h=2^{1/3}\ 3^{2/3}\ \epsilon^{4/3}\,,\qquad h_0^h=P^2g_s\ \epsilon^{-8/3}\ \times\ 0.056288(0)\,,\\
&K_{1,3}^h=\frac{4\sqrt{6}}{9\ \epsilon^2}\ P^2g_s\,,\qquad K_{2,2}^h=\frac{2^{2/3}}{3^{2/3}\ \epsilon^{4/3}}\,,\qquad 
K_{2,4}^h=-\frac{11\ 2^{1/3}\ 3^{2/3}}{45\ \epsilon^{8/3}}\,,\\
&K_{3,1}^h=\frac{4\sqrt{6}\ 2^{1/3}\ 3^{2/3}}{27\epsilon^{2/3}}\ P^2g_s\,,\qquad g_0^h=g_s\,,\qquad
f_{1,0}^h=1\,.
\end{split}
\eqlabel{susyir}
\end{equation}
\end{itemize}
The relation between $K_0$ and the strong coupling scale $\Lambda$ of the
cascading gauge theory  \eqref{lmg} is given by \eqref{defkolambda}:
\begin{equation}
K_0=P^2g_s\biggl(\frac53\ \ln2-\ln 3-\frac23-\frac43\ \ln\e\biggr)\,.
\eqlabel{defk0ep}
\end{equation}

\subsection{Case (B): Schwarzschild horizon in warped deformed conifold with fluxes}\label{apa2}

Using the metric ansatz \eqref{metpar} we derive from \eqref{5action} the following equations of motion: 
\begin{equation}
\begin{split}
&0=f''-\frac{f'}{2} \biggl(\frac6\rho-\frac{2 f_b'}{f_b}-\frac{2 f_a'}{f_a}-\frac{f_c'}{f_c}\biggr)+4 h \mu^2\,,
\end{split}
\eqlabel{b1}
\end{equation}
\begin{equation}
\begin{split}
&0=f_a''-\frac{1}{8 f_a} (f_a')^2-\frac{f_a}{8f_b^2} (f_b')^2+\frac{f_a}{8g^2} (g')^2+\frac{f_a}{8h^2} (h')^2
+\frac{g P^2 (K_2')^2}{36h f_b}-\frac{3f_a (K_1')^2}{32g h f_b^2 P^2}
\\
&+\frac{5(K_3')^2}{32g h f_a P^2}+\biggl(
\frac{3f_a'}{4f}-\frac{f_a h'}{8h f}-\frac{f_a f_b'}{4f_b f}
-\frac{f_a f_c'}{8f_c f}-\frac{5f_a}{4\rho f}\biggr) f'
+\biggl(
\frac{f_a'}{4f_c}-\frac{f_a f_b'}{4f_b f_c}\biggr) f_c'
+\frac{f_b'f_a'}{2f_b}\\
&+\frac{f_a h'}{h \rho}-\frac{3 f_a'}{\rho}
-\frac{9(K_3-K_1)^2}{64h f_b g f_c \rho^2 f P^2}
-\frac{g (K_2^2 (5 f_a^2-3 f_b^2)+12 f_b^2 (K_2-1)) P^2}{8h f_b^2 f_a f_c \rho^2 f}
\\&- \frac{K_2 (K_3-K_1) (K_2 (K_3-K_1)+4 K_1)}{32f_b^2 h^2 f_a f_c \rho^2 f}
-\frac{K_1^2}{8 f_b^2 h^2 f_a f_c \rho^2 f} +\frac{5 f_a}{\rho^2}
-\frac{45f_a^2}{16f_b f_c \rho^2 f}+\frac{9f_a}{8f_c \rho^2 f}\\&+\frac{3 f_a}{f_b \rho^2 f}
+\frac{27f_b}{16f_c \rho^2 f}-\frac{9}{\rho^2 f}+\frac{3 f_c}{f_b \rho^2 f}-\frac{5h f_a \mu^2}{2f}\,,
\end{split}
\eqlabel{b2}
\end{equation}
\begin{equation}
\begin{split}
&0=f_b''-\frac{f_b (f_a')^2}{8f_a^2}-\frac{(f_b')^2}{8f_b}+\frac{f_b (h')^2}{8h^2}
-\frac{3f_b (K_3')^2}{32h f_a^2 P^2 g}+\frac{5(K_1')^2}{32f_b h P^2 g}
+ \frac{P^2 g (K_2')^2}{36h f_a}
+\frac{f_b (g')^2}{8g^2}
\\&+\biggl(
\frac{3f_b'}{4f}- \frac{f_b f_a'}{4f_a f}-\frac{f_b h'}{8h f}
-\frac{5f_b}{4\rho f}-\frac{f_b f_c'}{8f_c f}\biggr) f'
+\biggl(
\frac{f_b'}{4f_c}-\frac{f_b f_a'}{4f_c f_a}\biggr) f_c'
+\frac{f_b' f_a'}{2f_a}-\frac{3 f_b'}{\rho}+\frac{f_b h'}{h \rho}
\\&
+ \frac{g (K_2^2 (3 f_a^2-5 f_b^2)+20 f_b^2 (K_2-1)) P^2}{8h f_c f_b f_a^2 \rho^2 f}
-\frac{K_2 (K_3-K_1) (K_2 (K_3-K_1)+4 K_1)}{32f_c f_b h^2 f_a^2 \rho^2 f}\\&
-\frac{9(K_3-K_1)^2}{64h f_a g f_c \rho^2 f P^2}
-\frac{K_1^2}{8f_c f_b h^2 f_a^2 \rho^2 f}+\frac{5 f_b}{\rho^2}
+\frac{27f_a}{16f_c \rho^2 f}+\frac{9f_b}{8f_c \rho^2 f}
-\frac{9}{\rho^2 f}-\frac{45f_b^2}{16f_a f_c \rho^2 f}\\&+\frac{3 f_b}{f_a \rho^2 f}
+\frac{3 f_c}{f_a \rho^2 f}-\frac{5h f_b \mu^2}{2f}\,,
\end{split}
\eqlabel{b3}
\end{equation}
\begin{equation}
\begin{split}
&0=f_c''-\frac{(f_c')^2}{2f_c}-\frac{f_c (f_a')^2}{8f_a^2}
-\frac{f_c (f_b')^2}{8f_b^2}
+\frac{f_c (h')^2}{8h^2}-\frac{P^2 g f_c (K_2')^2}{12f_b h f_a}
-\frac{3f_c (K_3')^2}{32h f_a^2 P^2 g}-\frac{3f_c (K_1')^2}{32f_b^2 h P^2 g}
\\&+\frac{f_c (g')^2}{8g^2}+
\biggl(
\frac{7f_c'}{8f}-\frac{f_c f_b'}{4f_b f}-\frac{h' f_c}{8h f}
- \frac{f_c f_a'}{4f_a f}- \frac{5f_c}{4\rho f}
\biggr) f'
+\biggl(
\frac{3f_b'}{4f_b}+\frac{3f_a'}{4f_a}-\frac3\rho\biggr) f_c'
-\frac{f_c f_b' f_a'}{2f_b f_a}
\\&+\frac{f_c h'}{h \rho}+\frac{27(K_3-K_1)^2}{64f_b h f_a g \rho^2 f P^2}
+\frac{3g (K_2^2 (f_a^2+f_b^2)-4 f_b^2 (K_2-1)) P^2}{8h f_b^2 f_a^2 \rho^2 f}
-\frac{K_1^2}{8f_b^2 h^2 f_a^2 \rho^2 f}
+\frac{5 f_c}{\rho^2}\\&+\frac{63f_a}{16f_b \rho^2 f}
-\frac{K_2 (K_3-K_1) (K_2 (K_3-K_1)+4 K_1)}{32f_b^2 h^2 f_a^2 \rho^2 f}
-\frac{63}{8 \rho^2 f}+\frac{3 f_c}{f_b \rho^2 f}+\frac{63f_b}{16f_a \rho^2 f}
\\&+\frac{3 f_c}{f_a \rho^2 f}-\frac{9 f_c^2}{f_a f_b \rho^2 f}-\frac{5h f_c \mu^2}{2f}\,,
\end{split}
\eqlabel{b4}
\end{equation}
\begin{equation}
\begin{split}
&0=h''+\frac{g P^2 (K_2')^2}{12f_b f_a}
+ \frac{3(K_1')^2}{32g f_b^2 P^2}+\frac{3(K_3')^2}{32g f_a^2 P^2}
-\frac{h}{8g^2} (g')^2-\frac{9}{8 h} (h')^2+\frac{h}{8f_a^2} (f_a')^2
\\&+\frac{h}{8f_b^2} (f_b')^2+\biggl(
 \frac{h f_b'}{4f_b f}+\frac{9h'}{8f}+\frac{h f_a'}{4f_a f}
+\frac{13h}{4\rho f}+ \frac{h f_c'}{8f_c f}\biggr) f'
+\biggl(
\frac{h f_b'}{4f_c f_b}+\frac{h'}{2f_c}
+\frac{h f_a'}{4f_c f_a}\\&+\frac{h}{f_c \rho}\biggr) f_c'
+\biggl(
\frac{f_a'}{f_a}+\frac{f_b'}{f_b}-\frac4\rho\biggr) h'
+\frac{h f_b' f_a'}{2f_a f_b}+\frac{2 h f_b'}{f_b \rho}
+\frac{2 h f_a'}{f_a \rho}+ \frac{45(K_3-K_1)^2}{64f_b f_a g f_c \rho^2 f P^2}
-\frac{13 h}{\rho^2}\\
&+\frac{9 K_1^2}{8 f_c f_b^2 h f_a^2 \rho^2 f}
+\frac{5g (K_2^2 (f_a^2+f_b^2)-4 f_b^2 (K_2-1)) P^2}{8f_c f_b^2 f_a^2 \rho^2 f}
+\frac{9K_1 K_2 (K_3-K_1)}{8f_c f_b^2 h f_a^2 \rho^2 f}
+\frac{h f_c}{f_a f_b \rho^2 f}\\
&+\frac{9f_a h}{16f_b f_c \rho^2 f}
+\frac{9(K_3-K_1)^2 K_2^2}{32f_c f_b^2 h f_a^2 \rho^2 f}
-\frac{9h}{8f_c \rho^2 f}-\frac{3 h}{f_b \rho^2 f}+\frac{9h f_b}{16f_a f_c \rho^2 f}
-\frac{3 h}{f_a \rho^2 f}
+\frac{13h^2\mu^2}{2f}\,,
\end{split}
\eqlabel{b5}
\end{equation}
\begin{equation}
\begin{split}
&0=K_1''+\biggl(
\frac{f'}{f}+\frac{f_c'}{2f_c}-\frac{f_b'}{f_b}-\frac{h'}{h}
-\frac{g'}{g}+\frac{f_a'}{f_a}
-\frac3\rho\biggr) K_1'
+\frac{9f_b (K_3-K_1)}{2f_a f f_c \rho^2}
\\&+\frac{g (K_2 (K_2 (K_3-K_1)-2 K_3)+4 K_1 (K_2-1)) P^2}{h f f_c f_a^2 \rho^2}\,,
\end{split}
\eqlabel{b6}
\end{equation}
\begin{equation}
\begin{split}
&0=K_2''+\biggl(
\frac{g'}{g}+\frac{f_c'}{2f_c}-\frac{h'}{h}
+\frac{f'}{f}-\frac3\rho\biggr) K_2'
-\frac{9(K_3-K_1) (K_2 (K_3-K_1)+2 K_1)}{8h f f_c f_a f_b g P^2 \rho^2}
\\&-\frac{9((f_a^2+f_b^2) K_2-2 f_b^2)}{2f f_c f_a f_b \rho^2}\,,
\end{split}
\eqlabel{b7}
\end{equation}
\begin{equation}
\begin{split}
&0=K_3''+\biggl(
\frac{f'}{f}-\frac{g'}{g}+\frac{f_b'}{f_b}-\frac{f_a'}{f_a}
+ \frac{f_c'}{2f_c}
-\frac{h'}{h}-\frac3\rho
\biggr) K_3'
-\frac{9f_a (K_3-K_1)}{2f_b f f_c \rho^2}
\\&-\frac{g K_2 (K_2 (K_3-K_1)+2 K_1) P^2}{h f f_c f_b^2 \rho^2}\,,
\end{split}
\eqlabel{b8}
\end{equation}
\begin{equation}
\begin{split}
&0=g''-\frac{(g')^2}{g}-\frac{g^2 P^2 (K_2')^2}{9h f_a f_b}
+\frac{(K_1')^2}{8h f_b^2 P^2}+\frac{(K_3')^2}{8h f_a^2 P^2}+\biggl(
\frac{f'}{f}+\frac{f_c'}{2f_c}+\frac{f_b'}{f_b}+\frac{f_a'}{f_a}-\frac3\rho
\biggr)g'
\\&+\frac{9(K_3-K_1)^2}{16f_af_bhff_c\rho^2 P^2}
-\frac{g^2 ((f_a^2+f_b^2) K_2^2+4 f_b^2 (1-K_2)) P^2}{2h f f_c f_a^2 f_b^2 \rho^2}\,.
\end{split}
\eqlabel{b9}
\end{equation}
Additionally we have the first order constraint
\begin{equation}
\begin{split}
&0=\frac{(K_1')^2}{P^2}+\frac{(K_3')^2 f_b^2}{f_a^2 P^2}
+ \frac{8g^2 P^2 (K_2')^2 f_b}{9f_a}
+\frac{4 h (g')^2 f_b^2}{g}-4 h g (f_b')^2+\frac{4 g f_b^2 (h')^2}{h}
\\&-\frac{4 h g (f_a')^2 f_b^2}{f_a^2}
+\biggl(
\frac{24 h g f_b^2}{f \rho}-\frac{4 h g f_b^2 f_c'}{f f_c}-\frac{8 h g f_b f_b'}{f}
-\frac{4 g f_b^2 h'}{f}-\frac{8 h g f_b^2 f_a'}{f f_a}\biggr) f'+\frac{64 h g f_b f_b'}{\rho}
\\&+\biggl(
\frac{32 h g f_b^2}{f_c \rho}-\frac{8 h g f_b f_b'}{f_c}
-\frac{8 h g f_b^2 f_a'}{f_c f_a}
\biggr) f_c'
-\frac{16 h g f_b f_b' f_a'}{f_a}+\frac{32 g f_b^2 h'}{\rho}
+\frac{64 h g f_b^2 f_a'}{f_a \rho}\\&
-\frac{4 g^2 (K_2^2 (f_a^2+f_b^2)+4 f_b^2 (1-K_2)) P^2}{f f_c f_a^2 \rho^2}
-\frac{g K_2 (K_3-K_1) (K_2 (K_3-K_1)+4 K_1)}{h f f_c f_a^2 \rho^2}
\\&-\frac{9f_b (K_3-K_1)^2}{2f_a f f_c \rho^2 P^2}
-\frac{4 g K_1^2}{h f f_c f_a^2 \rho^2}-\frac{96 h g f_b^2}{\rho^2}
-\frac{18 h f_a g f_b}{f f_c \rho^2}+\frac{36 h g f_b^2}{f f_c \rho^2}
+\frac{96 h g f_b}{f \rho^2}-\frac{18 h g f_b^3}{f f_c f_a \rho^2}
\\&+\frac{96 h g f_b^2}{f f_a \rho^2}-\frac{32 h f_c g f_b}{f f_a \rho^2}
+\frac{48 h^2 g f_b^2 \mu^2}{f}\,.
\end{split}
\eqlabel{bc}
\end{equation}
We explicitly verified that the constraint \eqref{bc} is consistent with \eqref{b1}-\eqref{b9}.
Thus the
second-order equation for $f_c$ \eqref{b4} can be eliminated in favor of the constraint equation
\eqref{bc} where $f_c'$ enters linearly. In total, we expect that a solution is
specified by $8\times 2+ 1\times 1=17$ parameters (8 second-order equations for
$f_{a,b}, K_{1,2,3}, h,g,f$ and a single first-order equation for $f_c$). 
 
The general UV (as $\rho\to 0$ ) asymptotic solution of \eqref{b1}-\eqref{bc} describing the black hole on 
the warped deformed conifold with fluxes and $S^3$ horizon (from the effective
5d perspective \eqref{5action}), \ie the thermal states of the cascading
gauge theory on $S^3$, takes the form
\begin{equation}
\begin{split}
&f_a=1+f_{a,1,0} \rho+\rho^2 \biggl(
\frac{3}{16} \mu^2 P^2 g_s+\frac14 f_{a,1,0}^2\biggr)+\rho^3 f_{a,3,0}+\calo(\rho^4\ln \rho)\,,
\end{split}
\eqlabel{bs1}
\end{equation}
\begin{equation}
\begin{split}
&f_b=1+f_{a,1,0} \rho+\rho^2 \biggl(
\frac{3}{16} \mu^2 P^2 g_s+\frac14 f_{a,1,0}^2\biggr)-\rho^3 f_{a,3,0}
+\calo(\rho^4\ln \rho)\,,
\end{split}
\eqlabel{bs2}
\end{equation}
\begin{equation}
\begin{split}
&f_c=1+f_{a,1,0} \rho+\rho^2 \biggl(
\frac14 \mu^2 P^2 g_s+\frac14 f_{a,1,0}^2\biggr)+\calo(\rho^4\ln \rho)\,,
\end{split}
\eqlabel{bs3}
\end{equation}
\begin{equation}
\begin{split}
&h= \frac18 P^2 g_s+\frac14 K_0-\frac12 P^2 g_s \ln\rho+\rho \biggl(
-\frac12 K_0 f_{a,1,0}+P^2 g_s  f_{a,1,0}\ln\rho\biggr)
+\rho^2 \biggl(
\frac19 P^4 g_s^2 \mu^2\\&-\frac{1}{24} K_0 P^2 g_s \mu^2-\frac14 P^2 g_s f_{a,1,0}^2 
+\frac58 K_0 f_{a,1,0}^2+\left(\frac{1}{12} P^4 g_s^2 \mu^2-\frac54 P^2 g_sf_{a,1,0}^2 \right) \ln\rho\biggr)
\\&+\rho^3 \biggl(-\frac38 f_{a,1,0} P^4 g_s^2 \mu^2+\frac18 K_0 f_{a,1,0} \mu^2 P^2 g_s
+\frac{11}{24} f_{a,1,0}^3 P^2 g_s-\frac58 K_0 f_{a,1,0}^3\\&+\biggl(
-\frac14 f_{a,1,0} P^4 g_s^2 \mu^2
+\frac54 f_{a,1,0}^3 P^2 g_s\biggr) \ln\rho\biggr)+\calo(\rho^4\ln^2 \rho)\,,
\end{split}
\eqlabel{bs4}
\end{equation}
\begin{equation}
\begin{split}
&K_1=K_0-2 P^2 g_s \ln\rho+f_{a,1,0} P^2 g_s \rho+\rho^2 \biggl(
\frac 14 P^2 g_s (3 P^2 g_s \mu^2+K_0 \mu^2-f_{a,1,0}^2)\\&-\frac12 P^4 g_s^2 \mu^2
\ln\rho\biggr)
+\rho^3 \biggl(
\frac{1}{12} P^2 g_s (f_{a,1,0}^3-6 P^2g_s f_{a,1,0}  \mu^2-3 K_0 f_{a,1,0} \mu^2
+8 f_{a,3,0}+8 k_{2,3,0})\\
&+\frac12 P^2 g_s (P^2 g_sf_{a,1,0}  \mu^2+4 f_{a,3,0}) \ln\rho\biggr)
+\calo(\rho^4\ln^2 \rho)\,,
\end{split}
\eqlabel{bs5}
\end{equation}
\begin{equation}
\begin{split}
&K_2=1+\rho^3 \biggl(k_{2,3,0}+3 f_{a,3,0} \ln\rho\biggr)+\calo(\rho^4\ln \rho)\,,
\end{split}
\eqlabel{bs6}
\end{equation}
\begin{equation}
\begin{split}
&K_3=K_0-2 P^2 g_s \ln\rho+P^2 g_s f_{a,1,0} \rho+\rho^2 \biggl(
\frac14P^2g_s(3P^2g_s\mu^2+K_0\mu^2-f_{a,1,0}^2)
\\&-\frac12 \mu^2 P^4 g_s^2 \ln\rho\biggr)+\rho^3 \biggl(
\frac{1}{12} P^2 g_s
(f_{a,1,0}^3-6 P^2 f_{a,1,0} g_s \mu^2-3 K_0 f_{a,1,0} \mu^2-8 f_{a,3,0}-8 k_{2,3,0})
\\&+\frac12 P^2 g_s (P^2 g_sf_{a,1,0}  \mu^2-4 f_{a,3,0}) \ln\rho\biggr)+\calo(\rho^4\ln^2 \rho)\,,
\end{split}
\eqlabel{bs7}
\end{equation}
\begin{equation}
\begin{split}
&g=g_s \biggl(1-\frac14\mu^2 P^2g_s\rho^2+\frac14\mu^2 f_{a,1,0}P^2g_s\rho^3+\calo(\rho^4\ln \rho)\biggr)\,,
\end{split}
\eqlabel{bs8}
\end{equation}
\begin{equation}
\begin{split}
&f=1+\rho^2 \biggl(
\frac18 \mu^2 P^2 g_s+\frac14 \mu^2 K_0-\frac12 \mu^2 P^2 g_s \ln\rho\biggr)
+\rho^3 \biggl(
\frac18 \mu^2 f_{a,1,0} P^2 g_s-\frac14 \mu^2 f_{a,1,0} K_0\\
&+\frac12 \mu^2 f_{a,1,0}
 P^2 g_s\ln\rho\biggr)+\rho^4\biggl(f_{4,0}-\left(
 \frac{13}{144} \mu^4 P^4 g_s^2+\frac{1}{12} \mu^4 K_0 P^2 g_s
 +\frac38 \mu^2 P^2 g_s f_{a,1,0}^2\right) \ln\rho
\\& +\frac{1}{12} \mu^4 P^4 g_s^2 \ln^2\rho\biggr)+\calo(\rho^5\ln^2 \rho)\,.
\end{split}
\eqlabel{bs9}
\end{equation}
It is characterized by the cascading gauge theory defining parameters \eqref{defuvcasea}. Additionally,
there are 9 normalizable coefficients, related to the diffeomorphism parameter $\alpha$ \eqref{leftover} and the expectation values of the various operators
in the boundary theory
\begin{equation}
\{f_{a,1,0}\,,\ f_{a,3,0}\,,\ k_{2,3,0}\,,\ f_{4,0}\,,\ f_{c,4,0}\,,\ g_{4,0}\,,\ f_{a,6,0}\,,\ k_{2,7,0}\,,\ f_{c,8,0}
\}\,.
\eqlabel{uvvevsb}
\end{equation}

Unlike the topologically distinct background geometries representing the $\calv_A$ or $\calv_B$ vacua
of the cascading gauge theories, the topology of the background geometry representing the deconfined
thermal states\footnote{Dual to bulk geometries with a regular Schwarzschild horizon, see \eqref{metpar}.} of the
cascading gauge theory $\calt_{decon}$ is unique: here, the Euclidean time direction smoothly shrinks to zero size,
with the boundary spatial $S^3$ and the conifold cycles remaining finite. 
Introducing
\begin{equation}
f_{a,b,c}^h\equiv y^2\ f_{a,b,c}\,,\qquad h^h\equiv y^{-4}\ h\,,
\eqlabel{bbredef}
\end{equation}
the IR asymptotic expansion takes form
\begin{equation}
\begin{split}
&f_{a,b,c}^h=\sum_{n=0} f_{a,b,c,n}^h\ y^{n}\,,\qquad h^h=\sum_{n=0} h^h_n\ y^{n}\,,\qquad K_{1,2,3}=\sum_{n=0} K_{1,2,3,n}^h\ y^{n}\,,
\\
&g=\sum_{n=0} g_{n}^h\ y^{n}\,,\qquad f=\sum_{n=1} f_{n}^h\ y^{n}\,,
\end{split}
\eqlabel{ircasebb}
\end{equation}
and is characterized by 9 parameters:
\begin{equation}
\{f_{a,0}^h\,,\, f_{b,0}^h\,,\,f_{c,0}^h\,,\,h_{0}^h\,,\,K_{1,0}^h\,,\,K_{2,0}^h\,,\,K_{3,0}^h\,,\,g_{0}^h\,,\ f_1^h\}\,.
\eqlabel{irvevsbb}
\end{equation}
A combination of these parameters is related to the Hawking temperature $T$ of the black hole:
\begin{equation}
T=\frac{f_1^h}{4\pi \sqrt{h_0^h}}\,.
\eqlabel{thaw}
\end{equation}
Note that given given \eqref{ircasebb},
\begin{equation}
\begin{split}
&\frac{1}{h^{1/2}\rho^2}\biggl(-f\ dt^2\biggr)+\frac{h^{1/2}}{f\rho^2}\ (d\rho)^2 \qquad
\underbrace{\longrightarrow}_{t\to i t_E}\qquad 
\frac{f}{\sqrt{h^h}}\ dt_E^2+
\frac{\sqrt{h^h}}{f} (dy)^2\\
&\underbrace{\longrightarrow}_{y\equiv z^2\to 0}\qquad \frac{4\sqrt{h_0^h}}{f_1^h}\biggl(z^2\ \frac{(f_1^h)^2}{4h_0^h}\ dt_E^2+(dz)^2\biggr)\,,
\end{split}
\eqlabel{irlimitb2}
\end{equation}
\ie the compactified Euclidean time direction $S^1$ indeed smoothly shrinks to zero size as $z\to 0$,
provided $t_E\sim t_E + \frac 1T$ with the temperature given by \eqref{thaw}. 
It is important to emphasize that
$\calt_{decon}$ deconfined thermal states defined by \eqref{ircasebb}
have either $U(1)$ or $\zet_2$ chiral symmetry ---
the chiral symmetry is unbroken in the former ($\calt_{decon}^s$ deconfined thermal states), and is
spontaneously broken in the
latter ($\calt_{decon}^b$ deconfined thermal states).  Specifically, unbroken chiral symmetry dictates \eqref{symsector}, leading to 
\begin{equation}
\begin{split}
&{\rm UV:}\qquad f_{a,3,0}=0\,,\qquad k_{2,3,0}=0\,,\qquad k_{2,7,0}=0\,;
\\
&{\rm IR:}\qquad f_b^h=f_a^h\,,\qquad K_{3,0}^h=K_{1,0}^h\,,\qquad K_{2,0}^h=1\,.
\end{split}
\eqlabel{uvirchiralt}
\end{equation}

\subsection{Holographic renormalization of the effective action \eqref{5action}}\label{apa3}

The holographic renormalization of the theory \eqref{5action} was discussed in \cite{Aharony:2005zr}
(see also appendix A.2 of \cite{Buchel:2018bzp}). We review here the expressions necessary to recover
various thermodynamic quantities of the black holes on the warped deformed conifold with fluxes, and the
Casimir energies of the $S^3$ conifold vacua.

The $SO(4)$ invariant five-dimensional metric ansatzes \eqref{5metrho} and \eqref{metpar}, convenient for
the numerical computations, differ depending whether or not the bulk geometry of $\calm_5$
has a horizon. To present the common expressions for the holographic renormalization we parameterize
the full ten-dimensional metric \eqref{10dmetric} as
\begin{equation}
ds_{10}^2=-c_1^2 dt^2 +c_2^2 \left(dS^3\right)^2+c_3^2 (d\rho)^2 + \Omega_1^2 g_5^2+\Omega_2^2(g_3^2+g_4^2)
+\Omega_3^2(g_1^2+g_2^2)\,,
\eqlabel{10dm2}
\end{equation}
where $c_i=c_i(\rho)$ and $\Omega_i=\Omega_i(\rho)$. Note that $\Omega_i$ are parameterized as in
\eqref{redef}, and
\begin{equation}
\begin{split}
&{\rm (A)\  [no\ horizon]:}\qquad c_1=h^{-1/4}\rho^{-1}\,,\qquad c_2=h^{-1/4}\rho^{-1}\mu^{-1}\ f_1\,,
\qquad c_3=h^{1/4}\rho^{-1}\,;
\\
&{\rm (B)\  [horizon]:}\qquad c_1=h^{-1/4}\rho^{-1} f^{1/2}\,,\qquad c_2=h^{-1/4}\rho^{-1}\mu^{-1}\,,
\qquad c_3=h^{1/4}\rho^{-1}\ f^{-1/2}\,.
\end{split}
\eqlabel{cis}
\end{equation}

Following \cite{Aharony:2005zr}, the renormalized five-dimensional effective action with a
cutoff $\rho=\hr$ takes form
\begin{equation}
S_{5,\hr}^{renom}=S_{5,\rho\ge \hr}^{bulk}+S_{GH,\rh}+S_{ct,\rh}\,.
\eqlabel{srenom}
\end{equation}
\nxt 
$S_{5,\rho\ge \hr}^{bulk}$ is the regularized bulk action \eqref{5action} with $\del\calm_5$ at
$\rho=\hr$, here  $'\equiv \del_\rho$,
\begin{equation}
\begin{split}
S_{5,\rho\ge \hr}^{bulk}=&\frac{108 }{16\pi G_5}\ {\rm vol}_{S^3}\ \int dt \int_{\hr}^{+\infty} d\rho \
c_1c_2^3c_3\ \times\ \Omega_1\Omega_2^2\Omega_3^2\ \times\ \biggl\{R_{10}+\cdots\biggr\}\\
=&\frac{108 }{16\pi G_5}\ {\rm vol}_{S^3}\ \int dt \int_{\hr}^{+\infty} d\rho \
\biggl[-\frac{2c_1'c_2^3\Omega_1\Omega_2^2\Omega_3^2}{c_3}\biggr]'\\
=&\frac{108 }{16\pi G_5}\ {\rm vol}_{S^3}\ \times\ \int dt  \ \times
\biggl[\frac{2c_1'c_2^3\Omega_1\Omega_2^2\Omega_3^2}{c_3}\biggr]_{+\infty}^{\hr}\,,
\end{split}
\eqlabel{s5reg}
\end{equation}
where in the second line we used the equations of motion \eqref{a1}-\eqref{ac} or \eqref{b1}-\eqref{bc}
to represent the bulk integral as a total derivative. 
\nxt $S_{GH,\rh}$ is the generalized Gibbons-Hawking term, evaluated at the regularization
boundary $\del\calm_5$,
\begin{equation}
\begin{split}
S_{GH,\rh}=&\frac{108}{16\pi G_5}\ \times\ 2\ \int_{\del\calm_5} {\rm vol}_{\del\calm_5}\
\Omega_1\Omega_2^2\Omega_3^2\ \biggl(\nabla_\mu n^\mu+n^\mu\nabla_\mu\ln
\left(\Omega_1\Omega_2^2\Omega_3^2\right)\biggr)\\
=&\frac{108 }{16\pi G_5}\ {\rm vol}_{S^3}\ \times\ \int dt\ \times\
\frac{2}{c_3}\biggl(c_1c_2^3\Omega_1\Omega_2^2\Omega_3^2\biggr)'\bigg|_{\rho=\hr}\,,
\end{split}
\eqlabel{ghreg}
\end{equation}
where $n^\mu$ is a unit space-like vector orthogonal to the four-dimensional boundary $\del\calm_5$. 
\nxt $S_{ct,\rh}$ is the counter-term action\footnote{
This counter-term action renormalizes the effective action, the boundary stress-energy tensor and the expectation values of all the relevant and the marginal operators of the boundary theory.
Additional counterterms are necessary
to remove power-law divergences of the irrelevant operators of the cascading gauge theory,
see \cite{Aharony:2005zr} for more details.},  evaluated at the regularization
boundary $\del\calm_5$,
\begin{equation}
\begin{split}
S_{ct,\rh}=&\frac{1}{16\pi G_5}\ \int_{\del\calm_5} {\rm vol}_{\del\calm_5}\ \call^{counter}\,,
\end{split}
\eqlabel{ctreg}
\end{equation}
where, in the "minimal subtraction scheme'' \cite{Aharony:2005zr},
\begin{equation}
\begin{split}
\call^{counter}=&\hK-2\hw_1^4-8\hw_2^4+\cala_4+\calr_\gamma\ \hw_1^2\biggl(-\frac{1}{12}\hK+\frac{1}{12}P^2 e^\Phi
-\frac16\hw_1^4+\calb_2\biggr)\\
&+\calr_\gamma^2\ \times\ \call_{\calr^2}^0+\calr_{ab\ \gamma}\calr^{ab}_\gamma\ \times\ \call_{{\calr ic}^2}^0\,,
\end{split}
\eqlabel{lct}
\end{equation}
\begin{equation}
X_a \equiv\left(1-\frac{\hw_2^2}{\hw_1^2}\right)\,,\qquad \cala_4=\frac{18}{5}\ X_a^2 \hw_1^4\,,\qquad
\calb_2=X_a\ \biggl(\frac 16 \hK-\frac{1}{30}P^2e^\Phi\biggr)\,,
\eqlabel{ap3def1}
\end{equation}
\begin{equation}
\begin{split}
\call_{\calr^2}^0=&-\frac{1}{144} P^4 e^{2\Phi} \ln^3\rho\ -
\frac{1}{96} P^2 e^\Phi \ln^2\rho\  \hK-\frac{1}{192} \ln\rho\  \hK^2\\
&+\left(\frac{1}{96}+4\kappa_1\right) P^4 e^{2\Phi} \ln^2\rho\ 
+\left(\frac{1}{96}+4\kappa_1\right) P^2 e^\Phi \ln\rho\  \hK\\
&+\left(\kappa_1+\frac{1}{1152}\right) \hK^2
+\left(2\kappa_2-\frac{43}{2304}\right) P^4 e^{2\Phi} \ln\rho\ 
\\ &+\left(\kappa_2
-\frac{13}{1152}\right) P^2 e^\Phi \hK+
\kappa_3 P^4 e^{2\Phi},\\
\end{split}
\eqlabel{resf11}
\end{equation}
\begin{equation}
\begin{split}
\call_{{\calr ic}^2}^0=&\frac{1}{48} P^4 e^{2\Phi} \ln^3\rho\ +
\frac{1}{32} P^2 e^\Phi \ln^2\rho\  \hK
+\frac{1}{64} \ln\rho\  \hK^2+\left(-\frac{1}{32}-12\kappa_1\right) 
P^4 e^{2\Phi} \ln^2\rho\ \\
&+\left(-\frac{1}{32}-12\kappa_1\right) P^2 e^\Phi \ln\rho\  \hK+
\left(-\frac{1}{256}-3\kappa_1\right) \hK^2
\\
&+\left(\frac{43}{768}-6\kappa_2\right) P^4 e^{2\Phi} \ln\rho\  
+\left(\frac{5}{192}-3\kappa_2\right) P^2 e^\Phi \hK\\
&+\left(\frac{541}{138240}-3 \kappa_3\right) P^4 e^{2\Phi},
\end{split}
\eqlabel{resf12}
\end{equation}
with \cite{Buchel:2018bzp}
\begin{equation}
\begin{split}
\hK=\frac 12\left(K_1+K_3\right)\,,\qquad \hw_1=3\Omega_1\,,\qquad
\hw_2=\frac{\sqrt{6}}{2}\left(\Omega_2+\Omega_3\right)\,,
\end{split}
\eqlabel{ap3defl}
\end{equation}
and, for a specific choice of the five-dimensional background metric in \eqref{10dm2}, 
\begin{equation}
\calr_\gamma=\frac{6}{c_2^2}\,,\qquad \calr_{ab\ \gamma}\calr^{ab}_\gamma=\frac{12}{c_2^4}\,,\qquad \square_\gamma\calr_\gamma=0\,.
\eqlabel{defrs}
\end{equation}

Notice that the holographic renormalization in the minimal subtraction\footnote{Additional ambiguities arise for generic $\del\calm_5$,
in particular when $\square_\gamma\calr_\gamma\ne 0$.}
has a 3-parameter ambiguity --- these are the coefficients $\kappa_1\cdots \kappa_3$ in \eqref{resf11} and \eqref{resf12}
parameterizing finite as $\rho\to 0$ counterterms
\begin{equation}
\begin{split}
\call_{finite}^{counter}=&\biggl(\kappa_1 \left(\hK+2 P^2e^\Phi\ln\rho\right)^2+\kappa_2 \left(\hK+2 P^2e^\Phi\ln\rho\right)P^2 e^\Phi+
\kappa_3 P^4 e^{2\Phi}\biggr)\\
&\times \left(\calr_\gamma^2-3 \calr_{ab\ \ga}\calr_\ga^{ab}\right)\,.
\end{split}
\eqlabel{lcfinite}
\end{equation}
The presence of the finite counterterms \eqref{lcfinite} is mandated \cite{Aharony:2005zr}
by the reparametrization of the radial coordinate $\rho\to \lambda \rho$ because of the explicit $\ln\rho$ dependence in
 \eqref{resf11} and \eqref{resf12}. Indeed, it is easy to see that the reparametrization $\ln\rho \to \ln\rho + \ln\lambda$
 is equivalent to 
\begin{equation}
\begin{split}
&\kappa_1\qquad \to\qquad \kappa_1-\frac{1}{192} \ln\lambda\,;
\\
&\kappa_2\qquad \to\qquad \kappa_2+4 \kappa_1\ \ln\lambda-\frac{1}{96} \ln^2\lambda+\frac{1}{96} \ln\lambda\,;
\\
&\kappa_3\qquad \to\qquad \kappa_3+2  \kappa_2\ \ln\lambda+4 \kappa_1\ \ln^2\lambda-\frac{1}{144} \ln^3\lambda+\frac{1}{96}
\ln\lambda-\frac{43}{2304} \ln\lambda\,.
\end{split}
\eqlabel{kappatr}
\end{equation}
The specific structure of the finite counter-term ambiguity, namely the combination $(\calr_\gamma^2-3 \calr_{ab\ \ga}\calr_\ga^{ab})$,
implies that the renormalized boundary stress-energy tensor  is ambiguity free when $\del\calm_5=R\times S^3$ \cite{Buchel:2011cc}. 
This might come as a surprise, as it is well-known \cite{Balasubramanian:1999re} that a finite counterterm (constant $\delta_{\calr^2}$)
\begin{equation}
\delta\call^{counter}_{finite}=\delta_{\calr^2}\ \times\ \calr_\gamma^2
\eqlabel{fcextra}
\end{equation}
would produce $\propto \delta_{\calr^2} R^2$ ambiguity in the boundary energy, as well as contribute $\propto \delta_{\calr^2} \square R$
ambiguity to the trace-anomaly. As explained in \cite{Aharony:2005zr}, a finite counterterm \eqref{fcextra} is absent
in the minimal subtraction since it reintroduces $\propto \delta_{\calr^2} \square R$ divergences in the one-point functions of the
irrelevant operators of the cascading gauge theory. If we restrict renormalization of the theory to the manifolds with
$\square_\gamma \calr_\gamma=0$, as in \eqref{defrs}, such a finite counter-term is allowed.

As usual, in the present of the Schwarzschild horizon in the bulk, we can analytically continue time
\begin{equation}
t\ \to\ i t_E\,,\qquad t_E\sim t_E+\frac 1T \,,
\eqlabel{te}
\end{equation}
where $t_E$ is periodic with inverse temperature $T$, and identify the free energy density $\calf$ of the black hole as
\begin{equation}
\frac 1T\ \times\ \frac{1}{\mu^3}{\rm vol}_{S^3}\ \times\ \calf=\lim_{\hr\to 0} S^{renom}_{5,\rh}\bigg|_{t\to i t_E}\,.
\eqlabel{deff}
\end{equation}

The renormalized effective action  $S^{renom}_{5,\rh}$ can further be used to compute the boundary stress-energy tensor as detailed in 
\cite{Aharony:2005zr}.

\subsection{Boundary stress-energy tensor}\label{listtmunu}

In this section, using the results of sections \ref{apa1}-\ref{apa3}, we collect expressions
for the energy density $\cale$ and pressure $\calp$, as well as some additional characteristics,
for the vacua and the thermal states of the
cascading gauge theory on $S^3$.  

Recall that,
\begin{itemize}
\item $\calv_A^b$ denotes vacua of the cascading gauge theory with spontaneously broken chiral symmetry,
with topologically trivial boundary $S^3$ --- see \eqref{as1}-\eqref{as9} for the UV asymptotics, and
\eqref{ircaseaa} (with \eqref{irlimitaa}) for the IR asymptotics;
\item $\calv_A^s$ denotes vacua of the cascading gauge theory with unbroken chiral symmetry,
with topologically trivial boundary $S^3$ --- see \eqref{uvirchiral} for the constraints on
UV/IR parameters;
\item $\calv_B$ denotes vacua of the cascading gauge theory with spontaneously broken chiral symmetry,
with topologically non-trivial boundary $S^3$ (the Klebanov-Strassler solution \cite{Klebanov:2000hb}
is a member of this class in the boundary $S^3$ decompactification limit)---
see \eqref{as1}-\eqref{as9} for the UV asymptotics, and
\eqref{ircaseb} (with \eqref{irlimitb}) for the IR asymptotics;
\item $\calt_{decon}^b$ denotes thermal deconfined states of the cascading gauge theory with spontaneously broken
chiral symmetry --- see \eqref{bs1}-\eqref{bs9} for the UV asymptotics, and
\eqref{ircasebb} (with \eqref{irlimitb2}) for the IR asymptotics;
\item $\calt_{decon}^s$ denotes thermal deconfined states of the cascading gauge theory with unbroken
chiral symmetry --- see \eqref{uvirchiralt} for the constraints on
UV/IR parameters.
\end{itemize}

\subsubsection{$\calv_A^b$ vacua}\label{lvas}

For the energy density $\cale$ and the pressure $\calp$ we find \cite{Aharony:2005zr}
\begin{equation}
\begin{split}
8\pi G_5\ \cale=&\biggl(
\frac{403}{1920} P^4 g_s^2+\frac{3}{32} K_0 P^2 g_s+\frac{1}{32} K_0^2\biggr) \mu^4
+\frac{9}{32} f_{a,1,0}^2 \mu^2 P^2 g_s-\frac32 f_{a,1,0} f_{a,3,0}\\
&-3 f_{a,4,0}+\frac32 f_{c,4,0}\,,\\
8\pi G_5\ \calp=&\biggl(
\frac{283}{5760} P^4 g_s^2+\frac{1}{16} K_0 P^2 g_s+\frac{1}{96} K_0^2\biggr) \mu^4
-\frac{5}{32} f_{a,1,0}^2 \mu^2 P^2 g_s+\frac12 f_{a,1,0} f_{a,3,0}\\&+f_{a,4,0}-\frac32 f_{c,4,0}\,.
\end{split}
\eqlabel{epab}
\end{equation}
Additionally, the expectation value of the dim-4 operator $\calo_{K_0}$, associated with the coupling $K_0$, is
\cite{Aharony:2005zr}
\begin{equation}
8\pi G_5\ \langle\calo_{K_0}\rangle=\biggl(-\frac{3}{32} K_0+\frac{1}{16} P^2 g_s\biggr) \mu^4
+\frac34 f_{a,1,0}^2 \mu^2+\frac{6 (f_{c,4,0}-f_{a,4,0})-3 f_{a,1,0} f_{a,3,0}}{P^2 g_s}\,.
\eqlabel{ok0}
\end{equation}
As explained in details in \cite{Aharony:2005zr}, the conformal anomaly of the cascading gauge theory reads
\begin{equation}
{\rm conformal\ anomaly}=\langle T_i^i\rangle + P^2 g_s\  \langle\calo_{K_0}\rangle=
-\cale+3\calp + P^2 g_s\  \langle\calo_{K_0}\rangle\,,
\eqlabel{confano}
\end{equation}
which vanishes here. Indeed, in any local field theory with the gravitational dual,
the conformal anomaly is a linear combination of terms
$I_4-E_4\propto (3 R_{\mu\nu}R^{\mu\nu}-R^2)$ (see \eqref{trace1} and \eqref{trace2})
and $\square R$ --- both of these terms vanish for the $\calm_4=R\times S^3$ boundary.

Note that any gravitational solution representing a $\calv_A^b$ vacuum can be interpreted as
a thermal confined state of the cascading gauge theory with spontaneously broken chiral symmetry,
$\calt^b_{con,A}$,
provided the Euclidean time direction is compactified with the inverse temperature period as in \eqref{te}.
In this case, it is easy to verify that the free energy density computed from \eqref{deff} is exactly
the same as the energy density,
\begin{equation}
\calt_{con,A}^b:\qquad \calf=\cale\,,
\eqlabel{fise}
\end{equation}
consistent with vanishing (in the supergravity or large-$N$ approximation) entropy density of the confined
states.

\subsubsection{$\calv_A^s$ vacua}\label{lvab}

These vacua are the special case of the $\calv_A^b$ vacua,
subject to constraints of the unbroken chiral symmetry \eqref{uvirchiral}.
Explicitly, 
\begin{equation}
\begin{split}
8\pi G_5\ \cale=&\biggl(
\frac{403}{1920} P^4 g_s^2+\frac{3}{32} K_0 P^2 g_s+\frac{1}{32} K_0^2\biggr) \mu^4
-\frac{3}{32} f_{a,1,0}^2 \mu^2 P^2 g_s
-3 f_{a,4,0}+\frac32 f_{c,4,0}\,,\\
8\pi G_5\ \calp=&\biggl(
\frac{283}{5760} P^4 g_s^2+\frac{1}{16} K_0 P^2 g_s+\frac{1}{96} K_0^2\biggr) \mu^4
-\frac{1}{32} f_{a,1,0}^2 \mu^2 P^2 g_s+f_{a,4,0}-\frac32 f_{c,4,0}\,,
\end{split}
\eqlabel{epas}
\end{equation}
and
\begin{equation}
8\pi G_5\ \langle\calo_{K_0}\rangle=\biggl(-\frac{3}{32} K_0+\frac{1}{16} P^2 g_s\biggr) \mu^4
+\frac{6 (f_{c,4,0}-f_{a,4,0})}{P^2 g_s}\,.
\eqlabel{ok0s}
\end{equation}

\subsubsection{$\calv_B$ vacua}\label{lvb}

While $\calv_B$ vacua are represented by topologically district solutions in the holographic dual
from those corresponding to $\calv_A^b$ vacua, and thus have different IR asymptotics
(compare \eqref{ircaseaa} with \eqref{ircaseb}), both vacua have the same UV asymptotics
\eqref{as1}-\eqref{as9}. This leads to the identical expressions for the energy density $\cale$,
the pressure $\calp$, and the expectation value of $\calo_{K_0}$ operators as in
\eqref{epab}-\eqref{ok0}.

Supersymmetric Klebanov-Strassler solution \cite{Klebanov:2000hb} is a decompactification
$\mu\to 0$ limit of $\calv_B$ vacua. Using the identifications \eqref{susyuv} we compute from
\eqref{epab}-\eqref{ok0}
\begin{equation}
\cale=0\,,\qquad \calp=0\,,\qquad \langle\calo_{K_0}\rangle=0\,,
\eqlabel{ksepk0}
\end{equation}
as expected.

\subsubsection{$\calt_{decon}^b$ thermal states}\label{lvtb}

For the energy density $\cale$ and the pressure $\calp$ we find \cite{Aharony:2005zr}
\begin{equation}
\begin{split}
8\pi G_5\ \cale=&\biggl(\frac{171}{640} P^4 g_s^2+\frac{17}{192} K_0 P^2 g_s+\frac{1}{32} K_0^2
\biggr) \mu^4+\biggl(
\frac{9}{32} K_0-\frac{21}{64} P^2 g_s\biggr) f_{a,1,0}^2 \mu^2-\frac32 f_{4,0}\\&+\frac32 f_{c,4,0}\,,
\\
8\pi G_5\ \calp=&\biggl(-\frac{7}{5760} P^4 g_s^2+\frac{1}{576} K_0 P^2 g_s+\frac{1}{96} K_0^2\biggr) \mu^4
+\biggl(
\frac{3}{32} K_0-\frac{7}{64} P^2 g_s\biggr) f_{a,1,0}^2 \mu^2\\&
-\frac12 f_{4,0}-\frac32 f_{c,4,0}\,.
\end{split}
\eqlabel{eptb}
\end{equation}
Additionally, the expectation value of the dim-4 operator $\calo_{K_0}$, associated with the coupling $K_0$, is
\cite{Aharony:2005zr}
\begin{equation}
8\pi G_5\ \langle\calo_{K_0}\rangle=\biggl(
\frac{1}{12} K_0+\frac{13}{48} P^2 g_s\biggr) \mu^4+\frac{6 f_{c,4,0}}{P^2 g_s}\,.
\eqlabel{ok0tb}
\end{equation}
Once again, the conformal anomaly of the cascading gauge theory reads
\begin{equation}
{\rm conformal\ anomaly}=\langle T_i^i\rangle + P^2 g_s\  \langle\calo_{K_0}\rangle=
-\cale+3\calp + P^2 g_s\  \langle\calo_{K_0}\rangle=0\,,
\eqlabel{confanot}
\end{equation}
\ie it vanishes here. 

Thermal deconfined states $\calt^b_{decon}$  carry the entropy density $s$,
\begin{equation}
s=\frac{1}{4G_5}\ f_{a,0}^hf_{b,0}^h\sqrt{f_{c,0}^h h_0^h}\,,
\eqlabel{enttb}
\end{equation}
with the equilibrium temperature $T$ given by \eqref{thaw}. The basic thermodynamic relation
\begin{equation}
\calf=\cale-s T
\eqlabel{femst}
\end{equation}
automatically holds, with the free energy density $\calf$ evaluated from \eqref{deff}.

The first law of the thermodynamics, \ie
\begin{equation}
0=\frac{d\cale}{T ds}-1\,,
\eqlabel{firstlaw}
\end{equation}
is verified numerically in all
solutions to an accuracy $\sim 10^{-7}$ for the Klebanov-Tseytlin black holes, and
 an accuracy $\sim 10^{-5}$ for the Klebanov-Strassler black holes --- see fig.~\ref{fl20}
 for a typical example.

\subsubsection{$\calt_{decon}^s$ thermal states}\label{lvts}

These thermal states are the special case of $\calt_{decon}^b$ thermal states, subject to
constraints of the unbroken chiral symmetry \eqref{uvirchiralt}. We have identical expressions
to \eqref{eptb} and \eqref{ok0tb} for the energy density $\cale$, the pressure $\calp$, and the
expectation value of $\calo_{K_0}$ because they do not depend on $\{f_{a,3,0},k_{2,3,0},k_{2,7,0}\}$.

The equilibrium temperature $T$ is still given by \eqref{thaw}, while for the entropy density
we have
\begin{equation}
s=\frac{1}{4G_5}\ \left(f_{a,0}^h\right)^2\sqrt{f_{c,0}^h h_0^h}\,.
\eqlabel{entts}
\end{equation}

\section{Fluctuations about $\calv_A^s$ vacua}\label{flucas}

Introducing
\begin{equation}
\begin{split}
&f_a\equiv f_3+\df\,,\ f_b\equiv f_3-\df\,,\ K_1\equiv K+\dk_1\,,\ K_3\equiv K-\dk_1\,,\
K_2\equiv 1+\dk_2\,,
\end{split}
\eqlabel{deffluc}
\end{equation}
we obtain from \eqref{a1}-\eqref{ac} decoupled linearized equations for fluctuations
$\df(\rho)$, $\dk_1(\rho)$, $\dk_2(\rho)$ about $\va^s$ vacua  (see section \ref{calvas}):
\begin{equation}
\begin{split}
&0=\df''+\biggl(\frac{3 f_1'}{f_1}+\frac{f_c'}{2f_c}-\frac3\rho\biggr)
\df'-\frac{K'}{2 f_3 h P^2g} \dk_1'+\biggl(
\frac{3h \mu^2}{2f_1^2}+\frac{(h')^2}{8h^2}-\frac{3h' f_1'}{4h f_1}
+\frac{(g')^2}{8g^2}
\\&-\frac{3(f_1')^2}{2f_1^2}-\frac{3 f_1' f_3'}{f_1 f_3}
-\frac{3f_1' f_c'}{4f_1 f_c}+\frac{(f_3')^2}{4f_3^2}
-\frac{f_3' f_c'}{2f_c f_3}+\frac{h'}{h \rho}
-\frac{3f_1'}{2f_1 \rho}+\frac{5}{\rho^2}-\frac{9}{f_c \rho^2}+\frac{6}{f_3 \rho^2}
+\frac{3 f_c}{\rho^2 f_3^2}\\
&-\frac{K^2}{8h^2 f_c \rho^2 f_3^4}-\frac{7(K')^2}{16h f_3^2  P^2g}-\frac{9P^2g}{4h f_3^2 f_c \rho^2}
\biggr) \df
-\frac{2  P^2g}{f_3 h f_c \rho^2}  \dk_2\,,
\end{split}
\eqlabel{vacasfl1}
\end{equation}
\begin{equation}
\begin{split}
&0=\dk_1''+\biggl(
\frac{f_c'}{2f_c}-\frac{h'}{h}+\frac{3 f_1'}{f_1}-\frac{g'}{g}-\frac3\rho
\biggr) \dk_1'
+\frac{2 K'}{f_3} \df'+\biggl(\frac{4  K P^2g}{h f_3^3 f_c \rho^2}
-\frac{2 K' f_3'}{f_3^2}\biggr) \df\\
&+\frac{2  K P^2g}{h \rho^2 f_c f_3^2} \dk_2-\frac{9}{f_c \rho^2} \dk_1\,,
\end{split}
\eqlabel{vacasfl2}
\end{equation}
\begin{equation}
\begin{split}
&0=\dk_2''+\biggl(
\frac{3 f_1'}{f_1}-\frac{h'}{h}+\frac{f_c'}{2f_c}
+\frac{g'}{g}-\frac{3}{\rho}
\biggr) \dk_2'
+\frac{9K}{2 h f_3^2 f_c P^2g \rho^2}  \dk_1-\frac{9}{f_c \rho^2} \dk_2-\frac{18}{f_3 f_c \rho^2} \df\,.
\end{split}
\eqlabel{vacasfl3}
\end{equation}
We turn on a finite source term\footnote{This is one of the two possible source terms. It is possible to turn on an independent source term for $\dk_1$; such an additional source term  produces conceptually
identical results.}  $m$,  (a non-normalizable coefficient for $\df$) explicitly breaking the
chiral symmetry $U(1)\to \zet_2$, leading to
\begin{equation}
\begin{split}
&\df=m\ \rho+\frac 12 f_{a,1,0} m\ \rho^2+\rho^3\ \biggl(
\df_{3,0}+\frac18 m \mu^2 (K_0-10 P^2 g_s) \ln\rho-\frac18 m \mu^2 P^2 g_s \ln^2\rho\biggr)
\\&+\calo(\rho^4\ln^2\rho)\,,
\end{split}
\eqlabel{uv1}
\end{equation}
\begin{equation}
\begin{split}
&\dk_1=-\frac12 P^2 g_s m\ \rho+\frac14 P^2 g_s m f_{a,1,0}\ \rho^2+\rho^3\ \biggl(
-\frac{1}{1728} P^2 g_s (131 P^2 g_s m \mu^2+330 K_0 m \mu^2\\
&-432 f_{a,1,0}^2 m-1152 \df_{3,0}
-1152 \dk_{2,3,0})-\frac{1}{288} P^2 g_s (263 P^2 g_s m \mu^2+168 K_0 m \mu^2\\&
-576 \df_{3,0}) \ln\rho
+\frac{1}{24} P^2 g_s m \mu^2 (-16 P^2 g_s+3 K_0) \ln^2\rho-\frac{1}{12} P^4 g_s^2 m \mu^2 \ln^3\rho\biggr)
\\&+\calo(\rho^4\ln^3\rho)\,,
\end{split}
\eqlabel{uv2}
\end{equation}
\begin{equation}
\begin{split}
&\dk_2=-\frac94 m\ \rho+\frac98 f_{a,1,0} m\ \rho^2+\rho^3\ \biggl(\dk_{2,3,0}
+\left(-\frac{133}{192} m \mu^2 P^2 g_s-m \mu^2 K_0+3 \df_{3,0}\right) \ln\rho
\\&+\left(-\frac78 m \mu^2 P^2 g_s
+\frac{3}{16} m \mu^2 K_0\right) \ln^2\rho-\frac{1}{8} m \mu^2 P^2 g_s \ln^3\rho\biggr)
+\calo(\rho^4\ln^3\rho)\,,
\end{split}
\eqlabel{uv3}
\end{equation}
as $\rho\to 0$, and  (compare with the definition of $f_{a}^h$ in \eqref{aredef})
\begin{equation}
\begin{split}
&\df^h\equiv y^2\ \df=\sum_{n=0} \delta{f}^h_n\ y^{2n}\,,\qquad \dk_1=\sum_{n=0}\dk^h_{1,n}\ y^{2n}
\,,\qquad \dk_2=\sum_{n=0}\dk^h_{2,n}\ y^{2n}\,,
\end{split}
\eqlabel{ir1}
\end{equation}
as $y\equiv \frac 1\rho\to 0$. 
It is characterized by 3 normalizable coefficients in the UV, and 3 normalizable coefficients in the
IR:
\begin{equation}
\begin{split}
{\rm UV}:&\qquad \{\df_{3,0}\,,\ \dk_{2,3,0}\,,\ \dk_{2,7,0}\}\,;
\\
{\rm IR}:&\qquad \{\df^h_0\,,\ \dk_{1,0}^h\,,\ \dk_{2,0}^h\}\,,
\end{split}
\eqlabel{uvirfluc}
\end{equation}
precisely as needed to identify a solution of a system of three second-order ODEs \eqref{vacasfl1}- \eqref{vacasfl3}.

\section{Fluctuations about $\calt_{decon}^s$ thermal states}\label{flucts}

Introducing
\begin{equation}
\begin{split}
&f_a\equiv f_3+\df\,,\ f_b\equiv f_3-\df\,,\ K_1\equiv K+\dk_1\,,\ K_3\equiv K-\dk_1\,,\
K_2\equiv 1+\dk_2\,,
\end{split}
\eqlabel{defflucts}
\end{equation}
we obtain from \eqref{b1}-\eqref{bc} decoupled linearized equations for fluctuations
$\df(\rho)$, $\dk_1(\rho)$, $\dk_2(\rho)$ about $\calt_{decon}^s$ states (see section \ref{techkt}):
\begin{equation}
\begin{split}
&0=\df''+\biggl(
\frac{f_c'}{2f_c}+\frac{f'}{f}-\frac{3}{\rho}\biggr) \df'
-\frac{K'}{2 f_3 h P^2g} \dk_1'+\biggl(
\frac{(h')^2}{8h^2}-\frac{h' f'}{8h f}+\frac{(g')^2}{8g^2}+\frac{(f_3')^2}{4f_3^2}
\\&-\frac{f_3' f_c'}{2f_3 f_c}-\frac{f' f_3'}{2f_3 f}
-\frac{f' f_c'}{8f f_c}+\frac{h'}{h \rho}
-\frac{5f'}{4\rho f}+\frac{5}{\rho^2}-\frac{7(K')^2}{16h f_3^2 P^2g} 
- \frac{9P^2g}{4h f_3^2 \rho^2 f f_c}
-\frac{9}{\rho^2 f f_c}\\
&+\frac{6}{f_3 \rho^2 f}
+\frac{3 f_c}{f_3^2 \rho^2 f}-\frac{K^2}{8h^2 f_3^4 \rho^2 f f_c}-\frac{5h \mu^2}{2f}\biggr) \df
-\frac{2   P^2g}{f_3 h \rho^2 f f_c}\dk_2\,,
\end{split}
\eqlabel{tsfl1}
\end{equation}
\begin{equation}
\begin{split}
&0=\dk_1''+\biggl(
\frac{f'}{f}+\frac{f_c'}{2f_c}-\frac{h'}{h}-\frac{g'}{g}-\frac{3}{\rho}\biggr)
\dk_1'
+\frac{2 K'}{f_3} \df'+\biggl(
\frac{4  K P^2g}{h f_3^3 f_c \rho^2 f}-\frac{2 K' f_3'}{f_3^2}\biggr) \df
\\&+\frac{2   K P^2g}{f_3^2 h f_c \rho^2 f}\dk_2-\frac{9 }{f f_c \rho^2}\dk_1\,,
\end{split}
\eqlabel{tsfl2}
\end{equation}
\begin{equation}
\begin{split}
&0=\dk_2''+\biggl(
\frac{g'}{g}+\frac{f_c'}{2f_c}-\frac{h'}{h}+\frac{f'}{f}-\frac{3}{\rho}\biggr)
\dk_2'
+\frac{9K}{2h f f_c f_3^2  P^2g \rho^2}\dk_1
-\frac{9 }{f f_c \rho^2}\dk_2\\&-\frac{18 }{f_3 f f_c \rho^2}\df\,.
\end{split}
\eqlabel{tsfl3}
\end{equation}
We find
\begin{equation}
\begin{split}
&\df=\df_{3,0}\rho^3 +\calo(\rho^4)\,,
\end{split}
\eqlabel{tsuv1}
\end{equation}
\begin{equation}
\begin{split}
&\dk_1= \biggl(\frac23 P^2 g_s (\df_{3,0}+\dk_{2,3,0})+2P^2 g_s \df_{3,0} \ln\rho \biggr)\rho^3+\calo(\rho^4\ln\rho)\,,
\end{split}
\eqlabel{tsuv2}
\end{equation}
\begin{equation}
\begin{split}
&\dk_2= \biggl(\dk_{2,3,0}+3 \df_{3,0} \ln\rho\biggr)\rho^3+\calo(\rho^4\ln\rho)\,,
\end{split}
\eqlabel{tsuv3}
\end{equation}
as $\rho\to 0$, and  (compare with the definition of $f_{a}^h$ in \eqref{bbredef})
\begin{equation}
\begin{split}
&\df^h\equiv y^2\ \df=\sum_{n=0} \delta{f}^h_n\ y^{2n}\,,\qquad \dk_1=\sum_{n=0}\dk^h_{1,n}\ y^{2n}
\,,\qquad \dk_2=\sum_{n=0}\dk^h_{2,n}\ y^{2n}\,.
\end{split}
\eqlabel{ir1ts}
\end{equation}
as $y\equiv \frac 1\rho\to 0$. 
It is characterized by 3 normalizable coefficients in the UV, and 3 normalizable coefficients in the
IR:
\begin{equation}
\begin{split}
{\rm UV}:&\qquad \{\df_{3,0}\,,\ \dk_{2,3,0}\,,\ \dk_{2,7,0}\}\,;
\\
{\rm IR}:&\qquad \{\df^h_0\,,\ \dk_{1,0}^h\,,\ \dk_{2,0}^h\}\,.
\end{split}
\eqlabel{uvirflucts}
\end{equation}
Because the fluctuations describing the {\it spontaneous} chiral symmetry breaking are linearized,
their overall amplitude is not fixed:
\begin{equation}
\{\df\,,\ \dk_1\,,\ \dk_2\}\ \sim\
\lambda\ \{\df\,,\ \dk_1\,,\ \dk_2\}\,,\qquad \lambda\equiv {\rm const}\,.
\eqlabel{ltsflres}
\end{equation}
We find it convenient to fix the amplitude setting $\df^h_0=1$.  Note that we do not have this
rescaling freedom when the chiral symmetry is broken {\it explicitly}, as in appendix \ref{flucas}:
such a rescaling would affect the source term $m\to \lambda m$.

\section{Conifold black holes at $\{T,\mu\}\gg \Lambda$}\label{perbh}

Using \eqref{persolbh}, a new radial coordinate $x$ and parameter $q$ \eqref{defqx}, we find from
\eqref{b1}-\eqref{bc} at $n=1$:
\begin{equation}
\begin{split}
&0=f_1''+\frac{4 q x^3+q x^2-10 x^2-10 x-3}{x (x+1) \Delta_1} f_1'-\frac{x (q x^2-2 q x+2 x^2-q)}{(x+1) \Delta_1} f_{c,1}'
\\&-\frac{4 x (q x^2-2 q x+2 x^2-q)}{(x+1) \Delta_1} f_{3,1}'-\frac{4 q (f_1-h_1)}{\Delta_1}\,,
\end{split}
\eqlabel{perteq1}
\end{equation}
\begin{equation}
\begin{split}
&0=f_{c,1}''-\frac{5}{4 x (x+1)} f_1'+\frac{9 q x^4-2 q x^3+18 x^4-5 q x^2-40 x^2-40 x-12)}{4x (x+1) \Delta_1} f_{c,1}'
\\&+\frac{x (q x^2-2 q x+2 x^2-q)}{(x+1) \Delta_1} f_{3,1}'+\frac{q x^4+6 q x^3+2 x^4+3 q x^2+16 x^3+24 x^2+16 x+4}{4x (x+1) \Delta_1} h_1'
\\&-\frac 34 (k_{1,1}')^2-\frac{4 (x+1)^2}{\Delta_1 x^2} k_{1,1}+\frac{5(q x^2+2 x^2+4 x+2)}{2\Delta_1 x^2} f_1-\frac{7 (x+1)^2}{\Delta_1 x^2} f_{c,1}
\\&+\frac{20 (x+1)^2}{\Delta_1 x^2} f_{3,1}
- \frac{5 q x^2-8 x^2-16 x-8}{2\Delta_1 x^2} h_1+\frac{3 (x+1)^2}{\Delta_1 x^2}\,,
\end{split}
\eqlabel{perteq2}
\end{equation}
\begin{equation}
\begin{split}
&0=f_{3,1}''-\frac{5}{4 x (x+1)} f_1'+\frac{x (q x^2-2 q x+2 x^2-q)}{4\Delta_1 (x+1)} f_{c,1}'
+\frac14 (k_{1,1}')^2
-\frac{4 (x+1)^2}{\Delta_1 x^2} k_{1,1}
\\&+\frac{3 q x^4-2 q x^3+6 x^4-2 q x^2-10 x^2-10 x-3}{x (x+1) \Delta_1} f_{3,1}'
+\frac{8 (x+1)^2}{\Delta_1 x^2} f_{3,1}-\frac{(x+1)^2}{\Delta_1 x^2}
\\&-\frac{5 q x^2-8 x^2-16 x-8}{2\Delta_1 x^2} h_1
+\frac{q x^4+6 q x^3+2 x^4+3 q x^2+16 x^3+24 x^2+16 x+4}{4x \Delta_1 (x+1)} h_1'\\
&+\frac{5(q x^2+2 x^2+4 x+2)}{2\Delta_1 x^2} f_1+\frac{5 (x+1)^2}{\Delta_1 x^2} f_{c,1}\,,
\end{split}
\eqlabel{perteq3}
\end{equation}
\begin{equation}
\begin{split}
&0=h_1''+\frac{13}{4 x (x+1)} f_1'-\frac{q x^4-10 q x^3+2 x^4-5 q x^2-16 x^3-24 x^2-16 x-4}{4x (x+1) \Delta_1} f_{c,1}'
\\&-\frac{q x^4-10 q x^3+2 x^4-5 q x^2-16 x^3-24 x^2-16 x-4}{x (x+1) \Delta_1} f_{3,1}'+\frac 34 (k_{1,1}')^2+\frac{36 (x+1)^2}{\Delta_1 x^2} k_{1,1}
\\&+\frac{7 q x^4-6 q x^3+14 x^4-7 q x^2-16 x^3-64 x^2-56 x-16}{4x (x+1) \Delta_1} h_1'-\frac{17 (x+1)^2}{\Delta_1 x^2} f_{c,1}\\
&-\frac{13(q x^2+2 x^2+4 x+2)}{2\Delta_1 x^2} f_1
+\frac{13 q x^2-72 x^2-144 x-72}{2\Delta_1 x^2} h_1-\frac{68 (x+1)^2}{\Delta_1 x^2} f_{3,1}\\
&+\frac{5 (x+1)^2}{\Delta_1 x^2}\,,
\end{split}
\eqlabel{perteq4}
\end{equation}
\begin{equation}
\begin{split}
&0=k_{1,1}''+\frac{(2 q x^4+4 x^4-q x^2-10 x^2-10 x-3)}{x (x+1) \Delta_1} k_{1,1}'-\frac{8 (x+1)^2}{\Delta_1 x^2}\,,
\end{split}
\eqlabel{perteq5}
\end{equation}
\begin{equation}
\begin{split}
&0=g_1''+\frac{2 q x^4+4 x^4-q x^2-10 x^2-10 x-3}{x (x+1)\Delta_1} g_1'-\frac{4 (x+1)^2}{\Delta_1 x^2}+(k_{1,1}')^2\,,
\end{split}
\eqlabel{perteq6}
\end{equation}

\begin{equation}
\begin{split}
&0=f_{c,1}'+4 f_{3,1}'+h_1'+\frac{3}{\Delta_2} (q x^2+2 x^2+2 x+1) (2 x+1) f_1'
+\frac{x}{\Delta_2} (x+1) (2 x+1) \\
&\times(q x^2+2 x^2+2 x+1) (k_{1,1}')^2-\frac{16 (x+1)^3}{x \Delta_2} k_{1,1}-\frac{6 (x+1) (q x^2+2 x^2+4 x+2)}{x \Delta_2} f_1
\\&+\frac{4 (x+1)^3}{x \Delta_2} f_{c,1}+\frac{16 (x+1)^3}{x \Delta_2} f_{3,1}+\frac{2 (x+1) (3 q x^2+8 x^2+16 x+8)}{x \Delta_2} h_1-\frac{4 (x+1)^3}{x \Delta_2}\,,
\end{split}
\eqlabel{perteqc}
\end{equation}
where
\begin{equation}
\begin{split}
&\Delta_1=(2 x+1) ((q+2) x^2+2 x+1)\,,\\
&\Delta_2=x^2 (x^2+6 x+3) q+2 x^4+32 x^3+48 x^2+32 x+8\,.
\end{split}
\end{equation}

Notice that the equation for $k_{1,1}$ decouples and can be solved analytically:
\begin{equation}
k_{1,1}=\frac{1}{q+1}\biggl(q\left(\ln( (2+q) x^2+2 x+1) -2 \ln x\right) -2 \ln x+2 \ln(x+1)\biggr)\,,
\eqlabel{k11solve}
\end{equation}
where we fixed, without the loss of generality, the additive integration constant so
that\footnote{Choosing $\rm const\ne 0$ simply modifies $\hK_0\to \hK_0+P^2 g_s \times {\rm const}$. }
\begin{equation}
k_{1,1}={\rm const}-2\ln x +\calo(x)\,,\qquad {\rm const}=0\,.
\eqlabel{defconst}
\end{equation}
The remaining equations must be solved numerically. We use the first order equation \eqref{perteqc}
to eliminate (algebraically) $f_{c,1}'$ from \eqref{perteq1}, \eqref{perteq3}-\eqref{perteq5}, keep
\eqref{perteq6},  and drop the redundant equation \eqref{perteq2}. Thus we end up with
4 second-order equations  and a single first-order equation for $\{f_1,f_{3,1},h_1,g_1,f_{c,1}\}$.
\nxt In the UV, \ie as $x\to 0$, we have
\begin{equation}
\begin{split}
&f_1=x^2 \left(
\frac12 q-2 q\ln x \right)+x^3 \left((1-\hf_{c,1,0}) q+4 q\ln x \right)+x^4 \left(\hf_{1,4,0}
+\left(\frac23 q^2-6 q\right) \ln x\right)\\
&+\calo(x^5\ln x)\,,
\end{split}
\eqlabel{uvper1}
\end{equation}
\begin{equation}
\begin{split}
&f_{3,1}=x \hf_{c,1,0}+\left(\frac34 q-\hf_{c,1,0}\right) x^2
+\left(\hf_{c,1,0}-\frac32 q\right) x^3+\left(\frac94 q
-\hf_{c,1,0}-\frac29 q^2\right) x^4+\calo(x^5)\,,
\end{split}
\eqlabel{uvper2}
\end{equation}
\begin{equation}
\begin{split}
&h_1=\frac12-2 \ln x+\left(2-2 \hf_{c,1,0}\right) x+\left(2 \hf_{c,1,0}-1-\frac23 q\right) x^2
+\left(\frac23+\frac43 q-2 \hf_{c,1,0}\right) x^3
\\&+\left(-\frac{17}{24}+2 \hf_{c,1,0}+\frac{11}{72} q^2-\frac{55}{24} q\right) x^4+\calo(x^5)\,,
\end{split}
\eqlabel{uvper3}
\end{equation}
\begin{equation}
\begin{split}
&g_1=-q x^2+2 q x^3+x^4 \left(\hg_{1,4,0}+(1-q^2-q) \ln x\right)+\calo(x^5\ln x)\,,
\end{split}
\eqlabel{uvper4}
\end{equation}
\begin{equation}
\begin{split}
&f_{c,1}=x \hf_{c,1,0}+\left(q-\hf_{c,1,0}\right) x^2
+\left(\hf_{c,1,0}-2 q\right) x^3+\left(\frac13-\hf_{c,1,0}-\frac59 q^2+\frac83 q\right) x^4\\&+\calo(x^5)\,,
\end{split}
\eqlabel{uvper5}
\end{equation}
where the additive integration constant in $g_1$ is set to zero --- it can be absorbed in
$\calo(P^2/\hK_0)$ redefinition of $g_s$. 
\nxt In the IR, \ie as $y\equiv\frac1x\to 0$, we have
\begin{equation}
\begin{split}
&f_1=\hf_{1,0}^h+\calo(y)\,,\qquad f_{3,1}=\hf_{3,1,0}^h+\calo(y)\,,\qquad h_1=\hh_{1,0}^h+\calo(y)\,,\\
&g_1=\hg_{1,0}^h+\calo(y)\,,\qquad f_{c,1}=\hf_{c,1,0}^h+\calo(y)\,.
\end{split}
\eqlabel{irper}
\end{equation}

Besides $q$, the solution for
 $\{f_1,f_{3,1},h_1,g_1,f_{c,1}\}$ is characterized by 10 parameters:
 \begin{equation}
 \{\hf_{c,1,0}\,,\, \hf_{1,4,0}\,,\, \hf_{c,6,0}\,,\, \hf_{c,8,0}\,,\, \hg_{1,4,0}\,,\,
 \hf_{1,0}^h\,,\, \hf_{3,1,0}^h\,,\, \hh_{1,0}^h\,,\, \hg_{1,0}^h\,,\, \hf_{c,1,0}^h \}\,.
 \eqlabel{pertpar}
 \end{equation}
One combination of these parameters determines the temperature at order $\calo(P^2 g_s/\hK_0)$,
thus, we have precisely $2\times 4+1=9$ parameters left, necessary to specify a unique
physical solution.

We set up numerics varying $q$, resulting in all parameters in \eqref{pertpar},
except for $\hf_{1,4,0}$, being the functions of $q$. A choice of  $\hf_{1,4,0}$ is equivalent to
fixing the overall mass scale at order $\calo(P^2 g_s)$, and does not affect results expressed
as dimensionless quantities. Our specific choice of $\hf_{1,4,0}$ is explained in section
\ref{perkt}.

\section{Numerical tests}\label{numtests}

The work presented in this paper is numerical. It is imperative that
we do as many tests as possible to confirm the reliability of the results.
In the rest of this section we highlight a small subset
of the obvious, and less obvious (accidental) tests that we performed. 

For the numerical integration --- solving the boundary value problem
for a system of coupled ODEs --- we use the ``shooting'' procedure developed in
\cite{Aharony:2007vg}. Numerical integration uses a finite-difference method,
thus, concerns that there could be a singularity in the solution
inside the integration range is less of the concern, as it would have been
for the spectral methods. Still, we would often plot the resulting functions
over the integration range to inspect that the solutions are indeed free
from the singularities. 

\subsection{The first law of thermodynamics of $\calt_{decon}^s$ and $\calt_{decon}^b$
states at $\mu=\mu_1$}

\begin{figure}[ht]
\begin{center}
\psfrag{e}[cc][][1][0]{{$\hat\cale$}}
\psfrag{l}[bb][][1][0]{{$d\cale/(T ds)-1$}}
\psfrag{u}[tt][][1][0]{{$d\cale/(T ds)-1$}}
\includegraphics[width=2.6in]{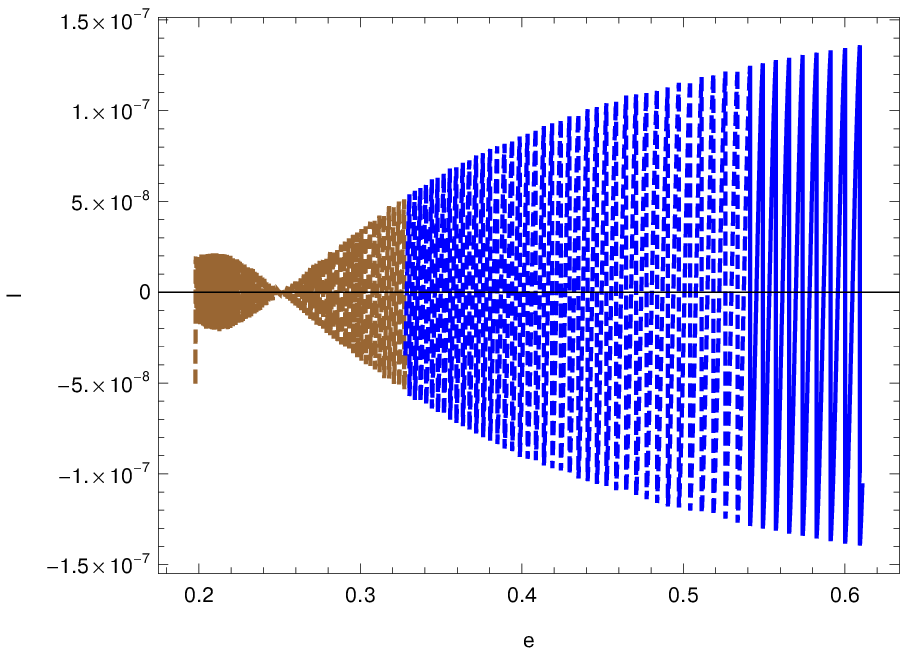}\ \  
\includegraphics[width=2.6in]{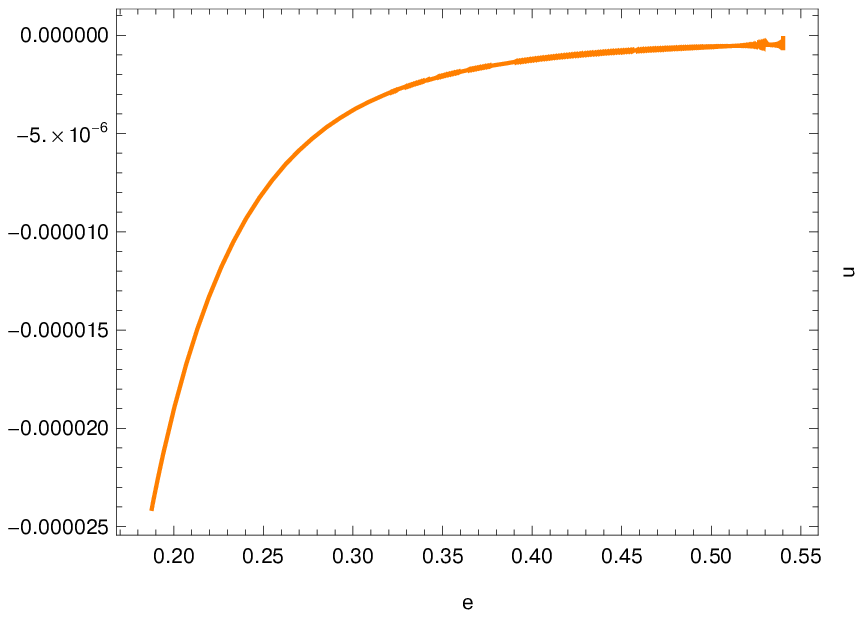} 
\end{center}
  \caption{The first law of thermodynamics for the Klebanov-Tseytlin (the left panel)
  and the Klebanov-Strassler (the right panel) black holes at $\mu=\mu_1$, see \eqref{mu1mu2}.
  The color coding as in fig.~\ref{micro20500}.
}\label{fl20}
\end{figure}

In order to study the phase diagram of the black holes on the conifold
we have to generate a large number of thermal spectra of these black holes.
These spectra at taken at fixed $\mu/\Lambda$, with $\mu$ begin
the compactification scale of the $S^3$ and $\Lambda$ the strong coupling
scale of the cascading gauge theory. We use the holographic renormalization of appendix
\ref{apa3} to compute the energy and free energy densities $\cale$ and $\calf$, the entropy density
$s$ and the temperature $T$. When both $\mu$ and $\Lambda$ kept first, the first law of
thermodynamics \eqref{firstlaw} must be satisfied. Note that \eqref{firstlaw}
is a differential constraint on our numerical data --- we use the default tools of
Wolfram \textit{Mathematica}$ ^{\rm \textregistered}$ to construct interpolating functions
for the collected data sets and verify that the first law of thermodynamics is satisfied
for the obtained interpolation functions (which can be easily differentiated).
In fig.~\ref{fl20} we present the verification of the first law for the Klebanov-Tseytlin
black holes (the left panel) and the Klebanov-Strassler black holes at
select value\footnote{We performed this check
for every single spectrum we collected.} $\mu=\mu_1$, see \eqref{mu1mu2}.  It is important to keep in mind that
not only the first law validates the numerics, but it also confirmed that
we correctly collected the black hole spectra using the defining parameters of the
theory fixed ($\mu,\Lambda$), and that we correctly implemented the holographic renormalization.

Results reported in fig.~\ref{fl20} are typical: 
\begin{itemize}
\item Klebanov-Tseytlin black holes are more symmetric than the Klebanov-Strassler black holes ---
the former ones require solving a coupled nonlinear system of 6 ODEs, while the latter
ones has 9 ODEs. The asymptotics of the KS black holes are more involved, and contain a gravitational
mode dual to a dimension-7 operator (parameter $k_{2,7,0}$ in \eqref{uvvevs}), which is absent in
KT black holes. It is thus not a surprise that numerical accuracy of KS black holes lags behind
that of KT black holes.
\item Numerics of black holes on the conifold, studied here,
is more challenging than that of the of the black branes discussed in
\cite{Aharony:2007vg,Buchel:2018bzp}. The reason is apparent from the UV asymptotic expansions
\eqref{bs1}-\eqref{bs9}. For concreteness, focus on the expansion for  $f$ in \eqref{bs9}:
when $\mu\ne 0$, there are additional terms, say $\propto \mu^4 \rho^4 \ln^2\rho$,
that are present for black holes,
and are absent for the black branes --- these additional
logarithms produce extra numerical difficulties.
\item The quality of the numerics deteriorates as the simulated system is pushed into more extreme
regime ---  the curvature of the background geometry increases, or we approach the state existence
boundary (as for the Klebanov-Strassler black holes as $\mu\to \mu_{KS}$ of \eqref{defmuks}),
or some of the parameters specifying the solution diverge (see the left panel of fig.~\ref{nonmon}).
\item Typically as in fig.~\ref{fl20} (more pronounced in the left panel),  the error has
a high-frequency noise --- this is a reflection of the discreteness in data collection, which was
used to produce interpolating functions. 
\end{itemize}

We stop trusting numerics when the first law constraint \eqref{firstlaw} violation exceeds
$\sim 10^{-5}-10^{-4}$. Mistakes in analytical results for the holographic renormalization,
or failures to keep mass scales defining the theory fixed,
lead to violations of \eqref{firstlaw} of order $\calo(1)$.

\subsection{Thermodynamics of $\calt_{decon}^s$
states at $\mu=\mu_2$}

The parameter state of the black holes on the conifold is 3-dimensional; we have:
\nxt $\mu$ --- the compactification scale of the boundary $S^3$;
\nxt $\Lambda$ --- the strong coupling scale of the theory;
\nxt $T$ --- the Hawking temperature of the black holes.\\
These scales have a nonlinear dependence on the numerical parameters defining the
solution \eqref{defuvcasea}. For example, from \eqref{defkolambda},
\begin{equation}
\frac{\mu^2}{\Lambda^2}=2^{-1/2}\mu^2\ P^2g_s\ e^{\frac{K_0}{P^2g_s}} \,.
\eqlabel{mulambdaee}
\end{equation}
The symmetries of the equations of motion describing the black holes, \ie  \eqref{betatransform},
\eqref{kssym1} and \eqref{kssym2}, allows for an inequivalent numerical schemes.
One such scheme is to set $\mu=1$,  $g_s=1$ and $K_0=1$, parameterizing 
\begin{equation}
\frac{\mu^2}{\Lambda^2}=2^{-1/2}\ b\ e^{\frac{1}{b}} \,,
\eqlabel{mulambdaee2}
\end{equation}
with $b\equiv P^2$, see \eqref{mulambdavs2}. Note that \eqref{mulambdaee2} is a decreasing function
of $b$ as $b\in [0,1]$, and then it increases again for $b>1$, retracing the same set of $\mu/\Lambda$
with $b\to \infty$, albeit with seemingly different set of defining parameters. Of course, any black
hole physical observable, say $\hat s(\hat\cale)$, in reality is represented as a parametric
dependence 
\begin{equation}
\{\hat\cale(b)\,,\, \hat s(b)\}\,,
\eqlabel{foldpar}
\end{equation}
with $\cale$ \eqref{eptb} and $s$ \eqref{enttb} depending on $b$ implicitly via parameters
\eqref{uvvevsb} and \eqref{irvevsbb}, 
must 'fold on itself' for $b>1$.  This is precisely what we observe in this work, and what was observed
for black branes in \cite{Aharony:2007vg,Buchel:2018bzp}. 
We call this 'an obvious' numerical test.

\begin{figure}[ht]
\begin{center}
\psfrag{e}[tt][][1][0]{{$\hat\cale$}}
\psfrag{s}[tt][][1][0]{{$\hat s$}}
\psfrag{k}[tt][][1][0]{{$K_0$}}
\psfrag{t}[bb][][1][0]{{$T/\Lambda$}}
\includegraphics[width=2.6in]{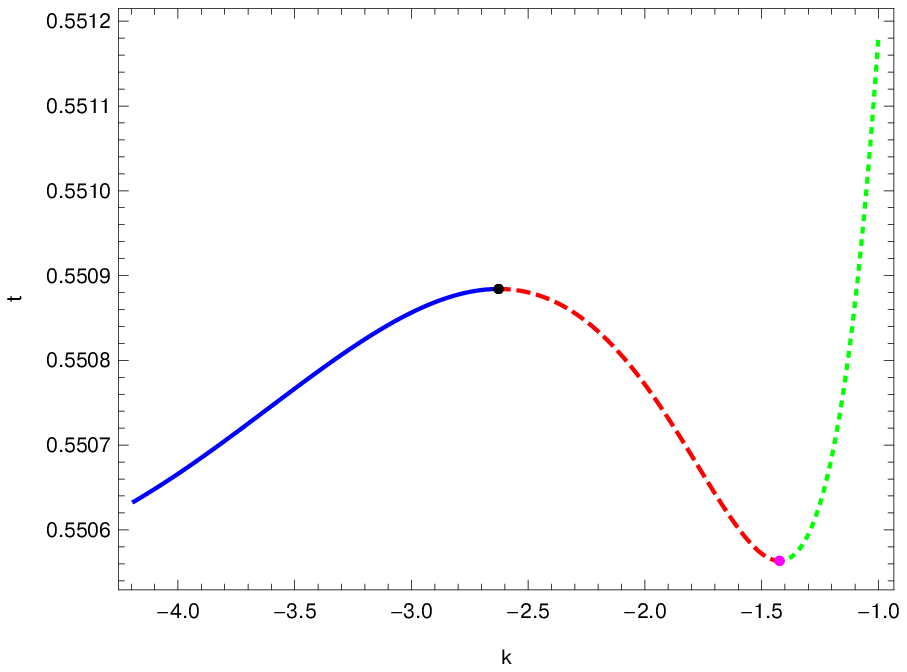}\ \  
\includegraphics[width=2.6in]{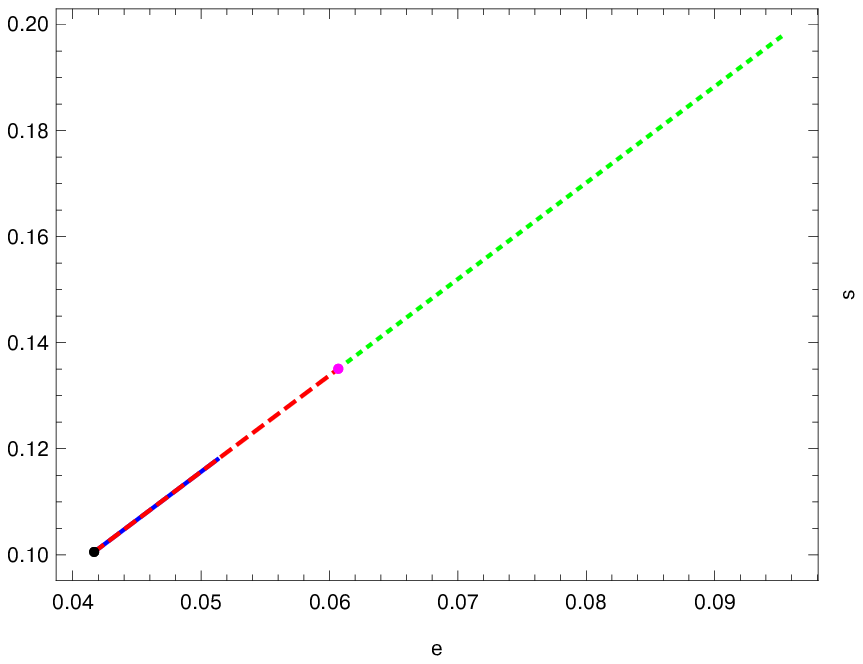} 
\end{center}
  \caption{Physical observables of conifold black holes, here the Klebanov-Tseytlin black hole
  at $\mu=\mu_2$ \eqref{mu1mu2}, can depend on the numerical parameters ($K_0$ in the left panel)
  non-monotonically. Some of this non-monotonicity, when translated to physical observables
  as in the right panel, results in retracing the same states of the black hole: the blue solid
  curve folds onto the red dashed curve in the right panel.
}\label{wierd}
\end{figure}

In fig.~\ref{wierd} we present an example of the 'accidental' numerical test.
The results reported relate to the Klebanov-Tseytlin black hole, \ie $\calt_{decon}^s$ thermal state,
at fixed $\mu/\Lambda=\mu_2/\Lambda$, see \eqref{mu1mu2}. We use numerical scheme with $P=1$, $g_s=1$.
We vary $K_0$ and $\mu$ along the line \eqref{k0muline} to enforce the condition
$\mu/\Lambda={\rm constant}$. The left panel presents the results for $T/\Lambda$ as a function of
$K_0$, and the right panel presents $\hat s$ as a function of $\hat\cale$. Notice that
$T/\Lambda$ is not a monotonic function of $K_0$. One aspect of this non-monotonicity is
actually physics: the magenta dot represents the terminal temperature $T_u(\mu_2/\Lambda)$
of the Klebanov-Tseytlin
black hole  separating the phases with the positive (dotted green curve) and negative (dashed red curve)
specific heat in the canonical ensemble\footnote{Of course, this feature is invisible
in the microcanonical ensemble, see the right panel of fig.~\ref{wierd}.}, see fig.~\ref{can20500}.    
On the contrary, the non-monotonicity highlighted by the black dot is a numerical artifact,
akin to 'folding' for $b>1$ discussed above --- the part of the blue curve retracing the same values
of $T/\Lambda$ as those to the right of the black dot, represented by the red dashed curve,
'folds on itself' when we use the dimensionless physical observables, see the right panel.
We did not show it, but the same 'folding' is true for the plot of $\hat\calf$ as a function of $T/\Lambda$.What is different in this folding, compare to the one discussed above, is that it could not have been
foreseen. 

A similar example of the accidental numerical test, due to unforeseen 'folding' is highlighted in
fig.~\ref{nonmon2}.

\subsection{First law of thermodynamics of near-conformal conifold black holes}

\begin{figure}[ht]
\begin{center}
\psfrag{c}[cc][][1.5][0]{{$\calc_{[1]}$}}
\psfrag{q}[bb][][1.5][0]{{$q$}}
\includegraphics[width=5in]{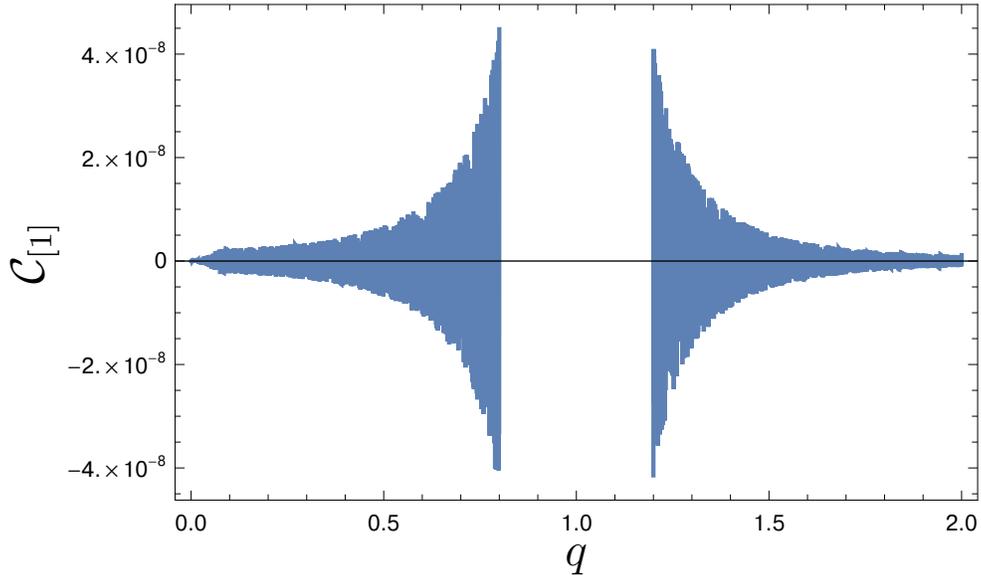}
\end{center}
  \caption{Verification of the first law of thermodynamics ---see \eqref{cdeef}  for
  a precise definition of $\calc_{[1]}$ ---  for the Klebanov-Tseytlin black holes
  to leading order in $1/\ln(\mu/\Lambda)$ as a function of $q$, equivalently $\frac{T}{\mu}$
  \eqref{zq}.
}\label{flpert}
\end{figure}

In section \ref{perkt} we discuss construction of $\calt_{decon}^s$ thermal states, \ie the
Klebanov-Tseytlin black holes, perturbatively in $z$ (see \eqref{defqx} and \eqref{mulambpert})
\begin{equation}
1\gg z\equiv\frac{P^2g_s}{\hK_0}\sim \frac{1}{2\ln \frac{\mu}{\Lambda}}\,.
\eqlabel{e3pert}
\end{equation}
The first law of thermodynamics in this case leads to a series of differential constraints
\begin{equation}
0\ \equiv\ \sum_{n=0}^\infty\ z^n\ \calf\call_{[n]}(q)\qquad \Longrightarrow\qquad \calf\call_{[n]}(q)=0
\eqlabel{econstkt}
\end{equation}
on parameters specifying the perturbative functions $f_n$, $f_{3,n}$, $f_{c,n}$, $h_n$, $k_{1,n}$, $g_n$.
Explicit expression for $\calf\call_{[1]}$ is given by \eqref{flpert1} and \eqref{flpert2}.
In fig.~\ref{flpert} we verify the first law of thermodynamics at order $n=1$ --- see \eqref{cdeef}
for a precise definition of $\calc_{[1]}$. We excluded the region around $q=1$, as in this
limit some of the parameters diverge --- see fig.~\ref{c1c2}. The results are presented as a function
of $q$; to leading order in $z$, see \eqref{tpert},
\begin{equation}
\frac{T}{\mu}=\frac{q+2}{2\pi q^{1/2}}\biggl(1+\calo(z)\biggr)\,.
\eqlabel{zq}
\end{equation}

\subsection{Additional numerical tests}

We list here additional some implicit and explicit numerical tests.
\nxt In multiple cases we cross checked numerics using different computational schemes,
see figs.~\ref{vbcompare} and \ref{vascompare}.
\nxt Black branes on the conifold are $\mu/\Lambda\to 0$
limiting cases of the black holes discussed here. Even in this limit, the computations
performed here differ from analysis in \cite{Aharony:2007vg,Buchel:2018bzp,Buchel:2010wp}.
Specifically, in the latter work a different radial coordinate was used, see section \ref{technical}.
In all cases we found a perfect agreement in the $\mu/\Lambda\to 0$ limit with the results reported
previously.

\bibliographystyle{JHEP}
\bibliography{kss3}

\end{document}